\documentclass{book}

\title{xMLC}
\newcommand{\booksubtitle}{A Toolkit for Machine Learning Control}
\newcommand{\seriestitle}{Machine Learning Tools in Fluid Mechanics}
\newcommand{\booklicense}{Creative Commons Attribution 4.0}

\author{Guy Y. Cornejo Maceda,\\[5pt]Fran\c{c}ois Lusseyran\\[5pt] \&  Bernd R.~Noack}
\newcommand{\authorsubtitle}{Shenzhen \& Paris}

\usepackage{Main}
\usepackage{wrapfig} %% Figure in text
\makeatletter
\newcommand{\booktitle}{\@title}
\newcommand{\bookauthor}{\@author}
\makeatother

\graphicspath{{Figures/}}
\newacronym{CGNS}{CGNS}{CFD General Notation System}
\newacronym{POD}{POD}{Proper Orthogonal Decomposition}
\newacronym{ROM}{ROM}{Reduced Order Model}

\newcommand{\Rey}{\mathrm{Re}} % Reynolds number
\newcommand{\bfKopt}{\boldsymbol{K}^*} % Optimal control law
\newcommand{\xMLC}{\texttt{xMLC}\ } % xMLC

\newcommand{\Ninstr}{N_{\rm Instr}} % Number of instructions
\newcommand{\Ninstrmax}{N_{\rm Instr, max}} % Maximum number of instructions
\newcommand{\Nvar}{N_{\rm Var}} % Number of variable registers
\newcommand{\Ncst}{N_{\rm Cst}} % Number of constant registers
\newcommand{\Ni}{N_{i}} % Total number of individuals
\newcommand{\Ntour}{N_{\rm Tour}} % Tournament selection size
\newcommand{\Ptour}{P_{\rm Tour}} % Selection probability
\newcommand{\Pmut}{P_{\rm Mut}} % Mutation probability
\newcommand{\Pcros}{P_{\rm Cros}} % Crossover probability
\newcommand{\Prep}{P_{\rm Rep}} % Replication probability
\newcommand{\Nb}{N_{\rm b}} % Number of inputs
\newcommand{\Nr}{N_{\rm R}} % Number of registers
\newcommand{\No}{N_{\rm O}} % Number of operators
\newcommand{\Nrho}{N_{\rho}} % Number of realizations
\newcommand{\Nps}{N_{\rm Pop. size}} % Population size
\newcommand{\Ng}{N_{\rm G}} % Number of generations
\newcommand{\Ne}{N_{\rm E}} % Elitism parameter
\newcommand{\Nh}{N_{\rm H}} % Number of time-dependent functions
\newcommand{\Ns}{N_{\rm S}} % Number of sensors

\usepackage{amsfonts}
\usepackage{amssymb}
\usepackage{multirow}
\usepackage{esint}
\usepackage{bigints}
\setcitestyle{open={(},close={)}}
\usepackage[section]{placeins}

\usepackage[bindingoffset=0.625in,
            left=.7in, right=.7in,
            top=0.9125in, bottom=1.0375in,
            paperwidth=8.3in, paperheight=11.7in]{geometry}
\usepackage{makeidx}
\makeindex % Initialize an index so we can add entries with \index

\begin{document}
\frontmatter

% No page numbers on the Frontispiece page
\thispagestyle{empty}
%\definecolor{covercolor}{RGB}{189, 28, 58}
\definecolor{covercolor}{RGB}{0, 64, 107}
\pagecolor{covercolor}
\color{yellow}

% ---- Title Page ----
% current geometry will be restored after title page
\newgeometry{top=1.75in,bottom=.5in}
\begin{titlepage}

\begin{flushleft}
\Large\bf Machine Learning Tools in Fluid Mechanics, Vol 2
\end{flushleft}

\begin{flushleft}
% Title
\textbf{\fontfamily{qcs}\fontsize{48}{54}\selectfont xMLC\\}

% Draw a line 4pt high
\par\noindent\rule{\textwidth}{4pt}\\

% Shaded box from left to right with Subtitle
% The text node is midway (centered).
\begin{tikzpicture}
\shade[top color=white,bottom color=lightgray]
    (0,0) rectangle (\textwidth, 2.5)
    node[midway] {\color{black}\Huge\bfseries{\begin{tabular}{l} A Toolkit \\ for Machine Learning Control \raisebox{-5pt}{\rule{0pt}{0pt}}\\ \end{tabular}}};
\end{tikzpicture}

% Edition Number
\begin{flushright}
\Large\bf  First Edition
\end{flushright}

\vspace{\fill}
\begin{center}
\includegraphics[width=0.5\textwidth]{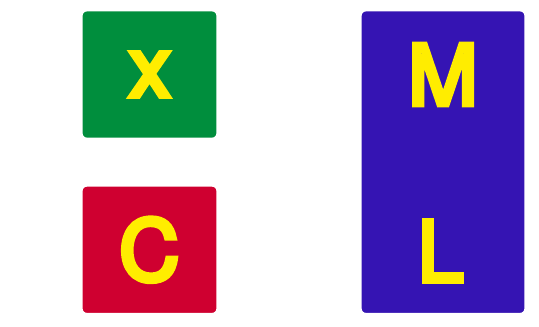}
\end{center}
\vspace{\fill}

\color{white}
% Author and Location
\textbf{\Huge \bookauthor}
\bigskip
\begin{flushright}
\textbf{\LARGE \textit{\authorsubtitle}}
\end{flushright}

\vspace{\fill}

% Self Publishing Logo. Free to use: CC0 license.
% The source file is book.svg. If you change the svg, you must then convert
% it to pdf. There are many online and offline tools available to do that.
\begin{center}\bfseries
Universit\"atsbibliothek\\ der Technischen Universit\"at Braunschweig\\
2022
\end{center}
\end{flushleft}
\end{titlepage}
\restoregeometry
% ---- End of Title Page ----
% Restore cover color 
\pagecolor{white}
\color{black}

% ---- Half Title Page ----
% current geometry will be restored after title page
\newgeometry{top=1.75in,bottom=.5in}
\begin{titlepage}
\begin{flushleft}

% Title
\textbf{\fontfamily{qcs}\fontsize{48}{54}\selectfont \seriestitle\\}

% Draw a line 4pt high
\par\noindent\rule{\textwidth}{4pt}\\

% Subtitle
% Shaded box from left to right. Text node is midway (centered).
%\begin{tikzpicture}
%\shade[bottom color=lightgray,top color=white]
%    (0,0) rectangle (\textwidth, 1.5)
%    node[midway] {\textbf{\large \textit{\booksubtitle}}};
%\end{tikzpicture}

% Edition Number
\begin{flushright}
\Large Series number: 2
\end{flushright}

% \vspace{\fill}
\vspace{\fill}

\end{flushleft}
\end{titlepage}
\restoregeometry
% ---- End of Half Title Page ----

% No page numbers on the Frontispiece page
\thispagestyle{empty}

% ---- Title Page ----
% current geometry will be restored after title page
\newgeometry{top=1.75in,bottom=.5in}
\begin{titlepage}
\begin{flushleft}

% Title
\textbf{\fontfamily{qcs}\fontsize{48}{54}\selectfont xMLC\\}

% Draw a line 4pt high
\par\noindent\rule{\textwidth}{4pt}\\

% Shaded box from left to right with Subtitle
% The text node is midway (centered).
\begin{tikzpicture}
\shade[bottom color=lightgray,top color=white]
    (0,0) rectangle (\textwidth, 1.5)
    node[midway] {\textbf{\large \textit{\booksubtitle}}};
\end{tikzpicture}

% Edition Number
\begin{flushright}
\Large First Edition
\end{flushright}

\vspace{\fill}

% Author and Location
\textbf{\large \bookauthor}\\[6.5pt]
\textbf{\large \textit{\authorsubtitle}}

\vspace{\fill}

% Self Publishing Logo. Free to use: CC0 license.
% The source file is book.svg. If you change the svg, you must then convert
% it to pdf. There are many online and offline tools available to do that.
%\begin{center}
%\includegraphics{booksvg.pdf}\\[4pt]
%\fontfamily{lmtt}\small{Self Publishers Worldwide\\
%Seattle San Francisco New York\\
%London Paris Rome Beijing Barcelona}
%\end{center}

\end{flushleft}
\end{titlepage}
\restoregeometry
% ---- End of Title Page ----

% Do not show page numbers on colophon page
\thispagestyle{empty}

\begin{flushleft}
\vspace*{\fill}
Series editors: Richard Semaan \& Bernd R.~Noack\\
\vspace{\fill}
Copyright \textcopyright{} \the\year{}  \bookauthor\\
License: \booklicense\\
DOI: \href{https://doi.org/10.24355/dbbs.084-202208220937-0}{10.24355/dbbs.084-202208220937-0}

\end{flushleft}

% A title page resets the page # to 1, but the second title page
% was actually page 3. So add two to page counter.
\addtocounter{page}{2}

%-----------------------------------------------------------------------
\newpage
\parbox[t]{75mm}{
\textbf{Dr.\ Guy Y. Cornejo Maceda}
\newline Postdoctoral fellow
\newline School of Mechanical Engineering
\newline and Automation
\newline Harbin Institute of Technology
\newline Shenzhen Campus, Building L, Room 319
\newline University Town, Xili
\newline Shenzhen 518055,  China
\newline Email: \texttt{Yoslan@hit.edu.cn}
\newline
\newline \textbf{Prof. Dr.~Fran\c{c}ois Lusseyran}
\newline Laboratoire Interdisciplinaire 
\newline des Sciences du Num\'{e}rique 
\newline Universit\'{e} Paris-Saclay, CNRS
\newline Campus universitaire B\^{a}t. 507
\newline Rue du Belved\`{e}re
\newline F - 91405 Orsay, France
\newline Email: \texttt{Francois.Lusseyran@limsi.fr}

}
\hfill
\parbox[t]{75mm}{\textbf{Prof.\ Dr.\ Bernd R. Noack}
\newline National Talent Professor
\newline School of Mechanical Engineering 
\newline and Automation
\newline Harbin Institute of Technology
\newline Shenzhen Campus, Building L, Room 2020
\newline University Town, Xili
\newline Shenzhen 518055,  China
\newline Email: \texttt{Bernd.Noack@hit.edu.cn}
}
\newpage

% The asterisk excludes chapter from the table of contents.
\chapter*{Preface}

xMLC is the second book of this 
`\textit{Machine Learning Tools in Fluid Mechanics}' Series 
and focuses on Machine Learning Control (MLC).
%In particular, the book gives access to a is an introduction to machine learning control \citet{Duriez2017book} law learning .
% --- Objectives of the book
The objectives of this book are two-fold:
First, provide an introduction to MLC for students, researchers, and newcomers on the field;
and second, share an open-source code, \texttt{xMLC}, 
to automatically learn open- and closed-loop control laws directly in the plant 
with only a few executable commands.

% --- Characteristics of the book/code
This presented MLC algorithm is based on genetic programming
and highlights the learning principles (exploration and exploitation).
The need of balance between these two principles is illustrated with an extensive parametric study 
where the explorative and exploitative forces are gradually integrated in the optimization process.
The provided software 
\xMLC is an implementation of MLC.
It builds on \texttt{OpenMLC} \citep{Duriez2017book} but replaces tree-based genetic programming 
but the  linear genetic programming framework \citep{Brameier2006book}.
The latter representation is preferred for its easier implementation 
of multiple-input multiple-output control laws and
of the genetic operators (mutation and crossover). 
The handling of the software is facilitated by a step by step guide 
that shall help new practitioners to use the code within few minutes.
We also provide detailed advice in using the code for other solvers and for experiments.
The code is open-source and a GitHub version is available for future updates, options and add-ons.

% --- Acknowledgement
\xMLC is the result of collaborations of many students, engineers and researchers 
in  Argentina, Canada, China, France, Germany, Japan, Jordan, Poland, Sweden, Spain, and USA
since almost one decade.
We deeply thank Marc Segond for initiating us into the world of genetic programming,
Mattias Wahde for alerting us to the advantages of the linear version
and Marc Schoenauer for advice regarding performance analysis and improvement.
We are indebted to Thomas Duriez and Ruiying Li for sharing 
their own implementation of MLC 
and for their valuable advice.
Special thanks are due to our numerous collaborators helping to improve MLC and 
to gain valuable experience in dozens of experiments and simulations:
Markus Abel,
Jean-Luc Aider,
Jacques Bor\'ee,
Jean-Paul Bonnet,
Steven L.\ Brunton, 
Sylvain Caillou, 
Rodrigo Castellanos, 
Camila Chovet, 
Laurent Cordier,
Christophe Cuvier,
Antoine Debien,
Jo\"el Delville,
Carine Fourment,
Hiroaki Fukumoto,
Ignacio de la Fuente,
Nan Deng, 
Dewei Fan,
Stefano Discetti,
Jean Marc Foucault,
Nan Gao,
Nicolas Gautier,
Fabien Harambat,
Andrea Ianiro,
Eurika Kaiser,
Laurent Keirsbulck, 
Azeddine Kourta,
Jean-Christophe Laurentie, 
Jean-Christophe Loiseau, 
Hao Li,
Songqi Li, 
Yiqing Li, 
Congjun Liu,
Yutong Liu,
Jiayang Luo,
Marc Lippert,
Robert Martinuzzi,
Lionel Mathelin, 
Nicolas Mazellier,
Marek, Morzy\'{n}ski, 
Philipp Oswald, 
Akira Oyama,
Vladimir Parezanovic,
Pierre-Yves Passaggia, 
Luc Pastur, 
C\'edric Raibaudo,
Richard Semaan, 
Tamir Shaqarin,
Ruixuan Shen,
Hongke Shi,
Michel Stanislas,
Jianguo Tan,
Kai, A.~F.~F.\ von Krbek and
Eliott Varon.
%Bingfu Zhang, and Zhi Wu.

\begin{flushright}
Shenzhen and Paris in July 2022\\[2pt]
	Guy Y. Cornejo Maceda \\
	Fran\c{c}ois Lusseyran\\
	Bernd R.~Noack 
\end{flushright}

%-----------------------------------------------------------------------

% Three-level Table of Contents
\setcounter{tocdepth}{3}
\tableofcontents

\mainmatter

\chapter{Introduction}
\label{ToC:Introduction}

This book shall facilitate a smooth application 
of machine learning control \citep{Duriez2017book}
to numerical and experimental plants.
Focus is placed on linear genetic programming control 
for nonlinear dynamics systems and, in particular, fluid flows.

Starting point of any control is the beautiful framework of linear theory  \citep{Doyle1992book,Astroem2010book}.
Linear theory has been the key enabler for stabilization of fluid flow instabilities
\citep{Rowley2006arfm,Kim2007arfm,Sipp2010amr}
and can provide mathematically rigorous performance guarantees.
Yet, linear dynamics has numerous implications which are uncommon for complex systems.
For instance, the unforced system either converges to a single fixed point
or diverges to infinity.
Or, the frequency of periodic actuation does not affect other frequencies of the system. 
Linear methods can be extended to weakly nonlinear dynamics,
e.g., via a nonlinear transformation to a linear system (input-output linearization)
or via a meaningful local linearization (linear parameter-varying systems).
Linear control theory is  inapplicable for
frequency crosstalk, i.e., if the actuation or system frequency effect other system frequencies.
Yet, frequency crosstalk is a defining feature 
and often key enabler for turbulence control
and many dynamical and complex systems.

Recently, machine learning has opened the new avenue of model-free nonlinear control.
\emph{Machine Learning Control (MLC)} can perform an automated control optimization
of multiple-input (multiple-actuator) multiple-output (multiple-sensor) plants.
MLC learns the control laws like mother nature---by clever trial and error.
More specifically, a regression problem of the second kind is formulated 
in which the cost is minimized via the MIMO control law.
Literature offers several approaches \citep{Brunton2015amr,Brunton2020arfm}.
The optimization of a parameterized control law, e.g., linear feedback,
invites all methods of optimization,
like Bayesian optimization \citep{Blanchard2022ams},
downhill simplex descent, Monte Carlo sampling, explorative gradient method \citep{LiY2022jfm}, 
genetic algorithm \citep{Benard2016ef},  
particle swarm optimization \citep{Shaqarin2022arxiv}, and 
numerous other biologically inspired algorithms \citep{Wahde2008book}.
Cluster-based control infers the structure of the nonlinear control law from flow data
and allows for downhill simplex optimization of simple feedbacks 
in only few dozen simulations or experiments \citep{Nair2019jfm}.

A spectrum of optimizing general nonlinear laws has been invented in the 1990's.
Neural network-based control \citep{Lee1997pof} has been developed by the team of J.\ Kim for wall-turbulence skin friction reduction.
Reinforcement learning (RL) \citep{Sutton1998book} has become popular in robotics
and has been pioneered in flow control by J. Rabault's team \citep{Rabault2019jfm,Tang2020pf,WangQL2022arxiv} and others \citep{Fan2020pnas}.
Genetic programming control (GPC) was discovered by \citet{Dracopoulos1997book} for stabilizing satellite motion
and was later re-invented for turbulence control \citep{Duriez2017book} under the name  \emph{Machine Learning Control (MLC)}.
First studies indicate that RL and GPC have overall similar performance with few application specific differentiations 
\citep{Castellanos2022pf}.
Numerous efforts to accelerate and improve the performance
of all approaches, e.g., gradient information has significantly reduced the learning time of GPC \citep{Cornejo2021jfm}.

In the sequel, the term MLC will be narrowly used for GPC only.
GPC is simple regression solver  
and seems the best tested version of MLC for turbulence control.
In dozens of experiments and simulations, 
GPC has consistently  outperformed other optimized control laws in their respective plant,
often with unexpected new actuation mechanisms, e.g.\ frequency crosstalk mechanisms.
GPC has even successfully learned a novel distributed actuation \citep{Zhou2020jfm}.
Successful applications include following:
\begin{enumerate}\setlength{\itemsep}{0pt}
		\item Stabilization of a noisy linear oscillator \citep{Duriez2017book}.
		\item Chaotization of the forced Lorenz system \citep{Duriez2014aiaa}.
		\item Stabilization of a generalized mean-field system (2 nonlinearly coupled oscillators as derived by \citet{Luchtenburg2009jfm}) \citep{Duriez2017book}
		\item Stabilization of a generalized mean-field system (3 nonlinearly coupled oscillators) \citep{LiR2018am}.
\item Stabilization of a 2D mixing layer in a direct numerical simulation (open loop) \cite{LiH2020pf}.
\item Energetization of a 2D mixing layer in a direct numerical simulation (open loop) \cite{LiH2020pf}.
\item Mixing layer energetization in a wind tunnel---feedback with 96 actuators in unison and 19 hot-wire sensors) \citep{Parezanovic2013tsfp,Parezanovic2016jfm}.
\item Mixing increase behind a backward-facing step in a wind-tunnel experiment---20 actuators (unison), 3 sensors \citep{Chovet2017ifac}.
\item Recirculation zone reduction behind a backward facing step in a water tunnel---feedback with single actuator, online PIV flow monitoring \citep{Gautier2015jfm}.
\item Separation control of a turbulent boundary layer over a ramp---feedback with 54 active vortex generators (unison) and 2 sensors \citep{Duriez2014aiaa,Debien2016ef}.
\item Smart skin separation control ---feedback with 2 $\times$ 30 actuators and 58 sensors [HIT 2022].
\item Lift-increase of a NACA0015 airfoil with an emulated plasma actuator
in a direct numerical simulation  [Joint work with H.~Fukumoto \& A.~Oyama].
\item Lift-increase of a high-lift configuration at high Reynolds number in experiment---Open-loop with 33 actuators and 196 sensors \cite{ElSayed2018aiaa}
\item Mitigation of dynamic bubble burst  for stall suppression ---feedback with 1 actuator and 1 sensor under transient conditions \citep{Asai2019aiaa}.
\item Open-loop stabilization of a fluidic pinball in a wind-tunnel experiment---open-loop with 3 actuators \citep{Raibaudo2017ifac}.
\item Stabilization of a fluidic pinball in a simulation---feedback with 3 actuators and 9 sensors \citep{Cornejo2021jfm}.
\item Ditto but with drag reduction \citep{CornejoMaceda2019pamm}.
\item Drag reduction of a D-shaped cylinder in a wind-tunnel experiment---feedback with 2 actuators and 5 sensors [TU Braunschweig].
\item Drag reduction of an Ahmed body in a wind tunnel---feedback with 4 actuators (unison) and 12 pressures sensors) \citep{LiR2017ef,LiR2018am}.
\item Symmetrization of a bi-modal Ahmed body wake control  in a wind tunnel---feedback with 4 actuators and 12 pressures sensors) \citep{LiR2016ef}.
\item Drag reduction and side wind stabilization of a yawed truck model---feedback with 5 actuators and 18 sensors under transients [TU Braunschweig].
\item Jet mixing optimization with a wind-tunnel experiment---multi-frequency open-loop with 1 actuator \citep{Wu2018ef}.
\item Ditto but with 6 actuators \citep{Zhou2020jfm}.
\item Vortex-induced vibration of a circular cylinder with 2 jets on the lee side working in unison and 1 sensor signal feeding back the transverse displacement \citep{Ren2019pof}.
\end{enumerate}
All examples are based on very similar meta parameters of GPC \citep{Duriez2017book}.
Typically, GPC is based on linear/tree-based genetic programming 
with a similar structure and
requires 10 generations with 100 individuals.
The genetic operations include
elitism (1 individual), replication (10\% probability), mutation (20\% probability) and crossover (70\% probability).
No meta parameter tuning seems necessary for typical applications.
The GPC can be taken `off the shelves'' and expected to work with the standard implementation.

This book is organized as follows.
Chapter~\ref{Cha:MLC} introduces the flow control problem as an optimization problem and describes the linear genetic programming methodology for function optimization.
Chapter~\ref{Sec:UG} constitutes the user guide. The chapter contains the main commands to get started with the code as well as a description of the content of the code and a list of useful commands. The commands are exemplified on the stabilization of a Landau oscillator.
The \xMLC code is then demonstrated on the net drag reduction of the fluidic pinball in chapter~\ref{Cha:Example}.
Three types of controllers are optimized: First, multi-frequency forcing; Second, feedback control; Thirdly, an hybrid control combining multi-frequency forcing and feedback control.
Finally, chapter~\ref{ToC:Conclusions} concludes this book and opens on the future of machine learning control.
%\colorbox{red}{\rule{20mm}{20mm}}

\chapter{Machine learning control}\label{Cha:MLC}

This chapter describes the machine learning control framework and the methodology for control law optimization.
First, the flow control problem is described as a regression problem (Sec.\ \ref{Sec2:OptimizationProblem}),
then the linear genetic programming control (LGPC) algorithm is detailed (Sec.\ \ref{Sec2:LGPC}),
finally, the main parameters for the optimization are summarized.

\section{Flow control as a regression problem}
\label{Sec2:OptimizationProblem}
A flow control problem consists on deriving the optimal control law $\bfKopt$ that fulfills a given objective, such as drag minimization, lift increase, noise reduction, mixing enhancement, etc. 
A control law is a mapping between the inputs ($\boldsymbol{b}$) and the outputs ($\boldsymbol{s}$) of the system to control:
\begin{equation}
\label{Eq:ControlLaw}
\boldsymbol{b}=\boldsymbol{K}(\boldsymbol{h}, \boldsymbol{s}).
\end{equation}
In equation~\eqref{Eq:ControlLaw}, the different quantities are:
\begin{itemize}
\item $\boldsymbol{b}$ is the actuation command, i.e., the signal sent to the actuators;
\item $\boldsymbol{s}$ is the sensor vector that comprises the measurements provided by the sensors;
\item $\boldsymbol{h}$ is a vector comprising time-dependent functions such as periodic functions (see Sec.\ \ref{sec:MF_optimization}).
\end{itemize}

A key enabler for employing machine learning techniques to derive the optimal control law $\bfKopt$ is to reformulate the problem as a regression problem.
In this framework, the control objective is translated in a \emph{cost function} or the \emph{loss function}, $J$, to be minimized.
The typical control problem can be expressed as a multi-objective minimization problem.
Let $J$ be the cost function to minimize according to the control law $\boldsymbol{K}$.
$J$ is then the sum of different control objectives that quantifies the control efficiency, it reads:
\begin{equation}
	J = J_a + \sum_{k}\gamma_k J_k
\end{equation}
where $J_a$ is the quantity to minimize, e.g. drag power, distance to given state, etc.
The $J_k$ are the secondary objectives to be interpreted as optimization constraints.
Their contribution is weighted by the coefficients $\gamma_k$.
The most common control problem comprises only two terms, $J_a$ and an actuation penalization term $J_b$ such as the cost function is reduced to:
\begin{equation}
	J = J_a + \gamma J_b
\end{equation}

The control problem can now be reformulated as an regression problem:
\begin{equation}
\label{Eq:ControlProblem}
\bfKopt = \underset{\boldsymbol{K}\in \mathcal{K}}{\operatorname{arg\,min}} \; J(\boldsymbol{K})
\end{equation}
where $\mathcal{K}: A \mapsto B$ is the space of all possible control laws, also referred as \emph{control law space}.
Here $A$ is the input space, e.g., the space of sensor signals and $B$ the actuation range.
The regression problem~\eqref{Eq:ControlProblem} is a hard, non-convex problem, possibly including many minima.
\begin{figure}[htb]
\centering
\includegraphics[width=0.5\textwidth]{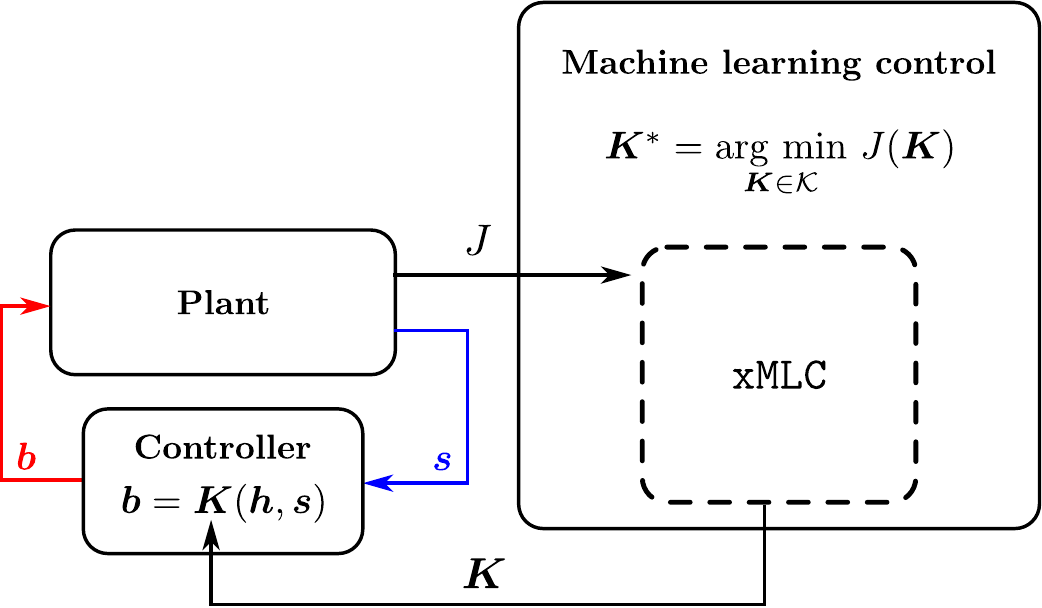}
\caption{Control problem overview.
	The plant is controlled by actuators and monitored by sensors.
	The control can be carried out in an open- or closed-loop manner.
	In closed-loop control, the controller sends the actuation commands $\boldsymbol{b}$ to the plant based on sensor signals $\boldsymbol{s}$.
	The software \xMLC builds a control law $\boldsymbol{K}$ that minimizes the cost function $J$.
	The control law is optimized through `trial and error' based on its performance.}
\label{Fig:LearningLoop}
\end{figure}
In this book, we introduce the software \xMLC based on a linear genetic programming to learn a control law $\boldsymbol{K}$ that minimize the cost function $J$.
Figure~\ref{Fig:LearningLoop} gives an overview of the system to control and the role of \xMLC in the learning loop.

\section{Linear genetic programming control}\label{Sec2:LGPC}
Machine learning control (MLC) based on genetic programming control \citep[GPC]{Dracopoulos1997book} is an evolutionary algorithm for function optimization.
MLC relies on biological-inspired mechanisms to build candidates solutions for the regression problem~\eqref{Eq:ControlProblem} in an iterative and stochastic manner.

The main idea of evolutionary algorithms is based on the evolution of a set of candidate solutions throughout generations thanks to selected recombinations.
Following the biological terminology, a candidate solution is also called an individual and a set of individuals, a population.
The optimization process relies mainly on three evolutionary principles:
\begin{description}
\item[The survival of the fittest:] it is the selection of the most fitting individual, or the most efficient individual according to the environment, to form the next generation. This mechanism allows that the features of the `best' individuals pass to the next generation to build, eventually, better individuals;
\item[Crossover:] it is one of the two forces of evolution that brings diversity to the population and gives opportunity to improve individuals; crossover is able to exploit the strengths of individuals by recombining two or more individuals and generating one or more offspring build from their `parents'.
\item[Mutation:] it is the second force of evolution; it is the force that brings novelty to the population; new and more better features are likely to appear thanks to mutation.
\end{description}
It is worth noting that both crossover and mutation are stochastic mechanisms.
Indeed, the recombination and the mutation of given individuals are random processes that, in general, give always different results.
When solving a regression problem with an evolutionary algorithm, the environment corresponds to the cost function $J$; It assess the performance/quality of an individual.
Mutation and crossover are referred as \emph{genetic operators}.

In the following, we describe the genetic programming control algorithm in the control framework.
First, we present the internal representation of the control laws and how we operate on them to generate new control laws.

\subsection{Control law representation}
To be able to combine and mutate the control laws throughout the generations, an internal representation of a mathematical function is needed.
The present software \xMLC is based on linear genetic programming (LGP) and adopts a matrix representation for the control laws \citep{Brameier2006book}.

In LGP, the individuals are considered as little computer programs,
using a finite number $\Ninstr$ of instructions,
a given register of variables
and a set of constants.
The instructions employ
 basic operations ($+$, $-$, $\times$, $\div$, $\cos$, $\sin$, $\tanh$, etc.)
using  inputs ($h_i$ time-dependent functions and $s_i$ sensor signals)
and yielding the control commands as outputs.
A matrix representation conveniently comprises the operations of each individual.
Every row describes one instruction.
The first two columns define the register indices of the arguments,
the third column the index of the operation
and the fourth column the output register.
Before execution, all registers are zeroed.
Then, the last registers are initialized with the input arguments,
while the output is read from the first registers after the execution of all instructions.
This leads to a $\Ninstr\times 4$ matrix
representing the control law $\boldsymbol{K}$.
The name `linear' refers to the sequential execution of the instructions.
If the operation to execute only requires one operand, only the first column is considered and the second one is ignored.
Each column of the matrix has its own range of values following what it codes.
For single input control, i.e. when there is only one controller, the control law is read in the first register.
For $\Nb$ controllers, the control laws are read in the first $\Nb$ registers.
Finally, to avoid definition problems, the operators such as division and logarithm are protected to be defined on $\mathbb{R}$ the space of all the real numbers, see \cite{Duriez2017book}.

The registers play the role of memory slots.
We distinguish two types of registers:
\begin{description}
\item[Variable registers:] they are registers that can be overwritten while executing an instruction. They help to store intermediate results.
\item[Constant registers:] they are registers that are protected during the reading of the matrix. They are used to store random constants or data of the problem. 
\end{description}
\begin{figure}[hbt]
\centering
\subfloat[]{\label{fig:instruction_representation}\includegraphics[width=0.2475\textwidth]{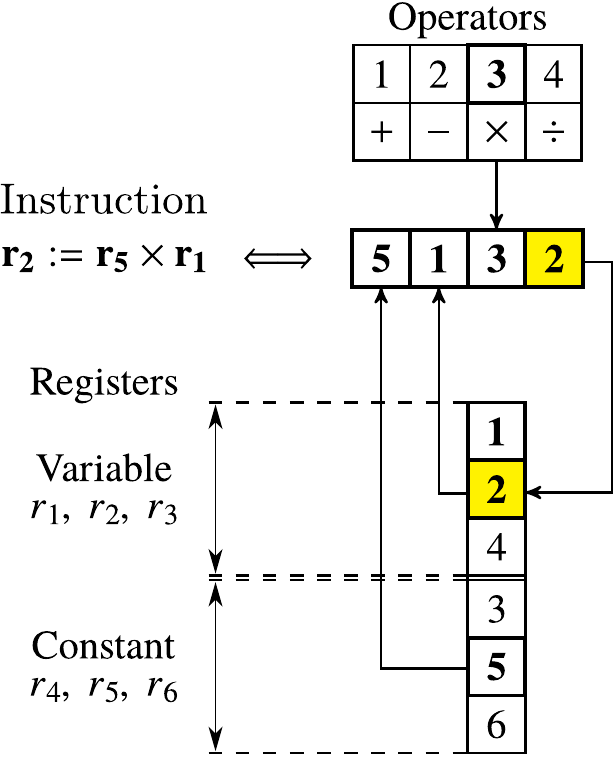}}%
\hfil
\subfloat[]{\label{fig:controllaw_representation}\includegraphics[width=0.45\textwidth]{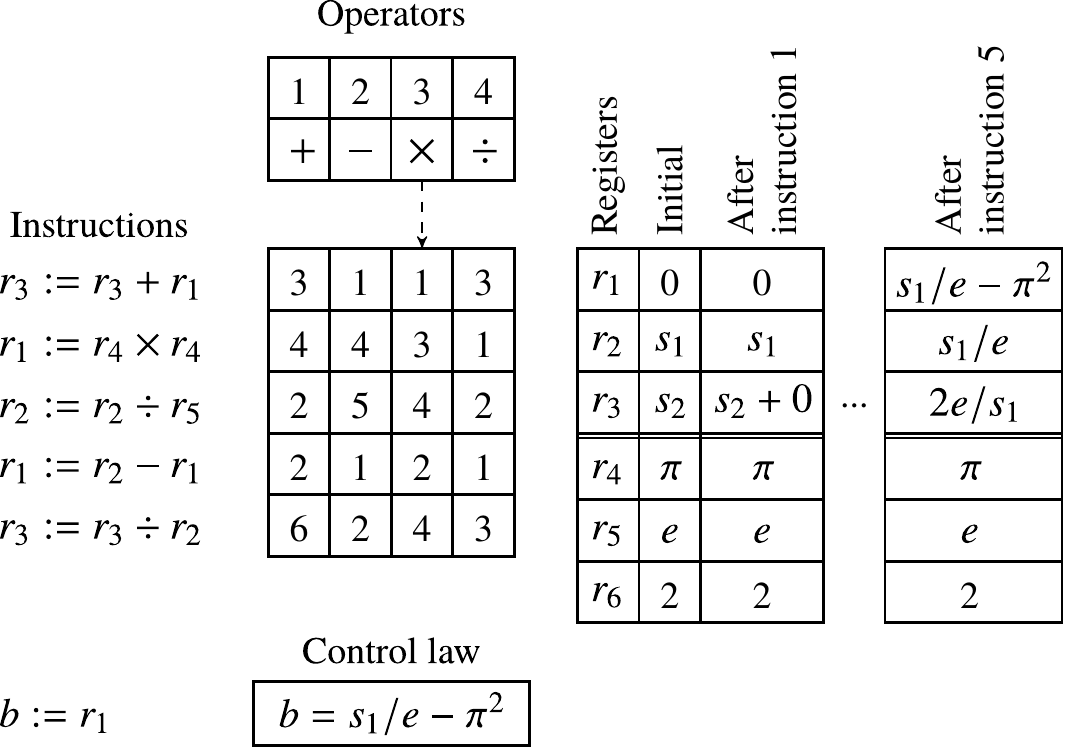}}%
\caption{(a) Matrix representation of one single operation.
The representation is inspired of reverse polish notation.
The vertical column are the indices of the registers.
The result of the instruction is stored in the variable register $r_2$, overwriting its previous value.
(b) Transcription of a matrix of instructions into a control law (expression framed in the bottom).
  The matrix (middle) has five instructions.
  The instruction are displayed on the left and the evolution of the registers after the execution of the first and fifth instruction on the right.
  The library of operators ($+$, $-$, $\times$, $\div$) and there corresponding index are displayed on the top.
  The first three registers ($r_1$, $r_2$, $r_3$) are the variable register and the last three ($r_4$, $r_5$, $r_6$) are the constant register.
  The control law is derived from the expression stored in the first register.
  In case of multiple input problems the others control laws would be derived from the following registers.}
\end{figure}
Figure~\ref{fig:instruction_representation}, illustrates how a single instruction is represented in matrix form.

It is worth noting that for a given mathematical expression, there is more than one matrix representation.
Indeed, as stated before, there are instructions in the matrix that have no impact in the output registers.
Also, the matrix representation takes into account the order of the operations even for operators that are commutative: the control laws $b = s_1+s_2$ and $b=s_2+s_1$ will be have different representation while being the same mathematical expression.
As consequence, several instances of the same individual can be present in the population.
In order to accelerate the learning, such individuals are removed.
To detect them, the control laws are all evaluated on the same random samplings.
If the result of the evaluation of one individual correspond  to a previous evaluation, then the individual is replaced.

Figure~\ref{fig:controllaw_representation} depicts how a matrix of instructions is read to build a control law.
We notice that in the instruction matrix, not all instruction lines are useful.
Indeed, if an instruction line does not affect one of the output registers then it is, in reality, useless.
However, in the process of recombination or mutation, these instructions lines can be `activated', changing the final control law.
Following \citet{Brameier2006book}, these `useless' instruction lines, also called \emph{introns}, play a major in the process of building relevant structures.

With enough instructions and operators, any function can be represented in matrix form.
LGP can, for example, reproduce the Taylor expansion of any function until an arbitrary order by deriving the coefficients of the power series.
Also, using the matrix representation, we do not constrain, a priori, the structure of the control laws.
Of course, the solutions built strongly depend on the library of operators and control inputs given to the algorithm.
Indeed, the richness of these libraries defines the complexity of the search space for the regression problem.
The choice of the function libraries is studied in Sec.\ \ref{sec:mlc_parametric_study} on a dynamical system whereas different sets of control inputs are tested for the control of the fluidic pinball in Sec.\ \ref{sec:fluidic_pinball_plant}.

Before giving the genetic algorithm in its final form, we first describe its first, the Monte Carlo step, as it is an optimization algorithm on its own.

\subsection{Monte Carlo sampling}
A starting point for genetic programming is the random generation of the first set of individuals.
This operation can be seen as a Monte Carlo sampling process.
In the LGPC framework, to define a control law is to chose a library for the operators ($+$, $-$, $\cos$, etc.), a library for the inputs ($a_1$, $a_2$, etc.), the maximum number of instruction $\Ninstrmax$, or the number of rows in the matrices,  the number of variable registers $\Nvar$ and the number of constant registers $\Ncst$.
From these parameters, we can generate random matrices that are then read sequentially to form control laws.
The number of instructions for each matrix is randomly drawn from an uniform distribution between 1 and $\Ninstrmax$.
In theory, a Monte Carlo process is enough to solve equation~\eqref{Eq:ControlProblem} but a very large number of individuals might be needed to reach the global optimum of the problem, especially for search spaces of infinite dimension.
For pragmatic reasons and also to emulate limited experiment time, we fixed the total number of individuals tested $\Ni$.
$\Ni$ can also be seen as the total number of cost function callings and also the total number of experiments to run.
\begin{figure}[htb]
  \centering
  \includegraphics[scale=0.8]{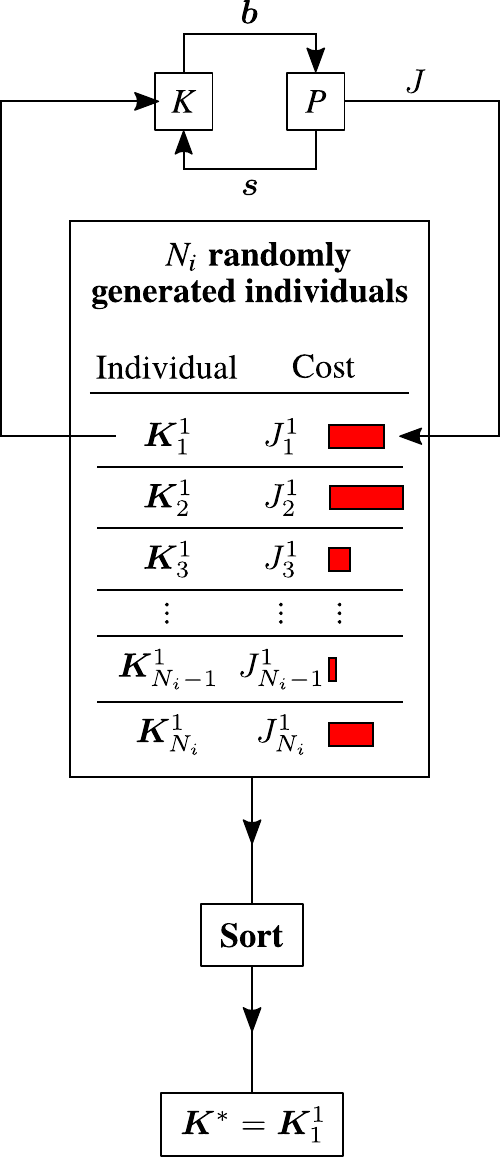}
  \caption{Monte Carlo sampling. 
  A set of randomly generated control laws is evaluated and sorted following their performances.
  The red bar on the right of each individual represents their performance.
  The smaller the bar, the better the individual performs.
	The superscript $1$ on the individuals $\boldsymbol{K}_i$ means that they belong to the first generation.
 The top part of the figure, represents the controller K, receiving the sensors signals $\boldsymbol{s}$ as input and giving back the actuation command $\boldsymbol{b}$ to the plant P.
  }
  \label{fig:monte_carlo}
\end{figure}
Figure~\ref{fig:monte_carlo} illustrates the Monte Carlo sampling process.
$\Ni$ individuals are generated randomly, they are all tested and sorted following their cost.
The final result of the algorithm is the individual with the lowest cost $J$, thus the most performing following the cost function criterion.
Figure~\ref{fig:monte_carlo} depicts the Monte Carlo process for controlling a plant P (framed in the figure).

In the next section, we describe how to create the next generations of individuals from a set of individuals generated thanks to a Monte Carlo process.

\subsection{The evolution process---Selection and genetic operators}

In the following sections, we describe how the natural selection is emulated and the implementation of the genetic operators.
In this section, we detail the steps carried out to create a new population based on a previous one.

\subsubsection{Selection}
To create the next generation of individuals, we need, first, to select the most performing individuals to be combined and mutated.
The operation of selection is carried out thanks to a tournament selection.
The idea of a tournament selection is to select $\Ntour$ individuals among the $\Ni$ individuals in the population.
Among the $\Ntour$ individuals selected, the best one is selected with a probability of $\Ptour$.
If the best one is not chosen, the second best is chosen with the same probability $\Ptour$ and so on for all the selected individuals.
At the end, if no other individual is chosen, the least performing among the $\Ntour$ is selected.
The choice of $\Ntour$ and $\Ptour$ influence the extent to which well-performing individuals are preferred over least-performing ones.
This feature is called \emph{selection pressure} and is developed in detail in \citet{Wahde2008book}.

\subsubsection{Crossover for exploitation}
\begin{figure}[htb]%
\centering
\subfloat[]{\label{fig:crossover}\includegraphics[scale=0.75]{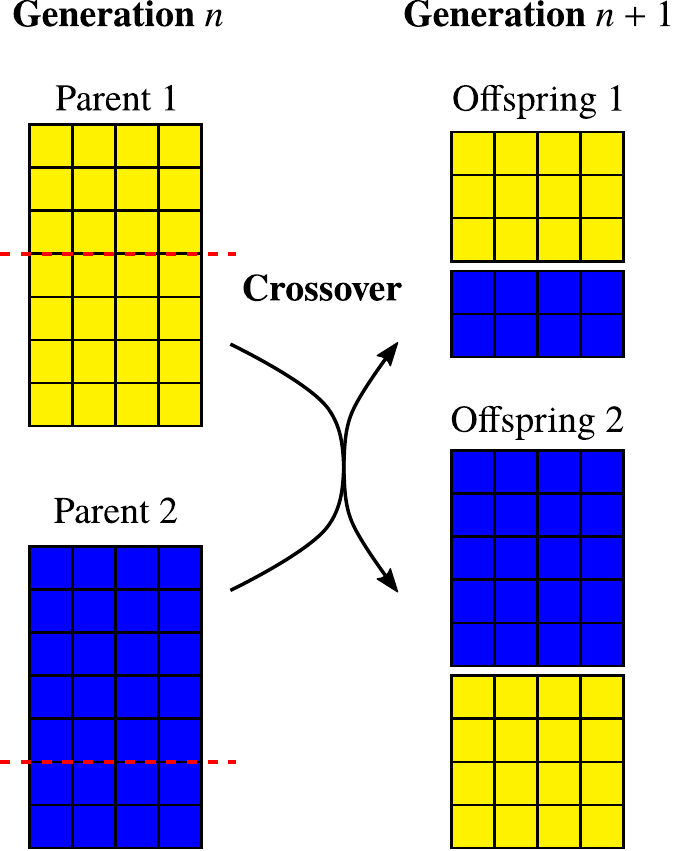}}%
\hfil
\subfloat[]{\label{fig:mutation}\includegraphics[scale=0.75]{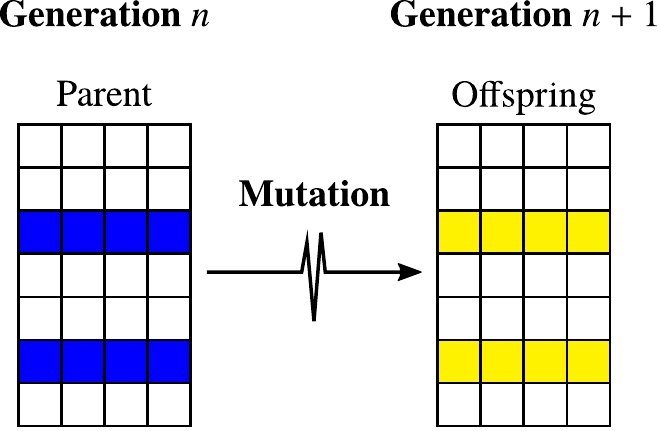}}%
\caption{(a) Exploitation of the genetic material by recombining two individuals. The parts of the matrices are switched to build new matrices.
(b) Exploration of the search space by creating new instructions and thus new structures.}
\end{figure}
Crossover is the operation of recombination of individuals.
It has the potential to extract and combine relevant structures in the individuals.
That is why, we refer this genetic operator as the \emph{exploitation operator}.
To combine the individuals, two individuals are selected in the population, and their matrices are split in two and the parts are swapped to generate two new individuals, also referred as \emph{offsprings}.
Figure~\ref{fig:crossover} illustrates the crossover operation between two individuals.

It is worth noting that, the crossover operation is defined such as the length of the matrices may increase or decrease.
To avoid that the size of the matrices explodes, we set a upper limit to the number of rows in the matrix.
In practice, this limit is the same $\Ninstrmax$.
If this limit is exceeded, then the operation is restarted until offspring with lesser instructions are built.

\subsubsection{Mutation for exploration}
Mutation is the operator that generates new sequences in the matrices.
The role of this operator is to find new structures, unknown to the population, to improve the solutions.
For the mutation of one individual, each row of the corresponding matrix representation has a probability of $\Pmut$ to be completely changed.
The parameter $\Pmut$ is chosen such as at least one row is changed in the matrix.
The change of one line can either have no consequences in the final output, if the instruction stays an intron, or it can also completely change the final output.
To improve our exploration potential, we choose to restart the mutation operation when the mutated individual is identical to the original one.
Figure~\ref{fig:mutation}, depicts the process of mutation for an individual.

Of course, there are several ways to define the crossover and mutation operators but we choose to realize the simplest implementation.

\begin{figure}[htb]
  \centering
  \includegraphics[scale=0.7]{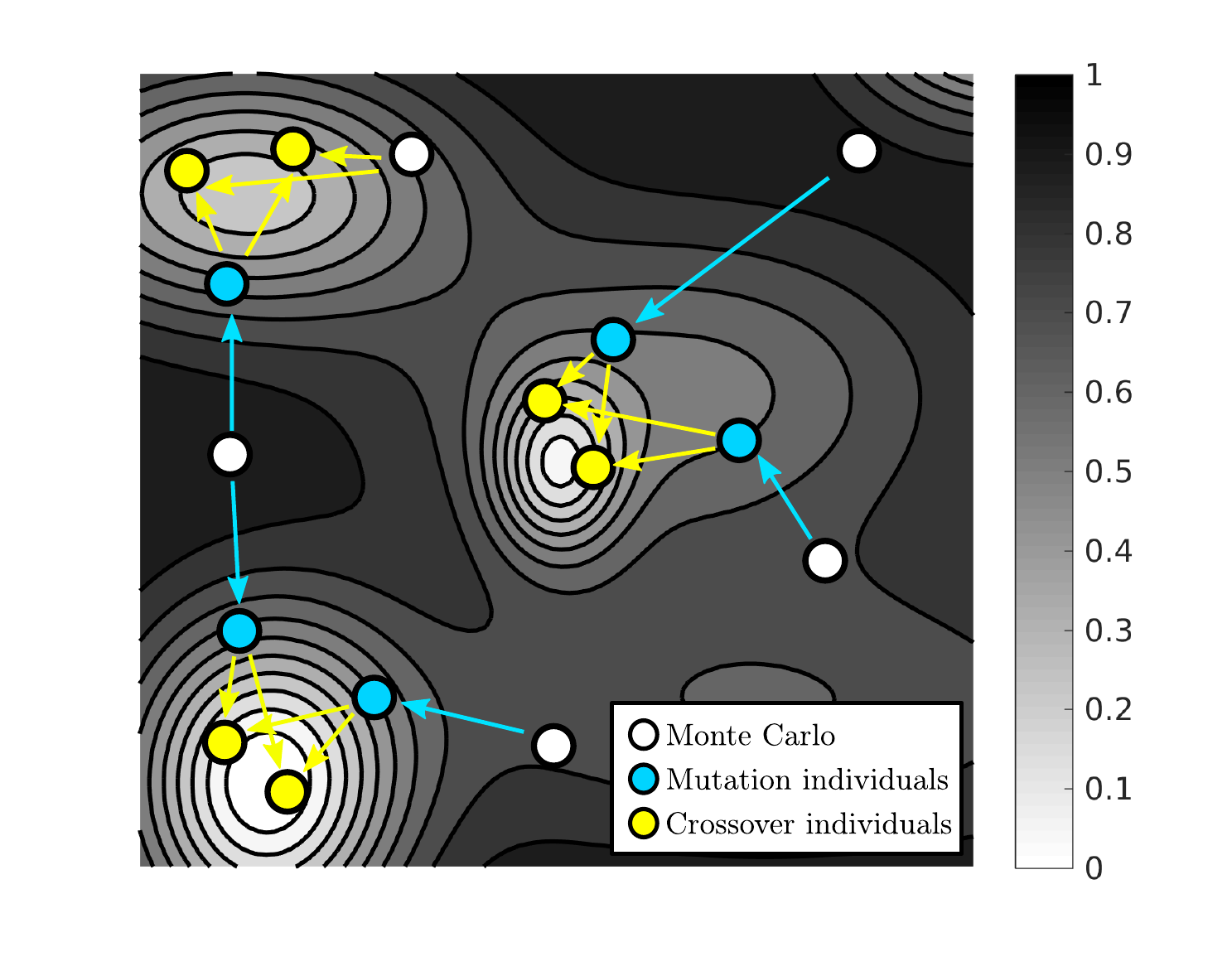}
  \caption{Principle sketch of the exploitation/exploration potential of crossover and mutation.
  The background is a 2D representation of a control landscape.
  White regions denoting good performances and dark regions poor performances.
  Three minima are depicted.
  After an a priori exploration of the control landscape with Monte Carlo, crossover and mutation improve the evaluated individuals.
  Crossover individuals have the potential to explore the neighborhood of a minimum whereas mutation discovers new minima by combining good individuals.}
  \label{fig:learning_principle}
\end{figure}
From an optimization point of view, crossover is the operator that improves existing solution.
Its role is to `explore' the neighborhood of a minimum, while mutation is the one that explores the control landscape to discover new minima.
The learning principles are illustrated in figure~\ref{fig:learning_principle}.

\subsubsection{Replication and elitism for memory}
In addition to crossover and mutation, we also consider two other operators: replication and elitism.
With replication an identical copy of one individual is copied to the next generation, assuring memory of good individuals and allowing future recombination.
This elitism operation assures that the bests individual are always in the latest generation so that `the winner does not get lost' throughout the generations.
In this case, the individuals are simply copied to the next generation.
The number of individuals selected by elitism is defined by the elitism parameter $\Ne$,
usually it is set to $\Ne=1$.

\begin{figure}[htb]
  \centering
  \includegraphics[scale=0.8]{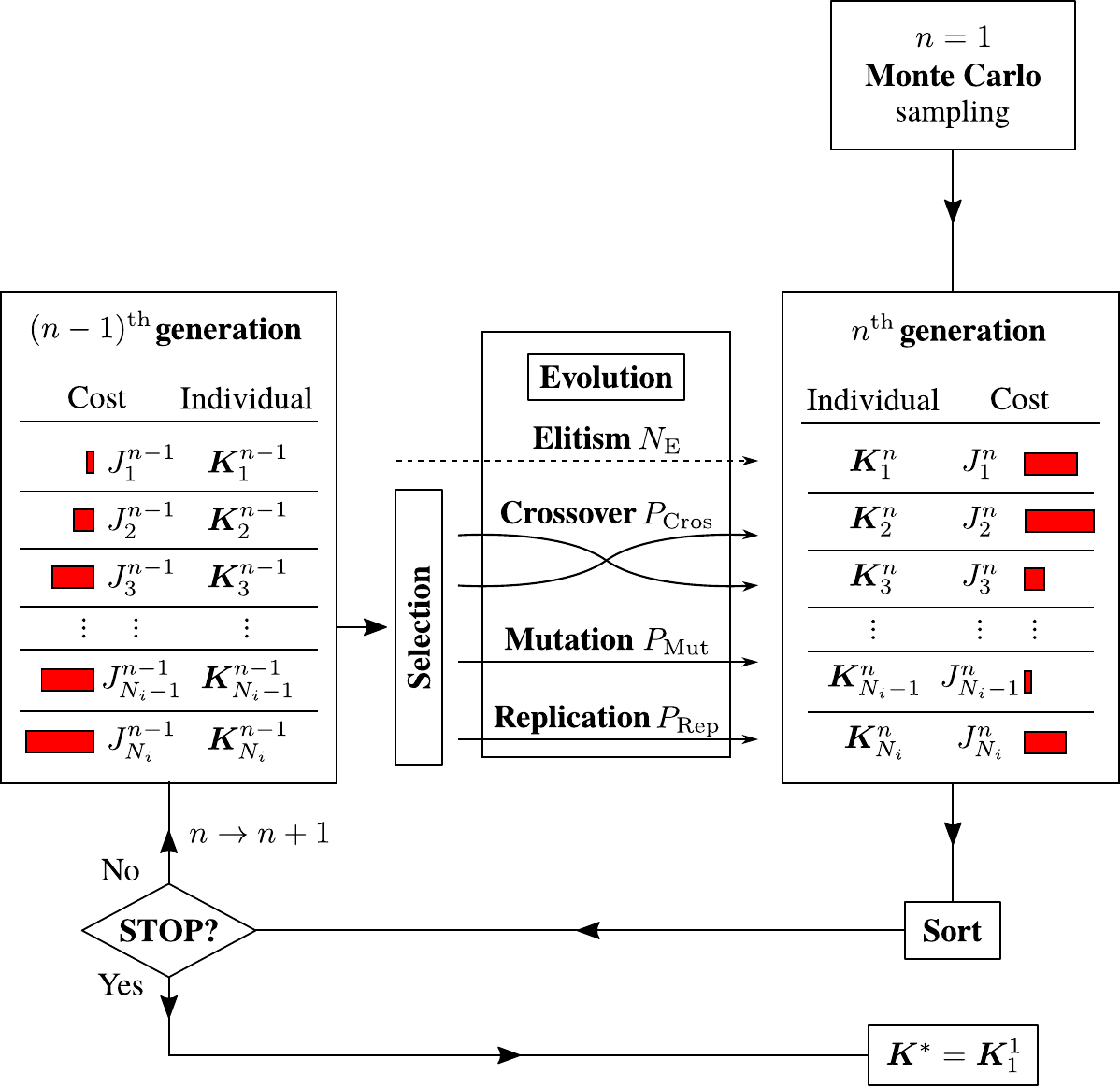}
  \caption{Linear genetic programming algorithm employed in \texttt{xMLC}.}
  \label{fig:LGPC_algo}
\end{figure}
Figure~\ref{fig:LGPC_algo}, illustrates the complete LGPC algorithm.
The first generation of individuals is generated thanks to a random sampling of the individuals (Monte Carlo method).
Then, from a generation $n$, the individuals are all evaluated and sorted following their performances.
The best individuals are then selected, thanks to a tournament method, to be modified and recombined with crossover and mutation.
Replication and elitism assures a memory of the good individuals.
The choice of crossover, mutation or replication to populate the next generation is controlled by the probabilities $\Pcros$, the crossover probability, $\Pmut$, the mutation probability and $\Prep$ the replication probability.
They are chosen such as $\Pcros + \Pmut + \Prep  = 1$.
The balance between crossover, mutation and replication is thoroughly analyzed in App.\ \ref{Sec:ParamStudyGenOp}.

There are several variations of the genetic programming algorithm, where the genetic operators are not separated but applied one after the other and where the offspring replaces the parent individual in the population only if it performs better.
In this study, we choose to follow the classical evolutionary algorithm described in \citet{Brameier2006book} and also employed by \citet{Duriez2017book}.

In the next section, we illustrate the learning process of LGPC by stabilizing a Landau oscillator with \texttt{xMLC}.

% -----------------------------------------------------------------------

\chapter{User guide}\label{Sec:UG}
This chapter contains all the elements to download, install and run the \xMLC software on a toy system.
This chapter is recommended for a quick start with \texttt{xMLC}.
First, the mean features of the software are described (Sec.\ \ref{Sec:Features}), then the requirements and installation are detailed (Sec.\ \ref{Sec:Installation}).
Thereafter, the main commands for a learning process are exemplified on the control of a toy system (Sec.\ \ref{Sec:QuickStart}.
Finally, an overview of the MLC MATLAB class is given and how to change the main parameters (Sec.\ \ref{Sec:MatlabClass}).
The user who wants to employ \xMLC to her/his own control problem is invited to follow the guide in App.~\ref{AppMyPlant} to write her/his own parameter file.
However, it is suggested to run the commands in this section first to get familiar with code.

\section{\xMLC features}\label{Sec:Features}
The \xMLC software is an implementation of the linear genetic programming algorithm for control optimization.
The software allows in particular for:
\begin{description}
\item[Open-loop control:] Control laws only depending on time-dependent functions can be optimized to derive, for example, a multi-frequency controller combining different periodic functions.
\item[Closed-loop control:] Of course, closed-loop control laws can be learned by including sensor signals in the input library.
\item[Hybrid control:] An hybrid optimization is also possible by combining sensor information and time-dependent functions.
\item[From SISO to MIMO control:] \xMLC is able to optimize control laws for single-input single-output (SISO) systems as well as multiple-input multiple-output (MIMO) systems.
There is no limit on the number actuators and sensors one can use.
\end{description}

In addition of the optimization process, \xMLC includes a variety of post-processing tools to analyze the learning process and visualize the distribution of evaluated individuals, such as learning curve, extraction of the best individual and Pareto front.
Some of these features are displayed in the next section  (Sec.\ \ref{Sec:QuickStart}).
Moreover, \xMLC includes a set of scripts and commands for a interfacing the optimization code with numerical simulations on computer clusters and experiments.
Finally, the code is not restricted to control problems as it can be adapted to solve any function regression problem.

\section{Download and installation}\label{Sec:Installation}
In this section we present the necessary steps to download and install the \xMLC software.

\subsubsection{Requirements}
\xMLC is available for both MATLAB and Octave.
It has been coded on MATLAB version 9.5.0.944444 (R2018b) and Octave version 4.2.2. any further version should be compatible with the software.
No particular MATLAB or Octave package is needed for the proper functioning of the software.

\subsubsection{Installation}
The \xMLC software can be download from the following links

\vspace{0.3cm}
 \url{https://doi.org/10.24355/dbbs.084-202208220937-0} or \url{https://github.com/gycm134/xMLC}
\vspace{0.3cm}
 
 \noindent
 under the MIT License (MIT).

Once downloaded, decompress the tar file and copy the \texttt{MLC/} folder where it is needed.
Installation is then complete.
For further information on the content of the \texttt{MLC/} folder please look at Sec.\ \ref{Sec:MatlabClass} and the README.md file.

\section{Quick start}\label{Sec:QuickStart}
In this section, we present the main commands to optimize control laws.
The process is illustrated by stabilizing the damped Landau oscillator, a dynamical system with a stable limit cycle.
The commands are executed on MATLAB but they can be also ran on Octave.
Only the methods containing MATLAB in their name needs to be replaced by their Octave counterparts.

% =====================================================================================
%To illustrate the linear genetic programming learning methodology, we optimize a controller for a damped Landau oscillator.
%First, we present the dynamical system used for this study.
%Then, we describe the learning mechanisms of LGPC by stabilizing the Landau oscillator with increasingly more complex algorithms.
%%We start with a Monte Carlo optimization then we progressively add the genetic operators until we reach the full-fledge LGPC algorithm.
%We start with a Monte Carlo sampling then we stabilize the oscillator with the LGPC algorithm described in Sec.\ \ref{fig:LGPC_algo}.
%In this section, we focus on the meta-parameters $(\Pcros,\Pmut,\Prep)$ to analyze the role of crossover, mutation and replication in the learning process.
%We reveal, among other things, that there is sweet spot in the meta-parameter space.
%
%All the simulations have been carried out thanks to our own MLC code LGPC code developed on MATLAB and also available for the free software GNU Octave.

\subsection{The damped Landau oscillator}\label{Sec:LandauOscillator}
\subsubsection{The controlled dynamical system}
The damped Landau oscillator is a system of two coupled ordinary differential equations with a nonlinear damping of the growth rate.
Despite its simplicity, it describes a fundamental oscillatory process at the heart of physical mechanisms such as the von Kármán vortex shedding behind a cylinder \citep{Luchtenburg2009jfm}.
To control the oscillator a forcing term $b$ is introduced in the second equation.
The systems reads:
\begin{equation}
    \left\{
    \begin{array}{lcl}
    \dot{a}_1 & = & \sigma a_1-a_2 \\
    \dot{a}_2 & = & \sigma a_2+a_1 + b(a_1,a_2)\\
    \sigma & = & (1-a_1^2-a_2^2)
  \end{array}
    \right.
  \label{eq:LandauOscillatorControlled}
\end{equation}

% Figure : Landau oscillator unforced
\begin{figure}[htb]
\centering
\subfloat[]{\label{Fig:Landau_unforced_1}\includegraphics[width=0.25\textwidth]{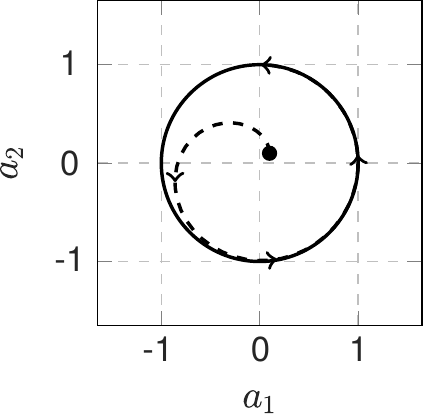}}%
\hfil
\subfloat[]{\label{Fig:Landau_unforced_2}\includegraphics[width=0.25\textwidth]{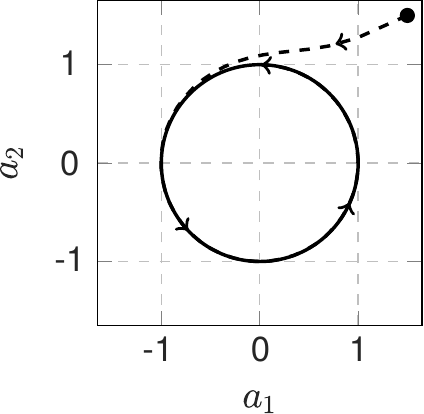}}
\caption{Phase portrait of the oscillator with no actuation
  ($b=0$) with $(a_1,a_2)_{t=0}=(0.1,0.1)$ as initial condition for (a)
  and $(a_1,a_2)_{t=0}=(1.5,1.5)$ as initial condition for (b).
  In both cases the system converges towards the limit cycle of radius $1$.}
\end{figure}
For $b=0$, we have an oscillator of growth rate $\sigma$, angular frequency $1$, period $T=2\pi$ and fixed point $(0,0)$.
For an initial condition close to the fixed point, the quadratic terms in $\sigma$ are negligible,
leading to an exponential growth.
When the system is far from the fixed point, the growth is damped due to the quadratic terms, stabilizing the oscillator to the limit cycle of radius $\sqrt{1} = 1$.
The same reasoning for an initial condition outside the circle of radius $1$ shows that the limit cycle is globally stable.
The uncontrolled dynamics are depicted in figure~\ref{Fig:Landau_unforced_1} and \ref{Fig:Landau_unforced_2}.
The control on the second equation has the effect of pushing the system upwards or downwards following the sign of $b$: Upwards if $b>0$ and downwards if $b<0$.

\subsubsection{Objective and cost function}
The control objective is to bring the system to the fixed point $(a_1,a_2)=(0,0)$ from the limit cycle.
Two terms are considered for the cost function: $J_a$, the averaged distance to the fixed point and $J_b$, an actuation penalization term.
\begin{equation}
  \begin{array}{rcl}
  J		& = &	J_a +\gamma J_b \\
J_a 		& = &	 \cfrac{1}{T_{max}} \bigintsss_{ 0}^{T_{max}} (a_1^2+a_2^2) d \tau \\
J_b 		& = &	\cfrac{1}{T_{max}} \bigintsss_{ 0}^{T_{max}} b^2 d \tau\\
  \end{array}
\label{Eq:CostFunction}
\end{equation}

The penalization parameter $\gamma$ is taken equal to $0.01$, such as the optimization process focuses first on the primary objective, i.e., stabilizing the oscillator.
%$J_a$ is the integral of the distance to the fixed point and $J_b$ measures the total energy delivered for the control.
Both are integrals quantities over 10 periods which corresponds to $T_{\rm max}=20\pi$ as we are interested not only in the final solution but also on the trajectory.
In order to assure a general solution, we consider four initial conditions on the limit cycle: $(1,0)$, $(0,1)$, $(-1,0)$, $(0,-1)$.
The cost function $J$ is then a mean value between these four initial conditions.
For the uncontrolled dynamics, we have $a_1^2 + a_2^2  = 1$ and $b^2 = 0$, therefore $ J_a=20\pi\approx 62.83$ and $J_b=0$.
The unforced cost is then $J_0 = 20\pi$.

In the \xMLC software, the damped Landau oscillator system and its resolution is implemented in the  \texttt{Plant/LandauOscillator/LandauOscillator\_problem.m} file.
See Sec.\ \ref{Sec:MatlabClass} for more information and the content of the code.

% ===================================================================================
\subsection{First run}
We now present the main commands to quickly use the \xMLC software.

\subsubsection{Initialization}
To use the MLC software launch a MATLAB session on the \texttt{MLC/} folder.
We then follow the steps of the \texttt{CheatSheet.m} file, providing the corresponding outputs and figures.
It is advised to not execute the whole \texttt{CheatSheet.m} script but rather execute it section by section.
The first step is to launch the \texttt{Initialization.m} script to load all the  necessary paths.
In the following insert, we include the command and the output.
\begin{lstlisting}
>> Initialization;
====================== xMLC v0.10 ====================
 Welcome to the xMLC software to solve function
 regression problems.
 In case of error please contact the author :
  Guy Y. Cornejo Maceda Website
 The MIT License (MIT)
 
 Start by creating a MLC object with : mlc=MLC;
=====================================================
\end{lstlisting}

The command output gives a short description of the \xMLC code and invites us to create a MLC object.
The MLC object is in fact a structure array containing the parameters, the database and information on the current generation.
Let's create such MLC object:
\begin{lstlisting}
>> mlc=MLC;
====================== xMLC v0.10 ====================
Name of the run : TestRun
Problem to solve : LandauOscillator
Problem type : MATLAB
   Number of actuators       : 1
   Number of control inputs  : 2

Parameters : 
   Population size : 10
   Elitism : 1
   Operator probabilities : 
      Crossover : 0.600
      Mutation : 0.300
      Replication : 0.100

To generate a population : mlc.generate_population;
To run N generations : mlc.go(N);
======================================================
\end{lstlisting}

The command output gives information on the problem and parameters in the MLC object.
The `Name of the run' is the name of the MLC object.
When an optimization is initiated, all its associated files are stored in a folder with the name of the run.
For this example, the folder \texttt{save\_runs/} will be automatically created along with the \texttt{save\_runs/TestRun/} folder.

The problem set by default is the stabilization of the Landau oscillator described in Sec.\ \ref{Sec:LandauOscillator};
it is implemented in the \texttt{Plant/LandauOscillator/LandauOscillator\_problem.m} file.
For this example, the system of equations~\eqref{eq:LandauOscillatorControlled} including the control is solved directly by MATLAB.
The \xMLC software can also be interfaced with other solvers or experiments, see App.\ \ref{AppInterface} for more information.

The output also displays the main parameters.
First, the number of actuators and control inputs are displayed
In this case, there is only one actuator, the forcing term $b$ in the second equation and its inputs are $a_1$ and $a_2$.
Then, the main optimization parameters, population size, elitism parameter and genetic operator probabilities.
All these parameters are defined in the \texttt{MLC\_tools/default\_parameters.m} file.
Here are some examples on how to modify these parameters:
\begin{lstlisting}
>> mlc.parameters.Name = 'AQuickTest';
>> mlc.parameters.PopulationSize=100;
>> mlc.parameters.Elitism = 1;
>> mlc.parameters.CrossoverProb = 0.6;
>> mlc.parameters.MutationProb = 0.3;
>> mlc.parameters.ReplicationProb = 0.1;
\end{lstlisting}
Note that these parameters are modified only for the current MLC object.
The user can also modify the initial parameters by creating her/his own parameter file, see App.\ \ref{AppMyPlant}.

\subsubsection{Control law optimization}
Once the MLC parameters are appropriately set, the optimization process can be begin.
There are two ways to advance the optimization process: either running the commands one after the other or using the \texttt{go} method.
Let's look at the first method.
For this, we follow the instructions displayed on the command prompt and generate a population with the appropriate command:
\begin{lstlisting}
>> % Create the first generation
>> mlc.generate_population;
Generating new population
    Population size = 100
    Generating new individual 1/10 Done. 
    Generating new individual 2/10 Done.
    Generating new individual 3/10 Done.
    ...
    Generating new individual 98/100 Done. 
    Generating new individual 99/100 Done. 
    Generating new individual 100/100 Done. 
End of generation : population generated in 3.4128 seconds.
No pre-evaluated individuals to be removed.

To evaluate the population: mlc.evaluate_population;
\end{lstlisting}

The output displays the creation process of the individuals.
Once an individual or a control law is created it is evaluated on a random set of inputs, if this evaluation returns \texttt{INF} or \texttt{NAN}, then the control law is discarded an a new is generated.
In this example, all the individuals have been properly evaluated thus the message: `No pre-evaluated individuals to be removed'.
Now, the population can be evaluated with the corresponding command:
\begin{lstlisting}
>> % Evaluate the first generation
>> mlc.generate_population;
Evaluation of generation 1 :
    Evaluation of individual 1/100  J= 62.831878
    Evaluation of individual 2/100  J= 75.394162
    Evaluation of individual 3/100  J= 59.252250
    ...
    Evaluation of individual 98/100  J= 90.154683
    Evaluation of individual 99/100  J= 46.342791
    Evaluation of individual 100/100  J= 78.925153
No bad individuals to be removed.

To create the next generation : mlc.evolve_population;

To create the next generation : mlc.evolve_population;
\end{lstlisting}

The cost of each individual is displayed next to its evaluation message.
The message `No bad individuals to be removed' means that all the individuals have been properly evaluated.
Otherwise,  the code attributes to them a high cost ($\approx 10^{36}$) so that the selection process automatically eliminates them.
The default parameters are such as the code replaces those `bad' individuals with new ones.
The replacement of bad individuals is repeated until the population does not include any of them.
This option is harmless when the evaluation of the individuals is `quick' but it is advised to deactivate it for time-consuming evaluations or experiments.
This option can be deactivated with the command before launching the optimization process:
\begin{lstlisting}
>> % Deactivate the recursive removal of individuals whose evaluation failed.
>> mlc.parameters.RemoveBadIndividuals = 0;
\end{lstlisting}

Let's generate the next generation and evaluate it with the corresponding commands:
\begin{lstlisting}
>> % Generate the next generation (Gen 2)
>> mlc.evolve_population;
Generation 2
    Generating new individual 1/10 Elitism.
    Generating new individual 2/10 Crossover.
    Generating new individual 3/10 Crossover.
    ...
    Generating new individual 98/100 Replication.
    Generating new individual 99/100 Mutation.
    Generating new individual 100/100 Mutation.
No pre-evaluated individuals to be removed.

To evaluate the population: mlc.evaluate_population;
>> % Evaluation of the new generation (Gen 2)
>> mlc.evaluate_population;
Evaluation of generation 2 :
    Evaluation of individual 1/100 already done (J= 32.039854)
    Evaluation of individual 2/100  J= 60.631446
    Evaluation of individual 3/100  J= 64.893979
    ...
\end{lstlisting}

Note the presence of the genetic operator that generated the new individual.
During the evaluation of the second generation, the individuals that have already been evaluated are not re-evaluated.
One can force the re-evaluation with the parameter \texttt{MultipleEvaluations}, see App.\ \ref{AppMyPlant}.
We could continue the optimization process by alternating the evolution and evaluation commands but there is a simpler way.
One can directly generate and evaluate the next generation with the \texttt{go} method and also directly indicate the final generation.
\begin{lstlisting}
>> % Generate and evaluate the next generation (Gen 3)
>> mlc.go;
Generation 3
    Generating new individual 1/100 Elitism.
    Generating new individual 2/100 Replication.
    Generating new individual 3/100 Crossover.
    ...
 Evaluation of generation 3 :
    Evaluation of individual 1/100 already done (J= 32.004032)
    Evaluation of individual 2/100 already done (J= 51.542092)
    Evaluation of individual 3/100  J= 53.512292
    ...
1 Generation(s) computed in 25.5563 seconds
 Cost of the best individual = 3.1794
 Its expression = 
    b = (tanh((s(1) - s(2))) - s(2))
\end{lstlisting}
  
When the \texttt{go} is employed it also gives back information on the best individual evaluated so far.
In this case, the best control law is \texttt{tanh(((s(1) - s(2)) - s(2)))} which after simplification corresponds to: $\tanh(a_1 - 2a_2 )$.
To continue the optimization until generation 10, run the command:
\begin{lstlisting}
>> % Generate and evaluate the next generation (Gen 3)
>> mlc.go(10);
    ...
Evaluation of individual 98/100 already done (J= 3.730651)
    Evaluation of individual 99/100  J= 5.254885
    Evaluation of individual 100/100  J= 22.423008
No bad individuals to be removed.

To create the next generation : mlc.evolve_population;
7 Generation(s) computed in 228.2561 seconds
 Cost of the best individual = 1.3074
 Its expression = 
    b = my_div(tanh(my_div(tanh(my_div(tanh((s(1) - s(2))),0.59655)),0.59655)),0.59655)
\end{lstlisting}

Once the process is over, the MLC object can be saved to continue the optimization or post-processing later.
\begin{lstlisting}
>> % Save the run
>> mlc.save_matlab;
\end{lstlisting}
 
For this example, the MLC object is saved in \texttt{save\_runs/AQuickTest/MLC\_Matlab.mat}.
To continue the optimization later or in a new MATLAB session, first create a MLC object then use the \texttt{load\_matlab} method with the name of the run.
\begin{lstlisting}
>> % To load an existing run on a new MATLAB session.
>> % First create a MLC object.
>> mlc=MLC;
>> % Then load the run with its name
>> mlc.load_matlab('AQuickTest');
\end{lstlisting}

For Octave, employ the corresponding methods \texttt{save\_octave} and \texttt{load\_octave}.
\xMLC also allows intermediate savings of the same MLC object, for more information see Sec.\ \ref{Sec:MatlabClass}.

 \subsubsection{Post-processing and analysis}
Once the  optimization process is done, the best individual can be accessed with the following method :
\begin{lstlisting}
>> mlc.best_individual;

Best individual after 10/10 generations (ID:3019):
   Its cost : 1.307
   Control law : my_div(tanh(my_div(tanh(my_div(tanh((s(1) - s(2))),0.59655)),0.59655)),0.59655) 
   Number of instructions (effective): 15 (7)
\end{lstlisting}

% Figure : Landau oscillator controlled
\begin{figure}[htb]
\centering
\subfloat[Phase space, actuation command, instantaneous cost function and radius for the controlled case.]{\label{Fig:BestCL}\includegraphics[width=0.45\textwidth]{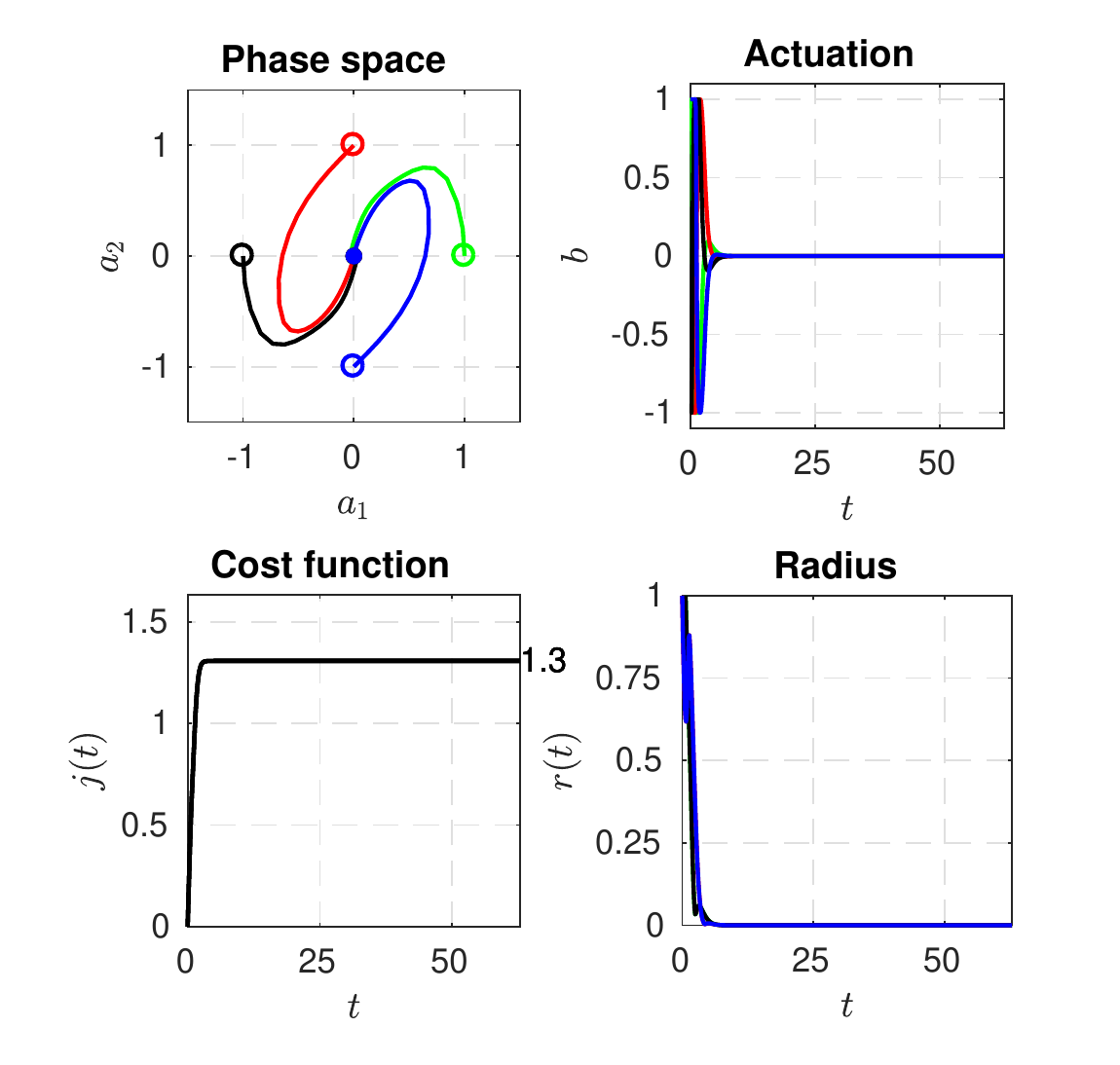}}%
\hfil
\subfloat[Visualization of the control law in the phase space.]{\label{Fig:BestCLVisu}\includegraphics[width=0.53\textwidth]{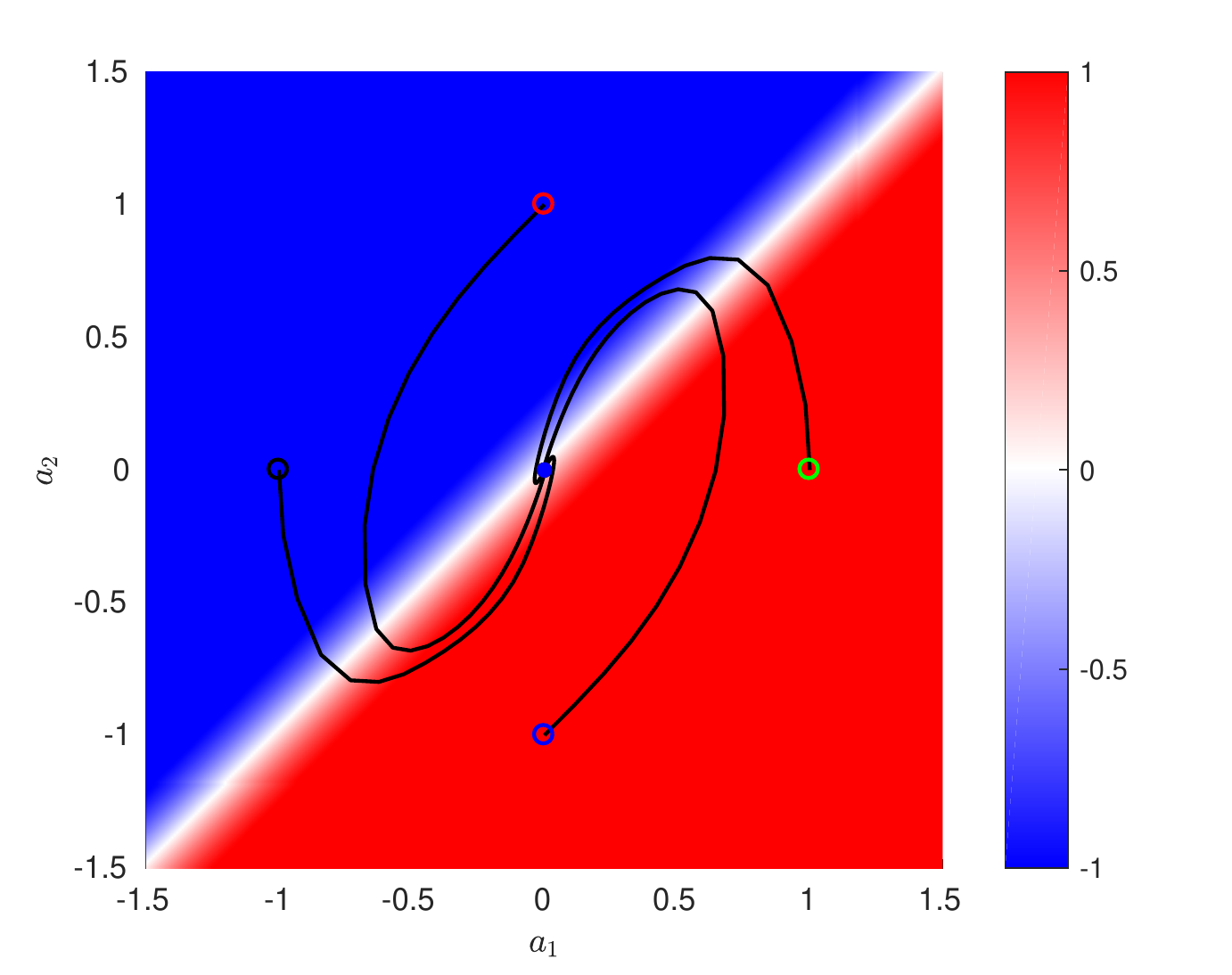}}%
\caption{\label{fig:BestControlLaw} Figures characterizing the control law optimized after 10 generations for the stabilization of the Landau oscillator.
The figures have been generated with the \texttt{best\_individual} method.}
\end{figure}
The output mentions that the best individual has been found in the $10^th$ generation over the 10 generations evaluated.
The ID number represents its index in the database.
The number of instructions corresponds to the number of lines in its matrix representation.
The number in parenthesis indicate that among the 15 lines of the matrix only 7 contribute to the final expression of the control law.
The \texttt{give} method also gives a figure displaying the performance of the best individual, see figure \ref{fig:BestControlLaw}.
The code for the figure is included in the \texttt{Plant/LandauOscillator/LandauOscillator\_problem.m} file.

For more information on the individual, run the command:
\begin{lstlisting}
>> mlc.give(3019);

3019-th Individual of the database:
   cost (mean over 1 evaluations): 1.307428
       1-st evaluation cost : 1.307428
   occurrences: 1
   control law:
      b1 = my_div(tanh(my_div(tanh(my_div(tanh((s(1) - s(2))),0.59655)),0.59655)),0.59655)
   ref:0
   
>> % Check the informations of the individual in the database.
>> % The matrix representation.
>> mlc.table.individuals(3019).chromosome;
>> % Cost of the individual: J, Ja, Jb.
>> mlc.table.individuals(3019).cost;
>> % Control law.
>> mlc.table.individuals(3019).control law;
\end{lstlisting}

% Figure : Landau oscillator controlled
\begin{figure}[htb]
\centering
\subfloat[]{\label{Fig:LearningProcess}\includegraphics[width=0.475\textwidth]{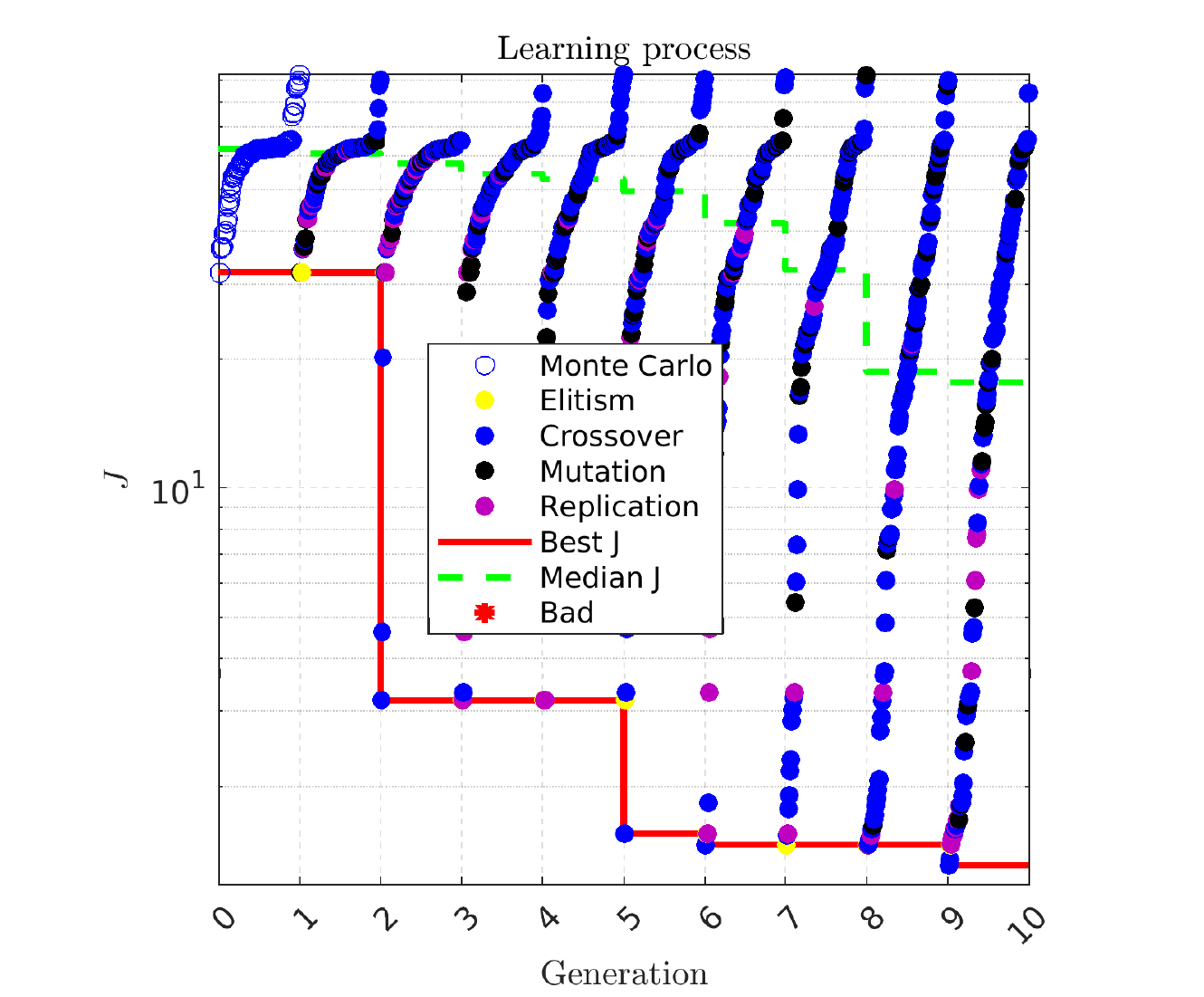}}%
\hfil
\subfloat[]{\label{Fig:ParetoDiagram}\includegraphics[width=0.475\textwidth]{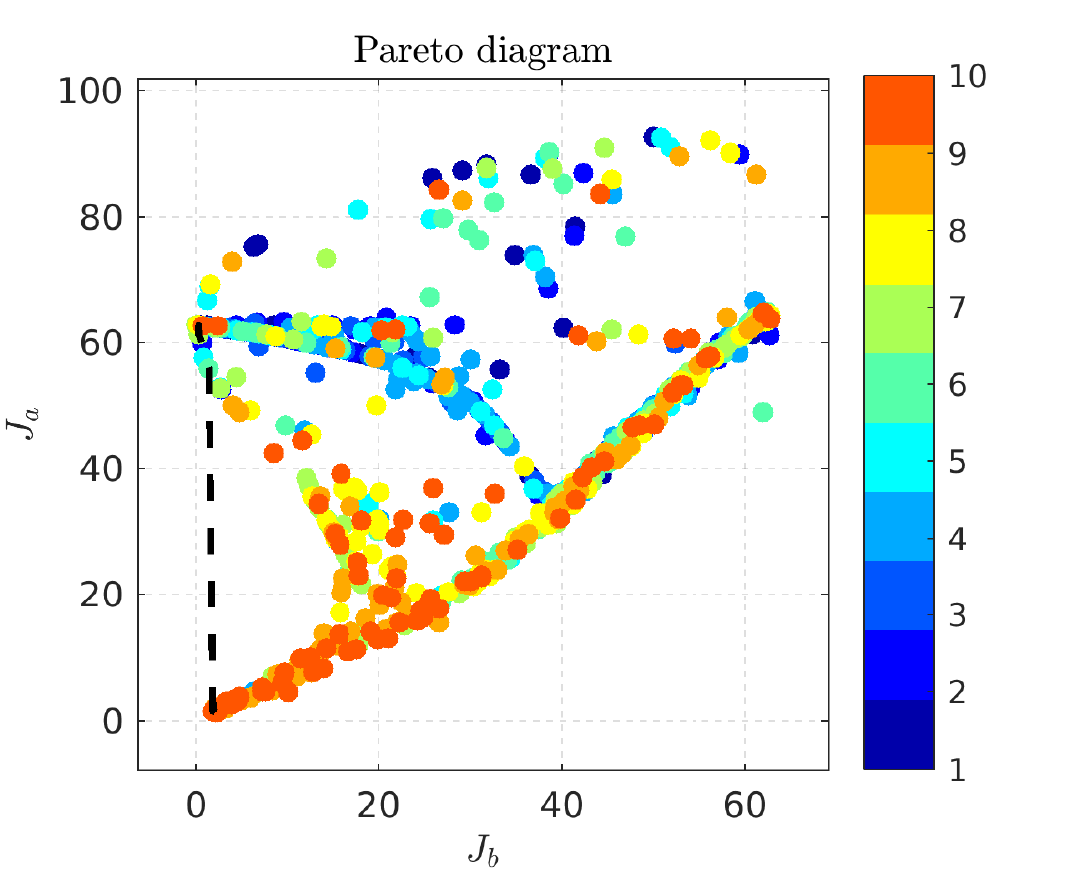}}%

\subfloat[]{\label{Fig:CostDistribution}\includegraphics[width=0.475\textwidth]{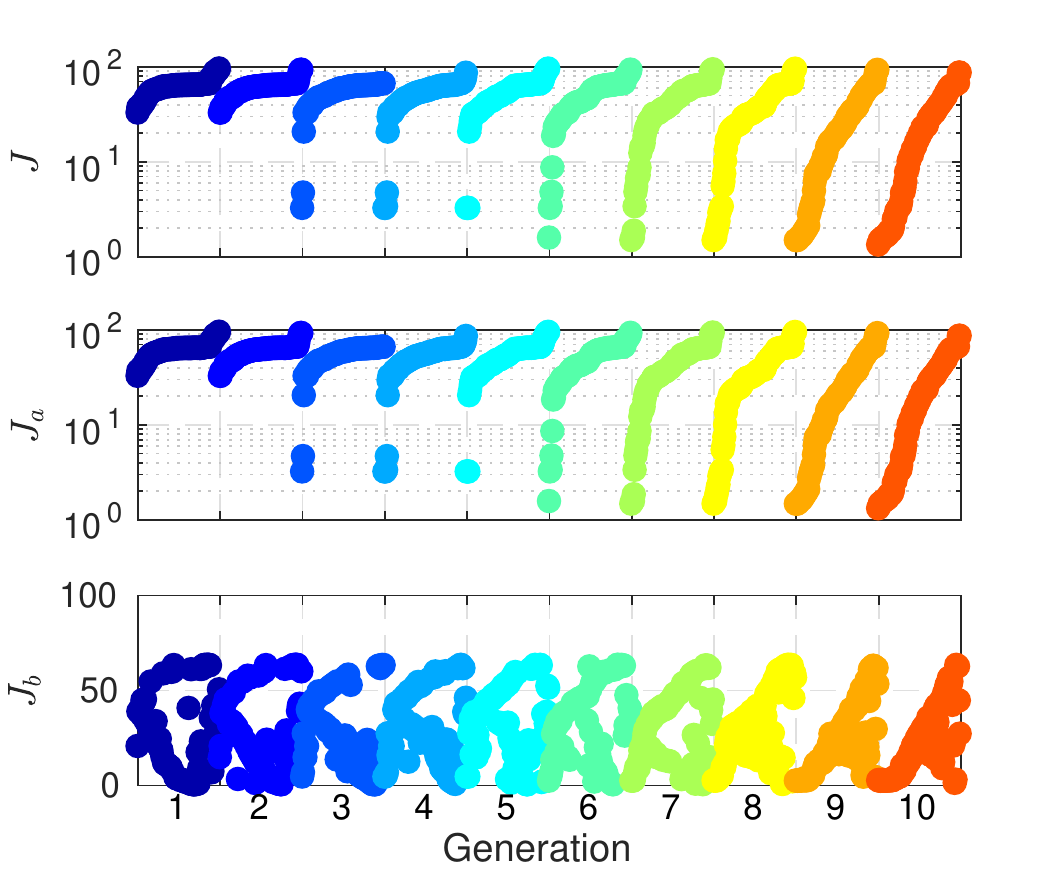}}%
\hfil
\subfloat[]{\label{Fig:Spectrogram}\includegraphics[width=0.475\textwidth]{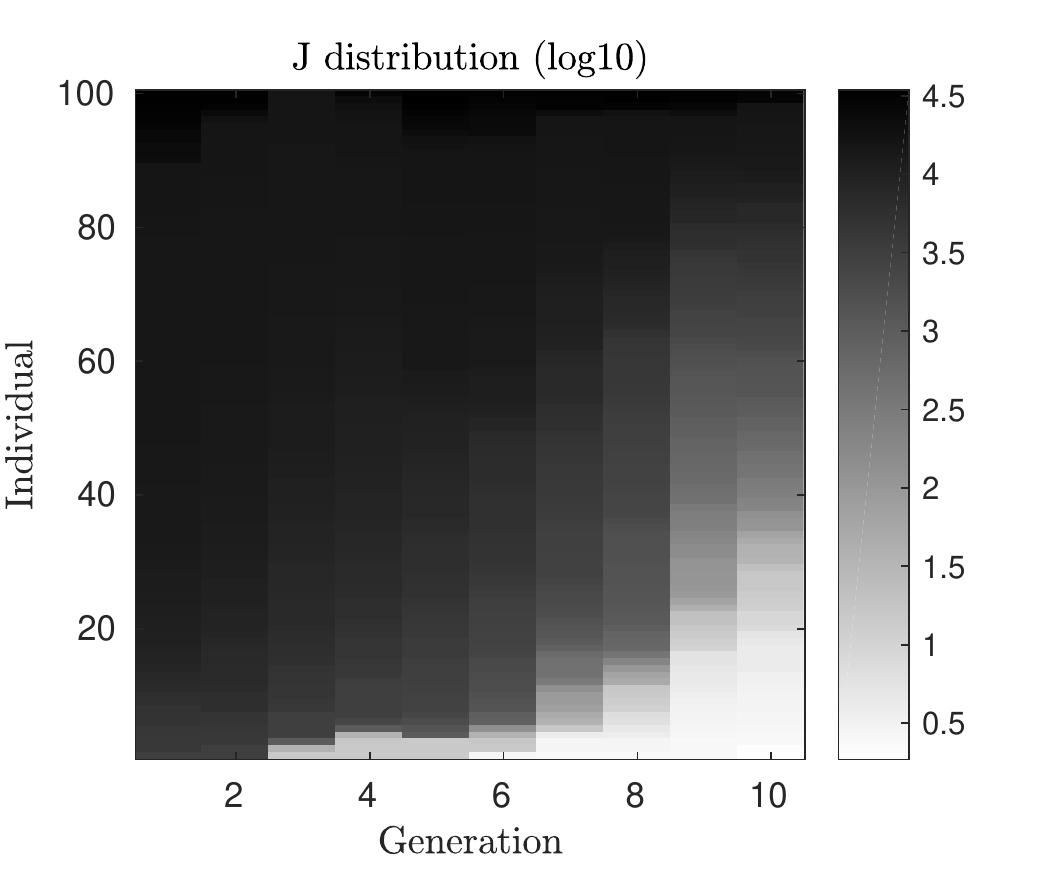}}%
\caption{Figures characterizing the optimization process and displaying the distribution of the control laws evaluated. Figure (b) and (c) share the same color code for the generations. The figures have been generated with specific methods: (a) \texttt{learning\_process}, (b) \texttt{Pareto\_diagram}, (c) \texttt{cost\_distribution} and (d) \texttt{spectrogram}.}
\end{figure}
Of course these commands can be executed for any ID in the database.
Note that the matrix is referred as chromosome in \xMLC to follow the evolutionary terminology.
Other features can be extracted such as the learning process (see figure~\ref{Fig:LearningProcess}), the Pareto diagram (see figure~\ref{Fig:ParetoDiagram}), the cost distribution (see figure~\ref{Fig:CostDistribution}) and the spectrogram (see figure~\ref{Fig:Spectrogram}) thanks to the following commands: 
\begin{lstlisting}
>> % Plot of the learning process.
>> mlc.learning_process;
>> % Plot of the Pareto diagram and front for the first components of the cost function.
>> mlc.Pareto_diagram;
>> % Plot of the cost distribution sorted by J.
>> mlc.cost_distribution;
>> % Plot of the distribution of costs in each generation.
>> mlc.spectrogram;
\end{lstlisting}

\section{Code description}\label{Sec:MatlabClass}
In this section, we give a brief description of the \texttt{xMLC}.
First, we detail the content of the code then we give an overview of the MLC class and its properties.
Finally, we provide a list of useful commands for the analysis and extraction of information from the MLC object.

\subsection{Content}
\begin{figure}[htb]
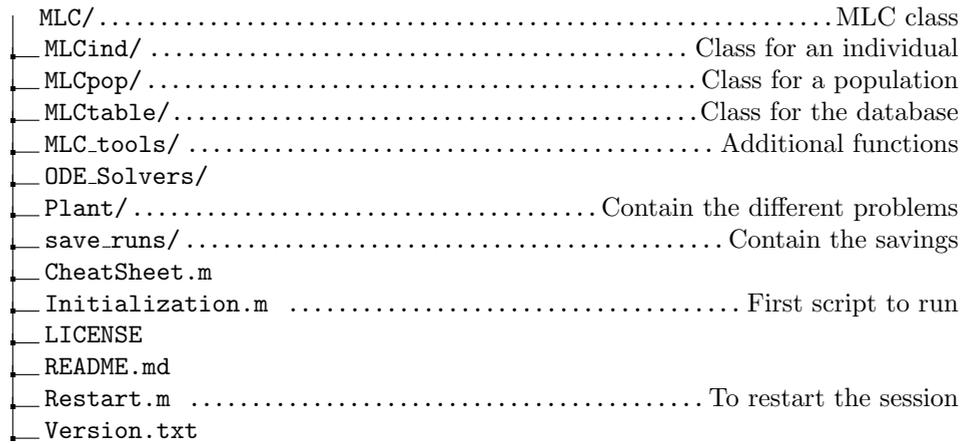

      \centering
        \begin{minipage}[b]{0.8\linewidth}
          \dirtree{%
            .1 \@MLC/\DTcomment{MLC class} .
            .1 \@MLCind/\DTcomment{Class for an individual} .
            .1 \@MLCpop/\DTcomment{Class for a population} .
            .1 \@MLCtable/\DTcomment{Class for the database} .
            .1 MLC\_tools/\DTcomment{Additional functions} .
            .1 ODE\_Solvers/ .
            .1 Plant/\DTcomment{Contain the different problems} .
            .1 save\_runs/\DTcomment{Contain the savings} .
            .1 CheatSheet.m .
            .1 Initialization.m \DTcomment{First script to run} .
            .1 LICENSE .
            .1 README.md .
            .1 Restart.m \DTcomment{To restart the session} .
            .1 Version.txt .
          }
      \end{minipage}
      \caption{File structure for the \texttt{xMLC/} folder.\label{fig:xMLCFolder}}
\end{figure}
Figure~\ref{fig:xMLCFolder} displays the content of the main folder \texttt{xMLC/}.
In the following, we give a description of each element of the folder.
\begin{itemize}
\item The folder \texttt{@MLC/} containing the file \texttt{MLC.m} that defines the MLC class with its properties and methods;
\item The folders, \texttt{@MLCind/}, \texttt{@MLCpop/}, \texttt{@MLCtable/}, defining three other classes that are employed in the MLC class: MLCind, MLCpop and MLCtable define an individual, a population and a database respectively;
\item The folder \texttt{MLCtools/} containing additional functions employed during the optimization;
\item The folder \texttt{ODE\_Solvers/} containing implementations of ODE solvers in particular Runge-Kutta methods from order 1 to 5;
\begin{figure}[htb]
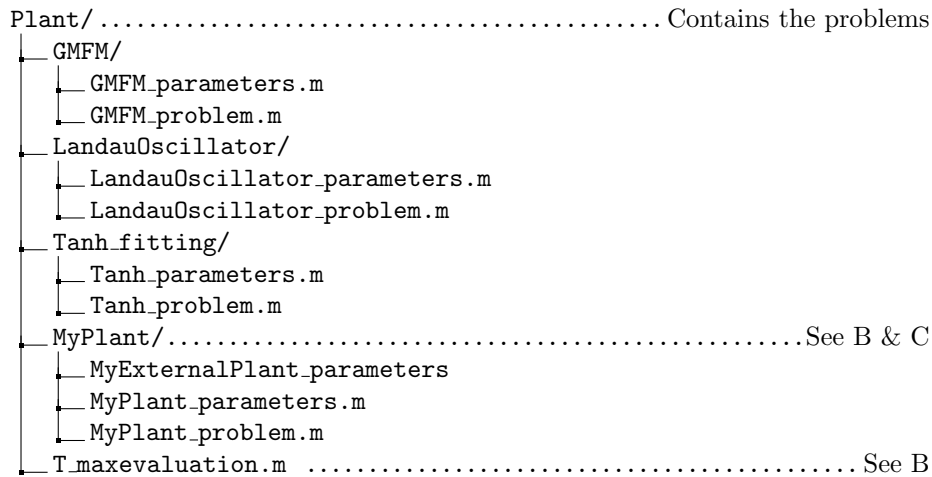

      \centering
        \begin{minipage}[b]{0.8\linewidth}
          \dirtree{%
            .1 Plant/\DTcomment{Contains the problems} .
              .2 GMFM/ .
               .3 GMFM\_parameters.m .
               .3 GMFM\_problem.m . 
              .2 LandauOscillator/ .
               .3 LandauOscillator\_parameters.m .
               .3 LandauOscillator\_problem.m .
              .2 Tanh\_fitting/ .
               .3 Tanh\_parameters.m .
               .3 Tanh\_problem.m .
              .2 MyPlant/\DTcomment{See\ \ref{AppMyPlant} \& \ref{AppInterface}} .
               .3 MyExternalPlant\_parameters .
               .3 MyPlant\_parameters.m .
               .3 MyPlant\_problem.m .
              .2 T\_maxevaluation.m \DTcomment{See\ \ref{AppMyPlant}} .
          }
      \end{minipage}
      \caption{File structure for the \texttt{Plant/} folder.\label{fig:PlantFolder}}
\end{figure}
\item The folder \texttt{Plant/} containing different parameter folders for each problem. Three toy problems are included, the stabilization of the Landau oscillator, the stabilization of the generalized mean-field model and the fitting of a $\tanh$ function (see figure~\ref{fig:PlantFolder}.
Each folder contains a parameter file and problem file.
The parameter file contains all the \xMLC parameters and can also include parameters for the problem.
The `problem' represents the plant, it includes the resolution of the corresponding dynamical.
This file is needed only when problems are solved by MATLAB.
It is not needed when the user employs an external solver or experiment.
To create a personalized problem, see App.\ \ref{AppMyPlant};
\item The folder \texttt{save\_runs/} containing the saving files and the outputs of the code. This folder does not exist on the original code but is created automatically if needed;
\item The file \texttt{CheatSheet.m} containing the main commands to start an optimization process, see Sec.\ \ref{Sec:QuickStart};
\item The file \texttt{Initialization.m} that loads all the paths necessary for the code. This script should be used at the beginning of each session;
\item The file \texttt{LICENSE} contains the license of the \xMLC code: \emph{The MIT License (MIT)}.
\item The file \texttt{README.md} gives a short description of the code along with the main commands to start an optimization process.
This file keeps also track of the updates/upgrades along the versions.
If you encounter any compatibility problem, contact the author at \href{mailto:Yoslan@hit.edu.cn}{Yoslan@hit.edu.cn};
\item The file \texttt{Restart.m} that clears all the variables and creates a new MLC object with the default parameters;
\item The file \texttt{Version.txt} containing the version of the code. The version should be communicated in case of contact with the author.
\end{itemize}

\subsection{Description of the MLC class}
Once the \texttt{Initialization.m} script is launched, a small description of the problem to be solved is printed.
It contains information about the number of inputs (controllers), the number of outputs (sensors), the population size and the strategy (genetic operators probabilities).
This is shown each time a new instance of a MLC class object is created.
To show it again use the \texttt{show\_problem} method by using the command : \texttt{mlc.show\_problem;}.
The \texttt{MLC} object  has 6 properties:
\begin{itemize}
\begin{figure}[htb]
      \centering
        \begin{minipage}[b]{0.8\linewidth}
          \dirtree{%
            .1 individuals \DTcomment{List of ID numbers sorted by their cost} .
            .1 costs \DTcomment{Cost of each individual} .
            .1 chromosome\_lengths \DTcomment{Number of instructions and effective instructions} .
            .1 parents \DTcomment{ID number of the parents} .
            .1 operation \DTcomment{Operation that generated the individual} .
            .1 generation .
            .1 evaluation \DTcomment{Status of the population: evaluated or not} .
            .1 CreationOrder \DTcomment{Order of creation of the individuals} .
          }
      \end{minipage}
      \caption{Properties of the MLCpop class.\label{fig:PopulationStructure}}
\end{figure}
\item \texttt{population} is an array of MLCpop.
Each one of then containing information of all the individuals of the a given generation.
In particular, the `ID' corresponds to the index of the individual in the database.
Note that the ID refers to a unique matrix representation.
Thus, two control laws that simplify to the same expression will have different ID numbers.
However, one of them may be replaced during the optimization process.
Figure~\ref{fig:PopulationStructure} gives more information on the properties of each population;
\item \texttt{parameters} is a structure object containing all the parameters for a problem. It contains in particular where all the parameters are defined, be it for the problem, the control law description or the MLC parameters; 
\begin{figure}[hbt]
      \centering
        \begin{minipage}[b]{0.8\linewidth}
          \dirtree{%
            .1 individuals \DTcomment{Array of MLCind objects} .
            .1 non\_redundant \DTcomment{ID list of all `unique' individuals} .
            .1 number \DTcomment{Total number of individuals explored} .
            .1 control\_points \DTcomment{See Sec.\ \ref{AppMyPlant}} .
            .1 costlist \DTcomment{List of costs for all individuals explored} .
          }
      \end{minipage}
      \caption{Properties of the MLCtable class.\label{fig:TableStructure}}
\end{figure}
\begin{figure}[!h]
      \centering
        \begin{minipage}[b]{0.8\linewidth}
          \dirtree{%
            .1 chromosome \DTcomment{Matrix representation of the individual} .
            .1 costs \DTcomment{Array of cell containing the cost $J$ and its components for all evaluations} .
            .1  control\_law \DTcomment{Array of cell containing the control law components} .
            .1 EI \DTcomment{Information on the effective instructions} .
            .1 occurrences .
            .1 evaluation\_time .
            .1 control\_points \DTcomment{See Sec\ \ref{AppMyPlant}} .
            .1 ref \DTcomment{ID of the reference individual if redundant individual} .
          }
      \end{minipage}
      \caption{Properties of the MLCind class.\label{fig:IndStructure}}
\end{figure}
\item \texttt{table} is the database, it contains all the individuals explored. This number may be superior to the number of individuals evaluated as some of them are removed due to failed evaluations or screening options.
Figure~\ref{fig:TableStructure} and \ref{fig:IndStructure} details the content of the MLCtable and MLCind classes respectively;
\item \texttt{generation} display the current generation.
It is set to $0$ by default when the population is empty.
One generation is counted only if it has been evaluated;
\item \texttt{version} contains the version of the MLC object. The structure of the parameter file may be modified with the new versions thus keeping track of the MLC object version is important.
\end{itemize}

\section{Handy commands}
We list now useful commands to extract individual information or the analyze the optimization process.
\begin{lstlisting}
>> % Information on the 5 best individuals evaluated so far.
>> % If an individual is evaluated several times (in case of stochastic problem),
>> % the command gives the best individuals based on an estimated performance.
>> % See the EstimatePerformance parameter in App. A.
>> mlc.list_best_individuals;
>> mlc.list_best_individuals(GEN); % Same as the previous command but up to generation GEN.
>> mlc.list_best_individuals(GEN,NI); % Same as the previous command but gives only the NI best individuals.

>> % Information on individual ID:
>> mlc.give(ID);

>> % Re-evaluation of individual ID and plot the control if possible.
>> % Only works for problems evaluated with MATLAB.
>> mlc.plotindiv(ID);

>> % Intermediate save of a run name AQuickTest:
>> mlc.save_matlab('Gen2');
>> % The saving file is then stored in save_runs/AQuickTest/Gen2_Matlab.mat
>> % Load an intermediate save of a run named AQuickTest:
>> mlc.load_matlab('AQuickTest','Gen2');

>> % Information on the sensor, constant and operation distribution for generation GEN.
>> mlc.CL_descriptions(GEN);

>> Distribution of genetic operators through the generations.
>> mlc.genoperatorsdistrib;

>> Relationship matrix between the parents and the offsprings.
>> mlc.relationship;

>> % Print the current figure.
>> % The figure is saved in the save_runs/AQuickTest/Figures/ folder in PNG and EPS format.
>> mlc.printfigure('AQuickTest')	
>> mlc.printfigure('AQuickTest',1) % To overwrite an existing figure.

>> % To initiate a problem different from the default one.
>> % For example, to create a MLC object wit the GMFM problem:
>> mlc=MLC('GMFM');

>> % One can also rename a run.
>> mlc.rename('AFastTest');
AQuickTest changed to AFastTest.

>> % If one's wish to duplicate a run, just copy-paste the run folder in save_runs/ and change it's name.
>> % The MLC object can then be loaded with the new name.
>> % The name in the parameters will be automatically updated.
\end{lstlisting}

\chapter{Example: Net drag power reduction of the fluidic pinball}\label{Cha:Example}

In this chapter, we reduce the net drag power of the fluidic pinball thanks to \xMLC software presented in Sec.\ \ref{Sec:UG}.
For this study, we explore three different search spaces: first we look for a multi-frequency forcing controller, second we look for feedback control laws and finally we investigate an hybrid search space comprising periodic functions and sensor signals.
Thus, in Sec.\ \ref{sec:fluidic_pinball_plant} we present the fluidic pinball and the regression problem.
In Sec.\ \ref{sec:pinball_openloop}, we detail an open-loop control study of the fluidic pinball for a control reference.
Lastly, we apply MLC to minimize the net drag power with multi-frequency forcing (Sec.\ \ref{sec:MF_optimization}), feedback control (Sec.\ \ref{sec:FB_optimization}) and a search space allowing both strategies (Sec.\ \ref{sec:HB_optimization}).

\section[The fluidic pinball]{The fluidic pinball---A benchmark flow control problem}
\label{sec:fluidic_pinball_plant}
In this section,
we describe the fluid system studied for the control optimization---the fluidic pinball.
First we present the fluidic pinball configuration and the unsteady 2D Navier-Stokes solver in Sec.\ \ref{sec:config_numerical_solver},
then the unforced flow spatio-temporal dynamics in Sec.\ \ref{sec:flow_characteristics}
and finally the control problem for the fluidic pinball in Sec.\ \ref{sec:control_objective}.
This section is largely inspired from section 2 of \citet{CornejoMaceda2021PhD}.

\subsection{Configuration and numerical solver}
\label{sec:config_numerical_solver}
The test case is a two-dimensional uniform flow past a cluster of three cylinders of same diameter $D$.
The center of the cylinders form an equilateral triangle pointing upstream.
The flow is controlled by the independent rotation of the cylinders along their axis.
The rotation of the cylinders enables the steering of incoming fluid particles,
like a pinball machine.
Thus, we refer this configuration as the fluidic pinball.
In our study, we choose the side length of the equilateral triangle equal to be $1.5D$.
Various side lengths have been explored numerically in \citep{ChenAlam2020jfm}, revealing a myriad of interesting regimes.

The flow is described in a Cartesian coordinate system,
where the origin is located midway between the two rearward cylinders.
The $x$-axis is parallel to the stream-wise direction.
The $y$-axis is orthogonal to the cylinder axis.
The velocity field is denoted by $\boldsymbol{u}=(u,v)$ and the pressure field by $p$.
Here, $u$ and $v$ are, respectively, the stream-wise and transverse components of the velocity.
We consider a Newtonian fluid of constant density $\rho$ and kinematic viscosity $\nu$.
For the direct numerical simulation,
the unsteady incompressible viscous Navier-Stokes equations
are non-dimensionalized with cylinder diameter $D$,
the incoming velocity $U_{\infty}$ and the fluid density $\rho$.
The corresponding Reynolds number is $\Rey_{D} = U_{\infty}D\slash \nu$.
Throughout this study, only $\Rey_D =100$ is considered.

The computational domain $\Omega$ is a rectangle bounded
by $[-6,20]\times[-6,6]$ excluding the interior of the cylinders:
\begin{equation*}
\Omega = \{[x,y]^{\rm T} \in \mathcal{R}^2 \colon [x,y]^{\rm T} \in [-6,20]\times[-6,6] \land (x-x_i)^2+(y-y_i)^2 \geq 1/4, i=1, 2, 3 \}.
\end{equation*}
Here,
$[x_i,y_i]^{\rm T}$ with $i=1,2,3$, are the coordinates of the cylinder centers,
starting from the front cylinder and numbered in mathematically positive direction,
\begin{equation*}
\begin{array}{clccl}
x_1 = & -3/2\cos(30^{\circ}) & &y_1= & ~0,\\
x_2 = & 0 & &y_2= & -3/4,\\
x_3 = & 0 & &y_3= & \quad 3/4.
\end{array}
\end{equation*}
The computational domain $\Omega$ is discretized on an unstructured grid comprising 4225 triangles and 8633 nodes.
The grid is optimized to provide a balance between computation speed and accuracy.
Grid independence of the direct Navier-Stokes solutions
has been established by \citet{Deng2020jfm}.

The boundary conditions for the inflow, upper and lower boundaries are $U_{\infty}=\boldsymbol{e}_x$ while a stress-free condition is assumed for the outflow boundary.
The control of the fluidic pinball is carried out by the rotation of the cylinders.
A non-slip condition is adopted on the cylinders:
the flow adopts the circumferential velocities of the front, bottom and top cylinder
specified by  $b_1 = U_F$, $b_2 = U_B$ and $b_3 = U_T$.
The actuation command comprises these velocities, $\boldsymbol{b} = [b_1,b_2,b_3]^{\rm T}$.
A positive (negative) value of the actuation command
corresponds to counter-clockwise (clockwise) rotation of the cylinders along their axis.
The numerical integration of the Navier-Stokes equations is carried by an in-house solver using a fully implicit Finite-Element Method \citep{Noack2003jfm, Noack2016jfm}.
The method is  third order accurate in time and space.
%% Figure : Snapshots---------------------------------------------------
\begin{figure}[htb]
\centering
\subfloat[Symmetric steady solution.]{\label{fig:steady_solution}\includegraphics[width=0.45\textwidth]{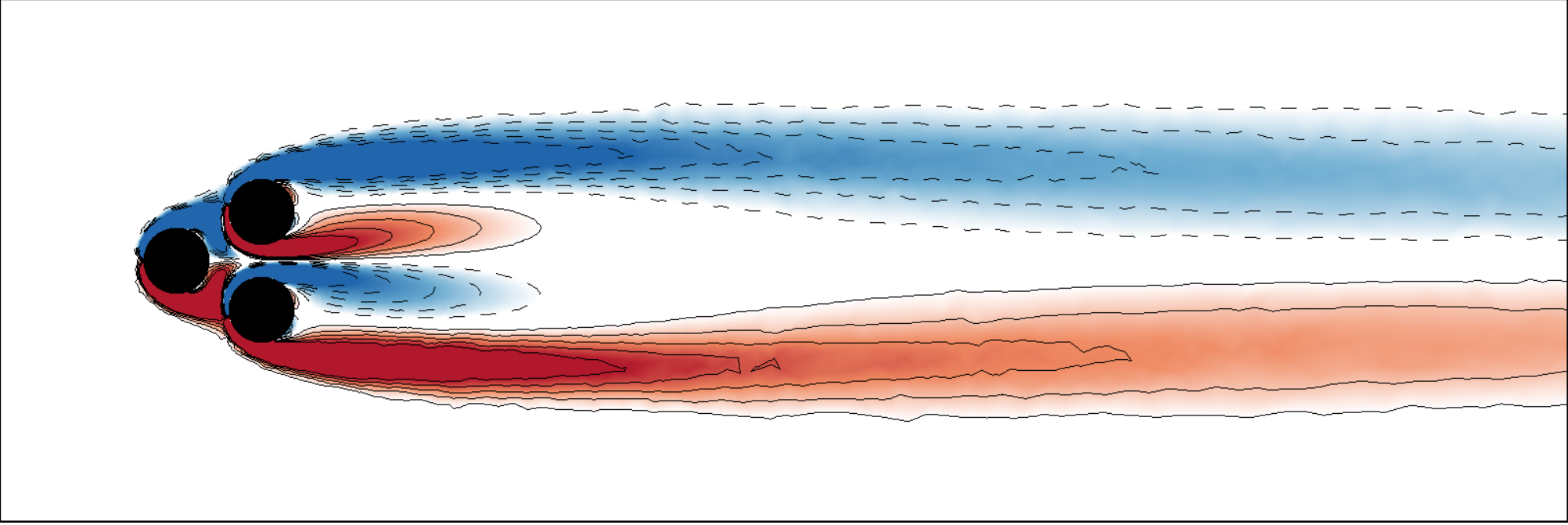}}%
\hfil
\subfloat[Unforced flow at $t=400$.]{\label{fig:natural_flow}\includegraphics[width=0.45\textwidth]{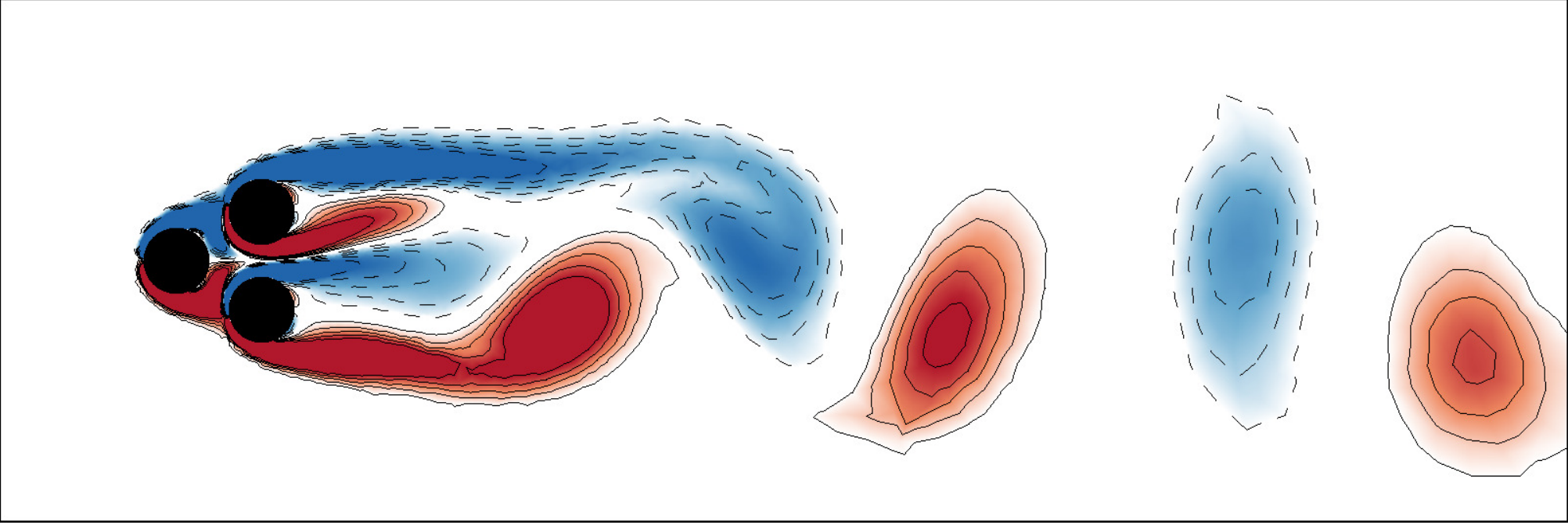}}%
\caption{\label{fig:unforced_flow}
Vorticity fields for the unforced fluidic pinball at $\Rey_D=100$.
Blue (red) regions bounded by dashed lines represent negative (positive) vorticity.
Darker regions indicate higher values of vorticity magnitude.
Figures reprinted from \citet{Cornejo2021jfm}.
}
\end{figure}
The initial condition for the numerical simulations is the symmetric steady solution, see figure~\ref{fig:steady_solution}.
The symmetrical steady solution is computed with a Newton-Raphson method on the steady Navier-Stokes.
An initial short, small rotation of the front cylinder is used to kick-start the transient to natural vortex shedding in the first period \citep{Deng2020jfm}.
The transient regime lasts around 400 convective time units.
Figure \ref{fig:unforced_flow} shows the vorticity field for the symmetric steady solution and the natural unforced flow after 400 convective units.
The snapshot at $t=400$ in figure \ref{fig:natural_flow} is the initial condition for all the following simulations.

% --- Description of the unforced flow
\subsection{Unforced reference}\label{sec:flow_characteristics}
%% Figures : Natural flow characteristics ------------------------------
\begin{figure}[htb]
\centering
\subfloat[]{\label{fig:CL_natural}\includegraphics[trim=0 14.25pt -20pt 30pt, clip,width=0.4\textwidth]{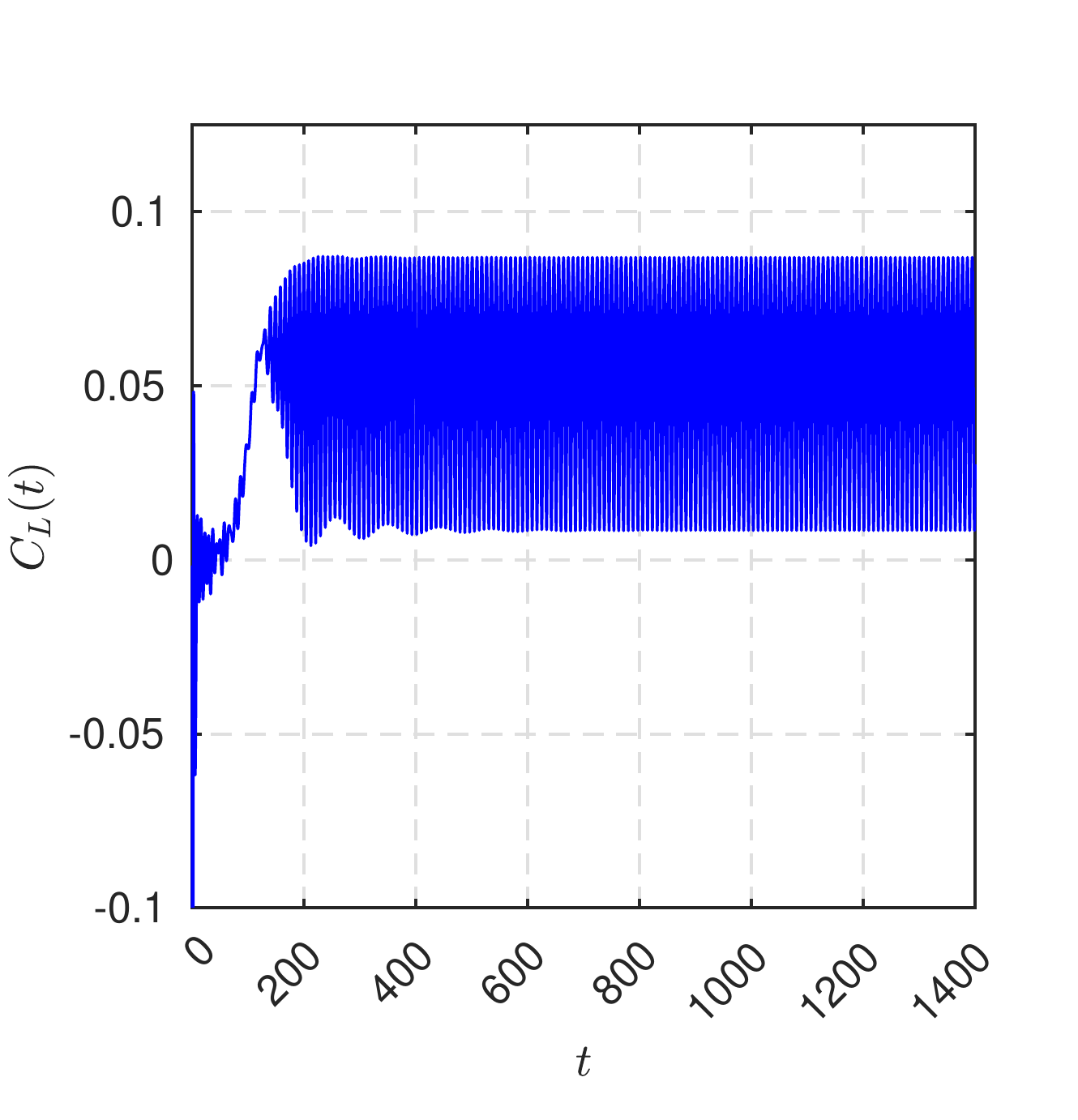}}%
\hfil
\subfloat[]{\label{fig:PP_natural}\includegraphics[trim=0 14.25pt -20pt 30pt, clip,width=0.4\textwidth]{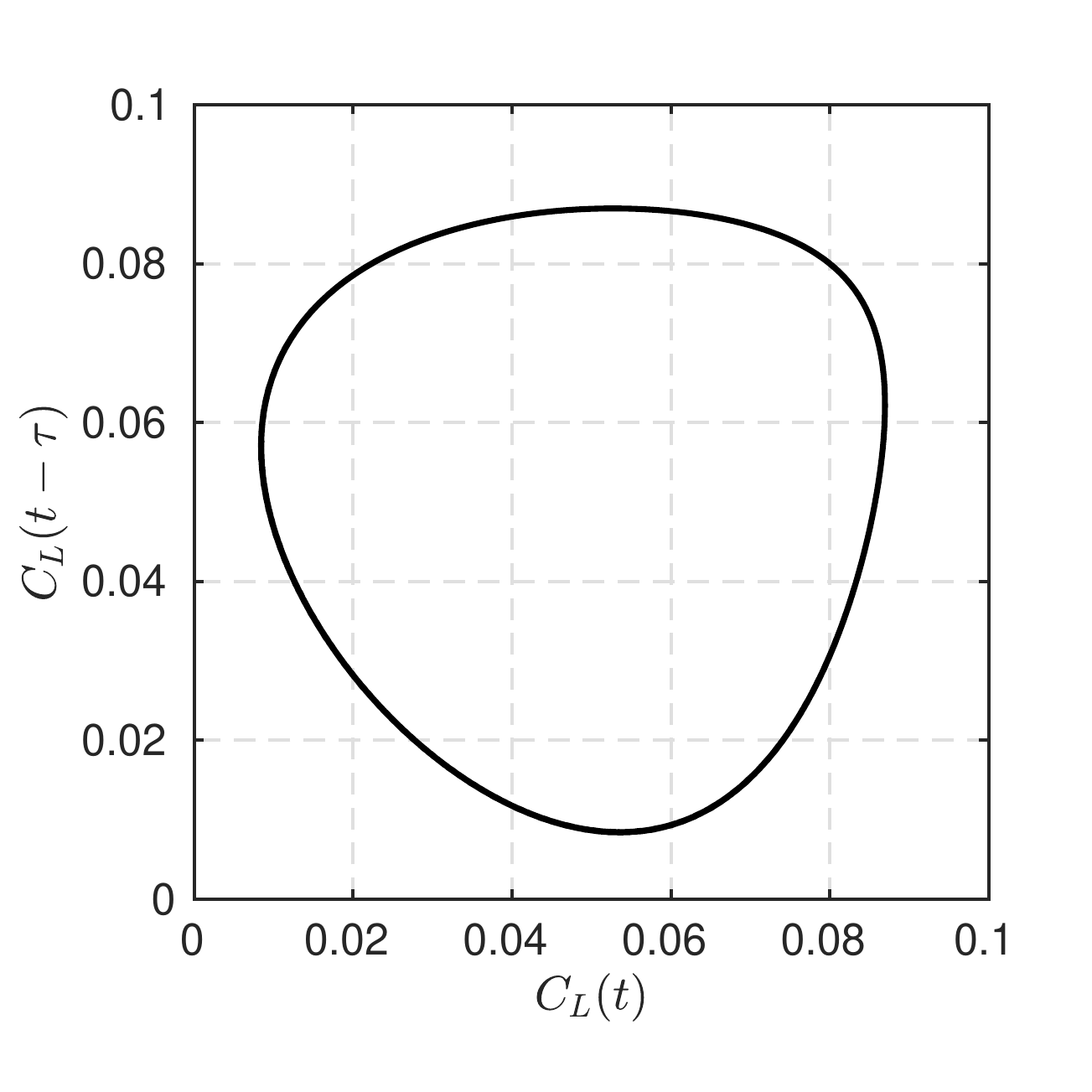}}%

\subfloat[]{\label{fig:DragPower_natural}\includegraphics[trim=0 14.25pt -20pt 30pt, clip,width=0.4\textwidth]{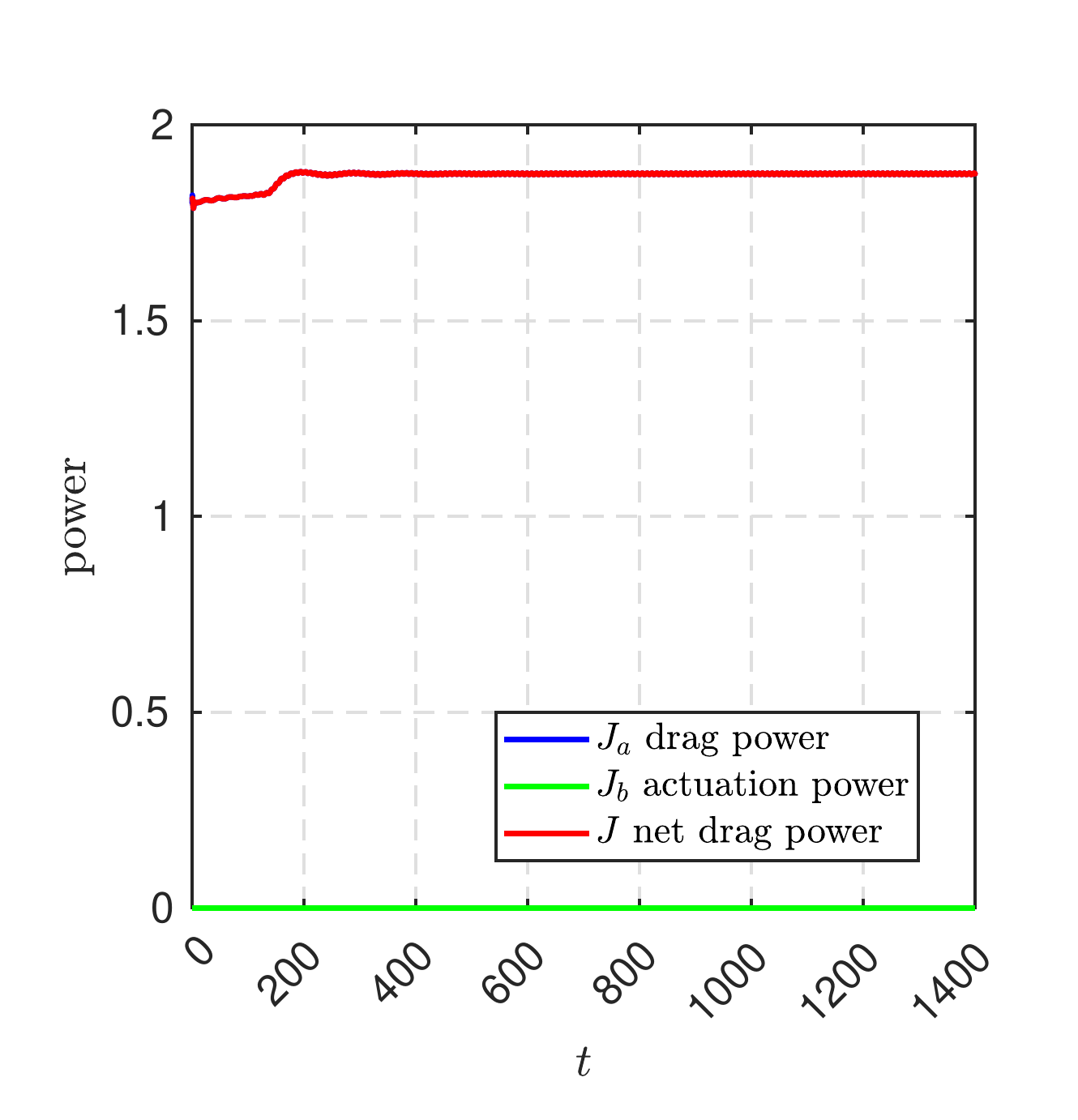}}%
\hfil
\subfloat[]{\label{fig:PSD_natural}\includegraphics[trim=0 14.25pt -20pt 30pt, clip,width=0.4\textwidth]{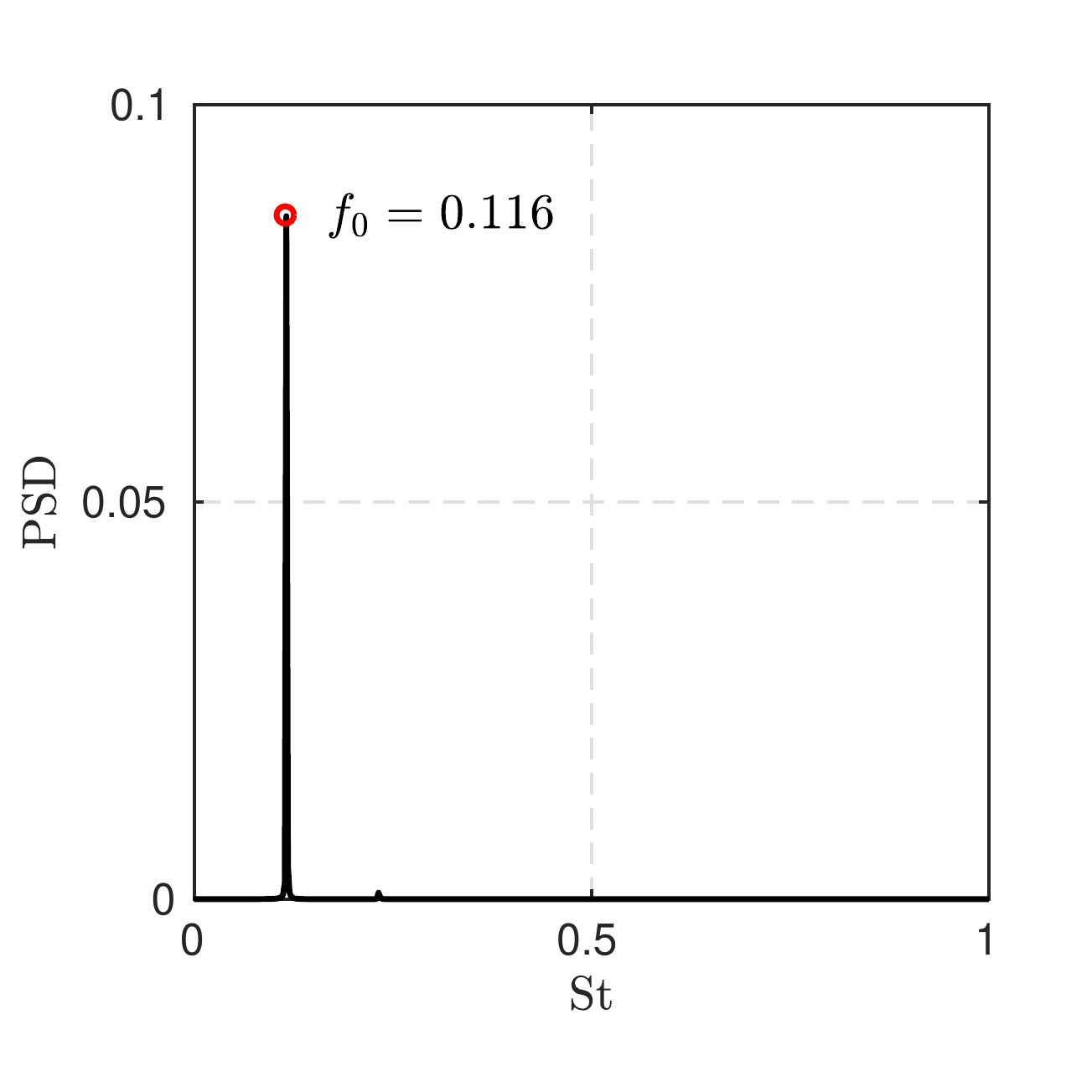}}%
\caption{\label{fig:natural_characteristics}
Characteristics of the unforced natural flow starting from the steady solution ($t=0$).
The transient spans until $t \approx 400$. (a) Time evolution of the lift coefficient $C_L$, (b) phase portrait, (c)  time evolution of the drag power $J_a$ (blue), actuation power $J_b$ (green) and net drag power $J$ (red) and (d) Power Spectral Density (PSD) showing the natural frequency $f_0=0.116$. The phase portrait is computed during the post-transient regime $t \in [900,1400]$ and the PSD is computed over the last 1000 convective time units, $t \in [400,1400]$.
}
\end{figure}
The fluidic pinball is a geometrically simple configuration
that comprises key features of real-life flows
such as successive bifurcations and frequency crosstalk between modes.
\citet{Deng2020jfm} shows that the unforced fluidic pinball undergoes successive bifurcations with increasing Reynolds number before reaching a chaotic regime.
The first Hopf bifurcation at Reynolds number $\Rey\approx 18$
breaks the symmetry in the flow and initiates the von K\'arm\'an vortex shedding.
The second bifurcation at Reynolds number $\Rey \approx 68$
is of pitchfork type and gives rise to a transverse deflection of jet-like flow appearing
 between the two rearward cylinders.
The bi-stability of the jet deflection has been reported by \citet{Deng2020jfm}.
At a Reynolds number $\Rey = 100$
the jet deflection is rapid and occurs before the vortex shedding is fully established.
Figure~\ref{fig:CL_natural} shows an increase of the lift coefficient $C_L$
before  oscillations set in and the lift coefficient converges
against a periodic oscillation around a slightly reduced mean value.
Those bifurcations are a consequence of  multiple instabilities
present in the flow: there are two shear instabilities,
on the top and bottom cylinder and a jet bi-stability originating from the gap between the two back cylinders.
The shear-layer instabilities synchronize to a von K\'arm\'an vortex shedding.
% Natural flow snapshots ---------------------------------------------
\begin{figure}[htb]%
\centering
\subfloat[$t+T_0/8$]{\label{fig:nat_T1}\includegraphics[width=0.45\textwidth]{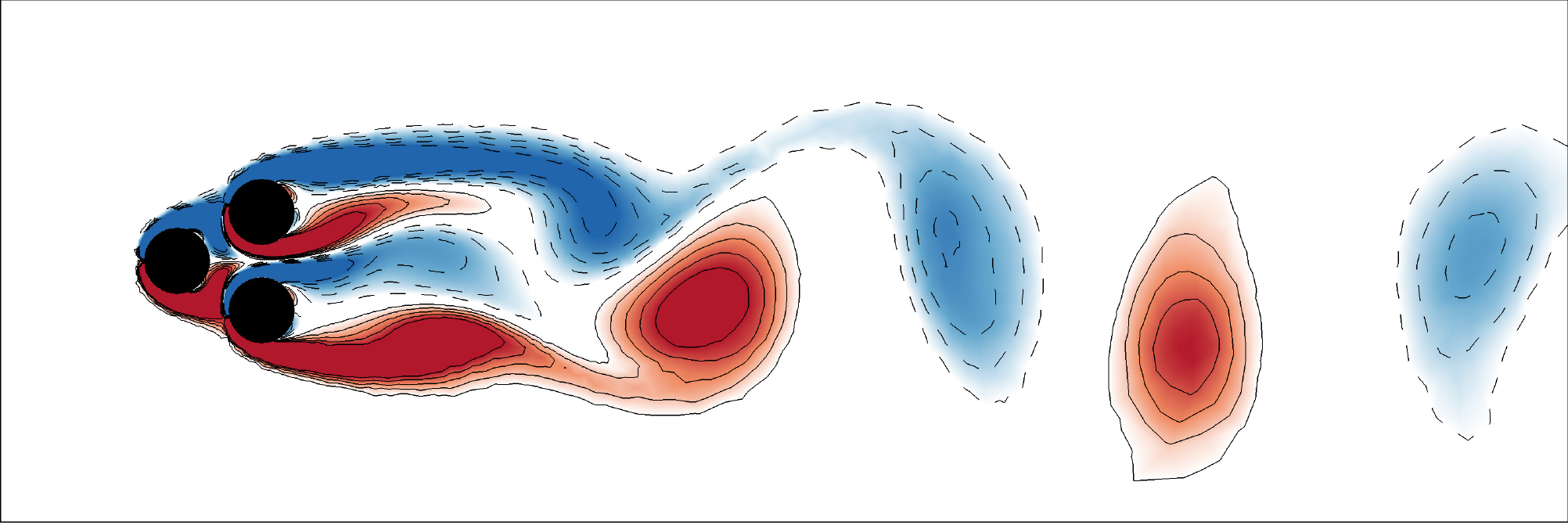}}%
\hfil
\subfloat[$t+2T_0/8$]{\label{fig:nat_T2}\includegraphics[width=0.45\textwidth]{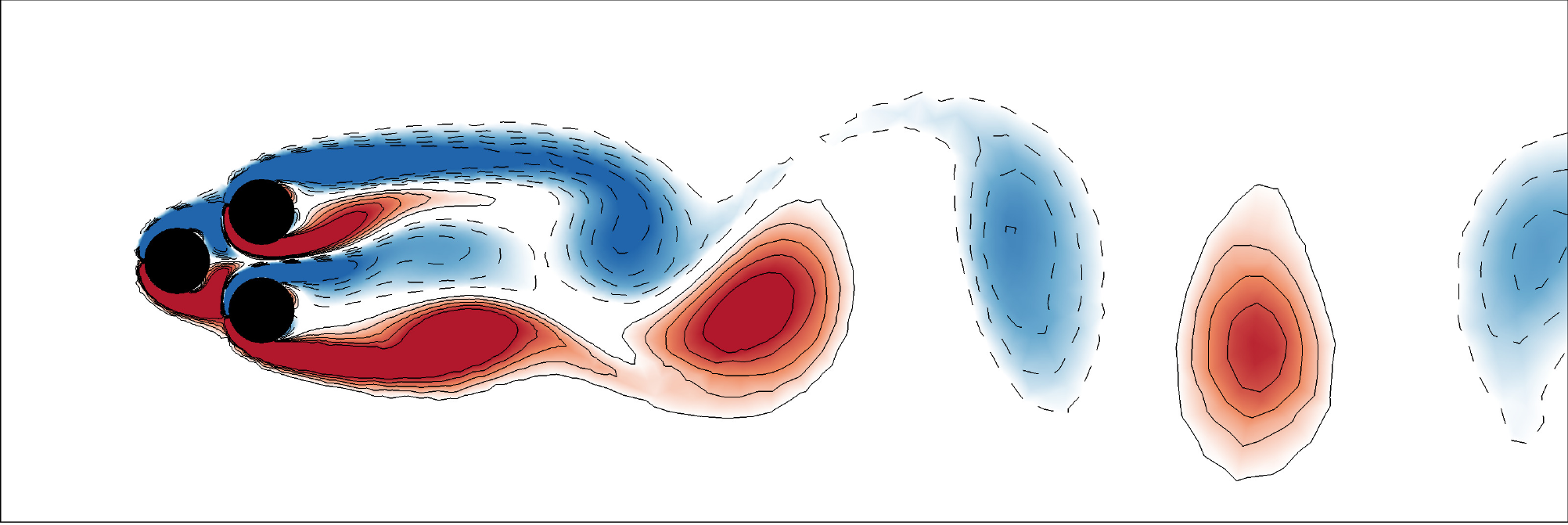}}%

\subfloat[$t+3T_0/8$]{\label{fig:nat_T3}\includegraphics[width=0.45\textwidth]{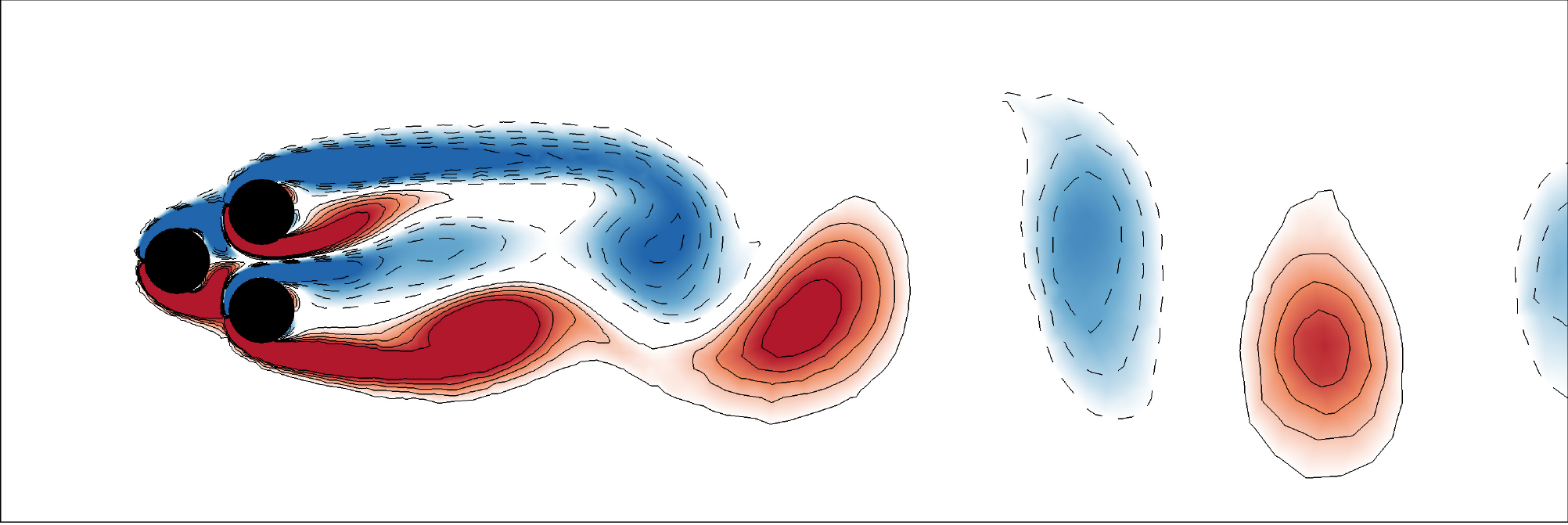}}%
\hfil
\subfloat[$t+4T_0/8$]{\label{fig:nat_T4}\includegraphics[width=0.45\textwidth]{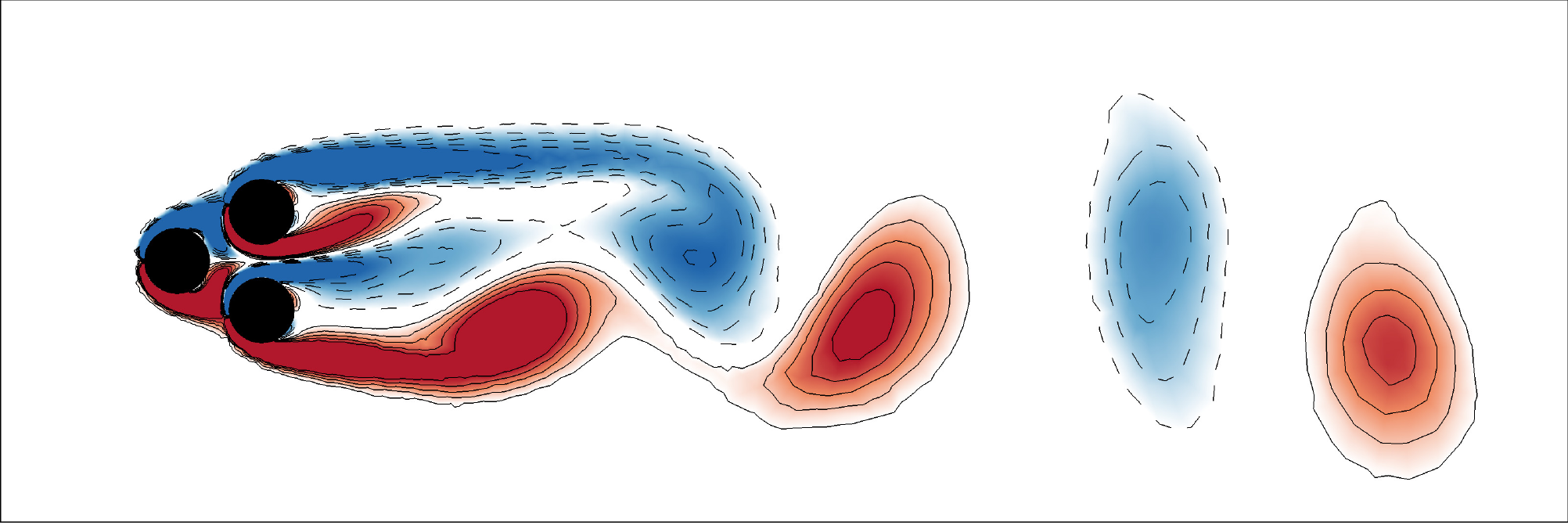}}%

\subfloat[$t+5T_0/8$]{\label{fig:nat_T5}\includegraphics[width=0.45\textwidth]{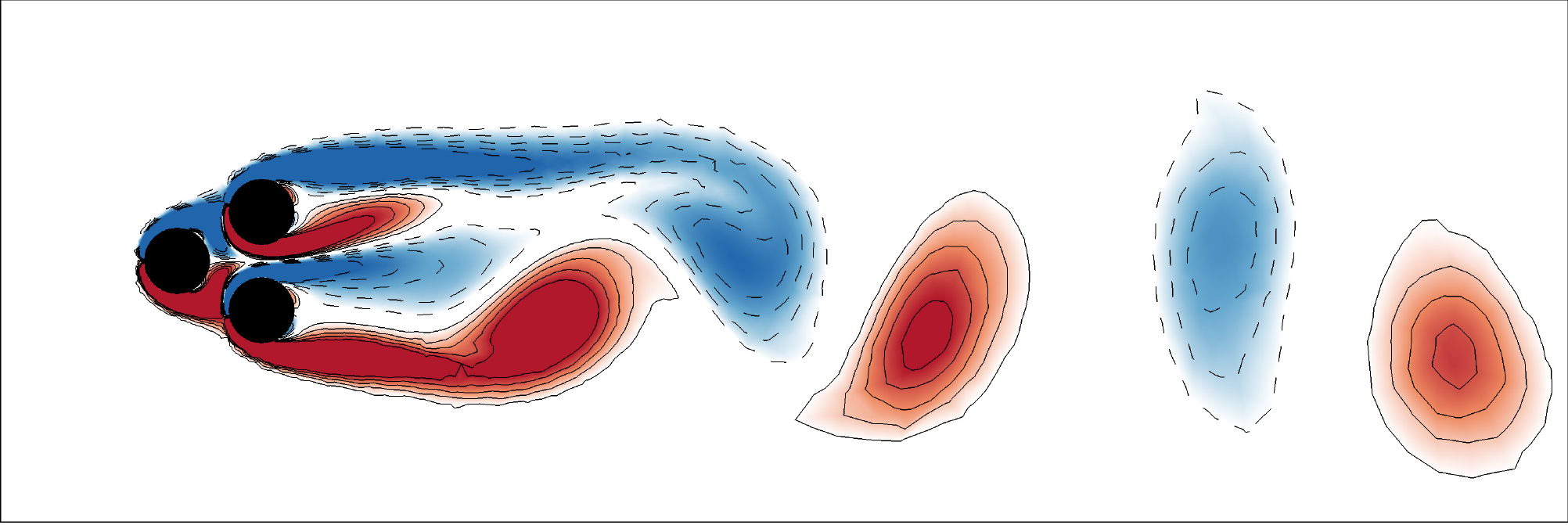}}%
\hfil
\subfloat[$t+6T_0/8$]{\label{fig:nat_T6}\includegraphics[width=0.45\textwidth]{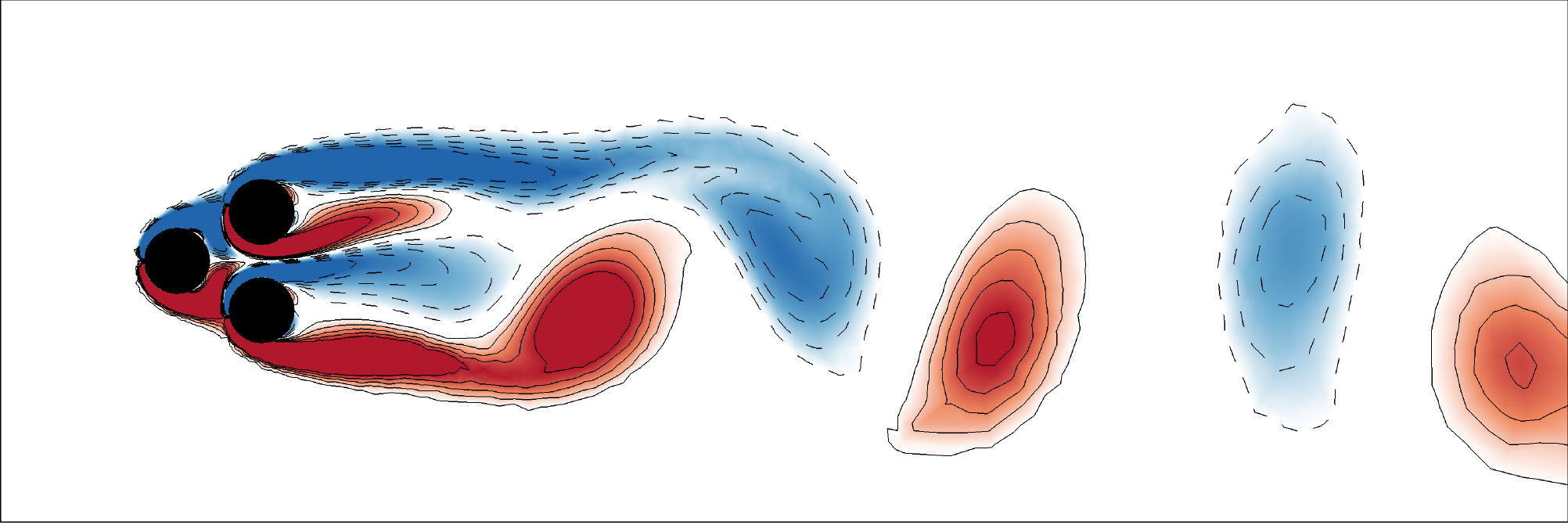}}%

\subfloat[$t+7T_0/8$]{\label{fig:nat_T7}\includegraphics[width=0.45\textwidth]{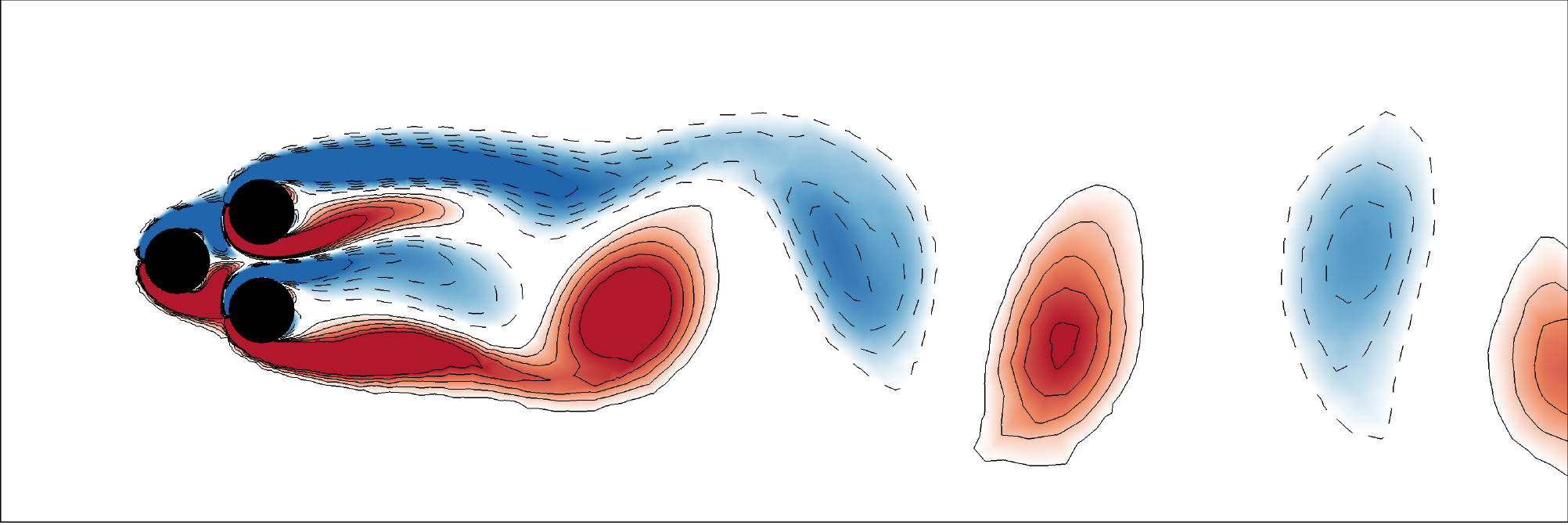}}%
\hfil
\subfloat[$t+T_0$]{\label{fig:nat_T8}\includegraphics[width=0.45\textwidth]{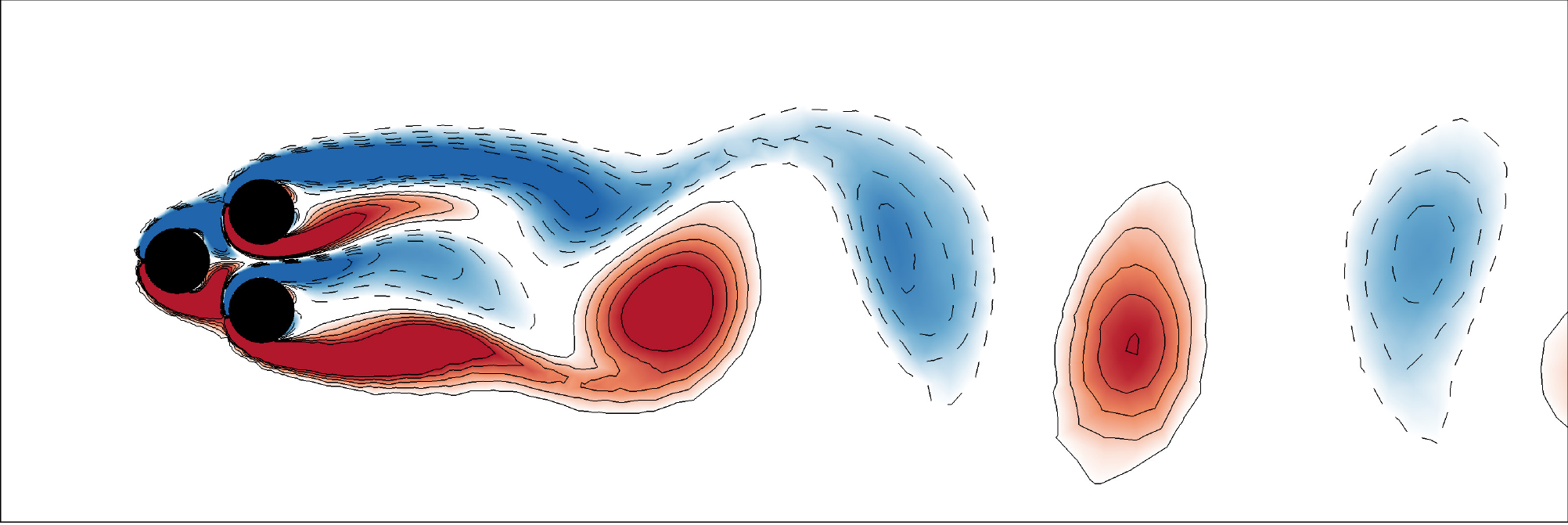}}%
\caption{\label{fig:natural_snap}Vorticity fields of the unforced flow. (a)-(f) Time evolution of the vorticity field in the last period of the  simulation.
The color code is the same as figure~\ref{fig:unforced_flow}.
$T_0$ is the natural period associated to the natural frequency $f_0$.
}
\end{figure}
Figure~\ref{fig:natural_characteristics} illustrates the dynamics of the unforced flow
from the unstable steady symmetric solution to the post-transient  periodic flow.
The phase portrait in figure~\ref{fig:PP_natural} and the power spectral density (PSD) in figure~\ref{fig:PSD_natural} show a periodic regime with frequency $f_0=0.116$ and its harmonic.
Figure~\ref{fig:CL_natural} shows that the mean value of the lift coefficient $C_L$ is not null.
This is due to the deflection of the jet behind the two rearward cylinders during the post-transient regime.
During this regime, the deflection of the jet stays on one side as it is illustrated in figure~\ref{fig:nat_T1}-\ref{fig:nat_T8} over one period and in figure~\ref{fig:natural_snap}j in the mean field.
This deflection explains the lift coefficient $C_L$ asymmetry.
Indeed, the upward oriented jet increases the pressure on the lower part of the top cylinder leading to an increase of the lift coefficient.
In figure~\ref{fig:CL_natural}, the initial downward spike on the lift coefficient is due to the initial kick.
The unforced natural flow is our reference simulation for future comparisons.

Thanks to the rotation of the cylinders,
the fluidic pinball is capable of reproducing six actuation mechanisms
inspired from wake stabilization literature and exploiting distinct physics.
Examples of those mechanisms can be found in \citet{Ishar2019jfm}.
First, the wake can be stabilized by shaping the wake region
more aerodynamically---also called fluidic boat tailing.
The shear layers are vectored towards the center region with passive devices,
like vanes \citep{Fluegel1930stg} or active control through Coanda blowing \citep{Geropp1995patent,Geropp2000ef,Barros2016jfm}.
In the case of the fluidic pinball,
we can mimic this effect by a counter-rotating rearward cylinders
which accelerates the boundary layers and delays separation.
This fluidic boat tailing is typically associated with significant drag reduction.
Second, the two rearward cylinders can also rotate oppositely ejecting a fluid jet on the centerline.
Thus, interaction between the upper and lower shear layer is suppressed,
preventing the development of a von K\'arm\'an vortex in the vicinity of the cylinders.
Such base bleeding mechanisms has a similar physical effect as a splitter plate behind a bluff body and has been proved to be an effective means for wake stabilization \citep{Wood1964jras,Bearman1967aq}.

Third, phasor control can be performed by estimating the oscillation phase
and feeding it back with a phase shift and gain \citep{Protas2004pf}.
Fourth, unified rotation of the three cylinders in the same direction
gives rise to higher velocities, and thus larger vorticity, on one side at the expense of the other side, destroying the vortex shedding.
This effect relates to the Magnus effect and stagnation point control \citep{Seifert2012review}.
Fifth, high-frequency forcing can be effected by symmetric periodic oscillation of the rearward cylinders.
With a vigorous cylinder rotation \citep{Thiria2006jfm}, the upper and lower shear layers are re-energized, reducing the transverse wake profile gradients and thus the instability of the flow.
Thus, the effective eddy viscosity in the von K\'arm\'an vortices increases, adding a damping effect.
Sixth and finally,
a symmetrical forcing at a lower frequency than the natural vortex shedding may stabilize the wake \citep{Pastoor2008jfm}.
This is due to the mismatch between the anti-symmetric vortex shedding and the forced symmetric dynamics whose clock-work is distinctly out of sync with the shedding period.
High- and low-frequency forcing lead to frequency crosstalk between actuation and vortex shedding
over the mean flows,
as described by  low-dimensional generalized mean-field model \citep{Luchtenburg2009jfm}.

We confirm therefore that the fluidic pinball is an interesting Multiple-Input Multiple-Output (MIMO) control benchmark.
The configuration
exhibits well-known wake stabilization mechanisms in physics.
From a dynamical perspective, nonlinear frequency crosstalk can easily be enforced.
In addition, even long-term simulations can easily be performed on a laptop within an hour.

%***********************************************************************
\subsection{Control objective and regression problem} 
\label{sec:control_objective}
Several control objectives related to the suppression or reduction of undesired forces
can be considered for the fluidic pinball.
We can increase the recirculation bubble length,
reduce lift fluctuations or even mitigate the total fluctuation energy.

In this study, we aim to reduce the net drag power at $\Rey_D=100$.
The associated objectives are $J_a$, the drag power and $J_b$, the actuation power.
The cost $J_a$ is defined as the temporal average of the drag power
of the controlled flow field:
\begin{equation}
J_a=\frac{1}{T_{ev}}\int_{t_0}^{t_0+T_{ev}}  j_a(t)  \:\mathrm{d} t
\end{equation}
with the  instantaneous cost function
\begin{equation}
 j_a(t) =\boldsymbol{F}_x(t) \cdot \boldsymbol{U}_{\infty}
\end{equation}
where $\boldsymbol{F}_x$ is the drag and $\boldsymbol{U}_{\infty}$ is the incoming velocity.
The control is activated at $t_0=400$ convective time units after the starting kick on the steady solution.
 Thus,  we have a fully established post-transient regime.
The cost function is evaluated until $T_{\rm ev}=525$ convective time units.
Thus, the time average is effected over 125 convective time units which corresponds to more than 10 periods $T_0$ of the unforced flow.

$J_b$ is naturally chosen as a measurement of the actuation energy investment.
Evidently, a low actuation energy is desirable.
The actuation power
is computed as the power of the torque applied by the fluid on the cylinders.
$J_b$ is the time-averaged actuation power over $T_{\rm ev} = 125$ time units:
\begin{equation}
J_b(\boldsymbol{b})=\frac{1}{T_{ev}}\int_{t_0}^{t_0+T_{\rm ev}}  \sum_{i=1}^{3}\mathcal{P}_{\rm{act},i} \:\mathrm{d} t
\end{equation}
where $\mathcal{P}_{\rm{act},i}$ is the actuation power supplied integrated over cylinder $i$:
$$\mathcal{P}_{\rm{act},i} = -  \oiint  b_i  F^{\theta}_{s,i} \:\mathrm{d} s $$
where $\left( F^{\theta}_{s,i} \mathrm{d}s \right)$ is the azimuthal component of the local fluid forces applied to cylinder $i$.
The negative sign denotes that the power is supplied and not received by the cylinders.

Thus, the cost function employed for the optimization is $J=J_a+\gamma J_b$.
$\gamma$ is the penalization parameter.
It allows to balance the terms of the cost function.
In this study as we aim to reduce the net drag power, we set $\gamma=1$ so both components $J_a$ and $J_b$ have the same weight.

The instantaneous values of $J_a$ and $J_b$ are plotted in figure~\ref{fig:DragPower_natural}.
Naturally, for the unforced flow, the actuation power is null, and the cost function is only the drag power.
We note that the drag power takes around 300 convective units to stabilize.
The cost of the post-transient regime is $J_0=1.87$.
$J_0$ serves as a reference for future comparisons.
The cost of the steady flow \ref{fig:steady_solution} is $J_{\rm steady}=1.80$ which is lower than $J_0$ but still high.
Therefore, we can assume that stabilizing the symmetric steady solution may not be the best strategy to reduce the net drag power.

In order to minimize the net drag power, the flow is forced by the rotation of the three cylinders.
The actuation command $\boldsymbol{b}=[b_1,b_2,b_3]^{\rm T}$
is determined by the control law $\boldsymbol{K}$.
This control law may operate open-loop or closed-loop with flow input.
Considered open-loop actuations are steady or harmonic oscillation around a vanishing mean.
Considered feedback  includes velocity sensor signals in the wake.
Thus, in the most general formulation, the control law reads the equation described in Sec.\ \ref{Sec2:OptimizationProblem}: $\boldsymbol{b}(t)=\boldsymbol{K}(\boldsymbol{h}(t), \boldsymbol{s}(t))$ 
with $\boldsymbol{h}(t)$ and $\boldsymbol{s}(t)$ being vectors comprising respectively time dependent harmonic functions and sensor signals.
The sensor signals include the instantaneous velocity signals
as well as three recorded values over one period
as elaborated in the result (Sec.\ \ref{sec:FB_optimization}).
In the following, $\Nb$ represents the number of actuators,
$\Nh$ for the number of time-dependent functions and $\Ns$ for the number of sensor signals.

\section[Symmetric steady actuation]{Symmetric steady actuation for net drag reduction}\label{sec:pinball_openloop}
\begin{figure}[htb]
  \centering
  \includegraphics[width=0.5\linewidth]{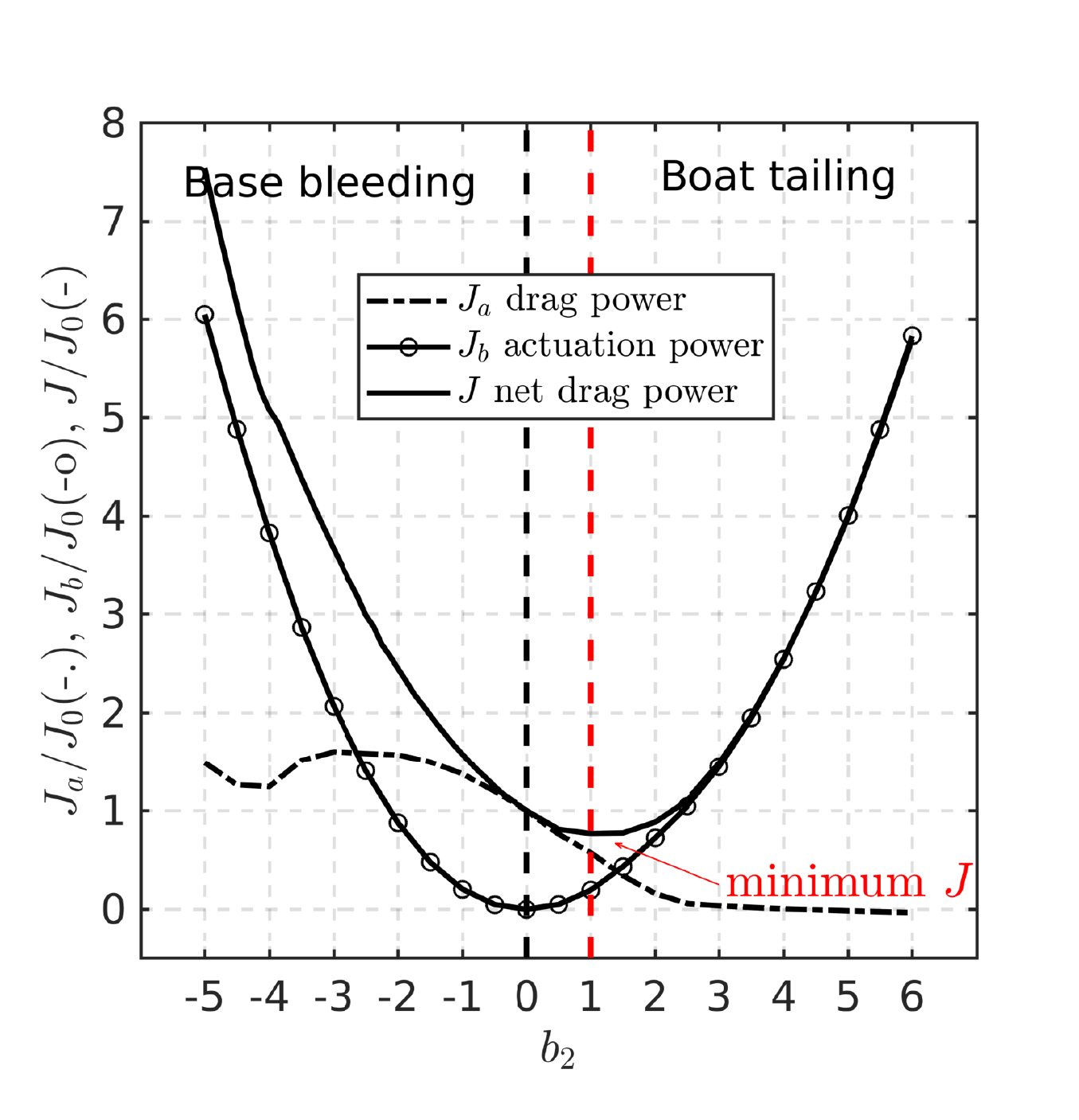}
  \caption{Cost function evolution regarding actuation intensity for symmetric constant actuation.
}
  \label{fig:constant_pinball}
\end{figure}
First, we carry out an open-loop parametric study to assess the effect of control on the drag power.
To achieve an exhaustive parametric study is a costly task as we need to operate in 3-dimensional parameter space.
That is why we restrict the search to the subspace of symmetric actuations: the front cylinder does not rotate and the two back cylinders rotate at the same speed but in opposite directions:
\begin{equation*}
\begin{array}{lll}
b_1 &=&0,\\
b_2  & =&-b_3.\\
\end{array}
\end{equation*}
Thus, we explore the effect of only one parameter $b_2$.
$b_2$ is defined so that when it is positive, the flow is vectored towards the centerline---boat tailing configuration---and when it is negative, the inner flow is accelerated---base bleeding configuration.
Figure~\ref{fig:constant_pinball} shows the evolution of $J_a$, $J_b$ and $J=J_a+J_b$ as a function of $b_2$.
Solely considering $J_a$, boat tailing is the best strategy to reduce the drag power.
Indeed, drag power decreases monotonously with increasing $b_2$.
For a strong actuation, $b_2>3$, $J_a$ even becomes negative and the fluidic pinball becomes a jet.
Base bleeding, on the other hand, is not a viable strategy to reduce the drag power.
There is a local minimum around $b_2=-4$ but its associated drag power is still higher than the natural unforced flow.
When adding the actuation power, there is only one minimum for the cost function $J$, around $b_2=1$.
Its associated cost is $J_{\rm BT}/J_0=0.77$.
%% Figures : Boat-tailing solution characteristics ------------------------------
\begin{figure}[htb]
\centering
\subfloat[]{\label{fig:CL_BT}\includegraphics[trim=0 14.25pt -20pt 30pt, clip,width=0.4\textwidth]{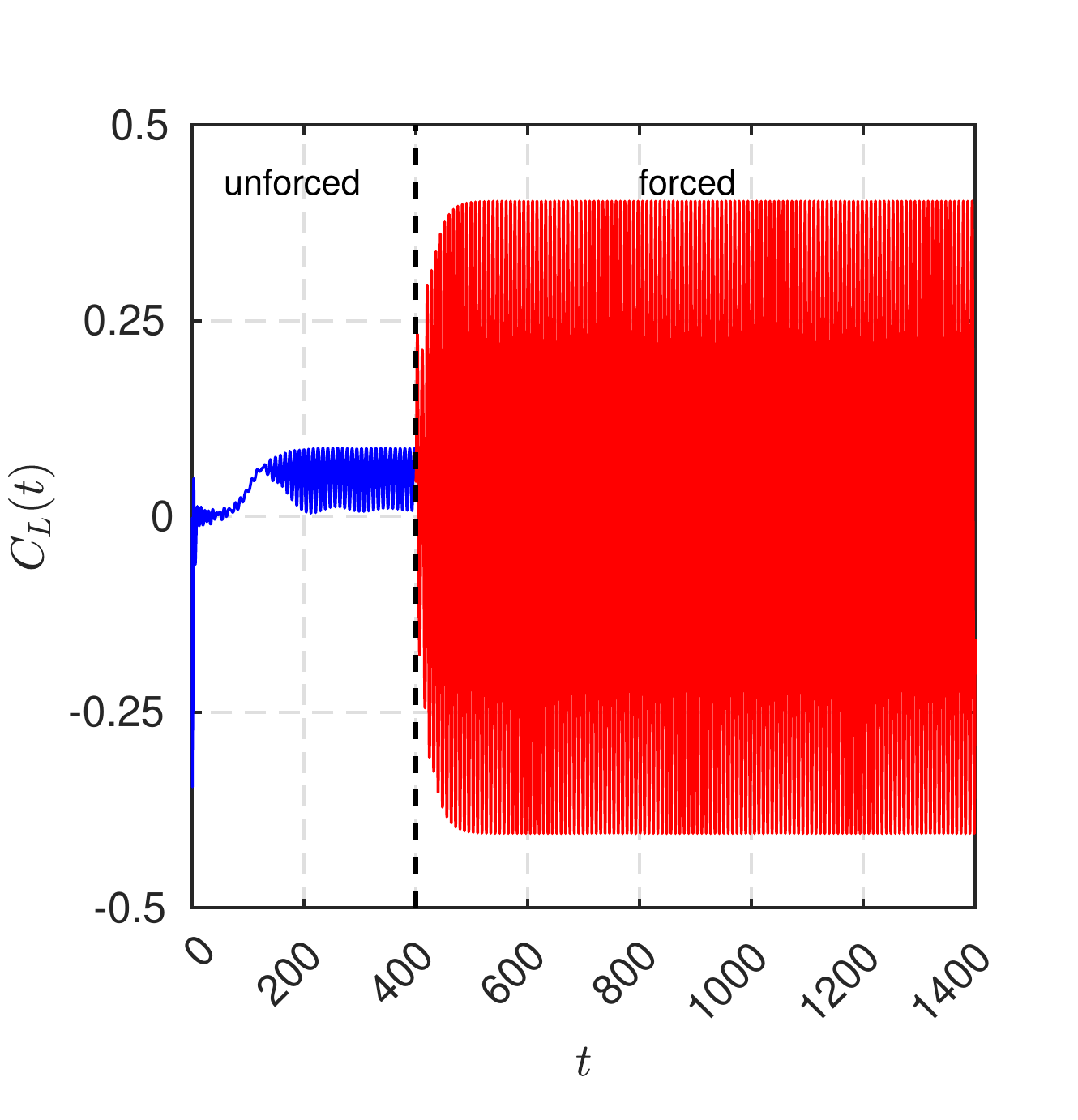}}%
\hfil
\subfloat[]{\label{fig:PP_BT}\includegraphics[trim=0 14.25pt -20pt 30pt, clip,width=0.4\textwidth]{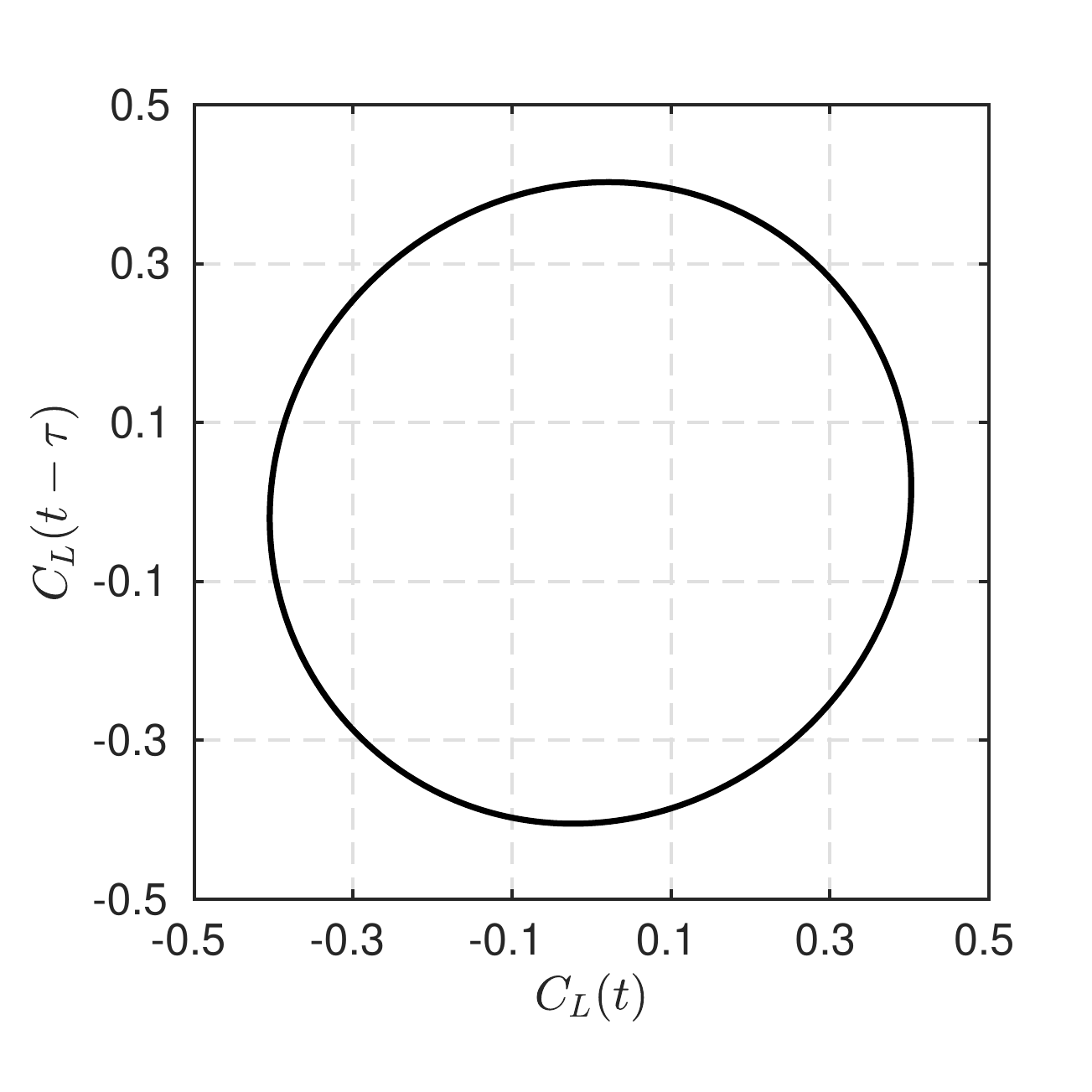}}%

\subfloat[]{\label{fig:DragPower_BT}\includegraphics[trim=0 14.25pt -20pt 30pt, clip,width=0.4\textwidth]{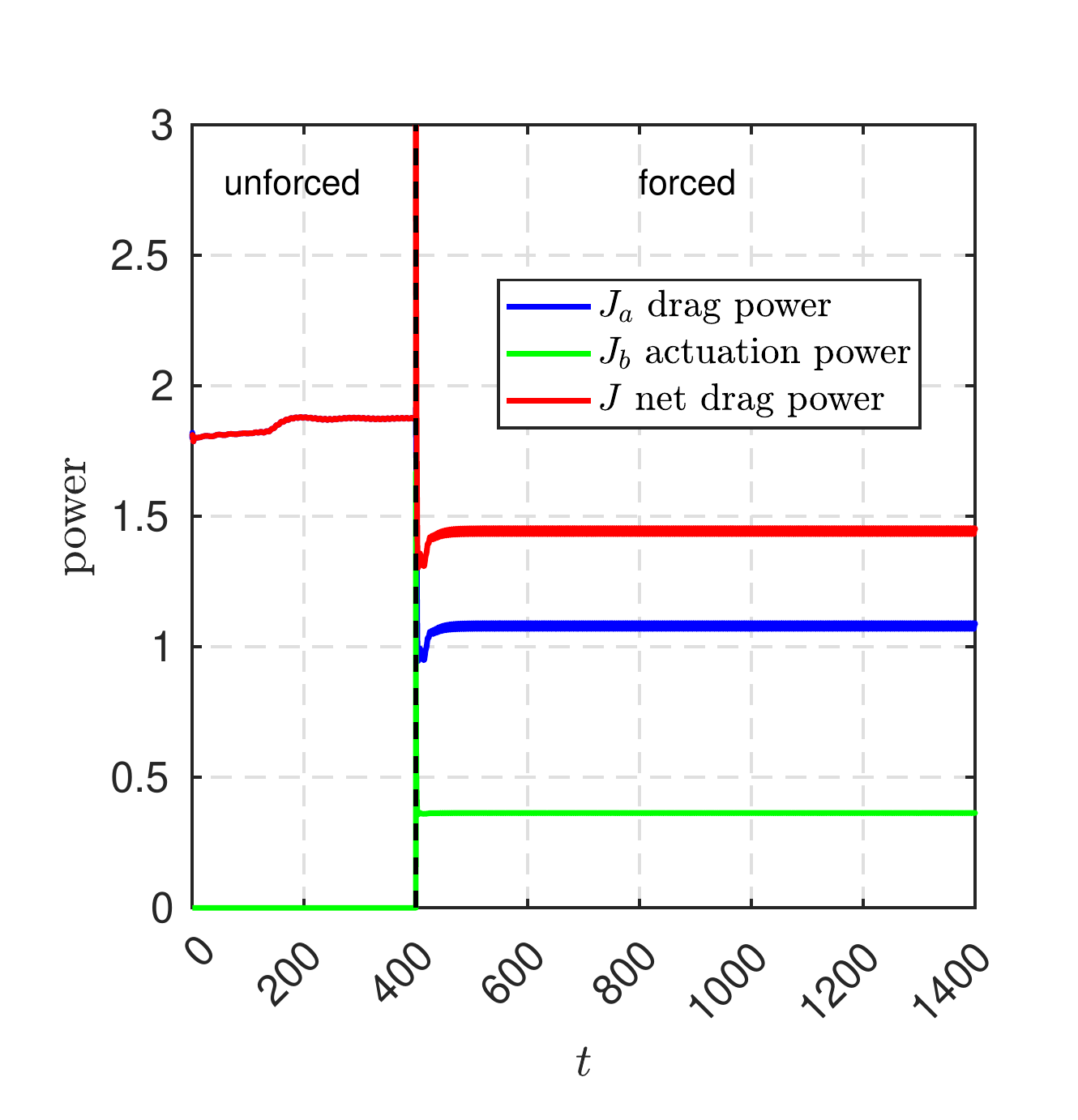}}%
\hfil
\subfloat[]{\label{fig:PSD_BT}\includegraphics[trim=0 14.25pt -20pt 30pt, clip,width=0.4\textwidth]{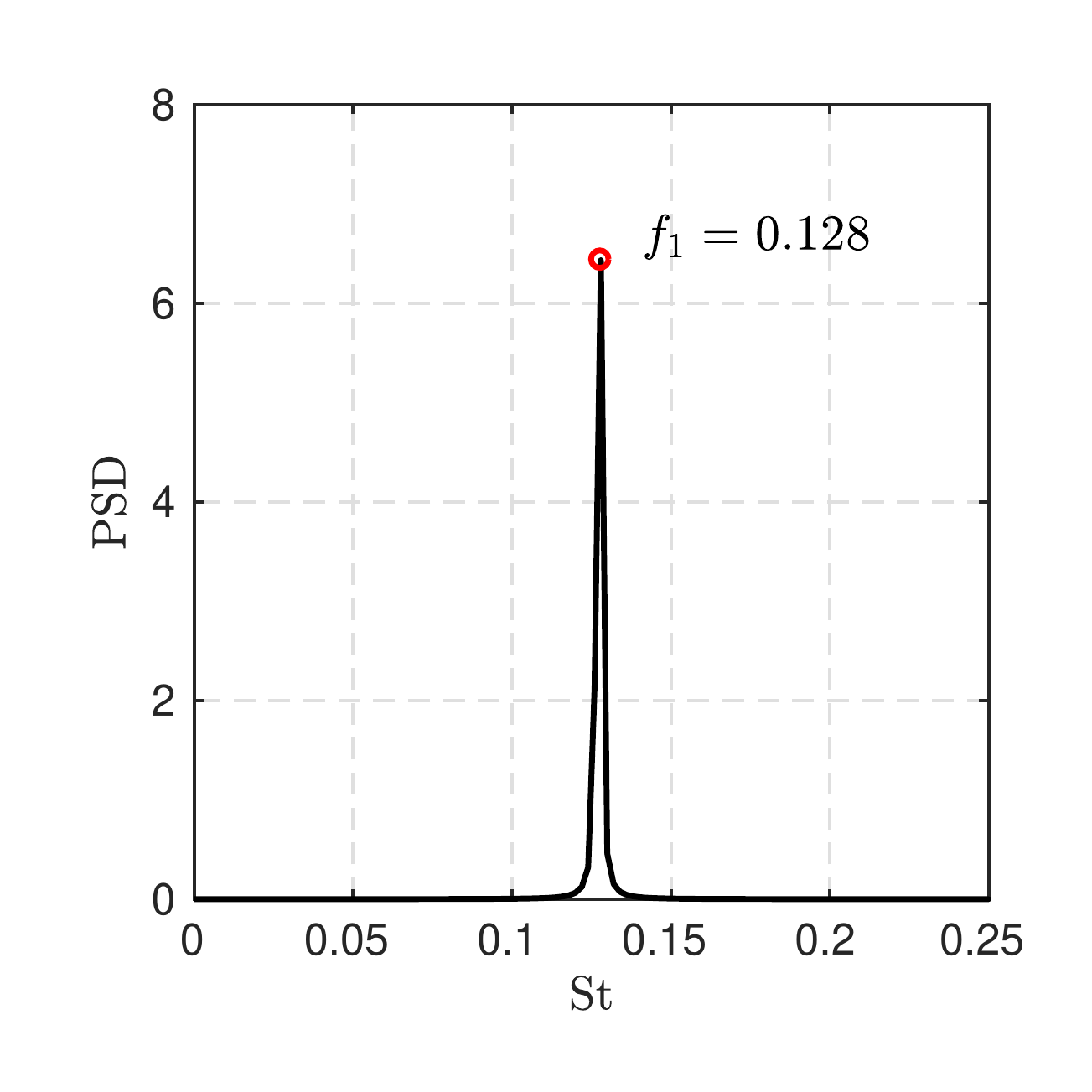}}%
\caption{\label{fig:BT_characteristics}Characteristics of the best boat tailing solution starting from the steady solution ($t=0$).
The transient spans until $t \approx 400$. (a) Time evolution of the lift coefficient $C_L$, (b) phase portrait, (c)  time evolution of the drag power $J_a$ (blue), actuation power $J_b$ (green) and net drag power $J$ (red) and (d) Power Spectral Density (PSD) showing the frequency $f_1=0.128$. 
}
\end{figure}
Figure~\ref{fig:PSD_BT} shows the characteristics of the controlled flow with the best boat tailing solution.
We note that the regime is purely harmonic with an increase of the main frequency from $f_0=0.116$, for the unforced flow, to $f_1=0.128$.
We also remark a symmetrization of the lift coefficient $C_L$ alongside a significant increase of the oscillations amplitude.
The loss of the mean value results in a symmetrization of the flow.
% ----------------------------- SNAPSHOTS -------------------------------
\begin{figure}[htb]
\centering
\subfloat[$t+T_1/8$]{\label{fig:BT_T1}\includegraphics[width=0.45\textwidth]{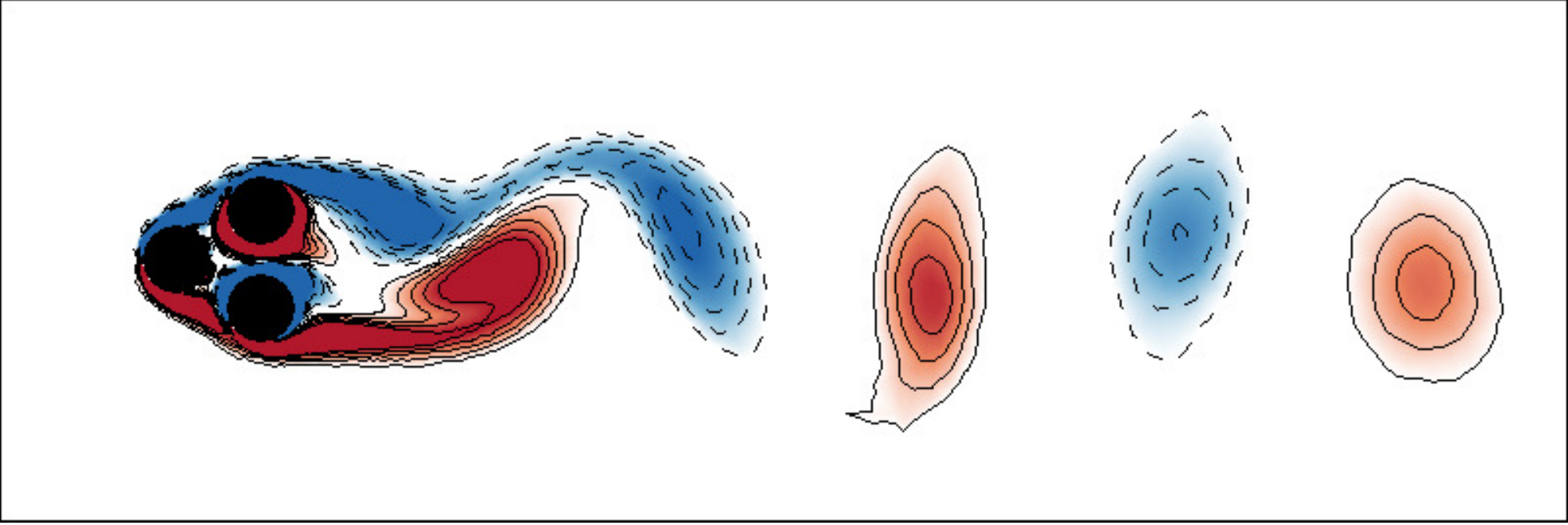}}%
\hfil
\subfloat[$t+2T_1/8$]{\label{fig:BT_T2}\includegraphics[width=0.45\textwidth]{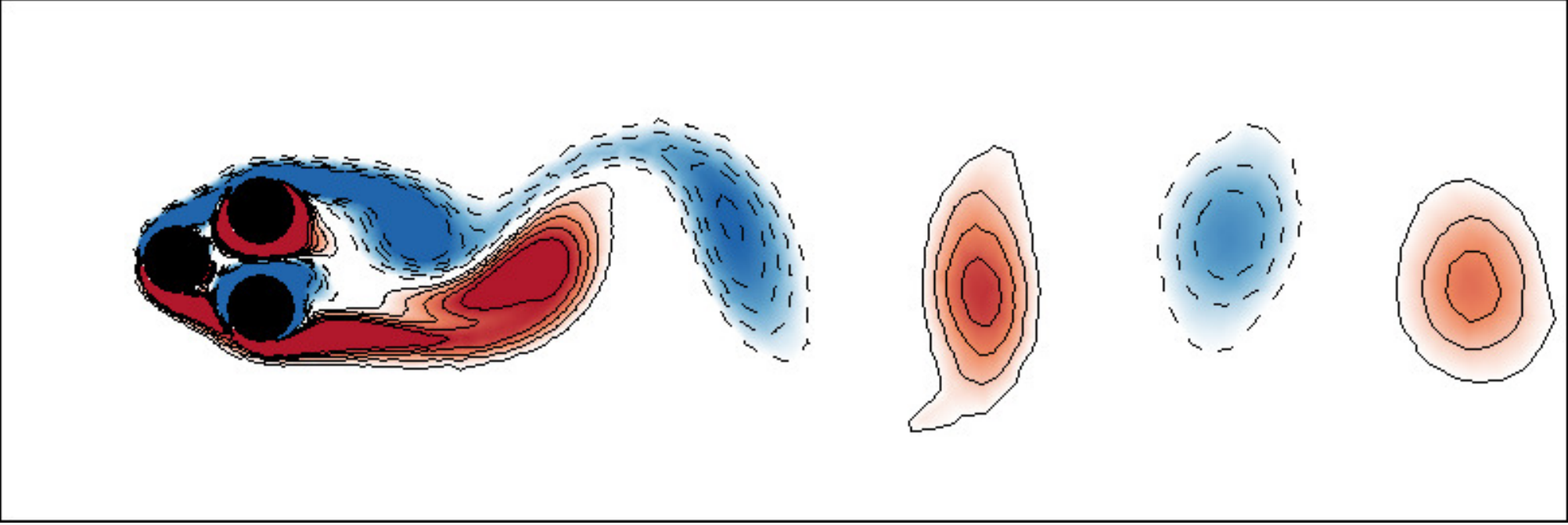}}%

\subfloat[$t+3T_1/8$]{\label{fig:BT_T3}\includegraphics[width=0.45\textwidth]{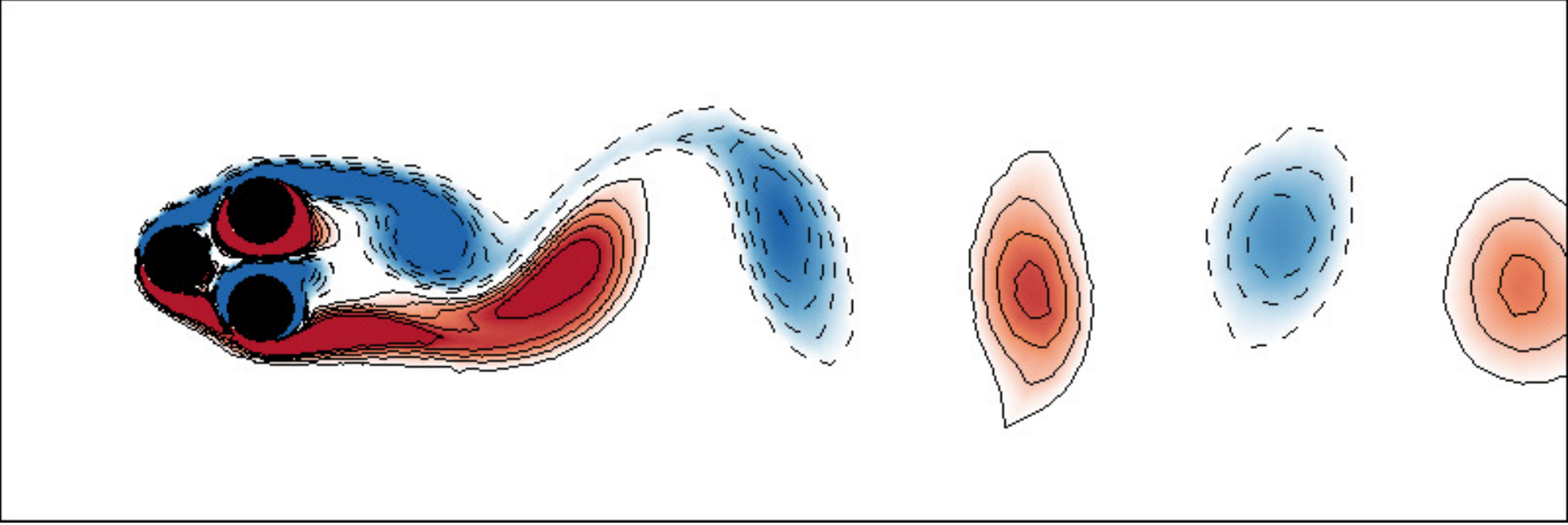}}%
\hfil
\subfloat[$t+4T_1/8$]{\label{fig:BT_T4}\includegraphics[width=0.45\textwidth]{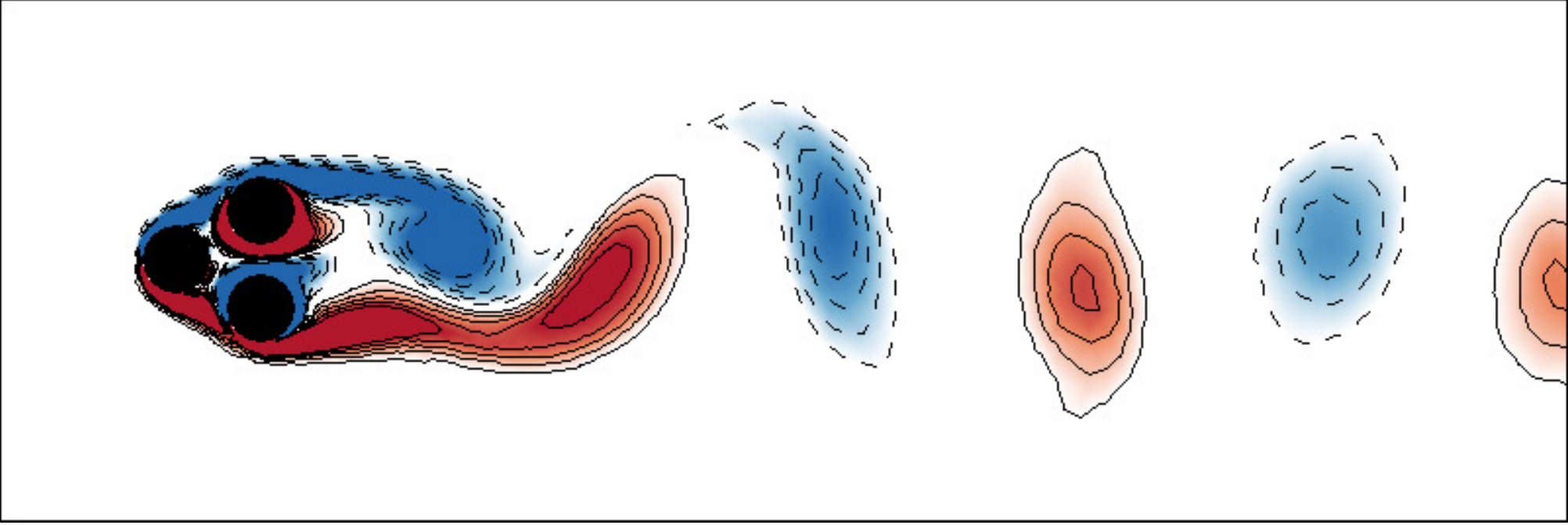}}%

\subfloat[$t+5T_1/8$]{\label{fig:BT_T5}\includegraphics[width=0.45\textwidth]{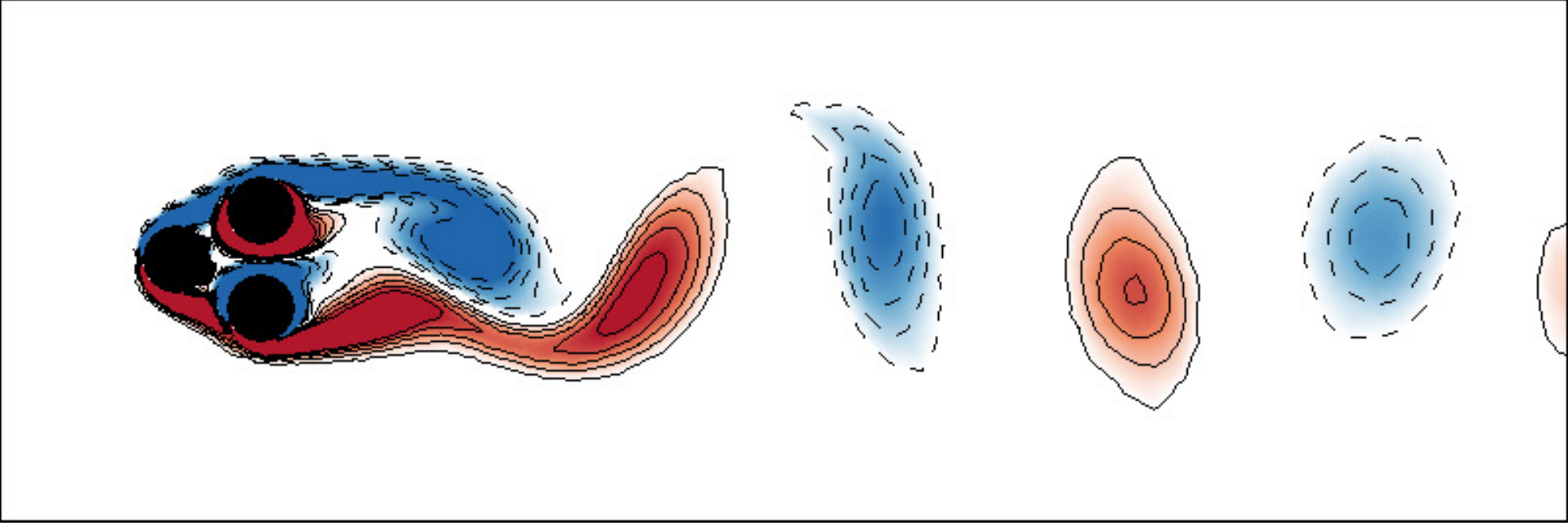}}%
\hfil
\subfloat[$t+6T_1/8$]{\label{fig:BT_T6}\includegraphics[width=0.45\textwidth]{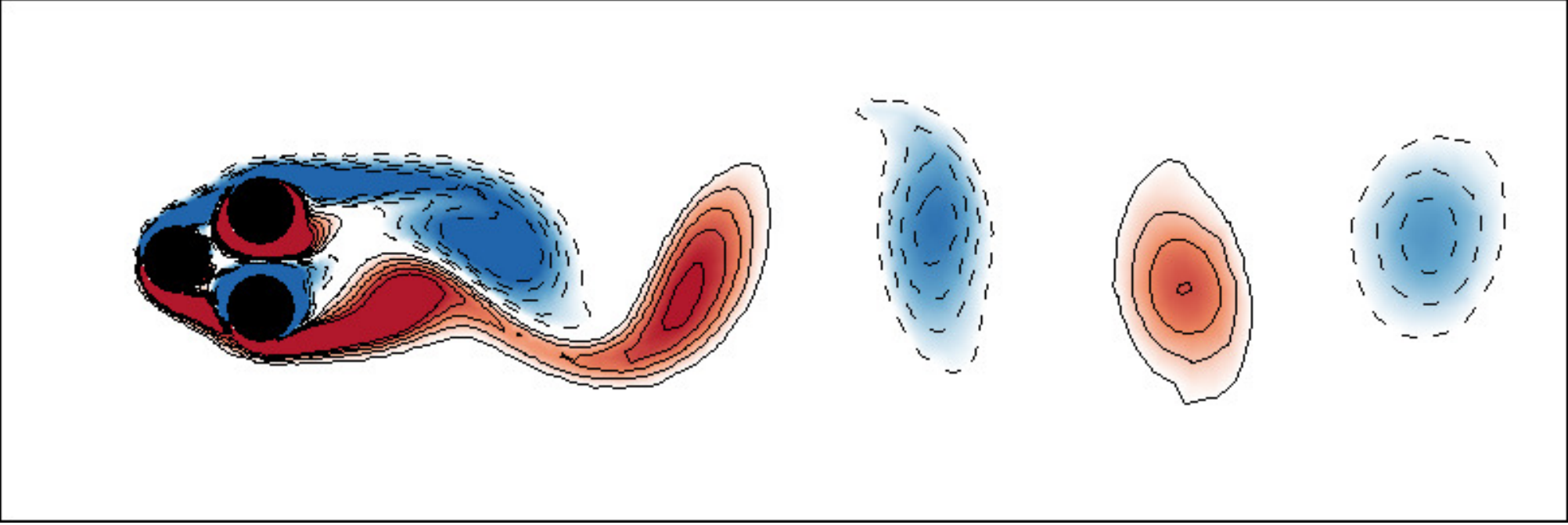}}%

\subfloat[$t+7T_1/8$]{\label{fig:BT_T7}\includegraphics[width=0.45\textwidth]{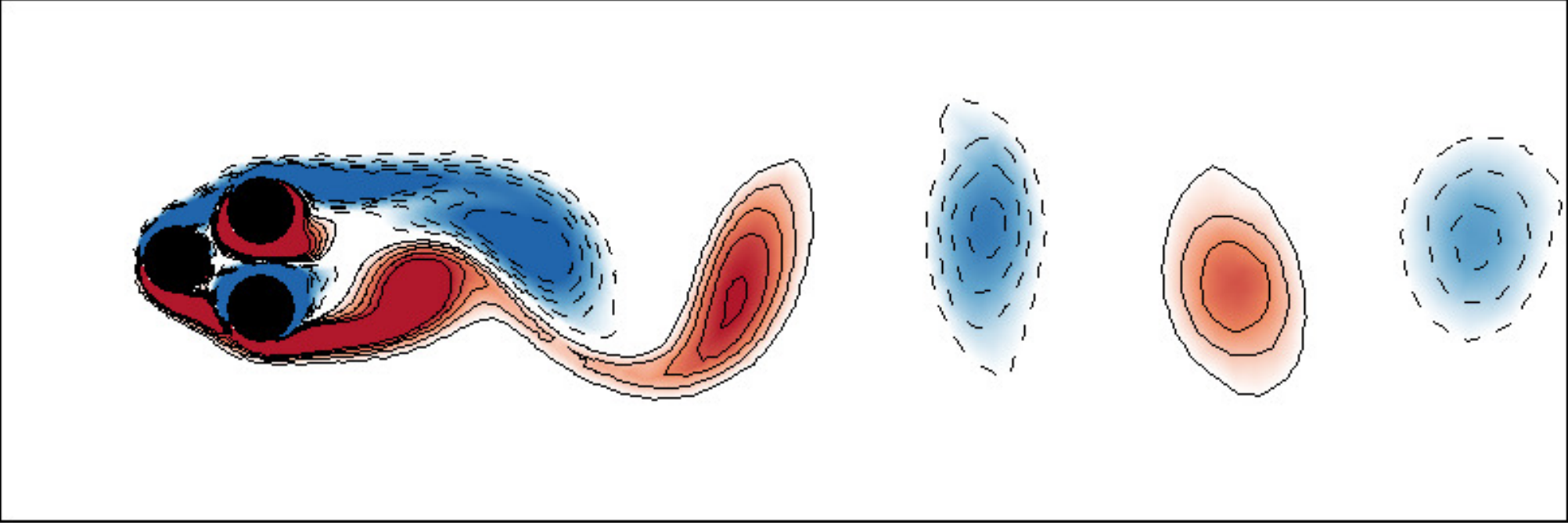}}%
\hfil
\subfloat[$t+T_1$]{\label{fig:BT_T8}\includegraphics[width=0.45\textwidth]{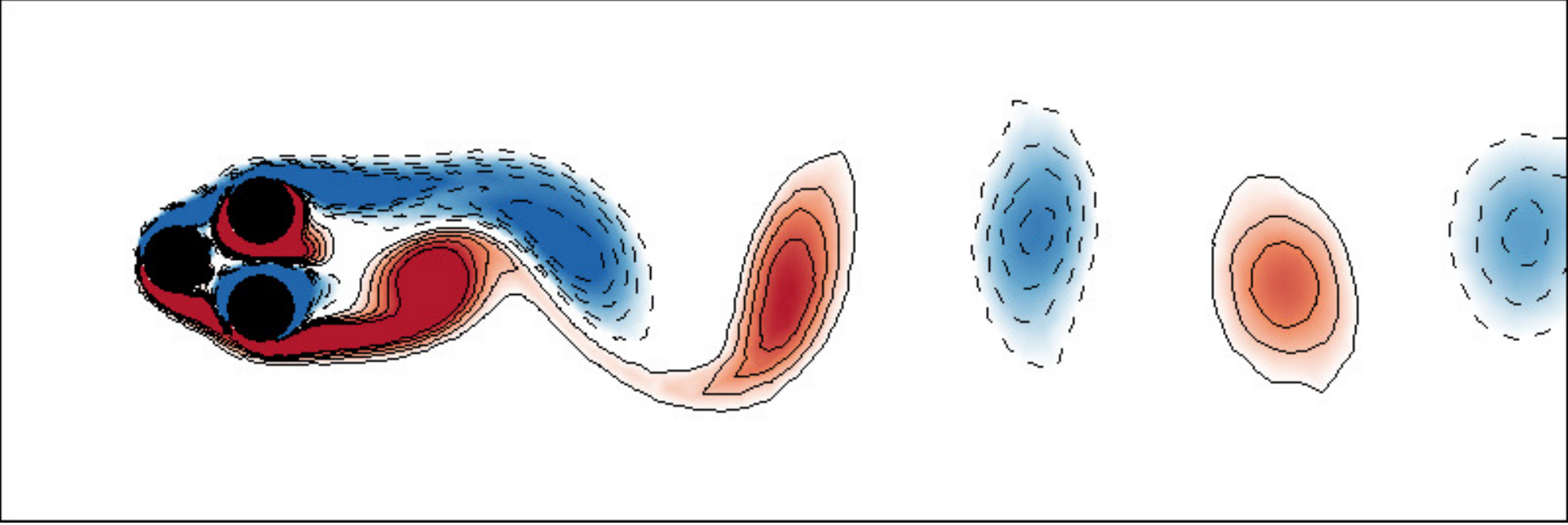}}%
\caption{\label{fig:BT_snap} Vorticity fields of the flow controlled by the best boat tailing solution. (a)-(f) Time evolution of the vorticity field in the last period of the 1000 time units simulation.
The color code is the same as figure~\ref{fig:unforced_flow}.
$T_1$ is the period associated to the frequency $f_1$.}
\end{figure}
Indeed, figure~\ref{fig:BT_snap} shows that the near jet completely disappeared.
We also observe a considerable reduction of the recirculation bubble, as a consequence, the base pressure behind the back cylinders increases and the total drag decreases.

\section{Multi-frequency optimization}\label{sec:MF_optimization}
In this section, we explore the space of open-loop controllers, in particular periodic control laws.
In order to do so, we consider cosine functions as inputs for the control laws.
Thus, equation~\eqref{Eq:ControlLaw} becomes: $\boldsymbol{b}(t) = \boldsymbol{K}(\boldsymbol{h}(t))$.
To enrich our search space and to avoid resonance effects, we choose eight periodic functions whose frequencies are incommensurable with the natural frequency $f_0$:
\begin{equation*}
\label{Eqn:h_definition}
\begin{array}{lllllll}
h_1 &=& \cos(2\pi \Phi^{-4} f_0 t ),&\hspace{1cm} &h_5 &=& \cos(2\pi \Phi^{1} f_0 t ), \\
h_2  & =&\cos(2\pi \Phi^{-3} f_0 t ),& &h_6 &=& \cos(2\pi \Phi^{2} f_0 t ),\\
h_3  & = & \cos(2\pi \Phi^{-2} f_0 t ),& &h_7 &=& \cos(2\pi \Phi^{3} f_0 t ),\\  
h_4 &=& \cos(2\pi \Phi^{-1} f_0 t ),& &h_8 &=& \cos(2\pi \Phi^{4} f_0 t ).\\
\end{array}
 \end{equation*}
The golden ration $\Phi$ assures that the periodic function is incommensurable with the natural frequency $f_0$, i.e. the natural frequency cannot be reconstructed thanks to the algebraic operators $(+,-,\times,\div)$.
\begin{table}[htb]
  \centering
   \begin{tabular}{>{\centering}p{2.5cm}>{\centering}p{6cm}>{\centering\arraybackslash}p{5cm}}
  Parameter & Description & Value\\
\midrule
   & \multirow{2}{*}{Function library} & $ F_2= \{+,-,\times,\div,$\\
   &  & $\mathrm{exp},\tanh,\sin,\cos,\mathrm{log}\}$\\
   $\boldsymbol{h}$  & Controller inputs & $h_i$, $i=-4,..,-1,1,..4$\\
   $\Nvar$ & Number of variable registers & $8+3=11$ \\
   $\Ncst$ & Number of constant registers & 10 \\
   $\Ninstrmax$ & Max. number of instructions & 50\\
  \hline
   $\Nps$ & Population size & 100\\
   $\Ng$ & Number of generations & 10\\
   $\Ntour$ & Tournament size & $7$\\
   $\Ne$ & Elitism & $1$\\
   $\Pcros$ & Crossover probability & 0.6 \\
   $\Pmut$ & Mutation probability & 0.3 \\
   $\Prep$ & Replication probability & 0.1 \\
\end{tabular}

  \caption{MLC parameters for multi-frequency forcing optimization.}
  \label{tab:MF_LGPCparameters}
\end{table}
The rest of the MLC parameters are summarized in table~\ref{tab:MF_LGPCparameters}.
We choose the operators probability $(\Pcros,\Pmut,\Prep)=(0.6,0.3,0.1)$ as explained in App.\ \ref{Sec:ParamStudyGenOp}.
As we optimize three controllers, we decided to increase the number of maximum instructions to 50.
To build complex control laws, we employ the function library $F_2= \{+,-,\times,\div,\mathrm{exp},\tanh,\sin,\cos,\mathrm{log}\}$.
Also, since we have eight inputs for the control laws, we need to increase the number of variable registers to include an instance of all inputs.
Thus, we also increase the number of constants.
\begin{figure}[htb]
 \centerline{\includegraphics[width=0.75\linewidth]{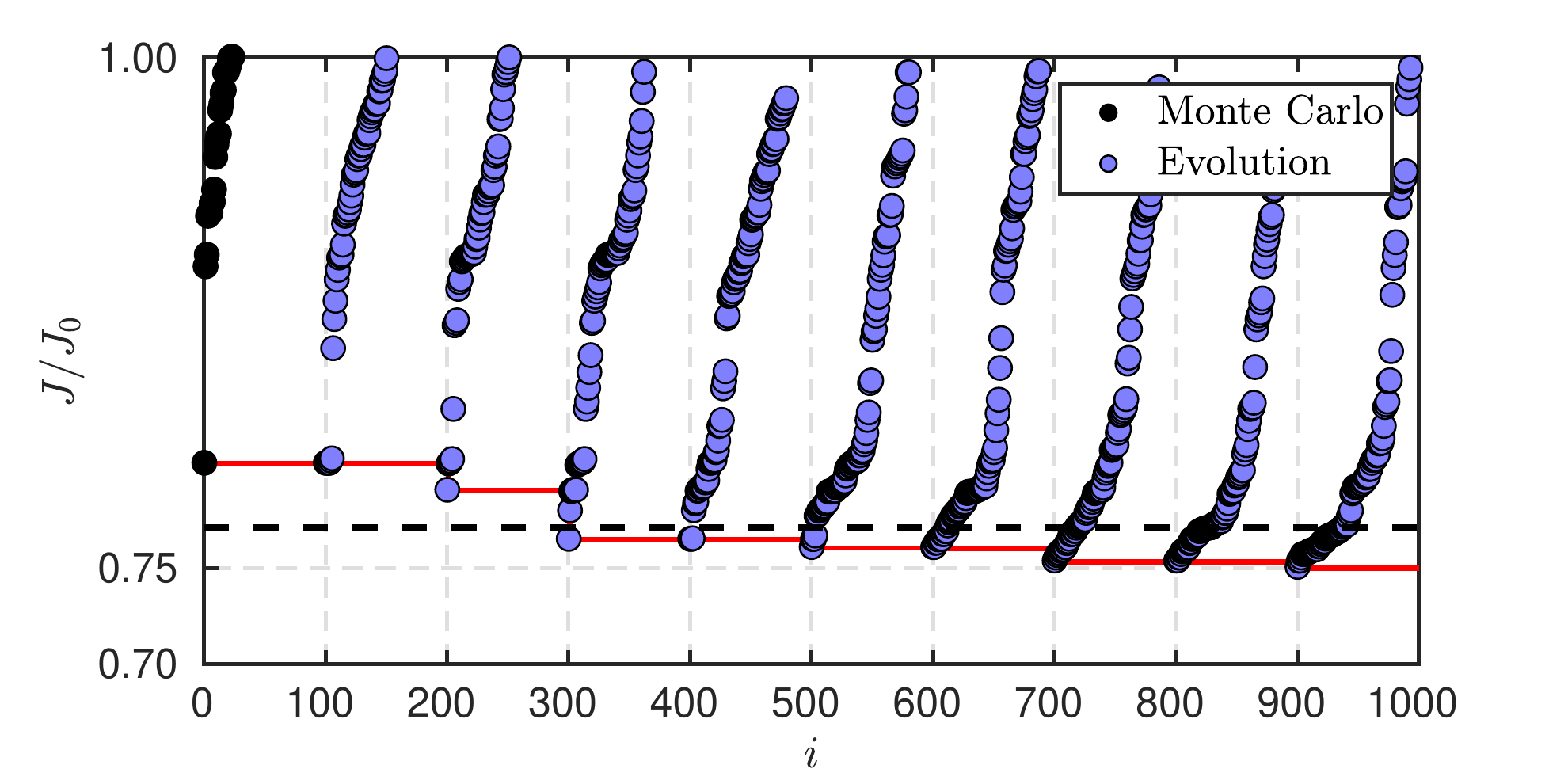}}
 \caption{Distribution of the costs during the multi-frequency forcing MLC optimization process.
 Each dot represents the cost $J_a/J_0$ of one individual.
 The color of the dots represents how the individuals have been generated.
 Black dots denote the individuals which are
 randomly generated (Monte Carlo).
 Blue dots refer to individuals which are generated from a genetic operator.
 The individuals from each generation have been sorted following their costs.
The red line shows the evolution of the best cost.
The dashed horizontal line corresponds to the best symmetric solution $b_2=1$ with a cost of $J_{\rm BT}/J_0=0.77$.
The vertical axis is in log scale and has been truncated to help the visualization of the best individuals.}
\label{fig:MF_Learning}
\end{figure}
We ran the optimization with a population of 100 individuals evolving through 10 generations.
The learning process is illustrated in figure~\ref{fig:MF_Learning}.
Most of the learning is done during the Monte Carlo step, indeed after 100 random evaluations the cost is equal to $J/J_0=0.80$.
Only small improvements are made until the fourth generation where a big jump manages to reduce the cost lower than the boat tailing solution.
From there, only small improvements are achieved.

The best control law $\boldsymbol{b}^{\rm MF}$ reads:
\begin{equation}
\label{Eqn:bMF}
\begin{array}{lll}
b^{\rm MF}_1 &=& -0.13732,\\
b^{\rm MF}_2  & =&0.982511,\\
b^{\rm MF}_3  & = & -1.1979,\\  
J_{\rm MF}/J_0 &=& 0.7476\\
\end{array}
 \end{equation}
 We note that the control is steady and does not contain any $h_i$, suggesting that periodic forcing is not a viable solution to reduce the net drag power.
 Indeed, periodic forcing must increase the gradient of the azimuthal speed, thus increasing the torque and actuation power.
$\boldsymbol{b}^{\rm MF}$ resembles a boat tailing configuration with a slight asymmetry as the bottom cylinders rotates faster than the top one and the front cylinder also presents a small rotation.
We suspect that this asymmetry is typical of pitchfork bifurcated flows as it was also reported by \citet{Raibaudo2020pf}.
%% Figures : Multi-frequency characteristics ------------------------------
\begin{figure}[htb]%
\centering
\subfloat[]{\label{fig:CL_MF}\includegraphics[trim=0 14.25pt -20pt 30pt, clip,width=0.4\textwidth]{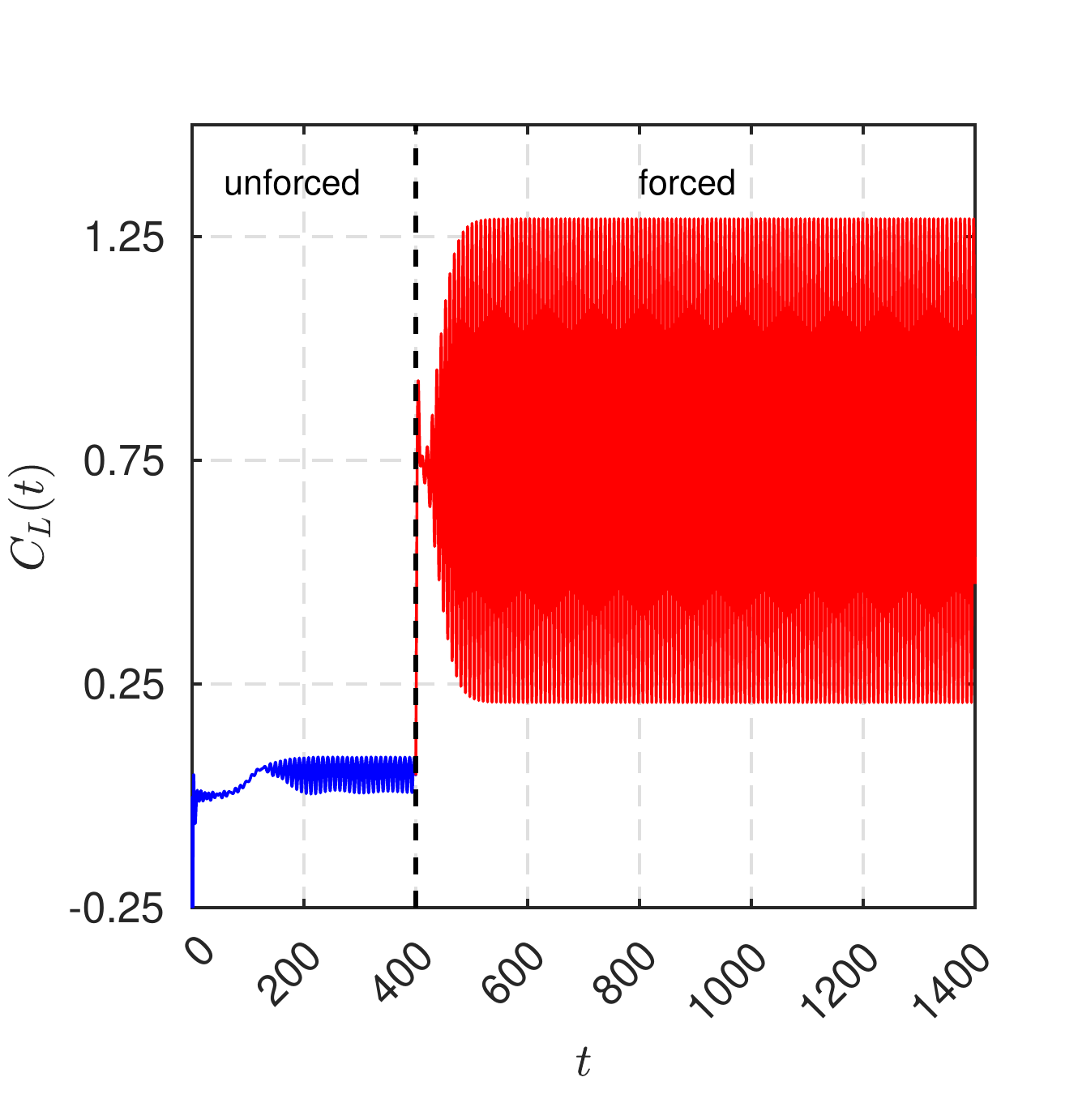}}%
\hfil
\subfloat[]{\label{fig:PP_MF}\includegraphics[trim=0 14.25pt -20pt 30pt, clip,width=0.4\textwidth]{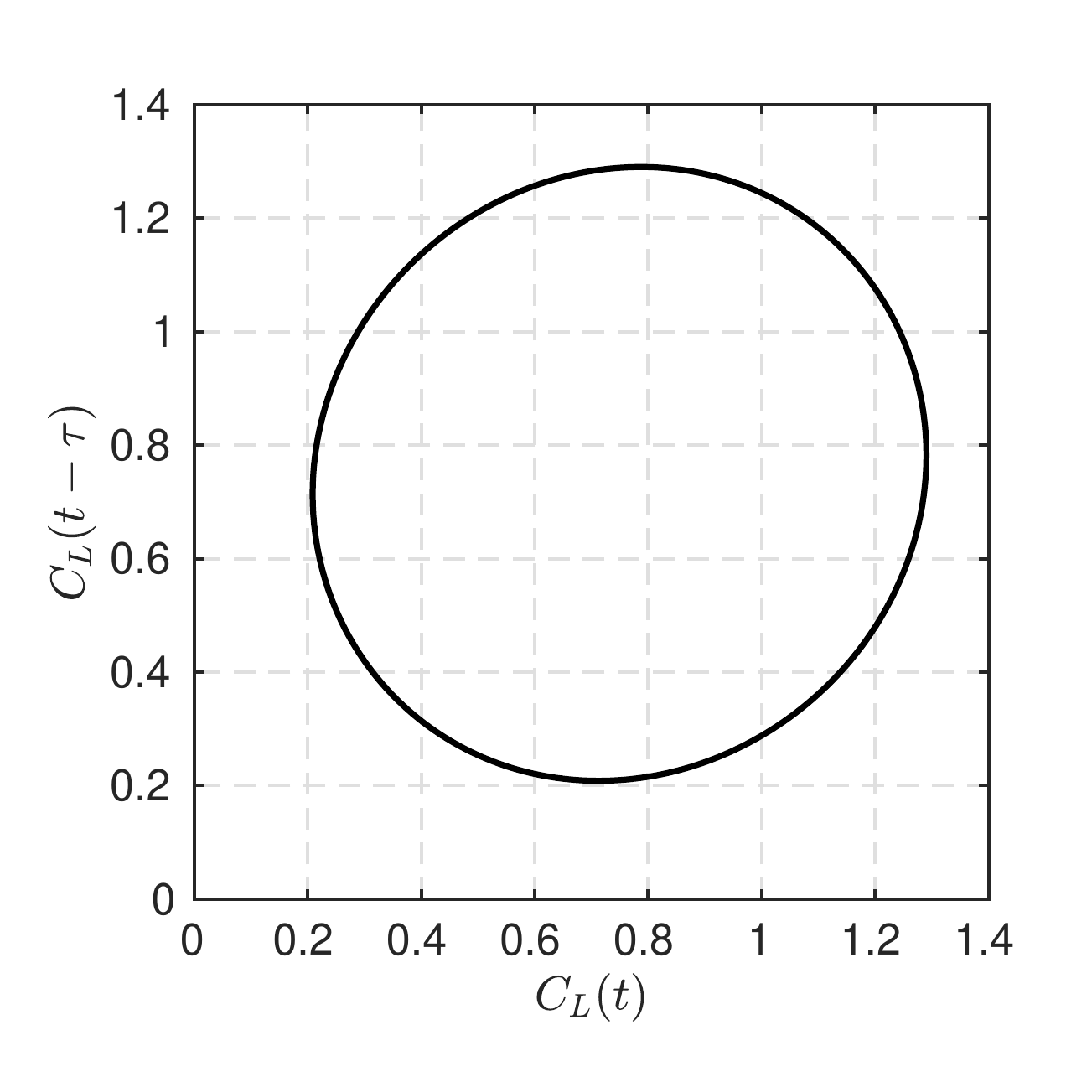}}%

\subfloat[]{\label{fig:DragPower_MF}\includegraphics[trim=0 14.25pt -20pt 30pt, clip,width=0.4\textwidth]{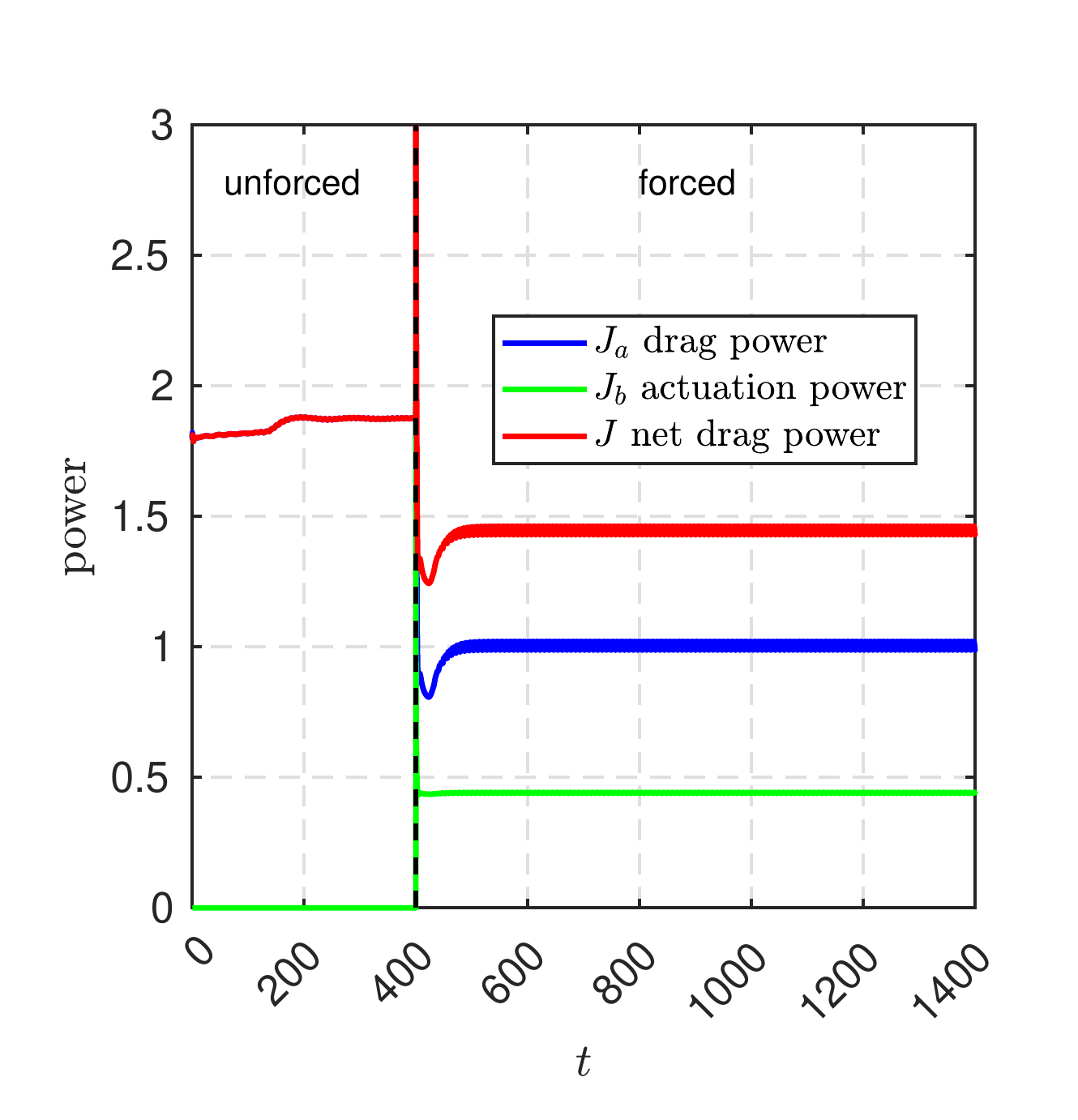}}%
\hfil
\subfloat[]{\label{fig:PSD_MF}\includegraphics[trim=0 14.25pt -20pt 30pt, clip,width=0.4\textwidth]{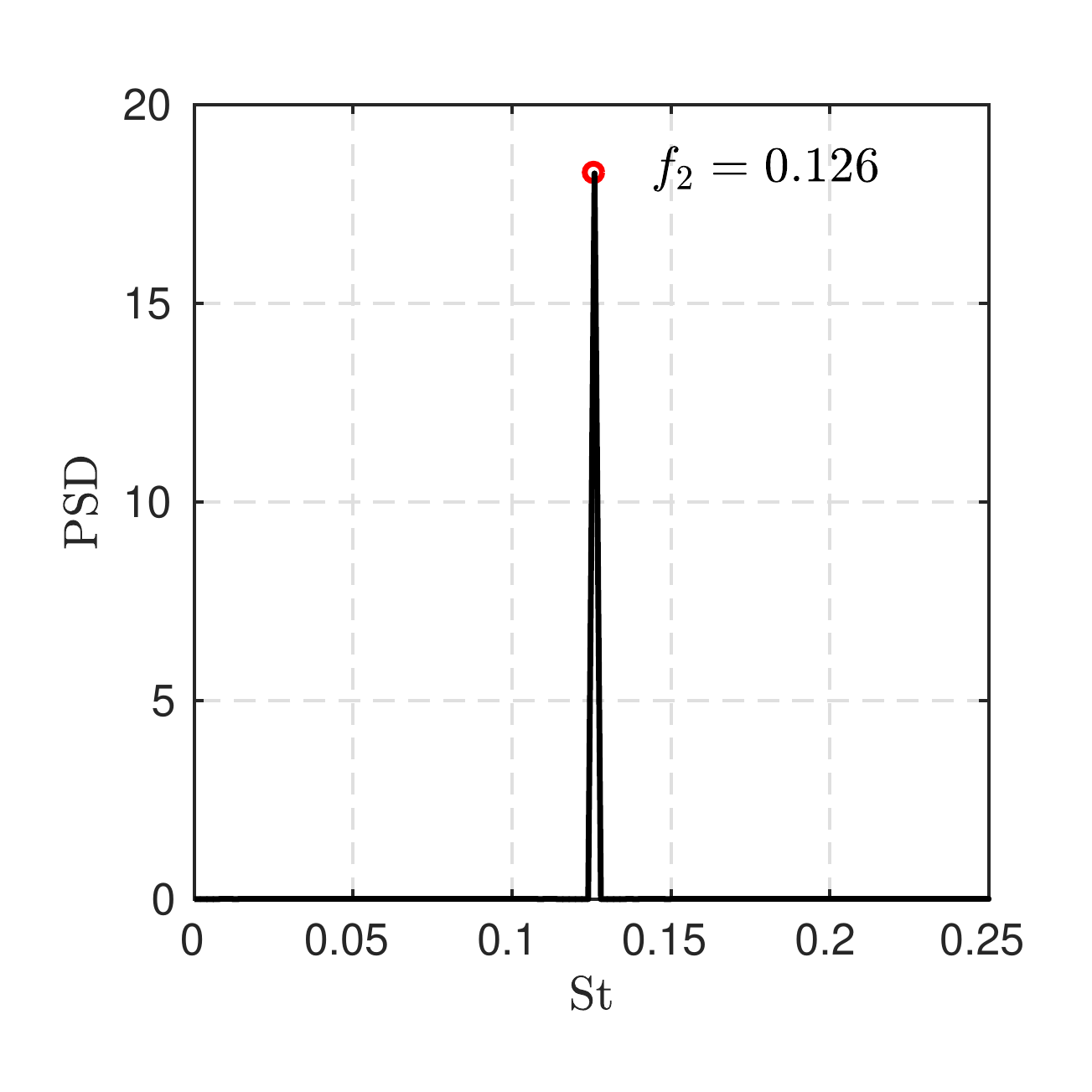}}%
\caption{\label{fig:MF_characteristics}Characteristics of the flow controlled by $\boldsymbol{b}^{\rm MF}$ starting from the steady solution ($t=0$).
The transient spans until $t \approx 400$. (a) Time evolution of the lift coefficient $C_L$, (b) phase portrait, (c)  time evolution of the drag power $J_a$ (blue), actuation power $J_b$ (green) and net drag power $J$ (red) and (d) Power Spectral Density (PSD) showing the frequency $f_2=0.126$. 
}
\end{figure} 
The characteristics of the flow controlled by $\boldsymbol{b}^{\rm MF}$ are shown in figure~\ref{fig:MF_characteristics}.
As for the best boat tailing solution, the controlled flow with $\boldsymbol{b}^{\rm MF}$ is purely harmonic with a slightly lower frequency $f_2=0.126$.
The amplitude of the oscillations of the lift coefficient has also increased.
However, the slight asymmetry in the control results in a significant increase of the mean value of the lift coefficient.
% Multi-frequency snapshots ---------------------------------------------
\begin{figure}%
\centering
\subfloat[$t+T_2/8$]{\label{fig:MF_T1}\includegraphics[width=0.45\textwidth]{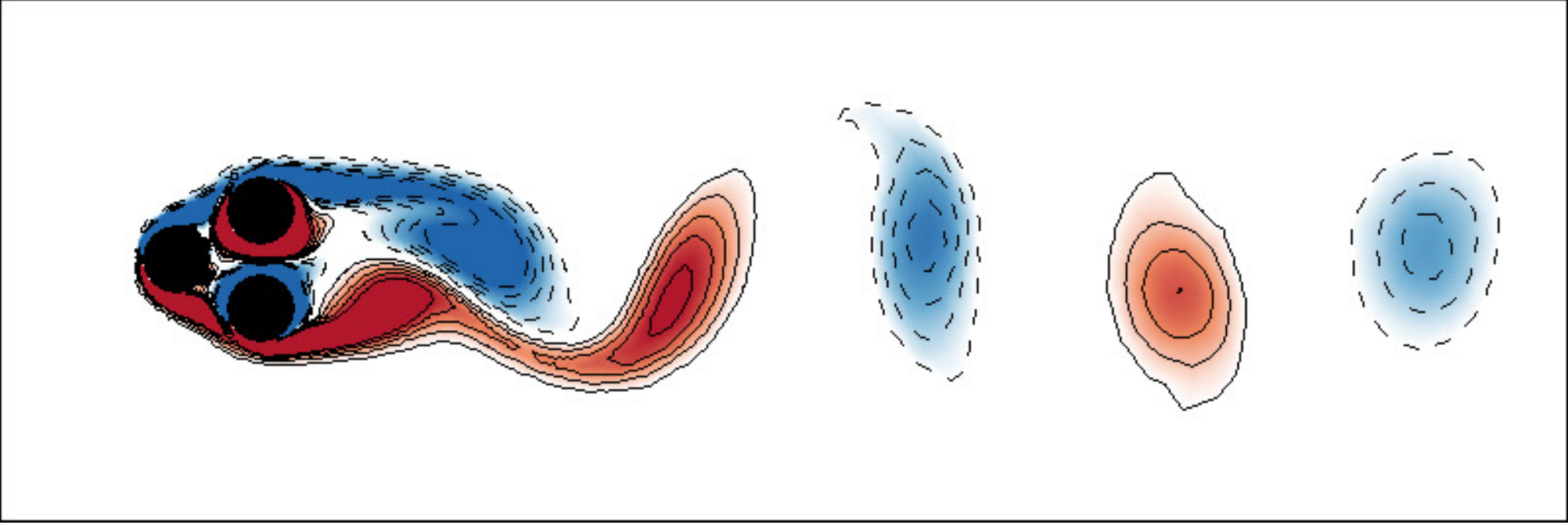}}%
\hfil
\subfloat[$t+2T_2/8$]{\label{fig:MF_T2}\includegraphics[width=0.45\textwidth]{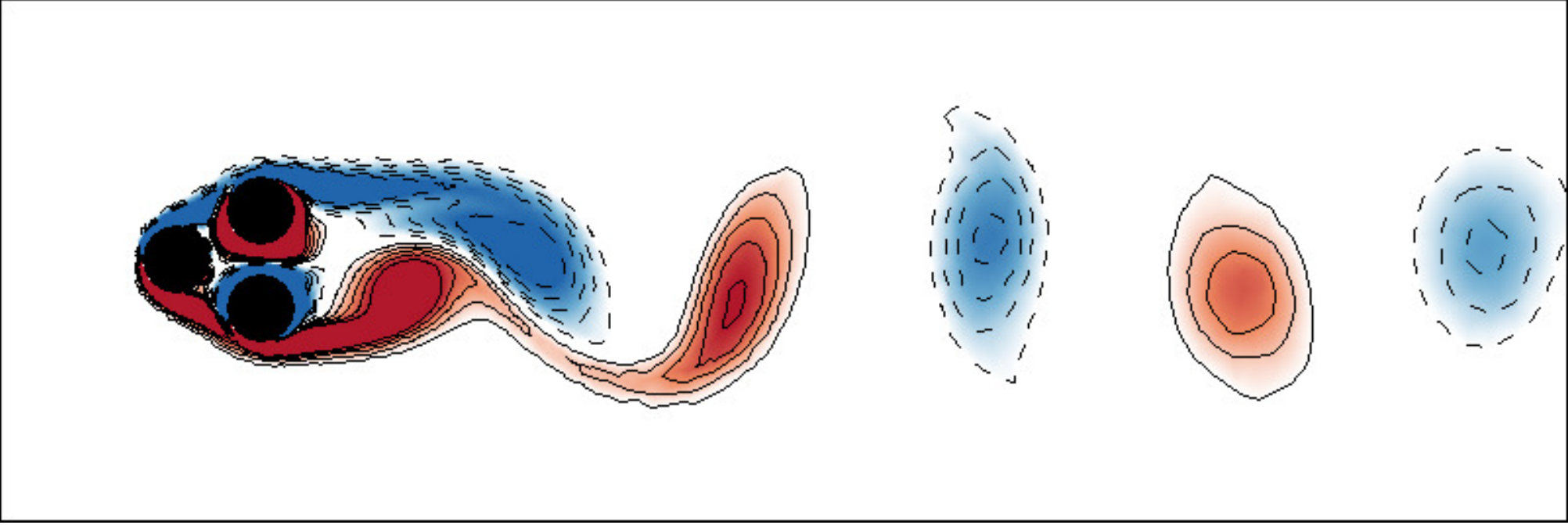}}%

\subfloat[$t+3T_2/8$]{\label{fig:MF_T3}\includegraphics[width=0.45\textwidth]{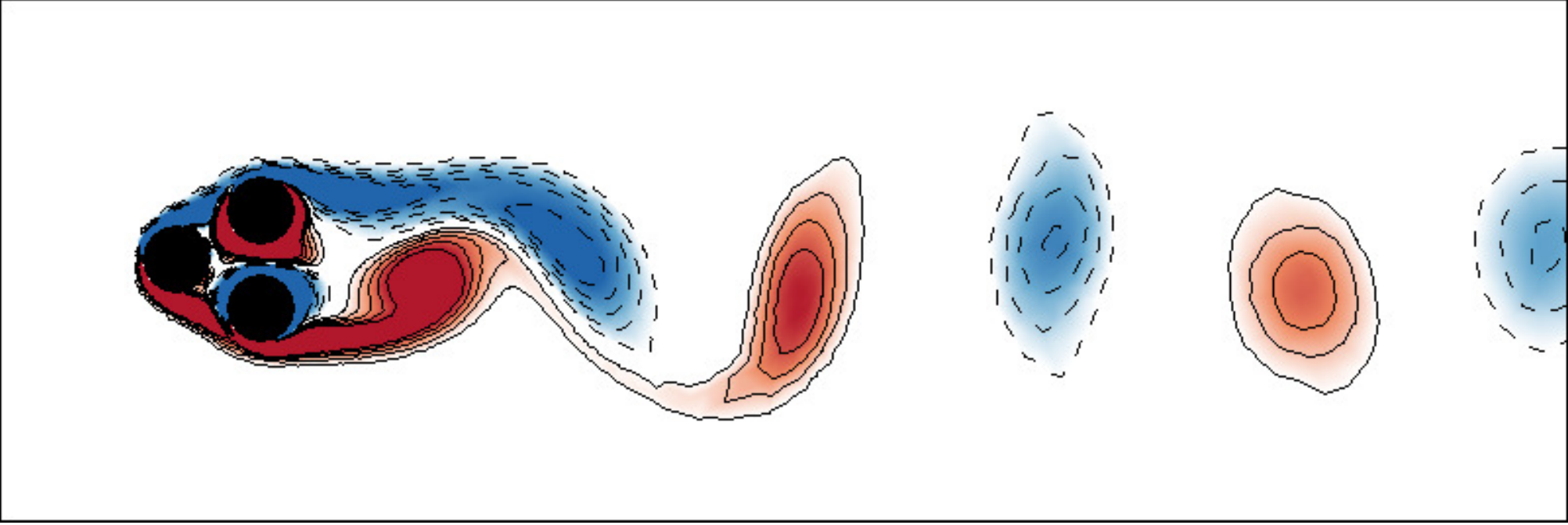}}%
\hfil
\subfloat[$t+4T_2/8$]{\label{fig:MF_T4}\includegraphics[width=0.45\textwidth]{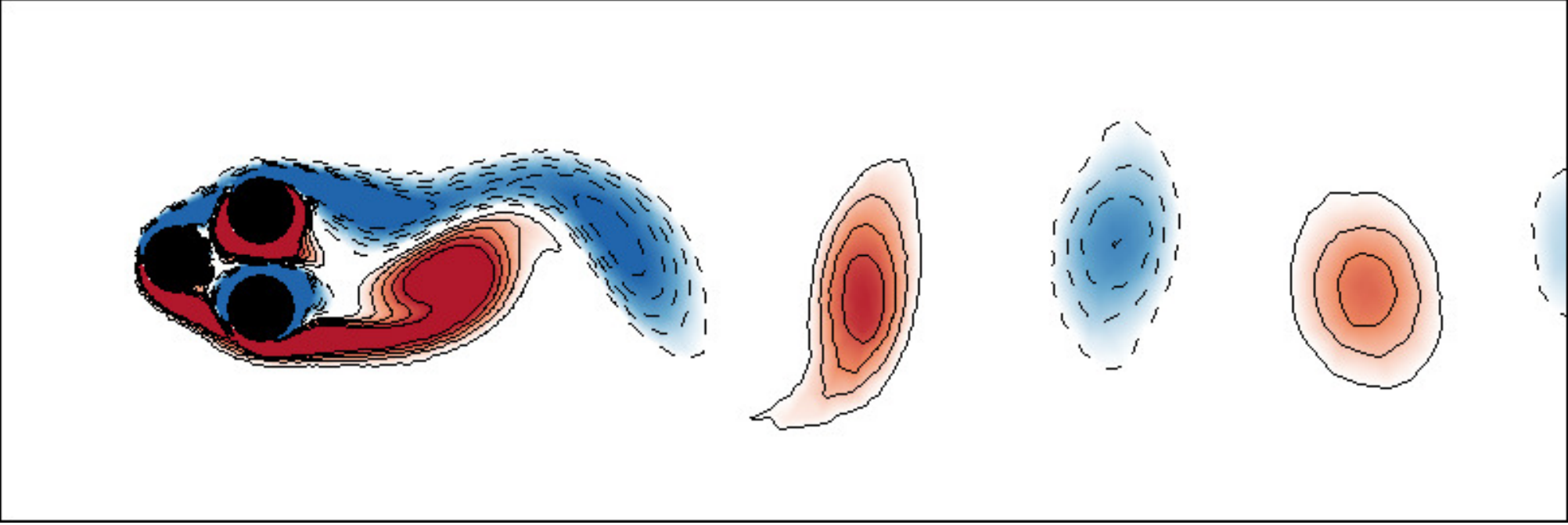}}%

\subfloat[$t+5T_2/8$]{\label{fig:MF_T5}\includegraphics[width=0.45\textwidth]{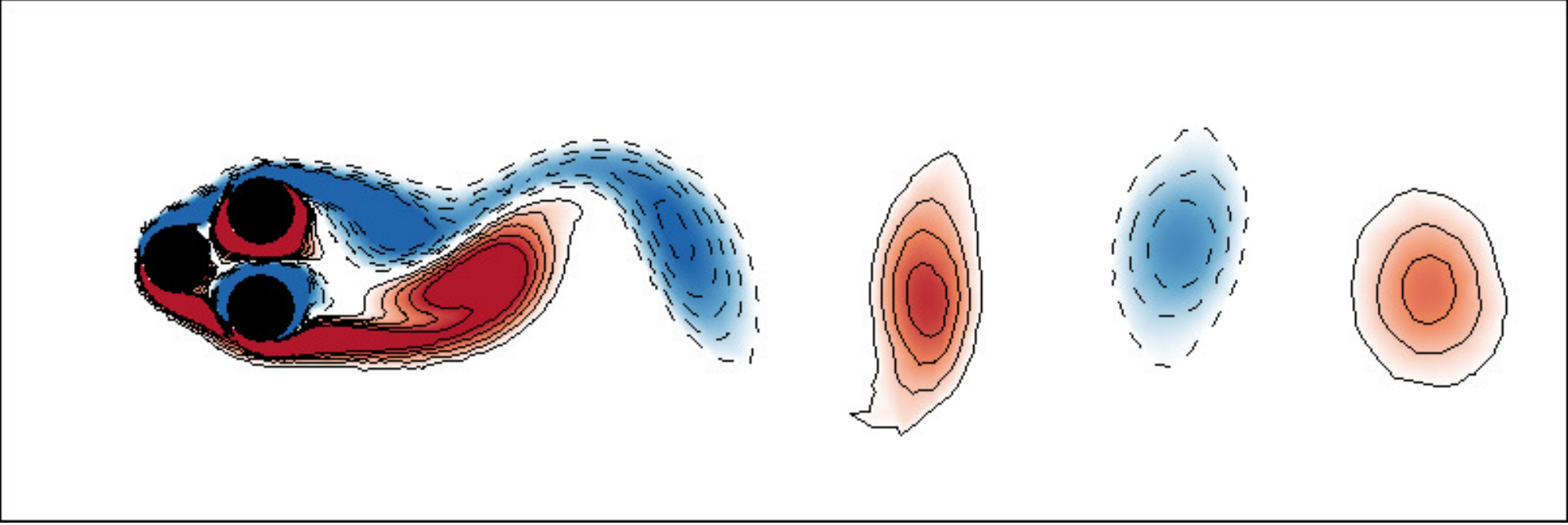}}%
\hfil
\subfloat[$t+6T_2/8$]{\label{fig:MF_T6}\includegraphics[width=0.45\textwidth]{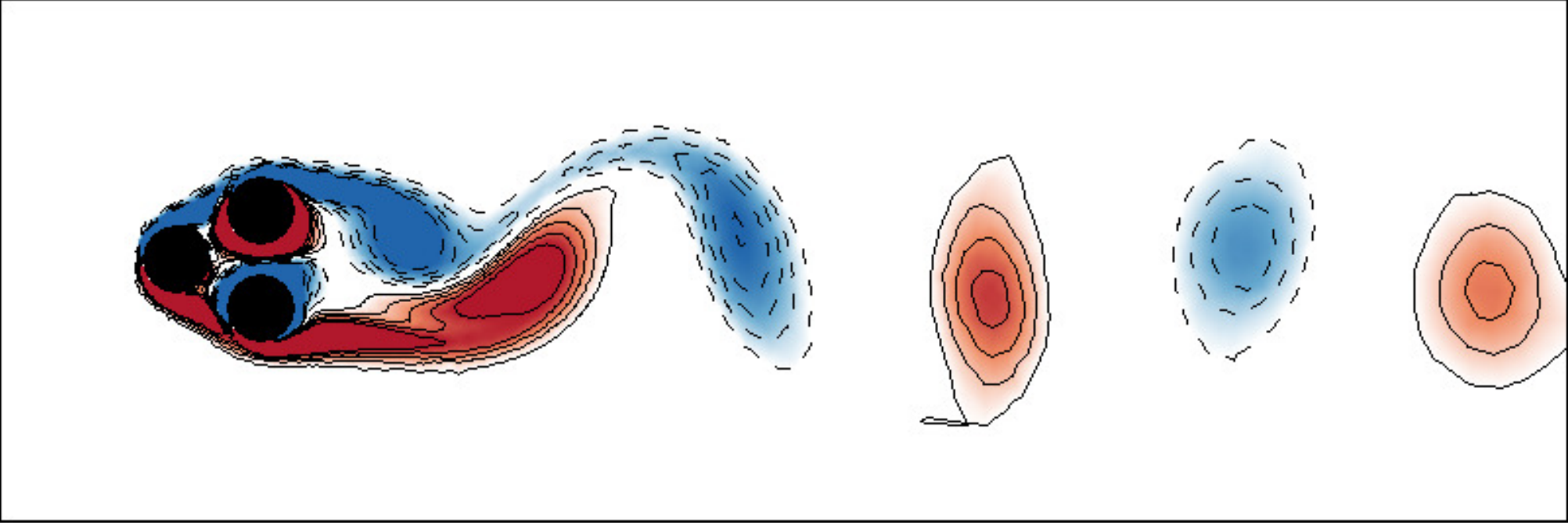}}%

\subfloat[$t+7T_2/8$]{\label{fig:MF_T7}\includegraphics[width=0.45\textwidth]{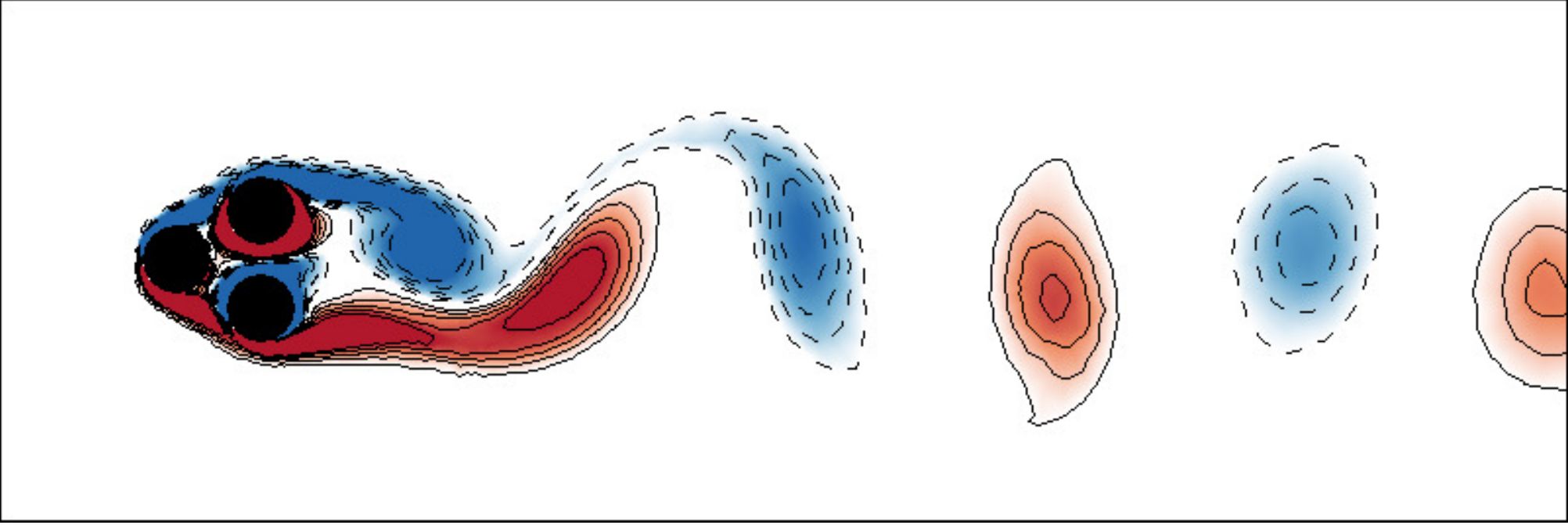}}%
\hfil
\subfloat[$t+T_2$]{\label{fig:MF_T8}\includegraphics[width=0.45\textwidth]{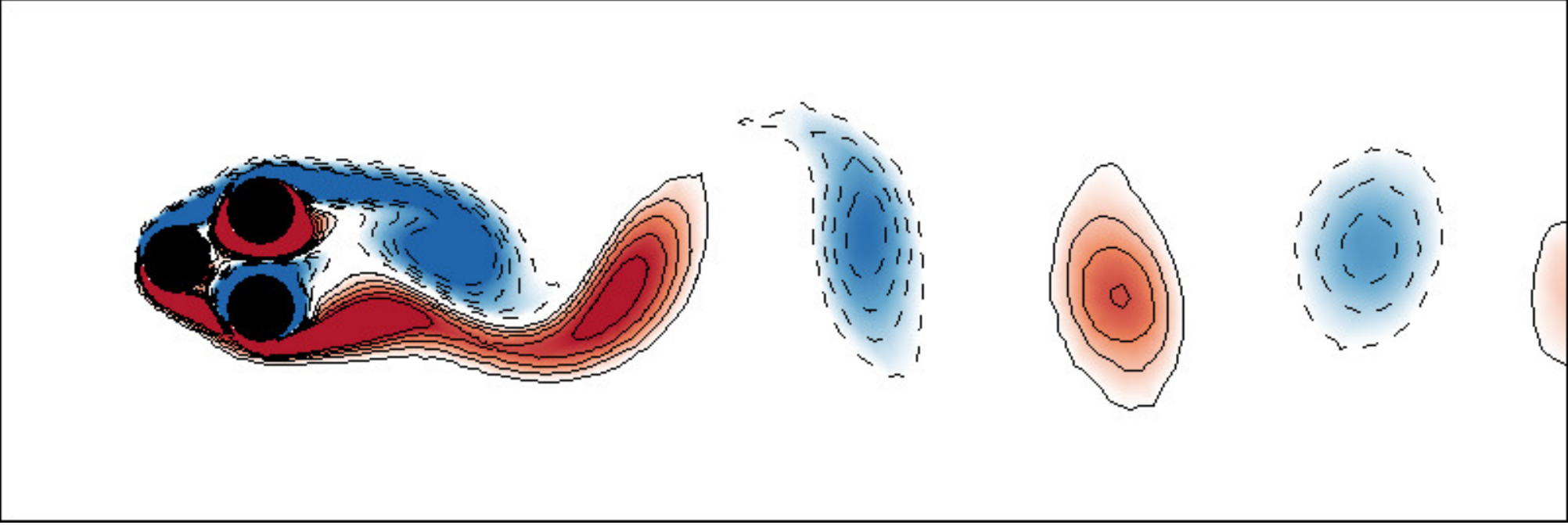}}%
\caption{\label{fig:MF_snap} Vorticity fields of the flow controlled by the best boat tailing solution. (a)-(f) Time evolution of the vorticity field in the last period of the 1000 time units simulation.
The color code is the same as figure~\ref{fig:unforced_flow}.
$T_2$ is the period associated to the frequency $f_2$.
}
\end{figure}
The mean value variation of the lift coefficient is, nonetheless, hardly visible on the snapshots in figure~\ref{fig:MF_snap}.
As the actuation is close to the best boat tailing control, the controlled flows are also similar.
By squinting ones eye, we notice that there is a small region, at the lower-back part of the front cylinder, with intense vorticity.
We can assume that this small vortex increased the local pressure and thus shifts the mean value of the lift positively.

\section{Feedback control law optimization}\label{sec:FB_optimization}
In this section, we allow feedback control laws by adding sensor signals as inputs.
Thus, equation~\eqref{Eq:ControlLaw} becomes: $\boldsymbol{b}(t) = \boldsymbol{K}(\boldsymbol{s}(t))$.
We choose a grid of nine sensor downstream measuring either $x$ or $y$ velocity components.
The coordinates of the sensors are $x =5, \>6.5, \> 8$ and $y=1.25,\> 0,\> -1.25$.
The six exterior sensors are $u$ sensors while $v$ sensors are chosen for the ones on the symmetry line $y=0$.
%-----------------------------------------------------------------------
\begin{table}[htb]
  \begin{center}
\def~{\hphantom{0}}
\begin{tabular}{cccc}
Sensor & $x$-coordinate & $y$-coordinate & Velocity component \\[3pt]
\midrule
$s_1$ & 5~ & ~1.25 & $u$ \\
$s_2$ & 6.5 & ~1.25 & $u$ \\
$s_3$ & 8~ & ~1.25 & $u$ \\
$s_4$ & 5~ & ~0~~ & $v$ \\
$s_5$ & 6.5 & ~0~~ & $v$ \\
$s_6$ & 8~ & ~0~~ & $v$ \\
$s_7$ & 5~ & -1.25 & $u$ \\
$s_8$ & 6.5 & -1.25 & $u$ \\
$s_9$ & 8~ & -1.25 & $u$ \\
\end{tabular}
\caption{\label{tab:sensor}Summary of sensor information.}
\end{center}
\end{table}
The information of sensors is summarized in table \ref{tab:sensor}.
Moreover, in order to take into account the convective nature of the flow,
we add time-delayed sensors as inputs of the control laws.
The delays are a quarter, half and three-quarters of the unforced natural period,
yielding following additional lifted sensor signals:
\begin{displaymath}
s_{i+9}(t) = s_i(t-T_0/4), \quad s_{i+18}(t)=s_i(t-T_0/2), \quad s_{i+27}(t) = s_i(t-3T_0/4).
\end{displaymath}
Hence, the dimension of the sensor vector $\boldsymbol{s}$ is  $9 \times 4 =36$ and $X \subset \mathbb{R}^{36}$.
The time-delayed sensors are also include as control inputs to mimic ARMAX-based control \citep{Herve2012jfm}.
\begin{table}[htb]
  \centering
   \begin{tabular}{>{\centering}p{2.5cm}>{\centering}p{6cm}>{\centering\arraybackslash}p{4cm}}
  Parameter & Description & Value\\
\midrule
   & \multirow{2}{*}{Function library} & $ F_2= \{+,-,\times,\div,$\\
   &  & $\mathrm{exp},\tanh,\sin,\cos,\mathrm{log}\}$\\
   $\boldsymbol{s}$  & Controller inputs & $s_i(t)$, $i=1,..,36$\\
   $\Nvar$ & Number of variable registers & $36+3=39$ \\
   $\Ncst$ & Number of constant registers & 10 \\
   $\Ninstrmax$ & Max. number of instructions & 50\\
  \hline
   $\Nps$ & Population size & 100\\
   $\Ng$ & Number of generations & 10\\
   $\Ntour$ & Tournament size & $7$\\
   $\Ne$ & Elitism & $1$\\
   $\Pcros$ & Crossover probability & 0.6 \\
   $\Pmut$ & Mutation probability & 0.3 \\
   $\Prep$ & Replication probability & 0.1 \\
\end{tabular}

  \caption{MLC parameters for feedback control optimization.}
  \label{tab:FB_LGPCparameters}
\end{table}
The same MLC parameters as for Sec.\ \ref{sec:MF_optimization} have been chosen.
The number of variable constants increased as we now have 36 sensor signals.
\begin{figure}[htb]
 \centerline{\includegraphics[width=0.75\linewidth]{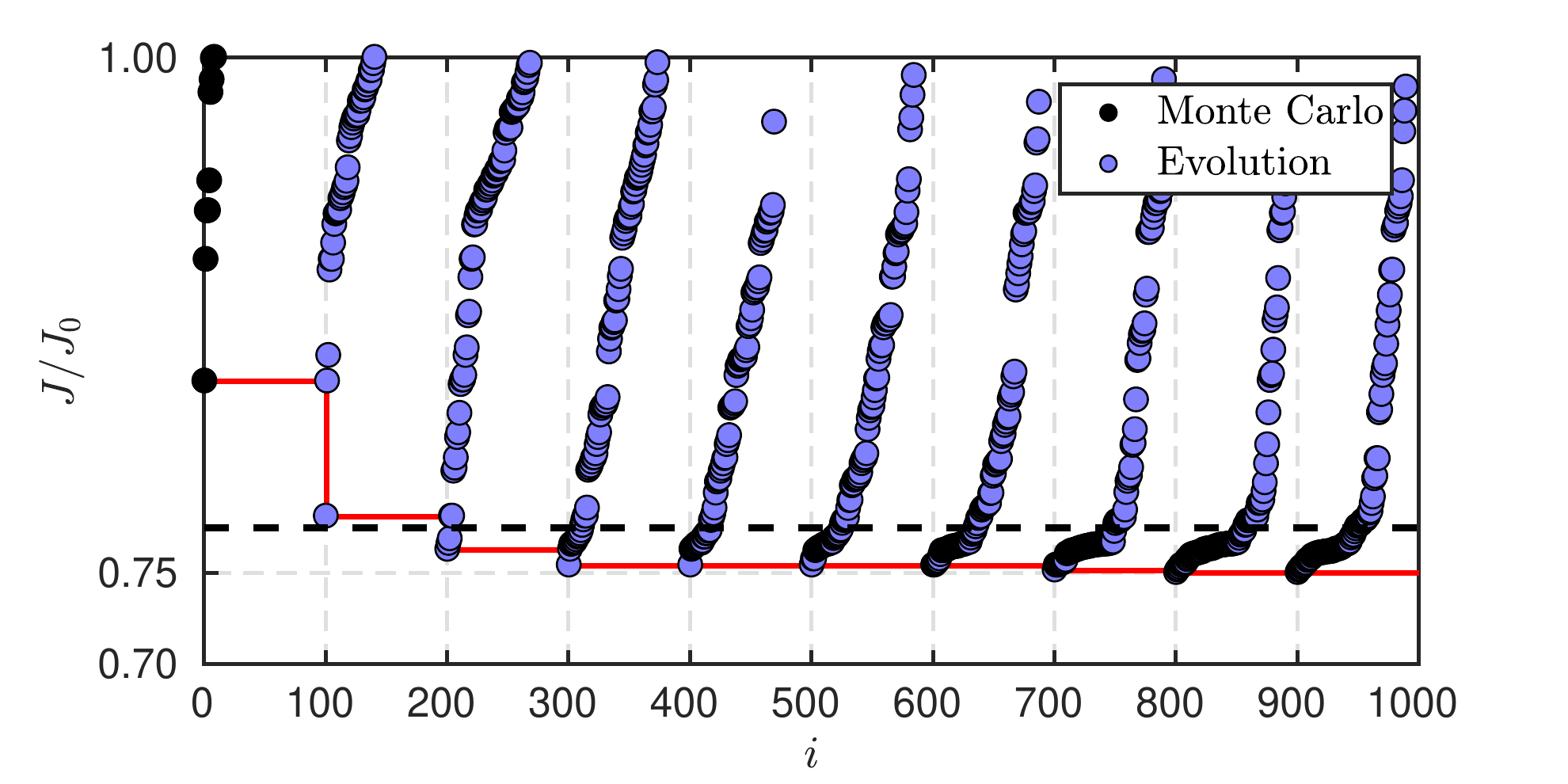}}
 \caption{Same as figure~\ref{fig:MF_Learning} but for the feedback optimization MLC.}
\label{fig:FB_Learning}
\end{figure}
Figure~\ref{fig:FB_Learning} shows cost of the individuals during the optimization process.
We note the same learning trend as for multi-frequency forcing optimization.
However in this case, the big jump appeared sooner, directly at the second generation.
From there, only small improvements are achieved.
We note also that the Monte Carlo step is less efficient as less individuals have reached a performance lower than the unforced one.
Moreover the best individual of the first generation has a cost of $J/J_0=0.84$ which is higher than in the multi-frequency forcing case.
This can be explained by the fact that as there are more inputs, the search space becomes larger thus the drop in performance of Monte Carlo.

The expression of the best control law $\boldsymbol{b}^{\rm FB}$ is:
\begin{equation}
\label{Eqn:bFB}
\begin{array}{lll}
b^{\rm FB}_1(t) &=&\log(s_{9}(t-T_0/2)),\\
b^{\rm FB}_2(t)  & =& \exp(0.18549\ \cfrac{s_{1}(t-4T_0/3)}{s_{9}(t)}) ,\\
b^{\rm FB}_3(t)  & = & -1.12724,\\     
J_{\rm FB} /J_0 &=& 0.7451.\\
\end{array}
 \end{equation}
MLC managed to successfully combined sensors signals, delayed sensor signals and nonlinear function to build a controller $b^{\rm FB}$ that reduces even further the cost function compared to the $b^{\rm MF}$ control law.
\begin{figure}[htb]
  \centerline{\includegraphics[width=0.75\textwidth]{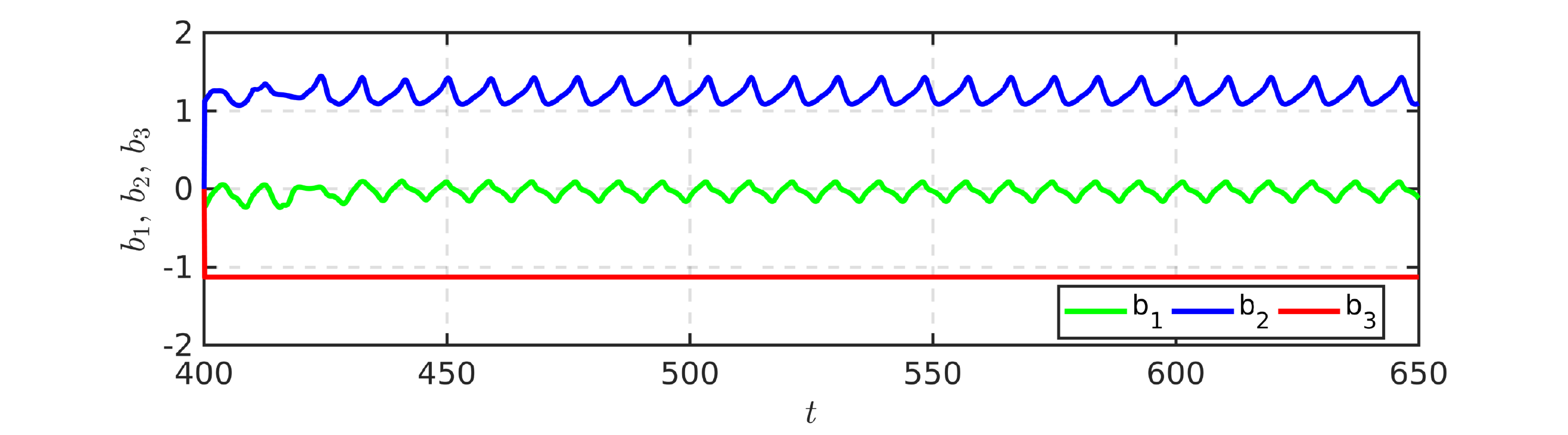}}
 \caption{Time series of the actuation command for the best feedback control law found with MLC.}
\label{fig:FB_Control}
\end{figure}
The actuation command resulting from $b^{\rm FB}$ is plotted in figure~\ref{fig:FB_Control}.
We note that the control resembles, again, the boat tailing solution, however this strategy is augmented by a phasor control for the front and bottom cylinder, meaning that the control of the front and bottom cylinder is directly related to the oscillatory dynamics of the flow \citep{Brunton2015amr}.
A spectral analysis shows that the main frequency of $b_2$ and $b_3$ are both $f_3=0.112$, suggesting that it is a direct feedback.

\begin{figure}[htb]
\centering
\subfloat[]{\label{fig:CL_FB}\includegraphics[trim=0 14.25pt -20pt 30pt, clip,width=0.4\textwidth]{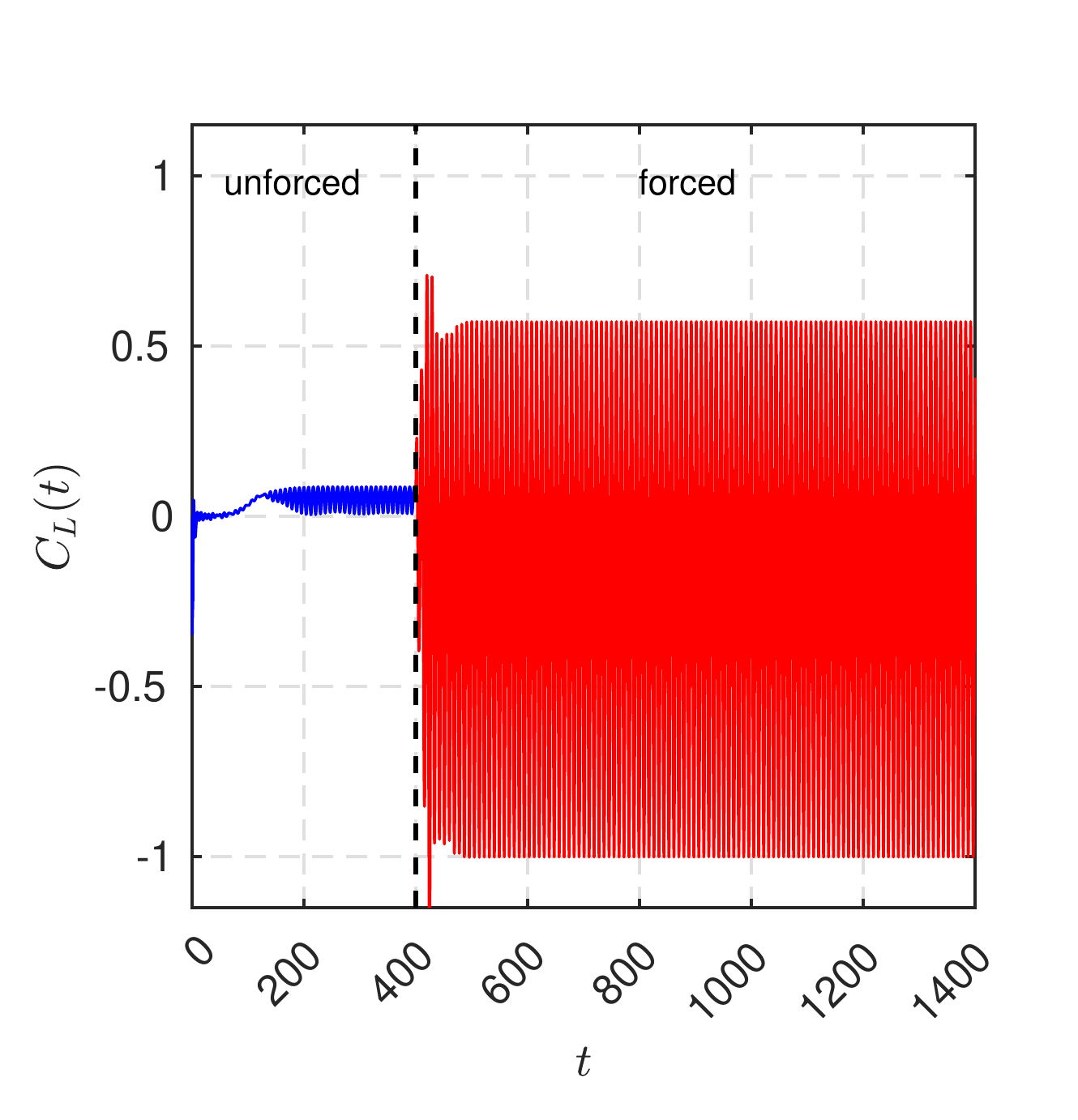}}%
\hfil
\subfloat[]{\label{fig:PP_FB}\includegraphics[trim=0 14.25pt -20pt 30pt, clip,width=0.4\textwidth]{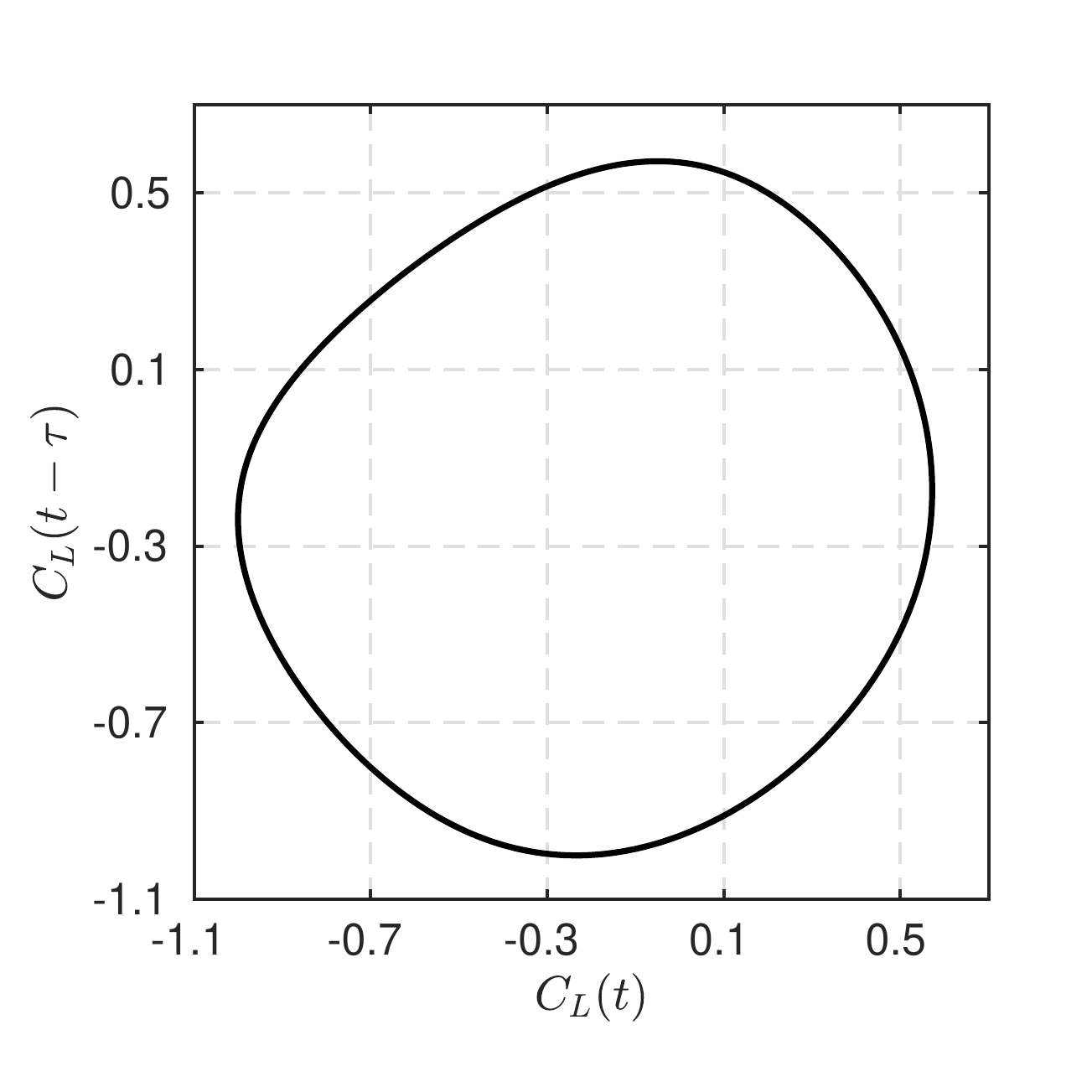}}%

\subfloat[]{\label{fig:DragPower_FB}\includegraphics[trim=0 14.25pt -20pt 30pt, clip,width=0.4\textwidth]{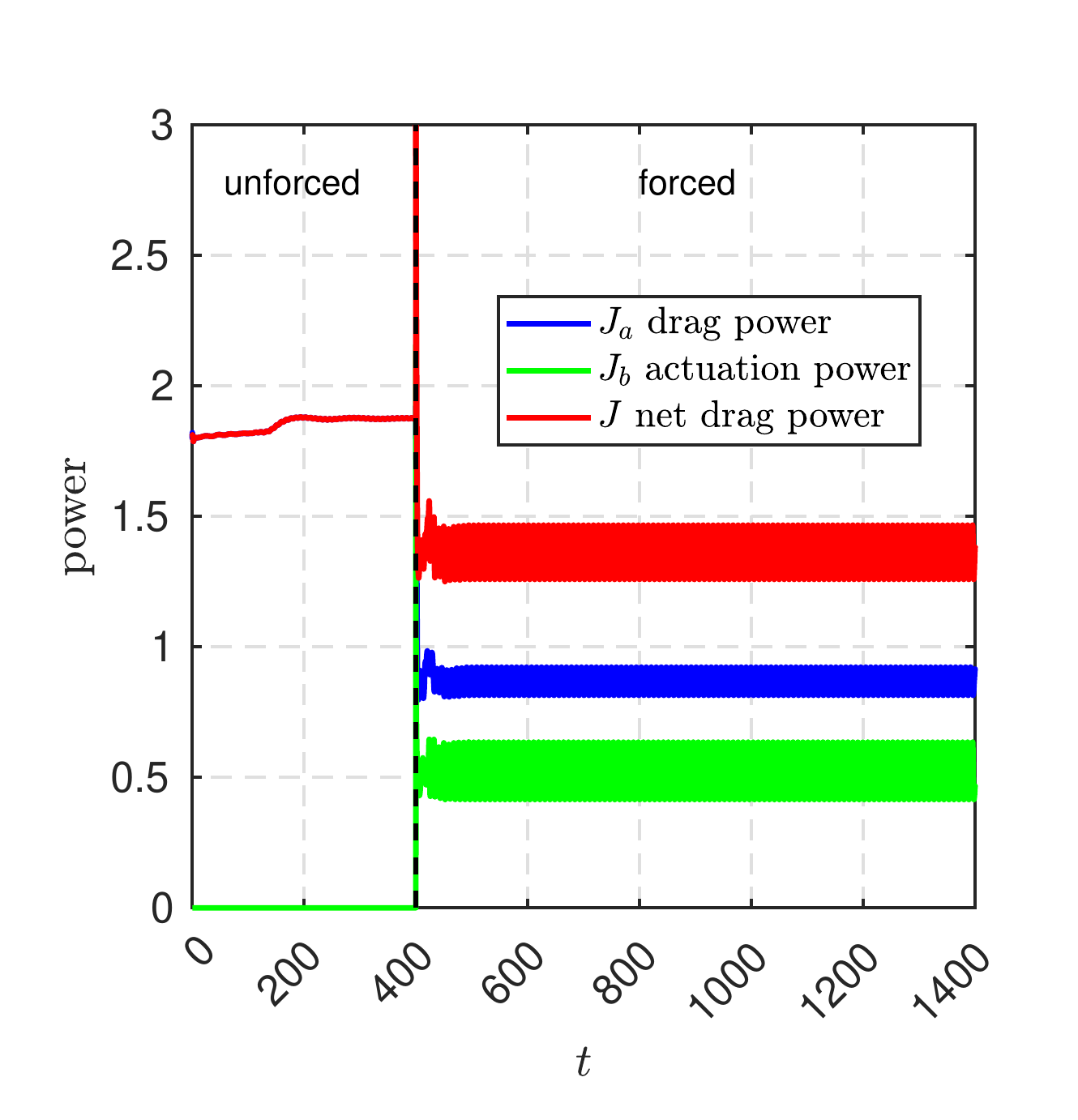}}%
\hfil
\subfloat[]{\label{fig:PSD_FB}\includegraphics[trim=0 14.25pt -20pt 30pt, clip,width=0.4\textwidth]{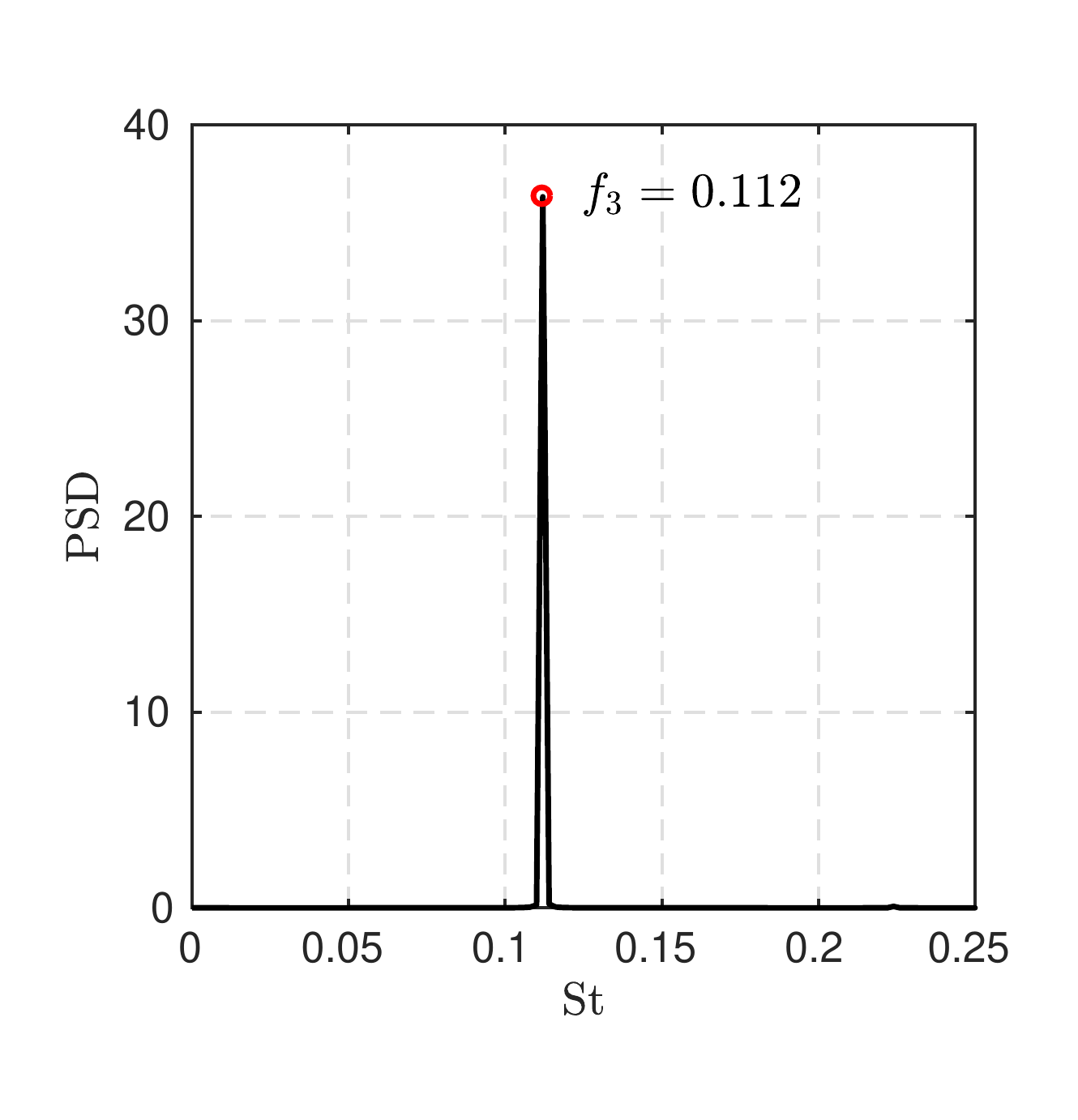}}%
\caption{\label{fig:FB_characteristics}Characteristics of the flow controlled by $\boldsymbol{b}^{\rm FB}$ starting from the steady solution ($t=0$).
The transient spans until $t \approx 400$. (a) Time evolution of the lift coefficient $C_L$, (b) phase portrait, (c)  time evolution of the drag power $J_a$ (blue), actuation power $J_b$ (green) and net drag power $J$ (red) and (d) Power Spectral Density (PSD) showing the frequency $f_3=0.112$. 
}
\end{figure}
Figure~\ref{fig:FB_characteristics} shows the characteristics of flow controlled by $b^{\rm FB}$.
The controlled flow is purely harmonic according to figure~\ref{fig:PSD_FB}, but the phase portrait, figure~\ref{fig:PP_FB}, is slightly deformed due to the third harmonic (not shown in the figure), even though its amplitude is imperceptible.
The amplitude oscillations of the lift coefficient also increased and the mean value is not null.
% Feedback control snapshots ---------------------------------------------
\begin{figure}[htb]%
\centering
\subfloat[$t+T_3/8$]{\label{fig:FB_T1}\includegraphics[width=0.45\textwidth]{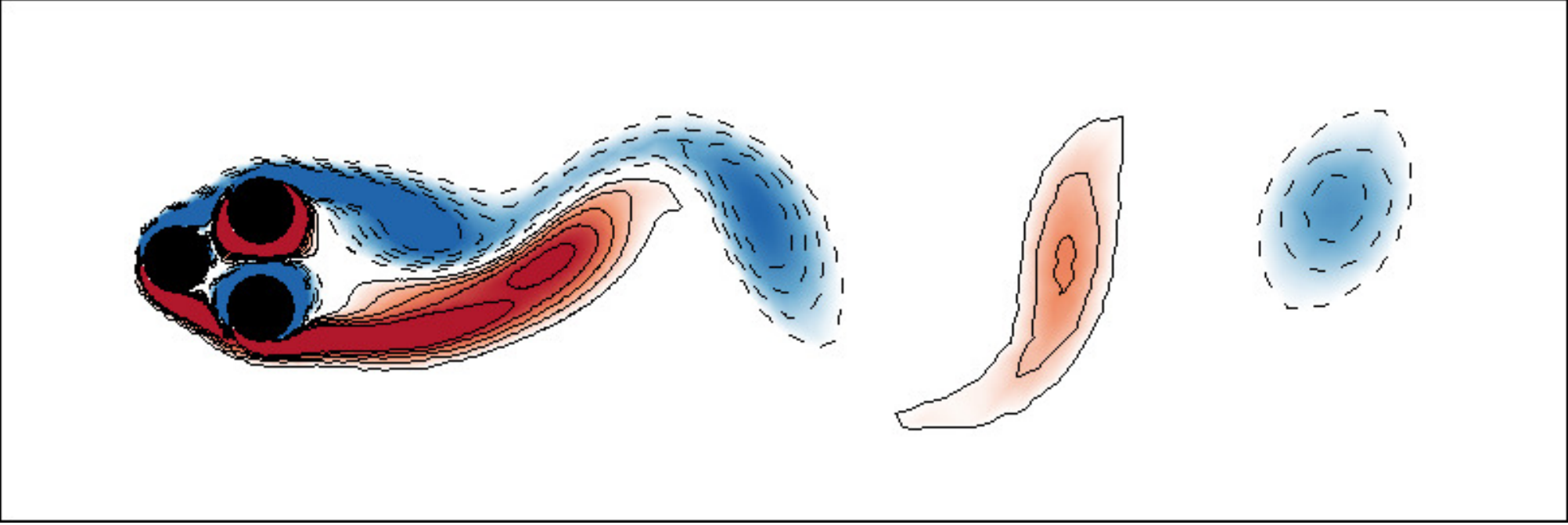}}%
\hfil
\subfloat[$t+2T_3/8$]{\label{fig:FB_T2}\includegraphics[width=0.45\textwidth]{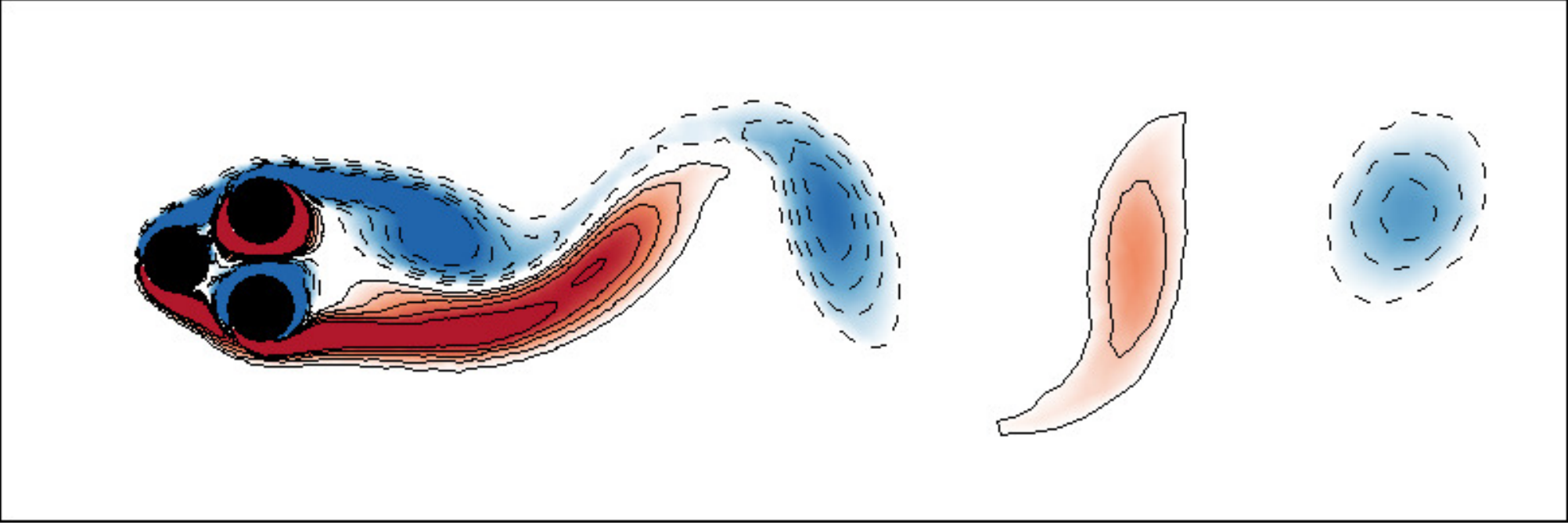}}%

\subfloat[$t+3T_3/8$]{\label{fig:FB_T3}\includegraphics[width=0.45\textwidth]{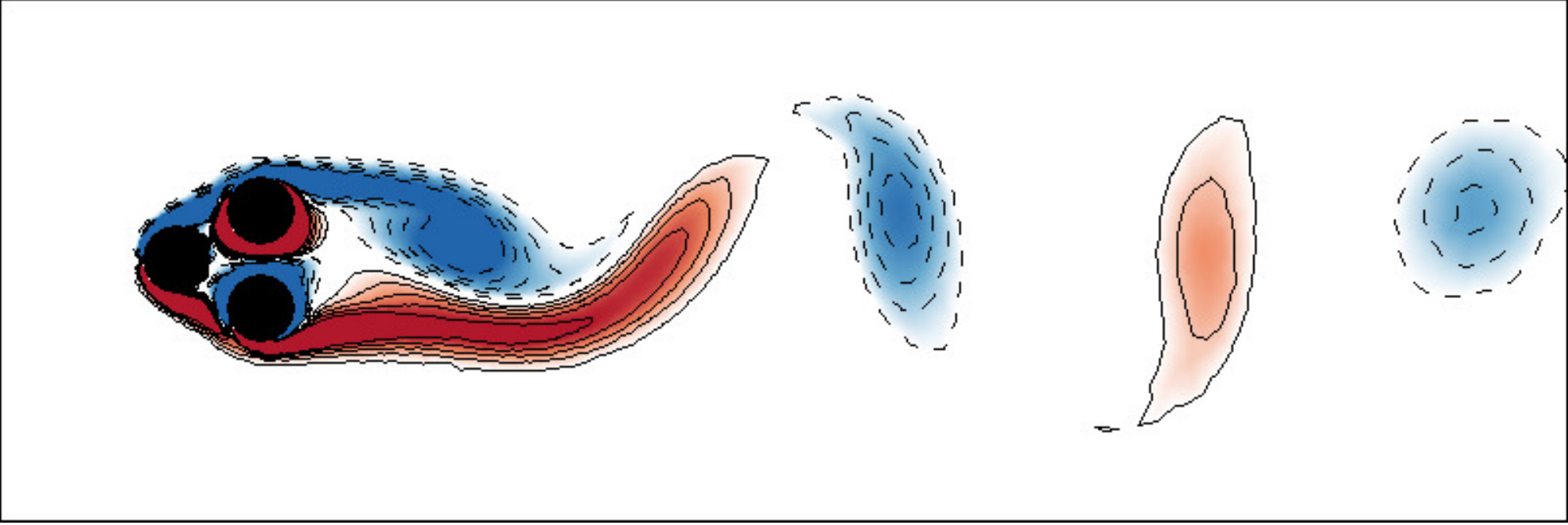}}%
\hfil
\subfloat[$t+4T_3/8$]{\label{fig:FB_T4}\includegraphics[width=0.45\textwidth]{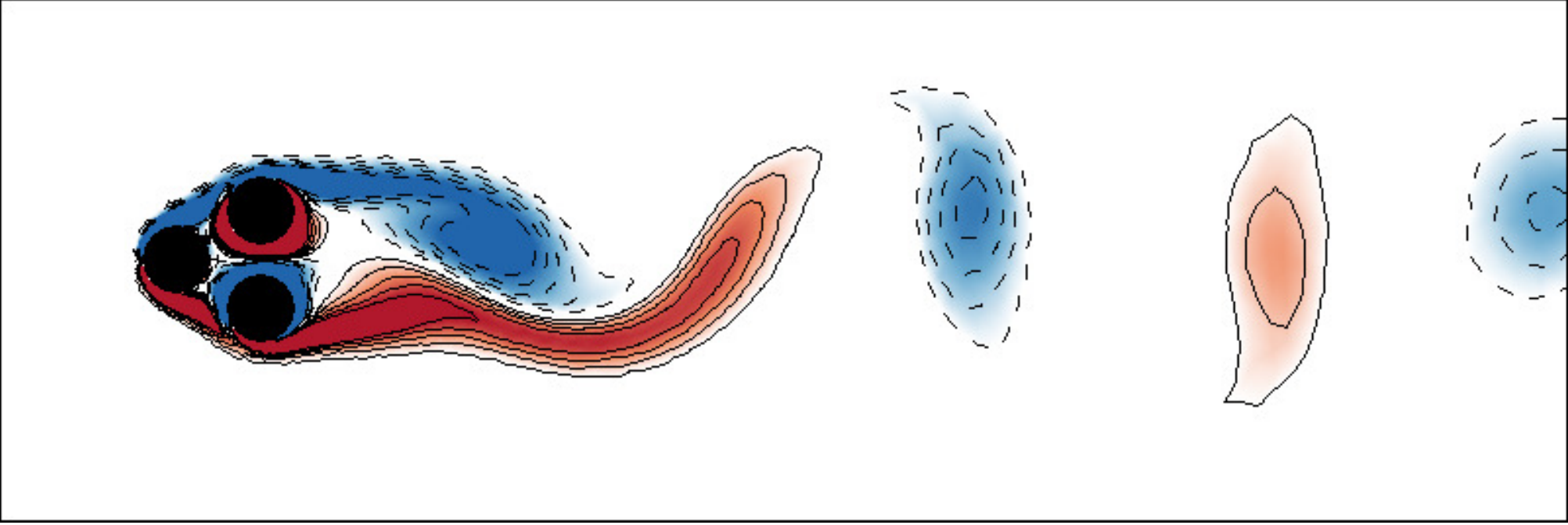}}%

\subfloat[$t+5T_3/8$]{\label{fig:FB_T5}\includegraphics[width=0.45\textwidth]{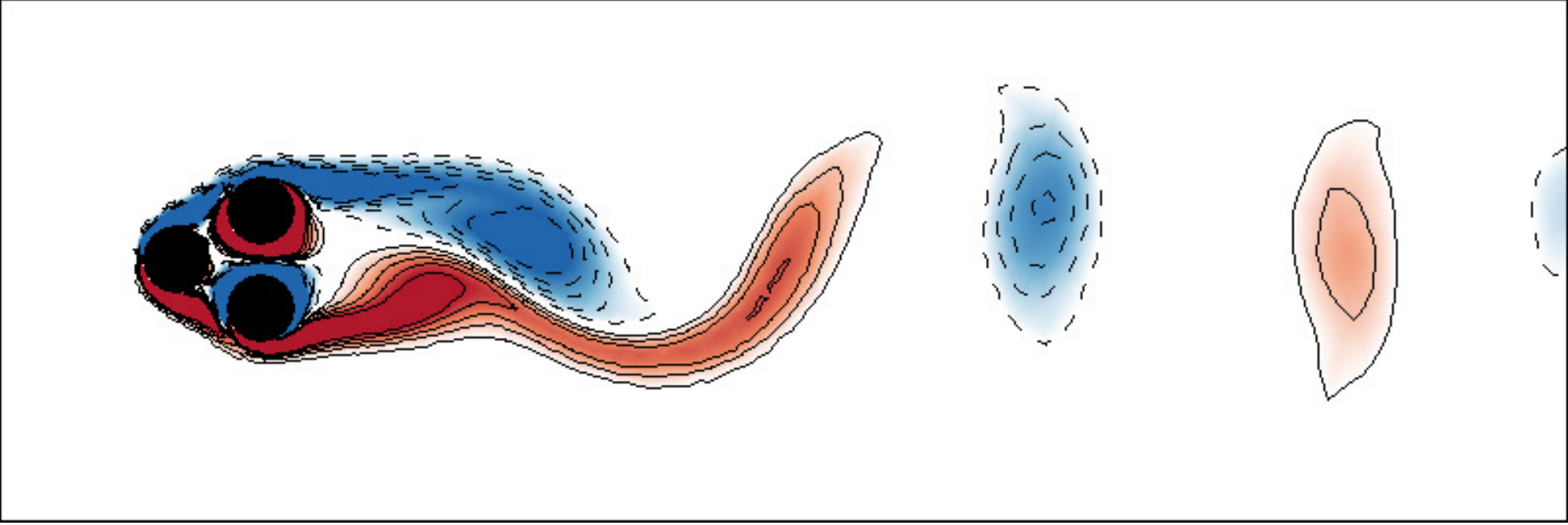}}%
\hfil
\subfloat[$t+6T_3/8$]{\label{fig:FB_T6}\includegraphics[width=0.45\textwidth]{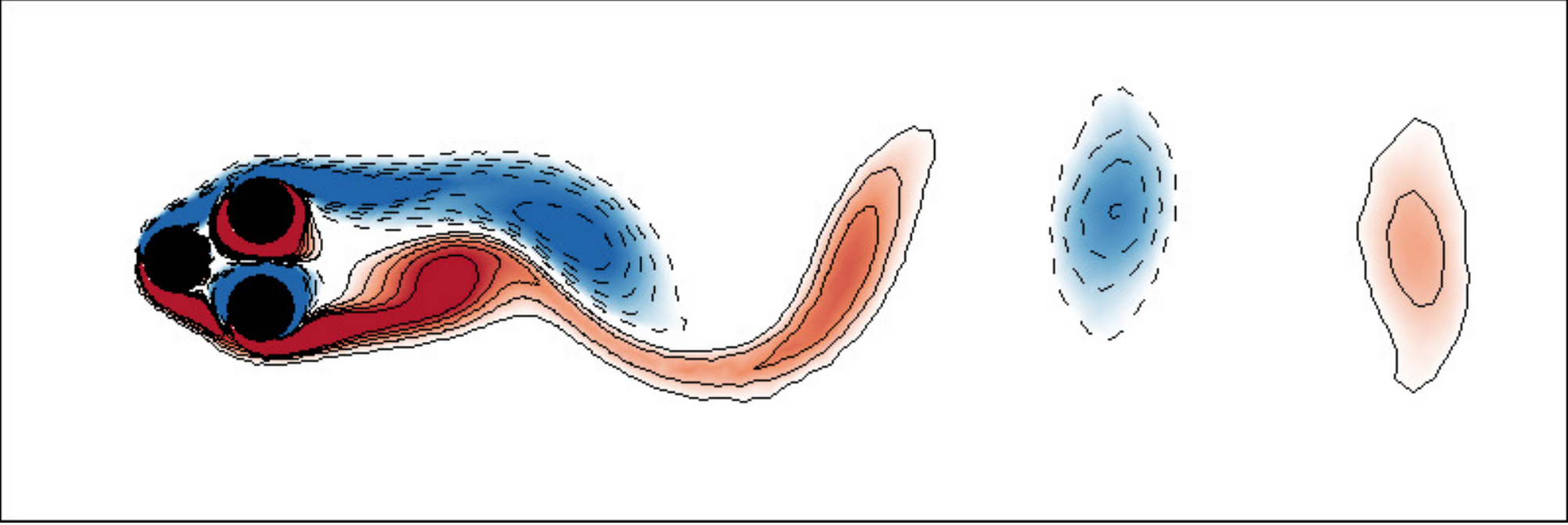}}%

\subfloat[$t+7T_3/8$]{\label{fig:FB_T7}\includegraphics[width=0.45\textwidth]{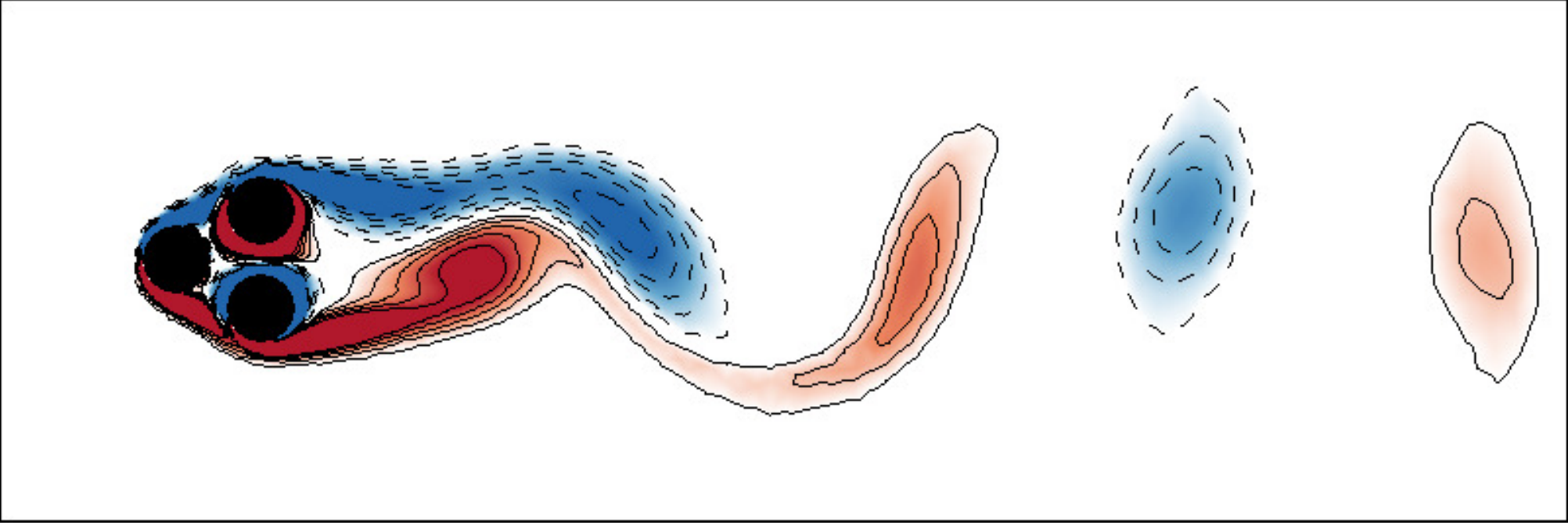}}%
\hfil
\subfloat[$t+T_3$]{\label{fig:FB_T8}\includegraphics[width=0.45\textwidth]{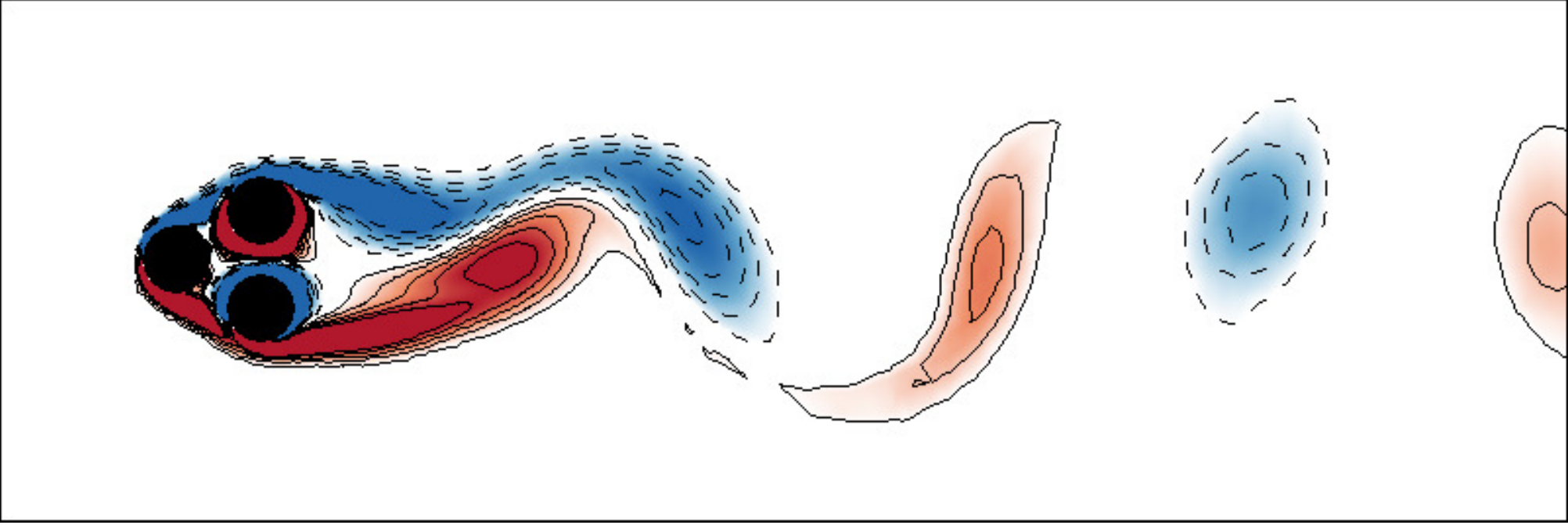}}%
\caption{\label{fig:FB_snap} Vorticity fields of the flow controlled by the feedback control law derived by MLC (a)-(f) Time evolution of the vorticity field in the last period of the 1000 time units simulation.
The color code is the same as figure~\ref{fig:unforced_flow}.
$T_3$ is the period associated to the frequency $f_3$.
}
\end{figure}
%% Figures : Feedback characteristics ------------------------------
The kinematics of the flow, figure~\ref{fig:FB_snap}, show that the near jet disappeared and the length  of the recirculation decreased like the two last control strategies.
However, we note that the recirculation bubble is a bit less reduced.
Also, the vortices stay attached longer before shedding.
This is especially true for the bottom part where the positive vortex stretches unusually, see figure~\ref{fig:FB_T3}, \ref{fig:FB_T4}, \ref{fig:FB_T5} and \ref{fig:FB_T6}.
The intensity of the positive vortices is also lesser than the previous control.
This can be explained by the re-energization of the shear layers, especially the bottom one, due to the periodic component of the forcing, like \citet{Protas2004pf}.
The rotation of the front cylinder has been reported in other studies, such as \citet{CornejoMaceda2019pamm}, but its effect is not yet fully understood.

\section[General control optimization]{General control law optimization: multi-frequency and feedback control}\label{sec:HB_optimization}
\begin{table}[htb]
  \centering
   \begin{tabular}{>{\centering}p{2.5cm}>{\centering}p{6cm}>{\centering\arraybackslash}p{5cm}}
  Parameter & Description & Value\\
\midrule
   & \multirow{2}{*}{Function library} & $ F_2= \{+,-,\times,\div,$\\
   &  & $\mathrm{exp},\tanh,\sin,\cos,\mathrm{log}\}$\\
    $\boldsymbol{s}$, $\boldsymbol{h}$  & \multirow{2}{*}{Controller inputs} & $s_i(t)$,  $i=1..36$\\
   &  & $h_i(t)$, $i=-4,..,-1,1,..4$\\   
   $\Nvar$ & Number of variable registers & $8+36+3=47$ \\
   $\Ncst$ & Number of constant registers & 10 \\
   $\Ninstrmax$ & Max. number of instructions & 50\\
  \hline
   $\Nps$ & Population size & 100\\
   $\Ng$ & Number of generations & 10\\
   $\Ntour$ & Tournament size & $7$\\
   $\Ne$ & Elitism & $1$\\
   $\Pcros$ & Crossover probability & 0.6 \\
   $\Pmut$ & Mutation probability & 0.3 \\
   $\Prep$ & Replication probability & 0.1 \\
\end{tabular}

  \caption{MLC parameters for general optimization including multi-frequency forcing and feedback control.}
  \label{tab:HB_LGPCparameters}
\end{table}
In this last section, we run a hybrid optimization allowing both multi-frequency forcing and feedback control.
We have seen in Sec.\ \ref{sec:MF_optimization}, that periodic forcing has not been selected to control the fluidic pinball, this may be related to the difficulty to build the proper frequency for control.
By adding flow information, we can expect MLC to build an open-loop periodic forcing modulated by a sensor signal and thus enable a richer control.
Such approach has been successfully employed to reduce the recirculation bubble of a back-ward facing step at Reynolds number $\Rey=31500$ in \citet{Chovet2017ifac}.
Then, we allow $\boldsymbol{s}$ and $\boldsymbol{h}$ as inputs of the controller and equation~\eqref{Eq:ControlLaw} becomes: $\boldsymbol{b}(t) = \boldsymbol{K}(\boldsymbol{s}(t),\boldsymbol{h}(t))$.
The MLC parameters are summarized in table~\ref{tab:HB_LGPCparameters}.
\begin{figure}[htb]
 \centerline{\includegraphics[width=0.75\linewidth]{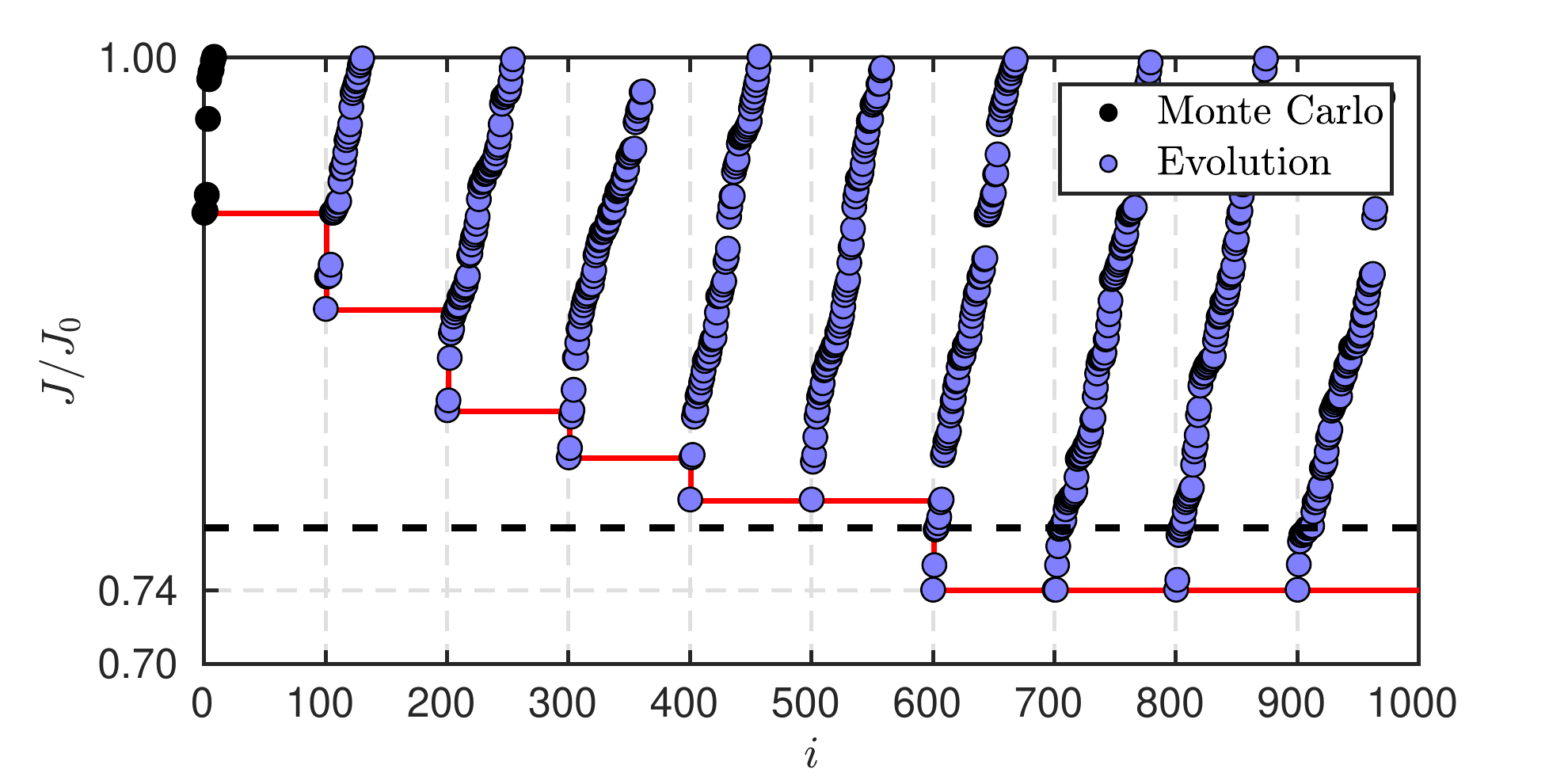}}
 \caption{Same as figure~\ref{fig:MF_Learning} but for the hybrid optimization MLC.}
\label{fig:HB_Learning}
\end{figure}
Figure~\ref{fig:HB_Learning} illustrates the learning process for the hybrid optimization.
First, Monte Carlo sampling struggles to find a good individual and after 100 random evaluations, the cost of the best individual is only $J/J_0=0.92$.
As detailed in Sec.\ \ref{sec:FB_optimization}, this can be explained by the adding of extra inputs, which enlarges the search space.
Contrary to the multi-frequency optimization and feedback control optimization, we note that significant and  regular improvements are made at each generation until reaching a plateau at $J_{\rm HB}=0.7363$ after seven generations.
For all generations the distribution of the individuals looks linear as opposed to the two previous optimizations where there was an accumulation of good individuals in the final generations.
This may be explained by the fact that as new and more efficient individuals are built at each generation, the is still a lot of diversity in the population.
When the learning is slowed down, then the best individual takes over the population thanks to replication.
From there, we enter a phase of fine tuning of the control law with only small improvements and thus an accumulation of good individuals in the generation.
Such behavior is likely to be observed if more generations were computed.

\begin{figure}[htb]
  \centerline{\includegraphics[width=0.75\textwidth]{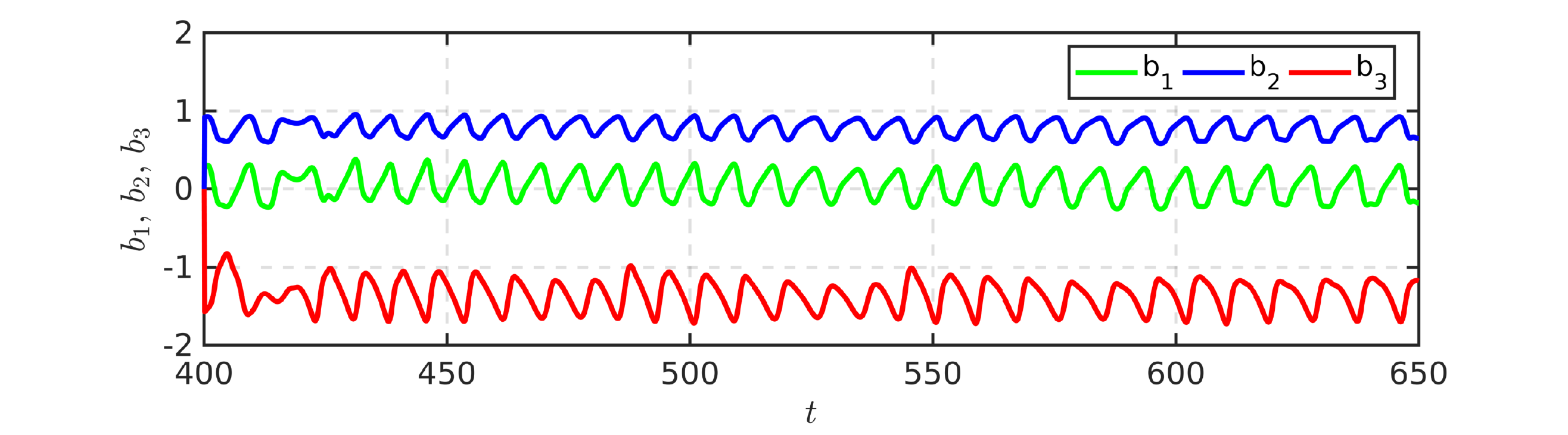}}
 \caption{Time series of the actuation command for the best feedback control law found with MLC.}
\label{fig:HB_Control}
\end{figure}
The final control law $\boldsymbol{b}^{\rm HB}$ reads:
\begin{equation}
\label{Eqn:bHB}
\begin{array}{lll}
b^{\rm HB}_1(t) &=& \cos(\cos(s_{8}(t-T_0/2)) + 0.88123),\\
b^{\rm HB}_2(t)  & =&\cos(\cos(s_{8}(t-T_0/2))) ,\\
b^{\rm HB}_3(t)  & = &-0.36574 - s_2(t) ,\\     
J_{\rm HB}/J_0 &=& 0.7363\\
\end{array}
 \end{equation}
The control built includes sensor information, a nonlinear function, $\cos$, but no open-loop periodic function.
The three components contain feedback information.
The time series of this control are plotted in figure~\ref{fig:HB_Control}.
\begin{figure}[htb]%
\centering
\subfloat[]{\label{fig:CL_HB}\includegraphics[trim=0 14.25pt -20pt 30pt, clip,width=0.4\textwidth]{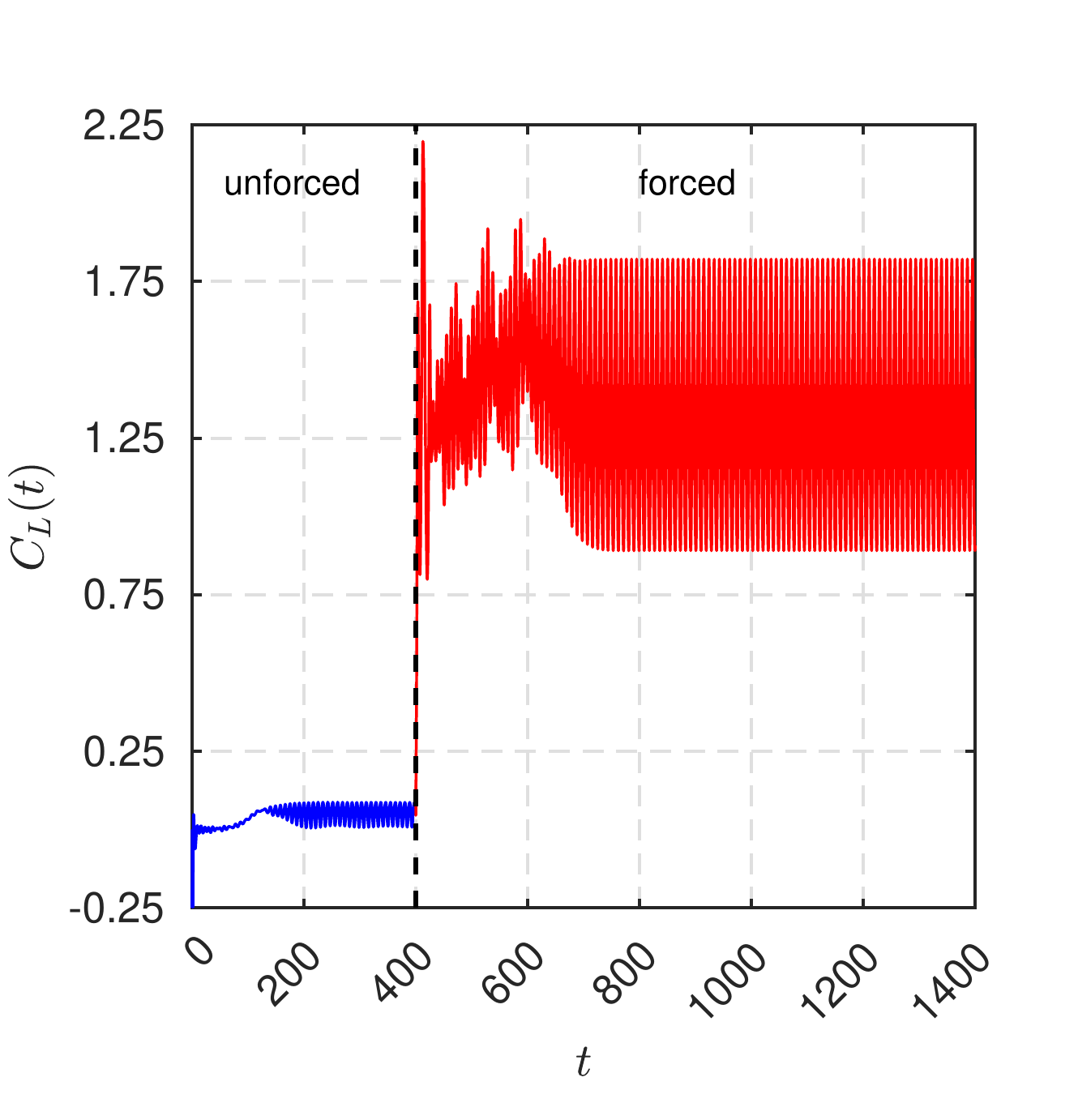}}%
\hfil
\subfloat[]{\label{fig:PP_HB}\includegraphics[trim=0 14.25pt -20pt 30pt, clip,width=0.4\textwidth]{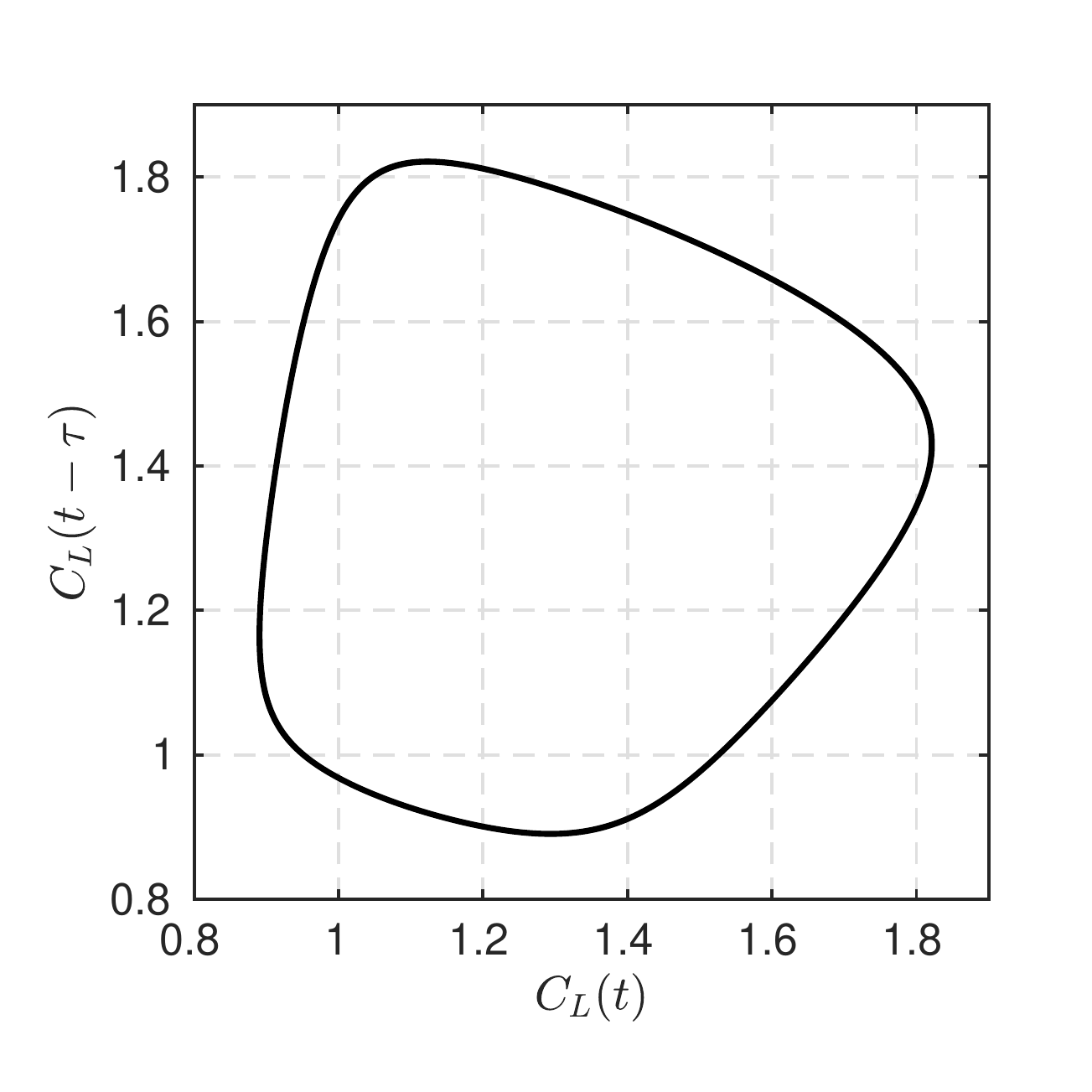}}%

\subfloat[]{\label{fig:DragPower_HB}\includegraphics[trim=0 14.25pt -20pt 30pt, clip,width=0.4\textwidth]{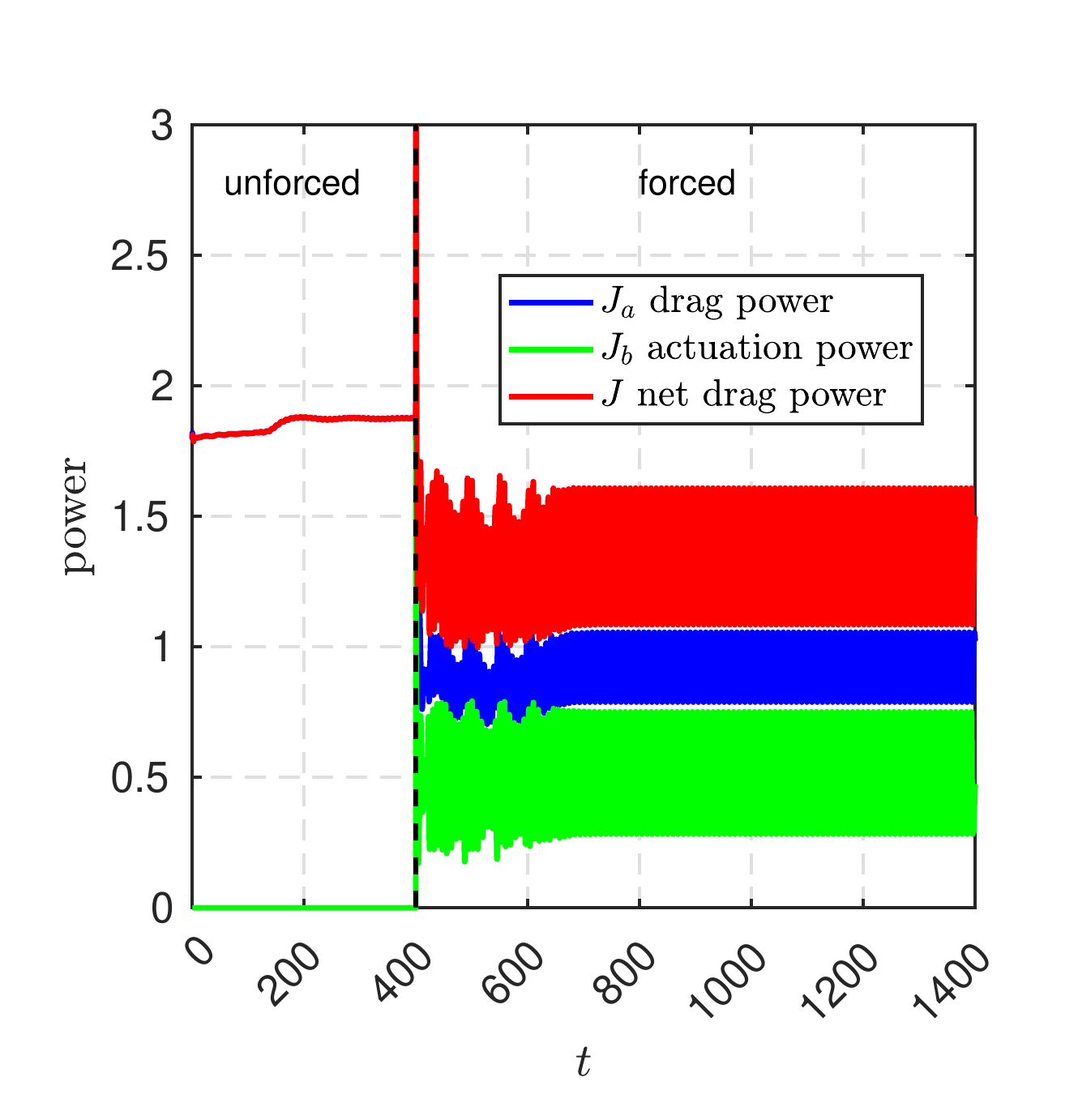}}%
\hfil
\subfloat[]{\label{fig:PSD_HB}\includegraphics[trim=0 14.25pt -20pt 30pt, clip,width=0.4\textwidth]{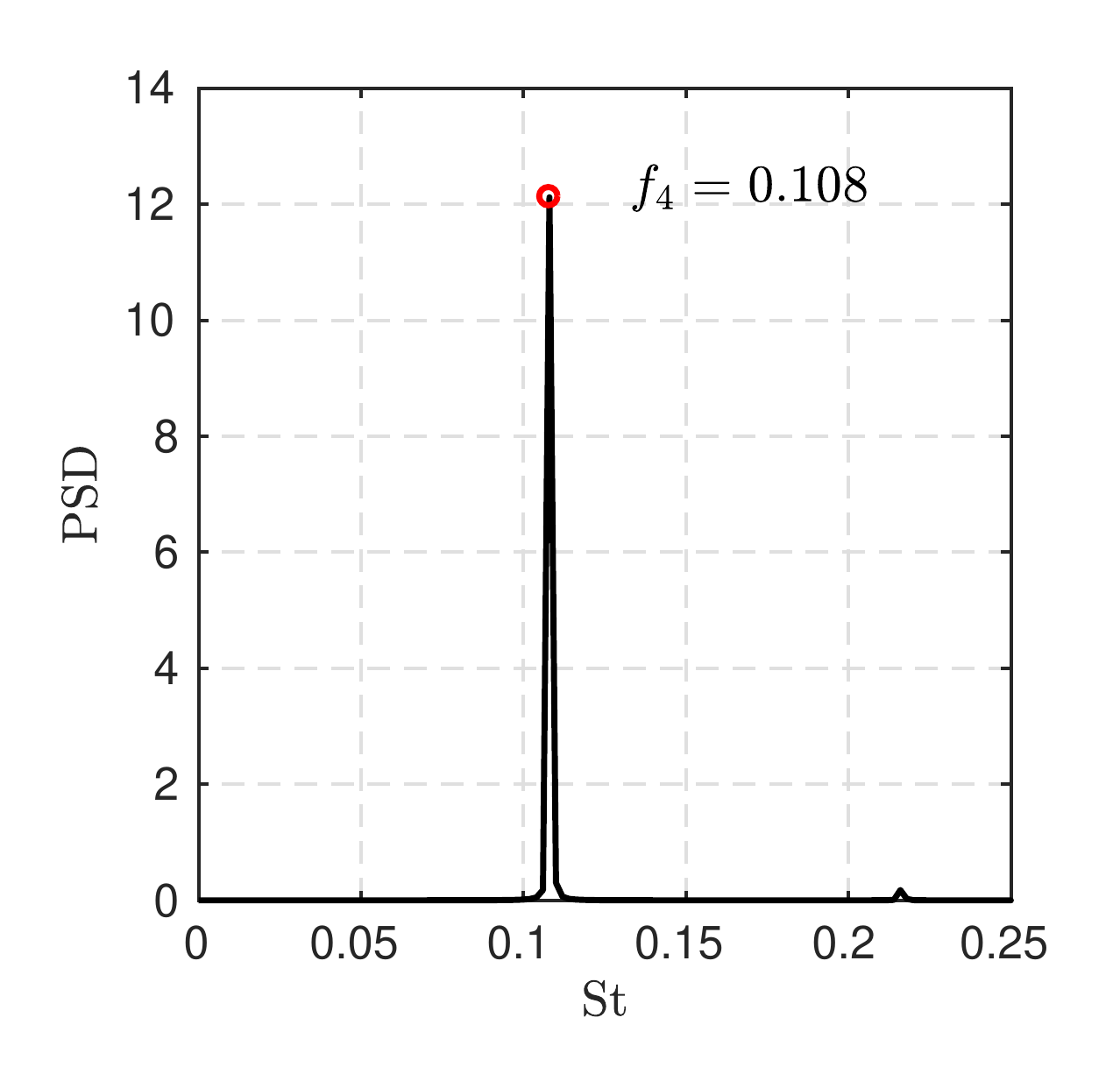}}%
\caption{\label{fig:HB_characteristics}
Characteristics of the flow controlled by $\boldsymbol{b}^{\rm HB}$ starting from the steady solution ($t=0$).
The transient spans until $t \approx 400$. (a) Time evolution of the lift coefficient $C_L$, (b) phase portrait, (c)  time evolution of the drag power $J_a$ (blue), actuation power $J_b$ (green) and net drag power $J$ (red) and (d) Power Spectral Density (PSD) showing the frequency $f_4=0.108$ and its first harmonic. 
}
\end{figure}
The mean values of all controllers are in line with a boat tailing configuration, however there is a non-negligible oscillatory component for all actuations.
The three cylinders are in direct feedback as the dominant frequency of the actuation commands are $f_4=0.108$, the main frequency of the flow.
So far, $b^{\rm HB}$ is the control that reduces the most the cost function with $J_{\rm HB}/J_0 = 0.7363$.
However, in figure~\ref{fig:CL_HB} and \ref{fig:DragPower_HB}, we notice that the transient extends until $\approx 700$ convective time units.
The cost of the controlled flow computed on the post-transient regime is $J/J_0=0.7369$, showing that the extended transient only brings a negligible improvement.
We note that taking into account a longer time-window may enable solutions with long transients.
% Hybrid control snapshots ---------------------------------------------
\begin{figure}[htb]
\centering
\subfloat[$t+T_4/8$]{\label{fig:HB_T1}\includegraphics[width=0.45\textwidth]{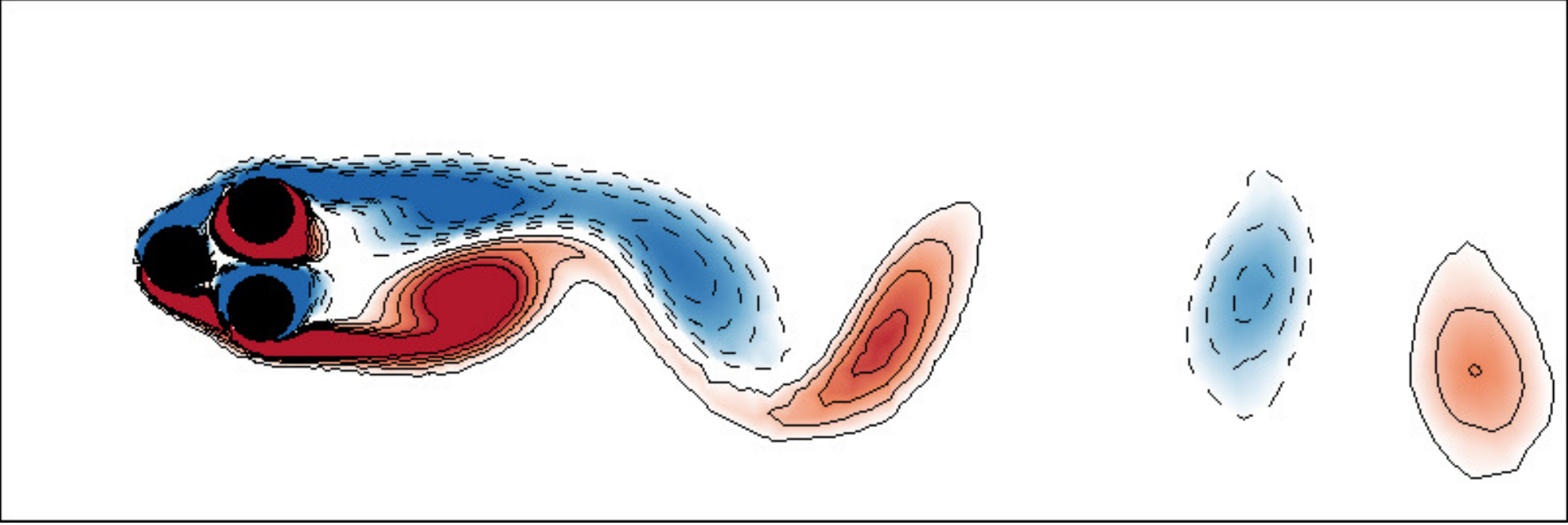}}%
\hfil
\subfloat[$t+2T_4/8$]{\label{fig:HB_T2}\includegraphics[width=0.45\textwidth]{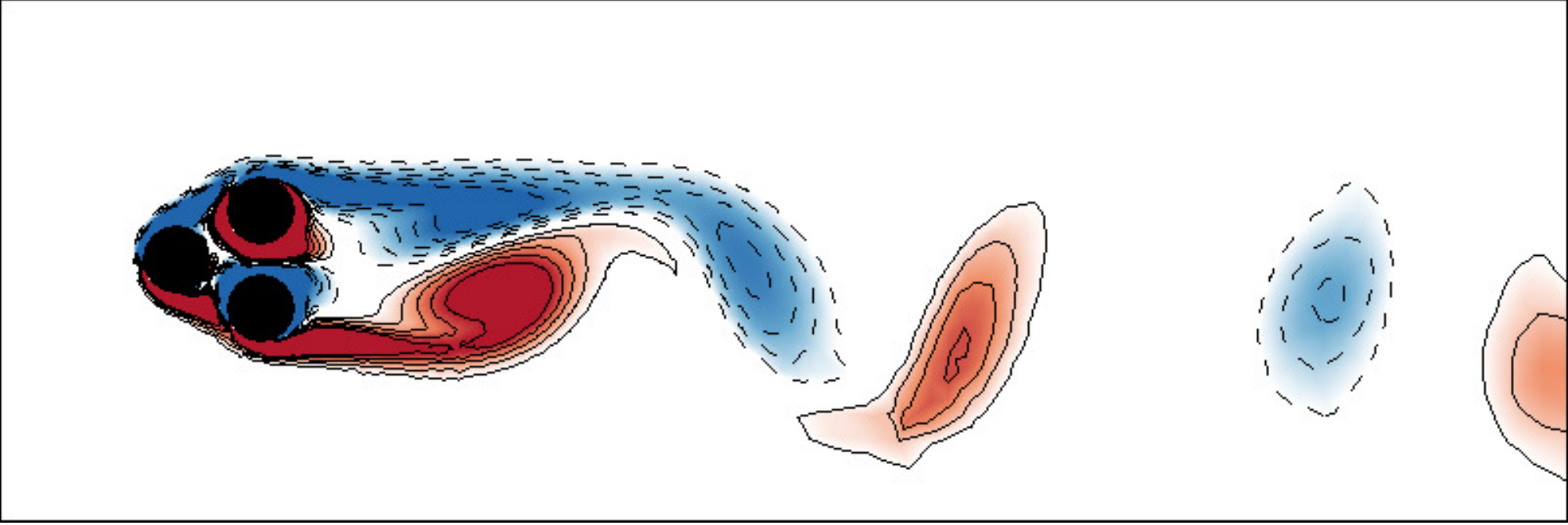}}%

\subfloat[$t+3T_4/8$]{\label{fig:HB_T3}\includegraphics[width=0.45\textwidth]{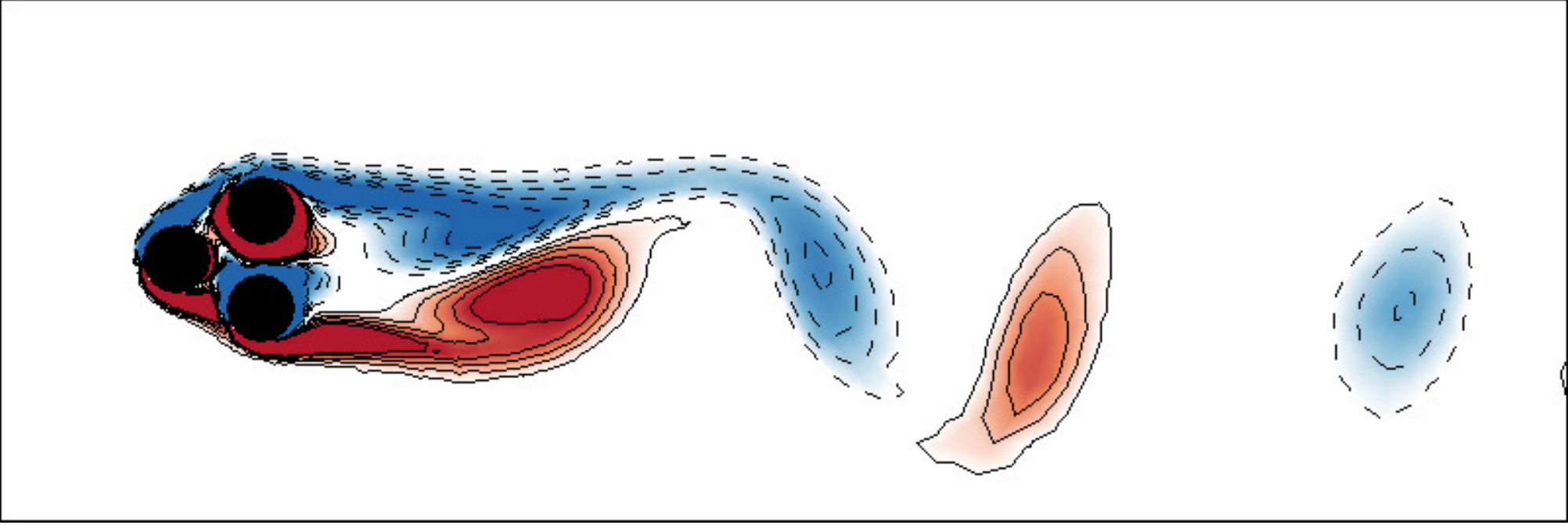}}%
\hfil
\subfloat[$t+4T_4/8$]{\label{fig:HB_T4}\includegraphics[width=0.45\textwidth]{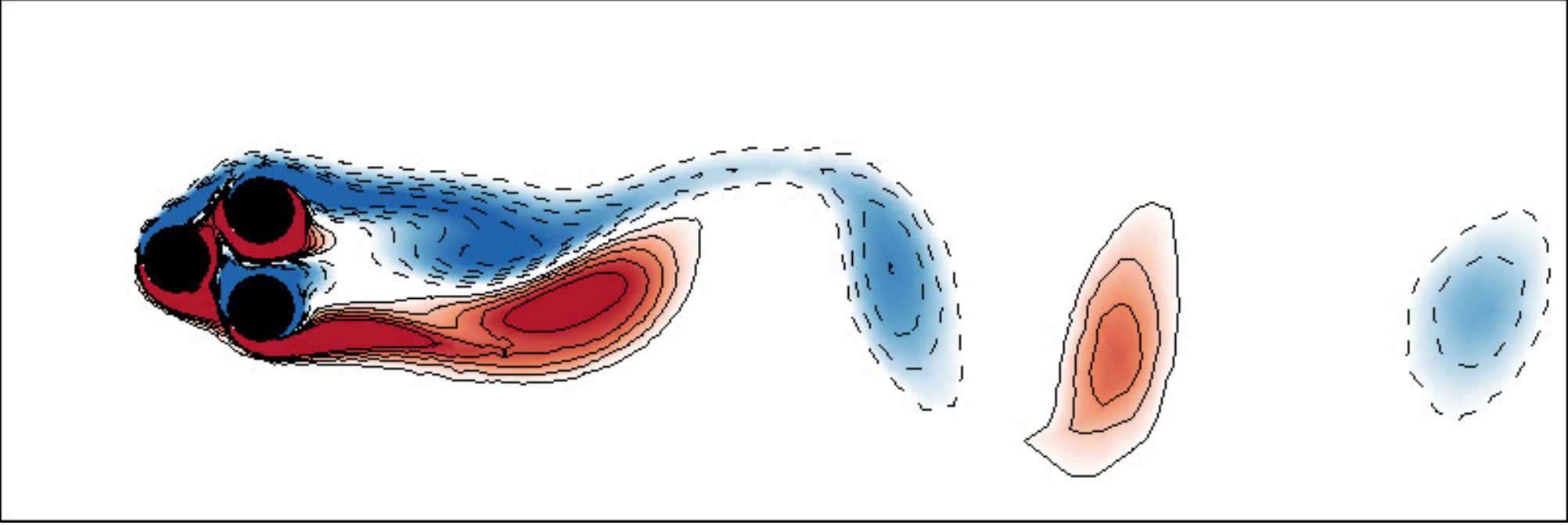}}%

\subfloat[$t+5T_4/8$]{\label{fig:HB_T5}\includegraphics[width=0.45\textwidth]{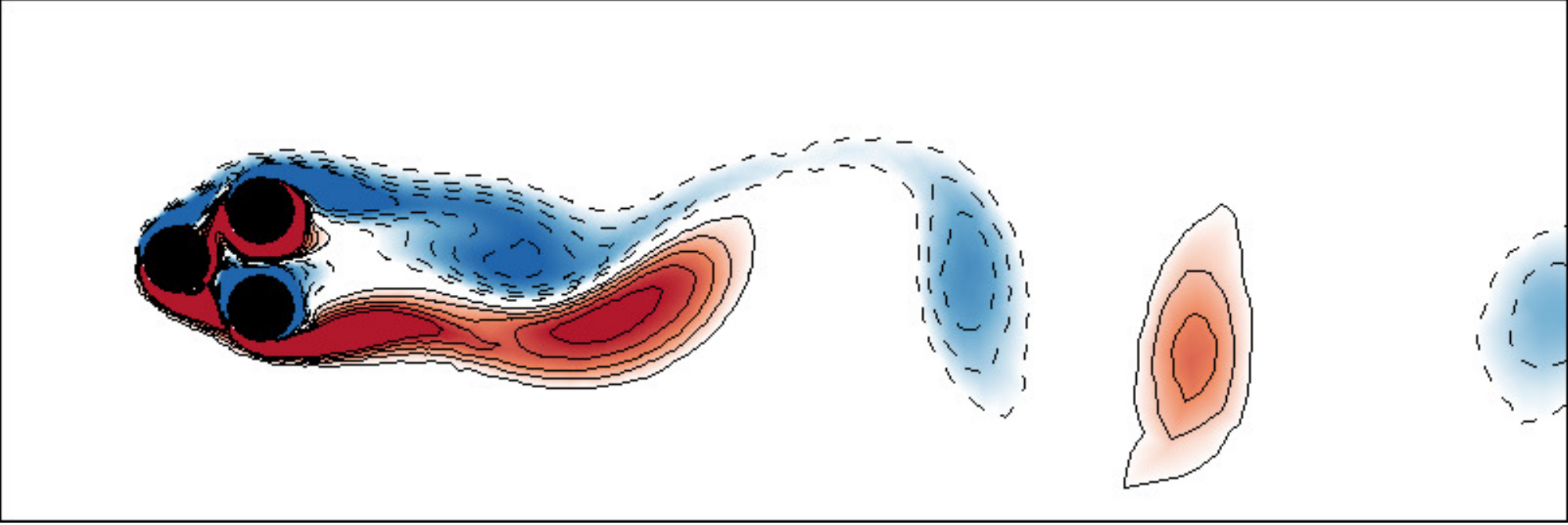}}%
\hfil
\subfloat[$t+6T_4/8$]{\label{fig:HB_T6}\includegraphics[width=0.45\textwidth]{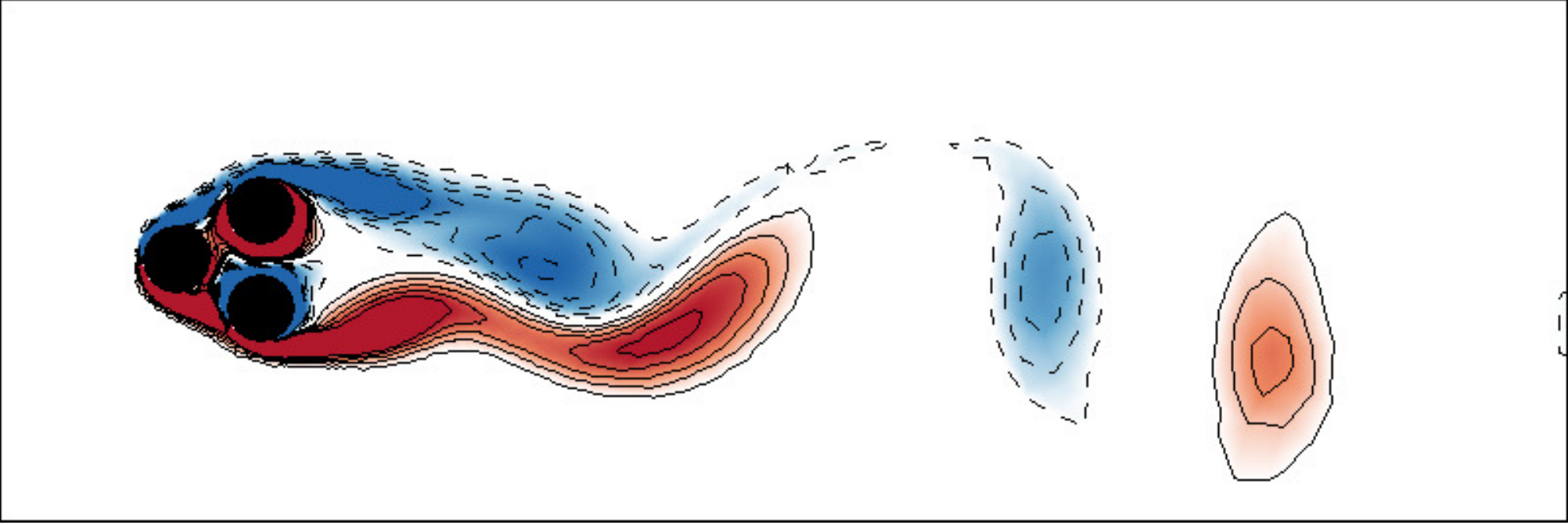}}%

\subfloat[$t+7T_4/8$]{\label{fig:HB_T7}\includegraphics[width=0.45\textwidth]{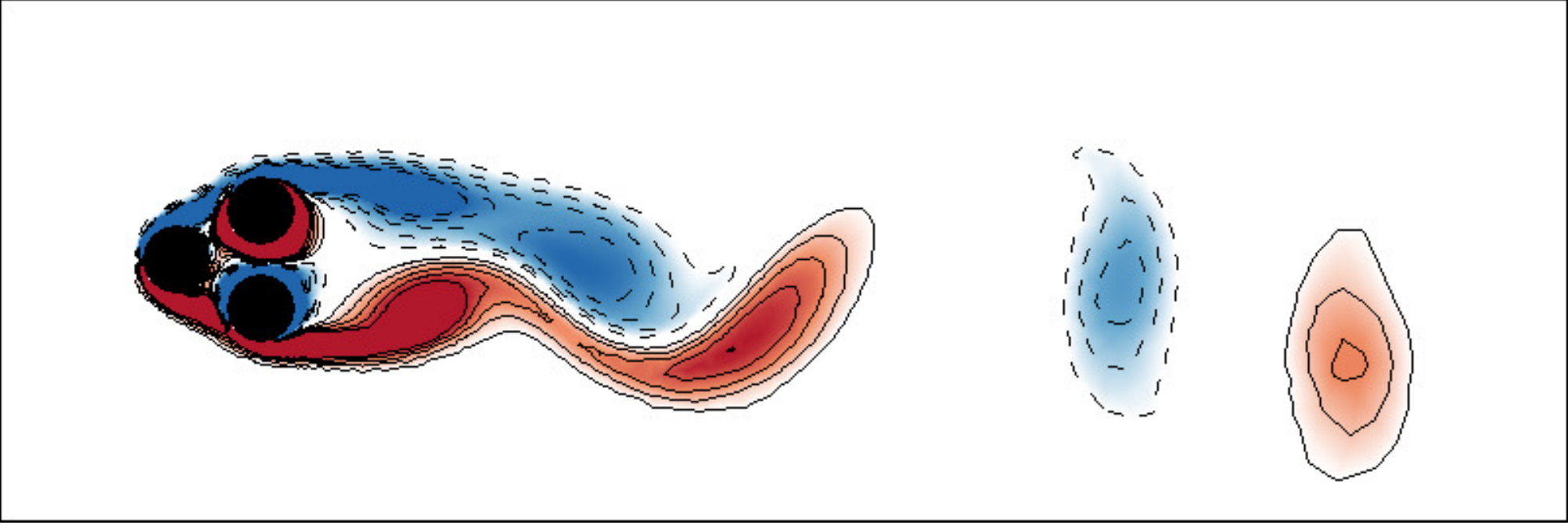}}%
\hfil
\subfloat[$t+T_4$]{\label{fig:HB_T8}\includegraphics[width=0.45\textwidth]{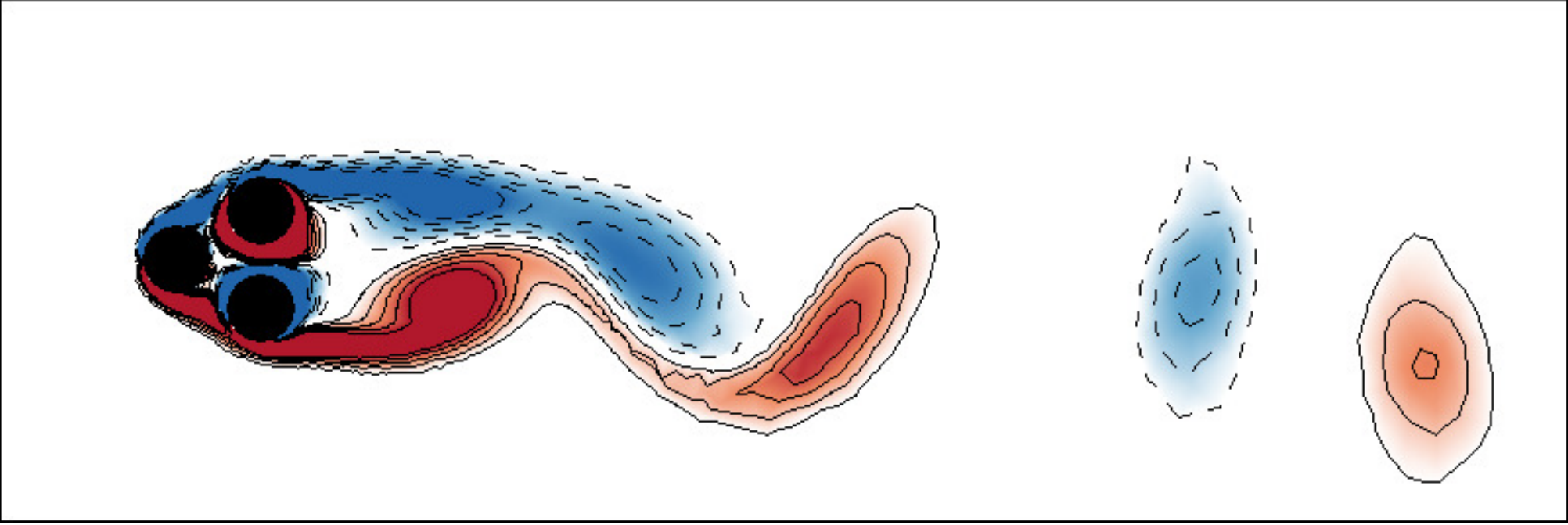}}%
\caption{\label{fig:HB_snap} Vorticity fields of the flow controlled by $b^{\rm HB}$. (a)-(f) Time evolution of the vorticity field in the last period of the 1000 time units simulation.
The color code is the same as figure~\ref{fig:unforced_flow}.
$T_4$ is the period associated to the frequency $f_4$.
}
\end{figure}
We notice in figure~\ref{fig:CL_HB} that the oscillations of the lift coefficient increased alongside with its mean value.
Yet, this asymmetry is not obvious in the figure~\ref{fig:HB_snap}.
Controlled with $b^{\rm HB}$, the flow is similar to the previous feedback control.
We note, nevertheless, that the recirculation bubble is slightly larger.
As for $b^{\rm FB}$, the vortices stay attached longer thanks to the closed-loop periodic forcing.
We notice in particular that the positive vortex is shed after a longer time interval than the negative vortex, indeed we notice a larger distance between a positive vortex and the previous negative vortex downstream than between a negative vortex and the previous positive vortex downstream.

\section{Summary}
\begin{table}[htb]
\begin{center}
\begin{tabular}{>{\centering}p{2.75cm}>{\centering}p{3cm}>{\centering}p{2cm}>{\centering}p{2cm}>{\centering\arraybackslash}p{2cm}}
  Search space & Control law & $J/J_0$ & $J_a/J_0$ & $J_b/J_0$
 \\
\midrule
Unforced natural & - & 1 & 1 & 0\\
 \rule{0pt}{5ex}
Symmetric steady & $b_1=0$, $b_2=-b_3= \rm cst$ & 0.7652 & 0.5695 & \textbf{0.1956} \\ 
\rule{0pt}{5ex}
Multi-frequency forcing & $\boldsymbol{b}(t)=K(\boldsymbol{h}(t))$ & 0.7476 & 0.5109 & 0.2368 \\ 
\rule{0pt}{5ex}
Feedback control & $\boldsymbol{b}(t)=K(\boldsymbol{s}(t))$ & 0.7451 & \textbf{0.4744} & 0.2706\\
\rule{0pt}{5ex}
General control & $\boldsymbol{b}(t)=K(\boldsymbol{s}(t),\boldsymbol{h}(t))$ & \textbf{0.7363} & 0.4845 & 0.2518 \\
%\rule{0pt}{5ex}
\end{tabular}
\end{center}
\caption{\label{tab:summary_mlc}Summary of the performances for the best solutions of each type of optimization.
The bold values are the best for each cost.}
\end{table}
In this chapter, we apply the MLC methodology previously described to minimize the net drag power of the fluidic pinball.
First, a parametric study on the subspace of symmetric steady forcing supports that the boat tailing configuration appears as a key strategy to reduce the drag power.
Then three search spaces are explored, first we allow for multi-frequency forcing, then we optimize a feedback control law, and finally, we allow both strategies for a hybrid optimization.
All three optimizations built control laws that include a boat tailing-like structure and discard the open-loop periodic functions when they are in the function library.
However, we notice that an asymmetry in the boat tailing is systematically favored.
Some improvement can be achieved with the addition of sensor information, reducing the cost from $J_{\rm MF}/J_0=0.7476$ to $J_{\rm HB}/J_0=0.7363$.
The costs of all runs are summarized in table~\ref{tab:summary_mlc}.
Thus, in less than 1000 evaluations, MLC managed to build a control combining asymmetric boat tailing and phasor control to reduce the net drag power in a model-free and with very few knowledge a priori.
MLC rediscovers, in particular, that to delay the vortex shedding, one can re-energize the shear layer with periodic forcing and that vectoring the flow towards the centerline helps to increase the base pressure.
The hybrid control built with MLC achieves the most net drag reduction so far.

In this part,
we unveiled the forces at play in the learning process of genetic programming and applied to the reduction of the net drag power of the fluidic pinball.
The parametric study of MLC, carried out in App.\ \ref{Sec:ParamStudyGenOp}, revealed the importance of key meta-parameters.
Such analysis serves as a guide to select adequate parameters for future MLC studies.
Applied to the fluidic pinball, MLC successfully managed to build control laws in different search spaces in less than 1000 evaluations, revealing key actuation mechanisms, without any prior knowledge, and combining them for further performance: the best control comprises a combination of asymmetric boat tailing and phasor control.

\chapter{Conclusions and outlook}
\label{ToC:Conclusions}

We have introduced  Machine Learning Control (MLC)
based on linear genetic programming
for a simple dynamical system.
This example is easily reproducible 
with the provided and documented Matlab code.
This code or small variations thereof
have been applied in dozens of plants:
nonlinear dynamical systems, direct numerical simulations
and turbulence control experiment \citep{Noack2019springer}.
The meta parameters have hardly been changed.
The algorithm has reliably and repeatability
converged to an assumed global minimum of the cost function.
Hence,  MLC can be expected to be successfully
applied in many other nonlinear dynamics systems 
and flow control simulations or experiments
without the need of change of meta parameters, used functions, etc.
While linear control theory provides unrivaled methods for linear dynamics,
MLC becomes more and more competitive with increasing control complexity,
e.g., by the degree of plant nonlinear, 
dimension of the dynamics, 
number of actuation commands and sensing signals 
and also nonlinearity of the control law.

Yet, a few words of caution are in order.
First, MLC is a stochastic optimization method for a nonlinear plant
and comes hence without any performance guarantees,
like convergence against a global minimum.
Second, the typical number of sensors
and actuators was $O(1)-O(10)$.
The learning time, i.e.\ the number of individuals to be tested before convergence,
is observed to slightly increase with the number of actuators and sensors.
The number of actuation commands is more critical than the number of sensors.
Third, a sensitive control problem where a small change of the control law
implies a large change in performance may not be best targets of vanilla MLC versions
and requires caution with any other control optimization/design as well.
Fourth, in engineering applications the performance of the optimized control law 
might be affected by a change of initial conditions,
measurements noise, plant uncertainty and changing operating conditions.
As rule of thumb, a control law which employs large-scale/dominant structures 
can be expected to be robust against such changes
while actuation involving small-scale/secondary structures may be fragile.

Machine learning control is an emerging 
rapidly evolving field overcoming some key challenges of linear control theory. 
Research needs to be done  on many fronts, inspired by the success stories of linear control:
\begin{enumerate}
\item \emph{Control landscape:} Visualization of the learning process.
A starting point is offered in the first book \citep{Duriez2017book}.
\item \emph{Human interpretable control laws.}
A cluster-based visualization \citep{Cornejo2021jfm,Castellanos2022pf} appears always doable and insightful.
\item \emph{Significant reduction of testing time.}
Here, myriad ideas come to mind, e.g., 
(1) surrogate modeling of the cost function as function of the control, 
(2) model identification from the MLC run as surrogate test plant, 
(3) pretesting of the control law to remove
the time consuming testing of similar or unpromising laws.
\item \emph{Robustness against measurement noise, plant uncertainty and changes of operating conditions.}
Currently, learning the control laws under varying operating conditions and with measurement noise
are a viable tested options.
\item \emph{Online adaptivity inspired by extremum seeking.}
The learning process for new operating conditions ideally requires only small changes in the operating plant.
Gradient-enriched MLC \citep{Cornejo2021jfm} 
offers one avenue with subplex optimization on a chosen subspace of control laws.
\item \emph{Transfer learning.} How can the control learning of a new operating conditions 
be shortened knowing the MLC run of another operating condition?
The task poses a big challenge which seems not currently addressed by any approach. 
\item \emph{Failure safety of actuators and sensors.} 
In praxi, actuators and sensors may fail, 
requiring a prompt alternative control logic with the new reduced hardware. 
This is also one of the big challenge problems formulated already 
in first book on genetic programming control \citep{Dracopoulos1997book}.
\end{enumerate}

An increasing community of researchers 
are working on all these research tasks.
Enjoy working with the MATLAB program \xMLC 
and stay tuned for the progress to come!

\subsubsection{Acknowledgements}
This work is supported
by the National Science Foundation of China (NSFC) through grants 12172109 and 12172111,
by the Natural Science and Engineering grant  2022A1515011492 of Guangdong province, P.R. China.

\appendix
\renewcommand\chaptername{Appendix}
\chapter{MLC parametric dependency}\label{Sec:ParamStudyGenOp}
In this appendix, we report the influence of some key parameters of MLC on the optimization process.
In particular, we study the influence of the genetic operator probabilities $(\Pcros,\Pmut,\Prep)$ (Sec.\ \ref{Sec:ParamStudyGenOp}) and the balance between population size and number of instructions (Sec.\ \ref{sec:mlc_parametric_study}).
The parametric study is carried on the stabilization of the damped Landau oscillator for its short evaluation time and its relatively simple dynamics.
The goal of this section is to establish some rules of thumb for the selection of these parameters.
First, we describe the optimal linear solution (Sec.\ \ref{Sec:OptLinSolution}), employed as reference for the future optimized control laws.

\section{Control benchmark and optimal linear solution}\label{Sec:OptLinSolution}
\subsection{Control benchmark}
The control benchmark for this parametric study is the stabilization of the damped Landau oscillator described in Sec.\ \ref{Sec:LandauOscillator}.
For this study, we employ the same cost function except that we choose the penalization parameter $\gamma$ equal to 1 to assure a fast convergence towards the fixed point.
Thus the cost function employed is: 
\begin{equation}
  \begin{array}{rcl}
  J		& = &	J_a + J_b \\
J_a 		& = &	 \cfrac{1}{T_{max}} \bigintsss_{ 0}^{T_{max}} (a_1^2+a_2^2) d \tau \\
J_b 		& = &	\cfrac{1}{T_{max}} \bigintsss_{ 0}^{T_{max}} b^2 d \tau\\
  \end{array}
\end{equation}

\subsection{Optimal linear solution}
We compute the optimal linear control of the control problem thanks to the \texttt{fminsearch} function of \textsc{MATLAB}.
\texttt{fminsearch} is a derivative-free method based on Nelder-Mead simplex method \citep{Nelder1965jc}.
We optimize the parameters of a linear control to stabilize the Landau oscillator.
The optimal linear control is detailed in equation~\eqref{eq:opt_lin}.
This solution has been found with a random initialization of the \texttt{fminsearch} function.
This local minimum may not be the global minimum of the minimization problem but for the following we will refer to it as the `optimal linear control'.
In fact, the control law learned in the following sections, reveal that this solution is not the optimal linear control as another linear solution performs better.
For now, we only focus on the solution in equation~\eqref{eq:opt_lin}, that is only a local minimum.
The associated cost is $J_{\rm opt} = 3.3403$ which corresponding to a $\Delta J_{\rm opt}/J_0=94.70\%$ reduction of the unforced cost.
\begin{equation}
      b_{\rm opt} = 2.4061a_1 -3.0984a_2
      \label{eq:opt_lin}
\end{equation}

\begin{figure}[htb]
\centering
\subfloat[Phase space, actuation command, instantaneous cost function and radius.]{\includegraphics[width=0.405\textwidth]{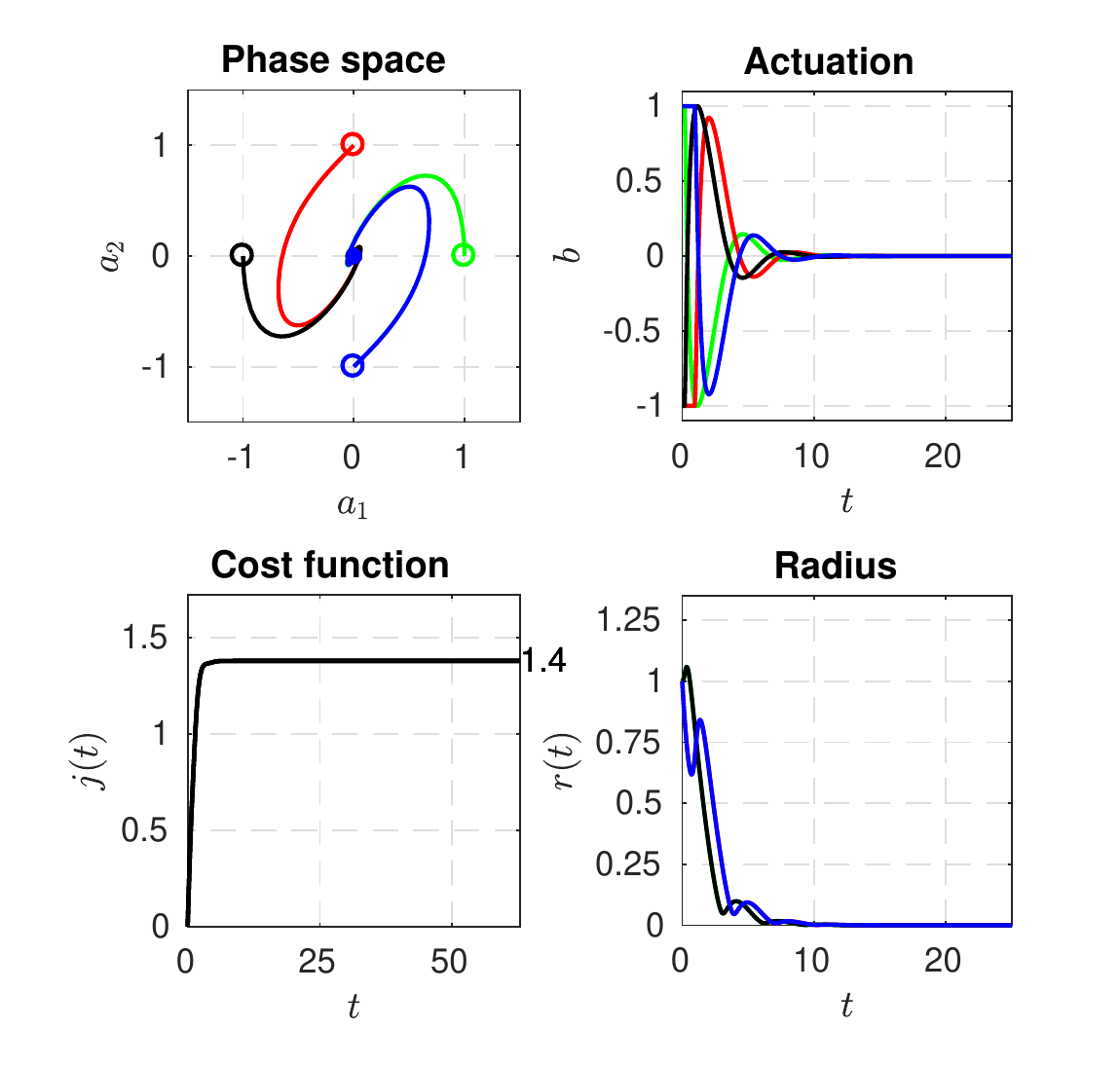}}%
\hfil
\subfloat[Visualization of the control law in the phase space. The limit cycle is depicted with a dashed line.]{\includegraphics[width=0.477\textwidth]{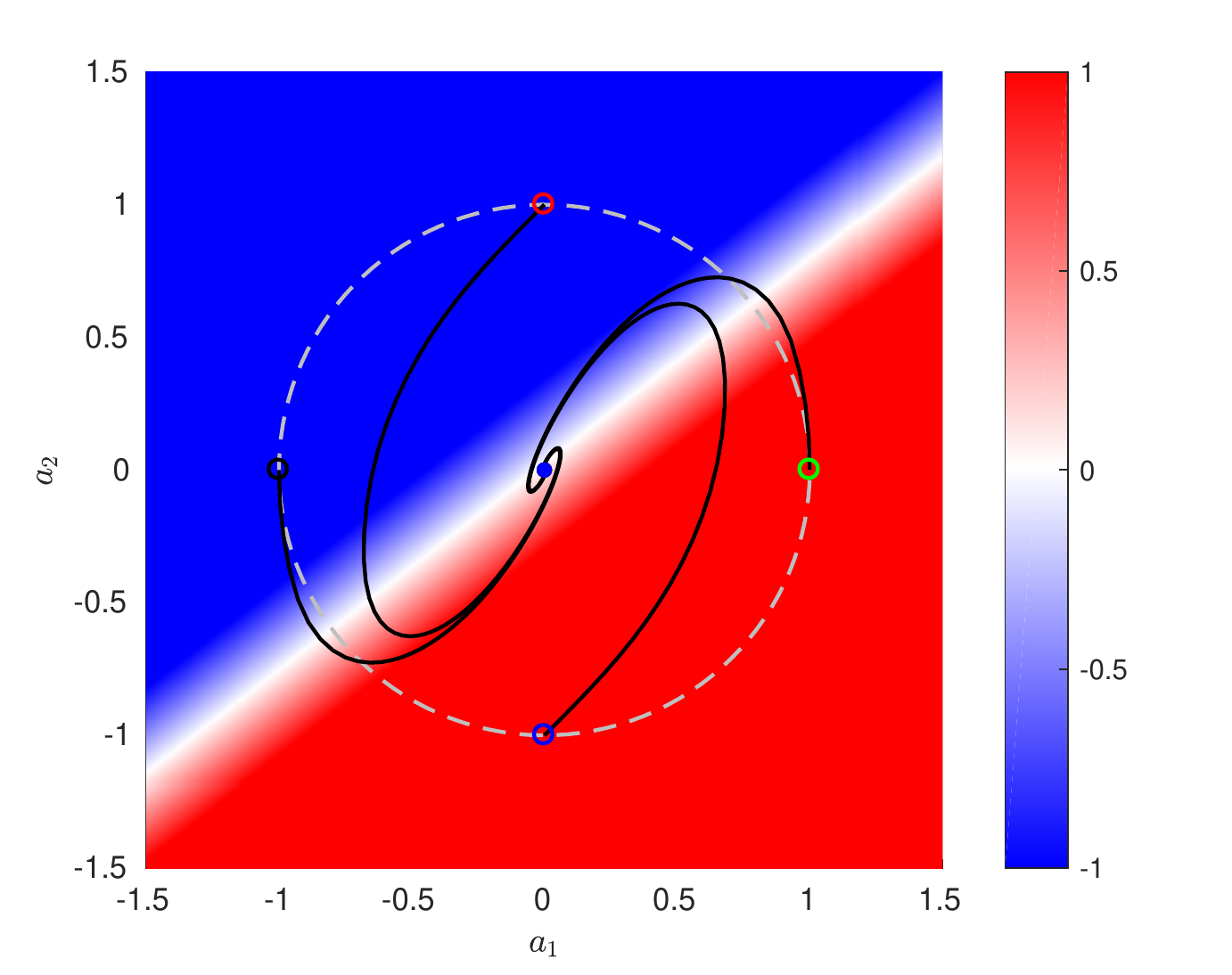}}%
\caption{\label{Fig:OptLinSolution} Figures characterizing the optimal linear solution.
}
\end{figure}
Figure~\ref{Fig:OptLinSolution} shows the controlled Landau oscillator and the actuation map for the control realized with equation~\eqref{eq:opt_lin}.

We notice that the symmetry of the problem is respected and that the system is successfully brought to the fixed point $(a_1,a_2)=(0,0)$ in less than two periods.
We recall that the actuation effect is to push the system upwards ($b>0$) or downwards ($b<0$).
The ratio of the coefficients in front of $a_1$ and $a_2$ define the angle for the separatrix dividing the space in two regions of positive and negative actuation.
The magnitude of the coefficients define the increase rate of the actuation level as the system moves away from the separatrix.
It is both the optimization of the separatrix angle and the actuation increase rate that brings the system faster towards the fixed point.
It is worth noting that the optimal control leads the system beyond the limit cycle for a short period of time.
Evidently, if the control would be on the first equation instead of the second, the results would be symmetrical to the $a_1=a_2$ axis.

Our goal is then to stabilize the Landau oscillator with MLC and to find a better solution than the optimal linear control.

\section{Influence of genetic operators}
\label{Sec:ParamStudyGenOp}

In this section, we stabilize the damped Landau oscillator with increasingly complex MLC algorithms.
The starting point is the special case of MLC, where only one generation is employed.
This case corresponds to a Monte Carlo sampling.
Then, we progressively include the genetic operators (crossover, mutation and replication) and analyze their influence on the algorithm's performance.

\subsection{Monte Carlo sampling}
As detailed in Sec.\ \ref{Sec2:LGPC}, a Monte Carlo sampling consists in generating random matrices for each individual.
Thus, we must first define the search space, i.e., the space of candidate solutions to the problem.

\subsubsection{The search space}
\begin{table}[htb]
  \begin{center}
  \def~{\hphantom{0}}
  \begin{tabular}{>{\centering}p{1.5cm}>{\centering}p{4.5cm}>{\centering\arraybackslash}p{4cm}}
Parameter & Description & Value\\
\midrule
    & Function library & $ F_1=\{+,-,\times,\div\}$\\
   $\boldsymbol s$ & Controller inputs & $a_1$, $a_2$ \\
   $\Nvar$ & Number of variable registers & $3$ \\
   $\Ncst$ & Number of constant registers & $3$ \\
   $\Ninstrmax$ & Max. number of instructions & $ 5$\\
\end{tabular}

    \caption{\label{tab:search_space_param}Parameters for the control ansatz.}
  \end{center}
\end{table}
In the linear genetic programming framework, the search space is defined by the function and inputs library.
The other parameters that can influence the exploration of the landscape are the number of variable registers $\Nvar$, the number of constant registers $\Ncst$ and the maximum number of instructions $\Ninstrmax$.
Indeed if $\Nvar$, $\Ncst$ and $\Ninstrmax$ are small then the  control laws are bound to contain only few operations, thus limiting the complexity of the accessible control laws.
This aspect is further studied in App.~\ref{sec:mlc_parametric_study}.
Table~\ref{tab:search_space_param} summarizes the parameters chosen for the Monte Carlo sampling.

Based on these parameters, we can compute the total number of control laws in the search space by listing all the possible combinations.
For this, we multiple the possible values for each columns of the instruction matrix and elevate the result at the power of the number of rows in the matrix.
As we allow matrices of different sizes, we need to add the results for each possible number of rows.
Expression~\eqref{eq:ss_size} gives the order of magnitude of the number of possible control laws with the chosen parameters.
\begin{equation}
  \label{eq:ss_size}
  \sum_{q=1}^{\Ninstrmax} [\Nr \times \Nr \times \No \times \Nvar]^{q} \approx 1.5 \times 10^{13}
\end{equation}
where

\begin{tabular}{rl}
    $\Ninstrmax$:& maximum number of instructions. \\
    $\Nr$:& total number of registers: $\Nr=\Nvar+\Ncst$. \\
    $\No$:& number of functions in the library or mathematical operators. \\
    $\Nvar$:& number of variable registers, where to store intermediate calculations. \\
\end{tabular}

This number $1.5 \times 10^{13}$ represents the total number of matrices, i.e., the total number of accessible control laws.
For this estimation, we assume that all registers are initialized with different values, which is often the case in practice, except for the output registers, i.e., the registers that store the control laws, that are initialed with zeros.
\begin{table}[htb]
  \begin{center}
  \def~{\hphantom{0}}
  \begin{tabular}{>{\centering}p{1cm}>{\centering\arraybackslash}p{1cm}}
  \multicolumn{2}{c}{Variable registers}\\
  \midrule
  $r_1$ & $0$ \\
  $r_2$ & $s_1$\\
  $r_3$ & $s_2$ \\
\end{tabular}
\hspace{1cm}
\begin{tabular}{>{\centering}p{1cm}>{\centering\arraybackslash}p{2.5cm}}
  \multicolumn{2}{c}{Constant registers}\\
  \midrule
  $r_4$ & $c_1 = -0.94$ \\
  $r_5$ & $c_2 = -0.05$ \\
  $r_6$ & $c_3 = -0.72$ \\
\end{tabular}

    \caption{\label{tab:reg_initialization}
Initialization of the registers. The constants $c_i$ have been randomly taken in the range $[-1,1]$.}
  \end{center}
\end{table}
Table~\ref{tab:reg_initialization} presents the initialization of the registers for the oscillator stabilization problem before the execution of the instruction matrices.

Among the $1.5 \times 10^{13}$ possible control law, there are, of course, a lot of identical and equivalent control laws, due to overwriting, useless instructions, multiplication by $0$, etc.
This constitutes a valuable feature as the expression of the global minimum is no longer unique and can be built in different ways, thus reducing the `size' of the search space.

\subsubsection{A randomly generated control law}
\begin{figure}[htb]
  \centering
  \includegraphics[width=0.65\textwidth]{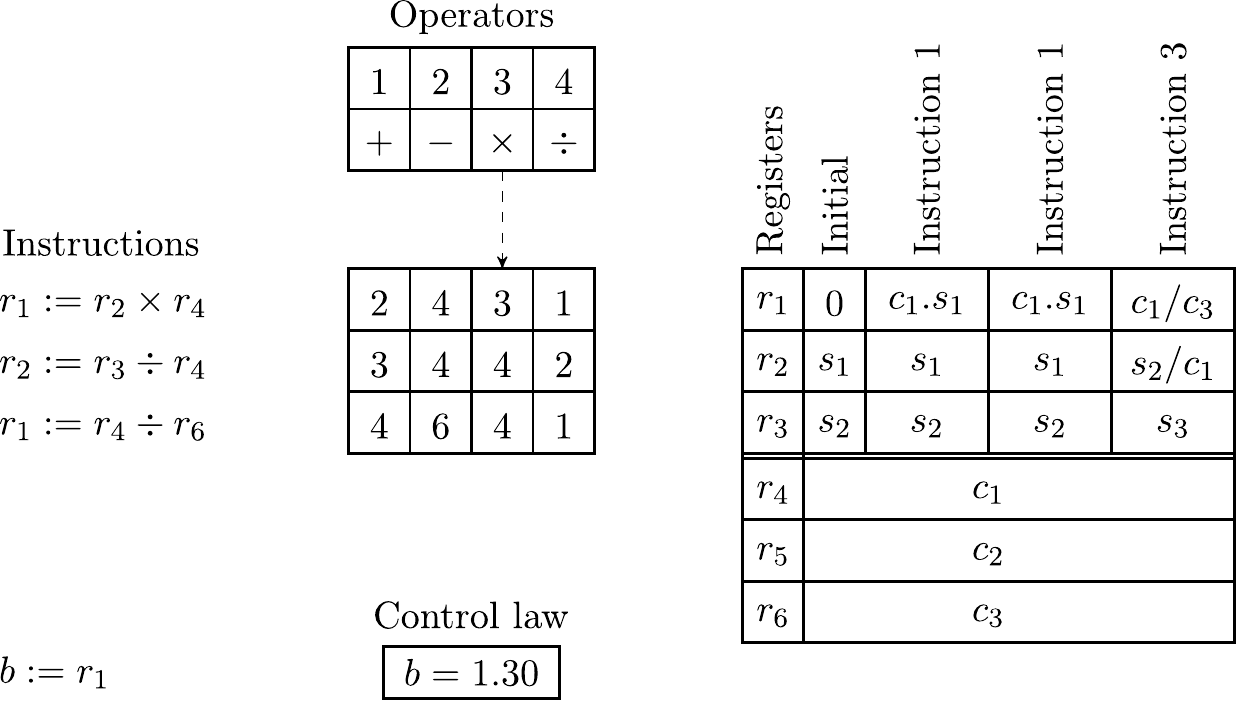}
  \caption{Example of randomly generated matrix and its translation to a control law.}
  \label{fig:controllaw_random_matrix}
\end{figure}
Figure~\ref{fig:controllaw_random_matrix} shows an instruction matrix randomly generated and depicts the process of translation.
The resulting control is a constant beyond the actuation limits, thus this control law is equivalent to a constant forcing at $b=1$.
\begin{figure}[htb]
\centering
\subfloat[Phase space, actuation command, instantaneous cost function and radius.]{\includegraphics[width=0.405\textwidth]{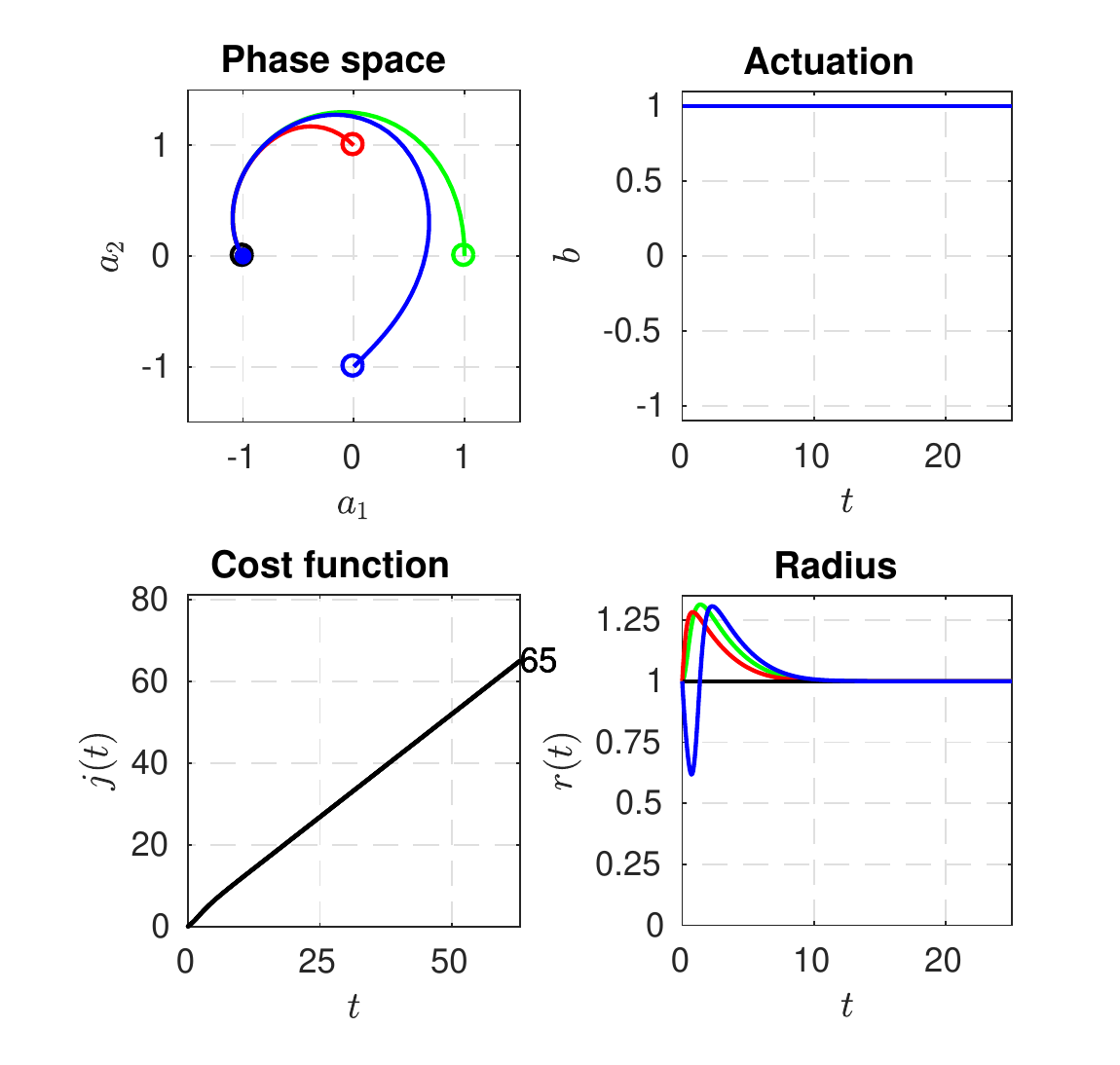}}%
\hfil
\subfloat[Visualization of the control law in the phase space. The limit cycle is depicted with a dashed line.]{\includegraphics[width=0.477\textwidth]{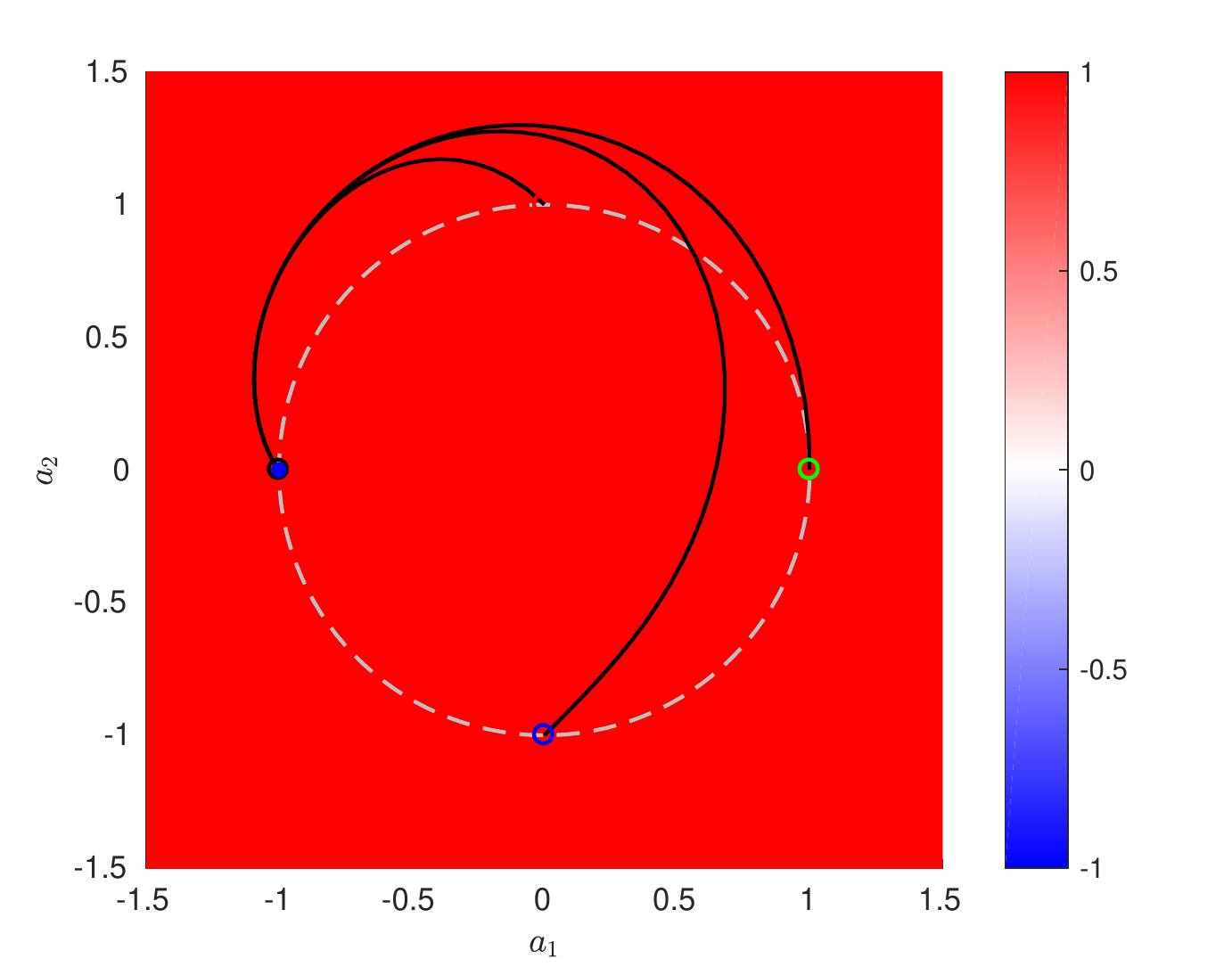}}%
\caption{\label{fig:random_controller}Visualization of the randomly generated control law $b=1.30$.
The control being limited between -1 and 1, this control is equivalent to 1.
}
\end{figure}
Before performing a Monte Carlo sampling, let's first generate a random control law.
Figure~\ref{fig:random_controller} gives a visualization of the control law.
To avoid controls with too strong actuations, we limit the actuations between -1 and 1 with a clip function.
We notice that with such action, the point $(a_1,a_2)=(-1,0)$ becomes a fixed point.
Indeed, without control this point is the unique point where $(\dot{a}_1,\dot{a}_2)  = (0,-1)$, thus when adding a constant forcing $b=1$, this point becomes a fixed point.

% ############################# MONTE CARLO ##############################
\subsubsection{Monte Carlo sampling}
Now, let's perform a Monte Carlo sampling with the parameters of \ref{tab:search_space_param}.
In order to emulate experimental conditions, we only perform $\Ni=1000$ evaluations of the cost functions.
Thus, we only generate 1000 control laws to be evaluated.
Our Monte Carlo implementation is optimized with the screening of redundant individuals.
More informations are provided in App.~\ref{sec:accelerators}.
% --- Monte Carlo performance
\begin{figure}[htb]
\centering
\subfloat[]{\label{fig:MC_one}\includegraphics[width=0.45\textwidth]{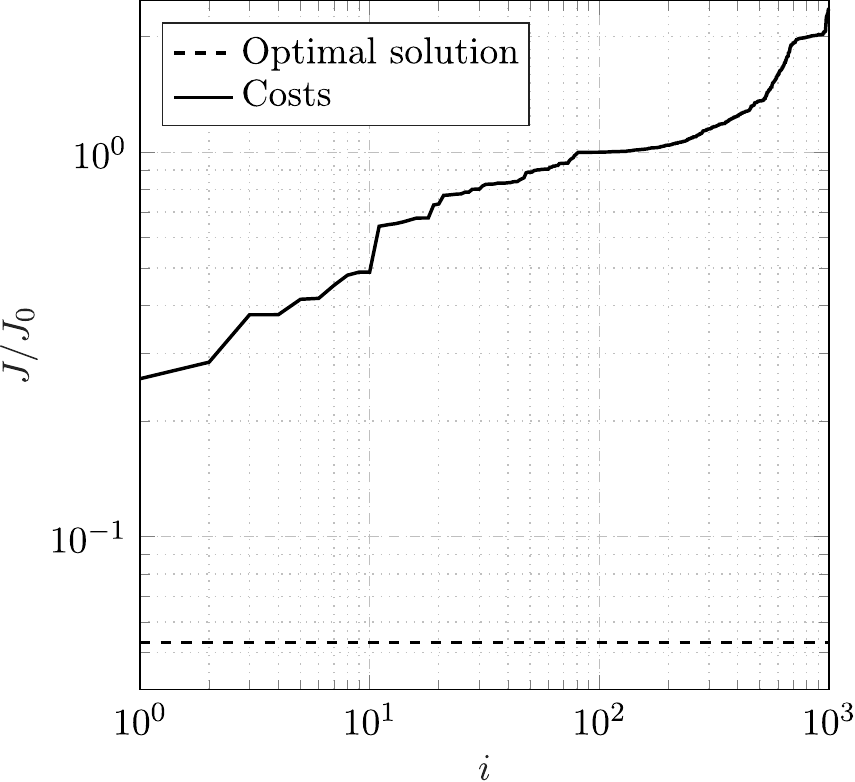}}%
\hfil
\subfloat[]{\label{fig:MC_envelop}\includegraphics[width=0.45\textwidth]{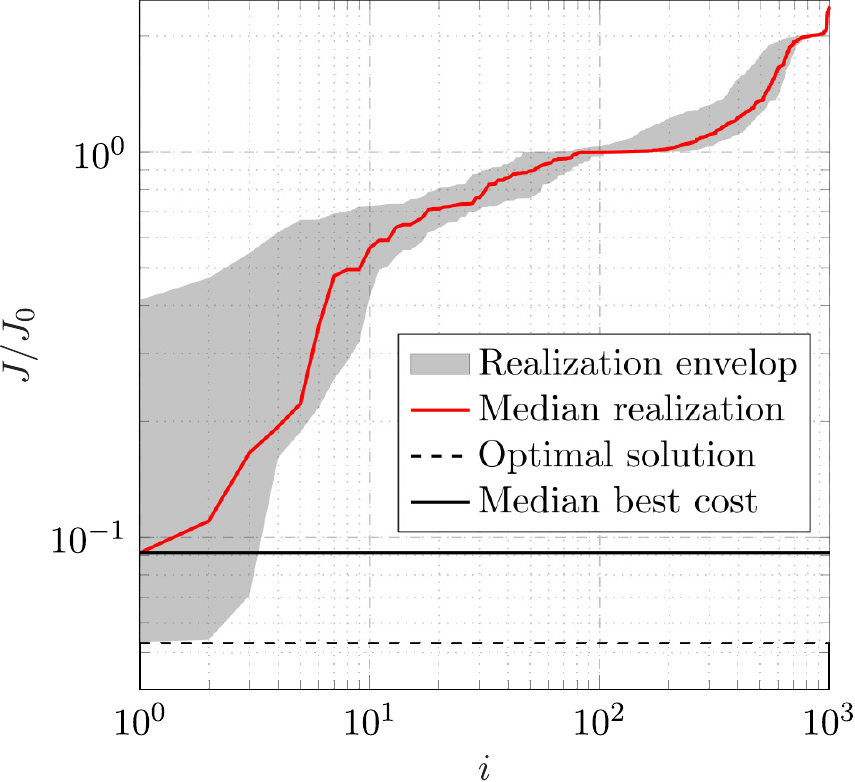}}
\caption{\label{fig:MC_optimization}(a) Distribution of the cost of 1000 randomly generated control laws for the stabilization of the damped Landau oscillator. 
(b) Envelop of 100 realizations of Monte Carlo sampling for the same problem.
  The median realization ($\rho=50^{\rm th}$ best) is depicted in red.
  The black horizontal line represents the cost of the best individual of the median realization.
  The associated cost reduction is $\Delta J /J_0 = 91.17\%$.
  The dashed line represents the cost of the optimal linear solution ($\Delta J_{\rm opt}/J_0=94.70\%$).
  The  vertical and horizontal axis are in $\log$ scale.
   In both (a) and (b) the individuals are sorted following their cost.}
\end{figure}
Figure~\ref{fig:MC_one} shows the distribution of costs in log-log scale for one realization of Monte Carlo sampling.
The values are sorted and normalized with the unforced cost $J_0$.
We notice in figure~\ref{fig:MC_one}, that the performance of the best solution after 1000 evaluated control laws is not as good as the optimal linear solution.
The associated cost reduction is $\Delta J/J_0=74\%$.

However, as we rely on a stochastic process, only one realization of the optimization process is not enough to estimate the performance of the method.
That is why, we perform $\Nrho=100$ realizations of Monte Carlo sampling to have statistically relevant realizations.
Figure~\ref{fig:MC_envelop} shows the envelop of the one hundred realizations of Monte Carlo.
The cost reduction of the best individuals in the realization envelop goes from $58.55\%$ to $94.64\%$, and the median performance reduces the cost by $\Delta J /J_0 = 91.17\%$.
In the following the realizations are indexed by $\rho$ and
the median performance is defined as the $\rho= 50^{\rm th}$ best realization.
\begin{figure}[htb]
\centering
\subfloat[]{\label{fig:MC_median}\includegraphics[width=0.45\textwidth]{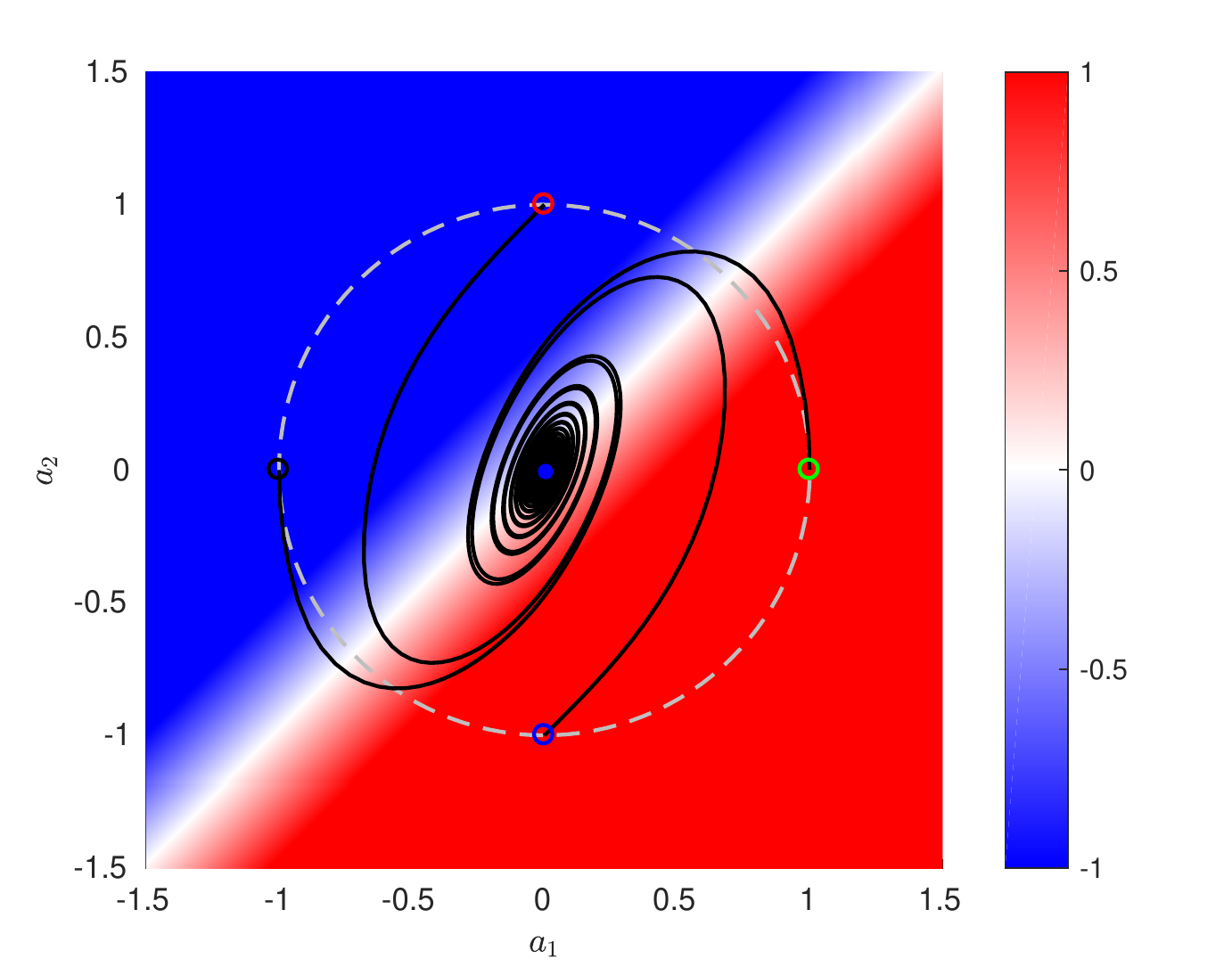}}%
\hfil
\subfloat[]{\label{fig:MC_best}\includegraphics[width=0.45\textwidth]{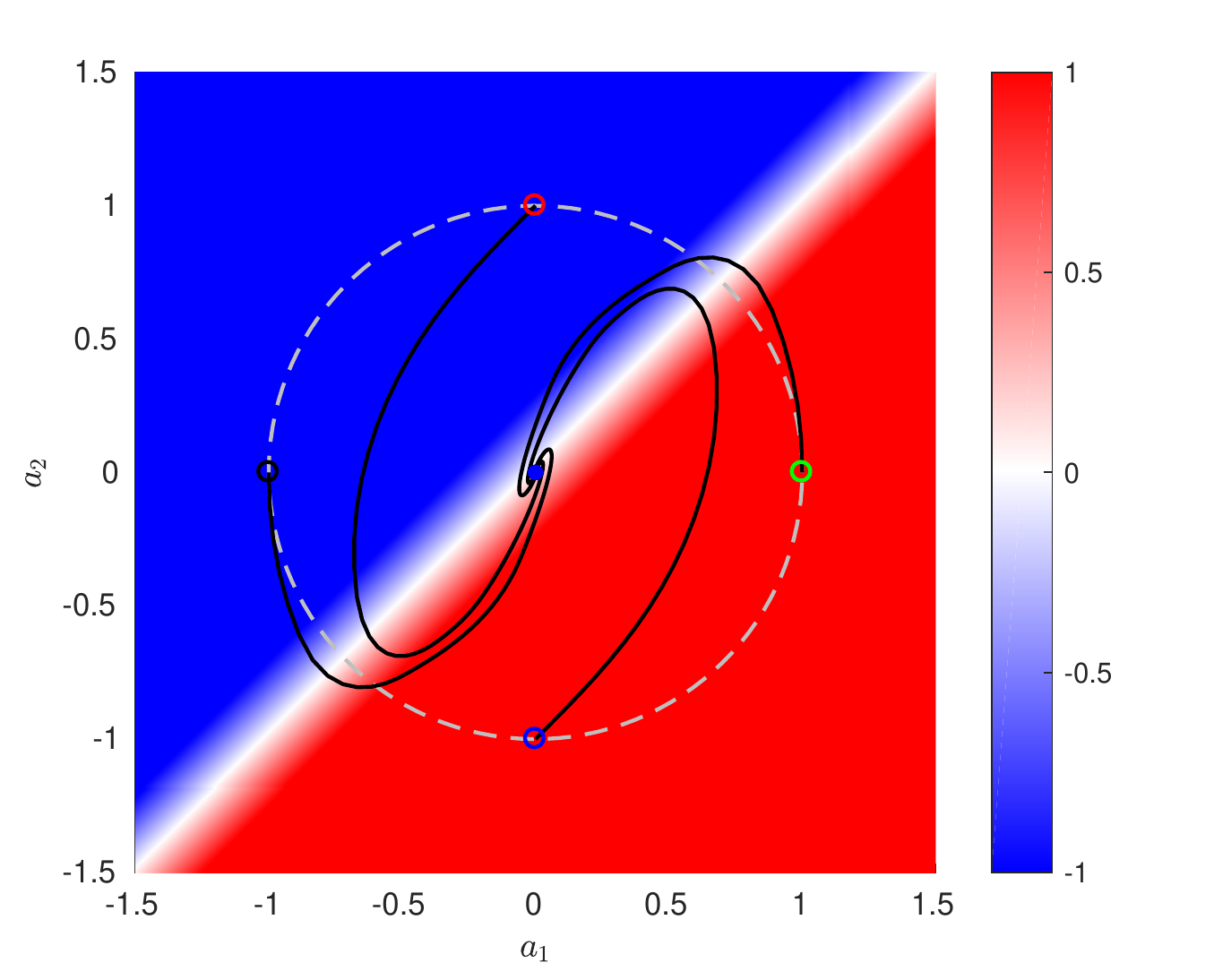}}%
\caption{Visualization of control laws in the phase space.
(a) Best control law of the median Monte Carlo realization \eqref{eq:MC_median}.
(b) Best control law of the best Monte Carlo realization \eqref{eq:MC_best}.
The limit cycle is depicted with a dashed line.
}
\end{figure}
The best control law of the median Monte Carlo realization is
\begin{equation*}
  b_{\rm MC, median }=2.0907(a_1 - a_2)
  \label{eq:MC_median}
\end{equation*}
and the best control law of the best Monte Carlo realization is 
\begin{equation*}
  b_{\rm MC, best}= 3.7747(a_1 - a_2);
  \label{eq:MC_best}
\end{equation*}
Their associated cost reduction are $\Delta J/J_0=91.17\%$ and $\Delta J/J_0=94.64\%$ respectively.
The expressions have been simplified compared to the expressions computed by \texttt{xMLC}.
The two controls have the same form but with a different coefficient.
The coefficients are related to the width of the control band and thus at the speed of convergence of the oscillator towards the fixed point.
This difference can be appreciated in figures~\ref{fig:MC_median} and \ref{fig:MC_best}
\begin{figure}[htb]
  \centering
  \includegraphics[width=0.7\textwidth]{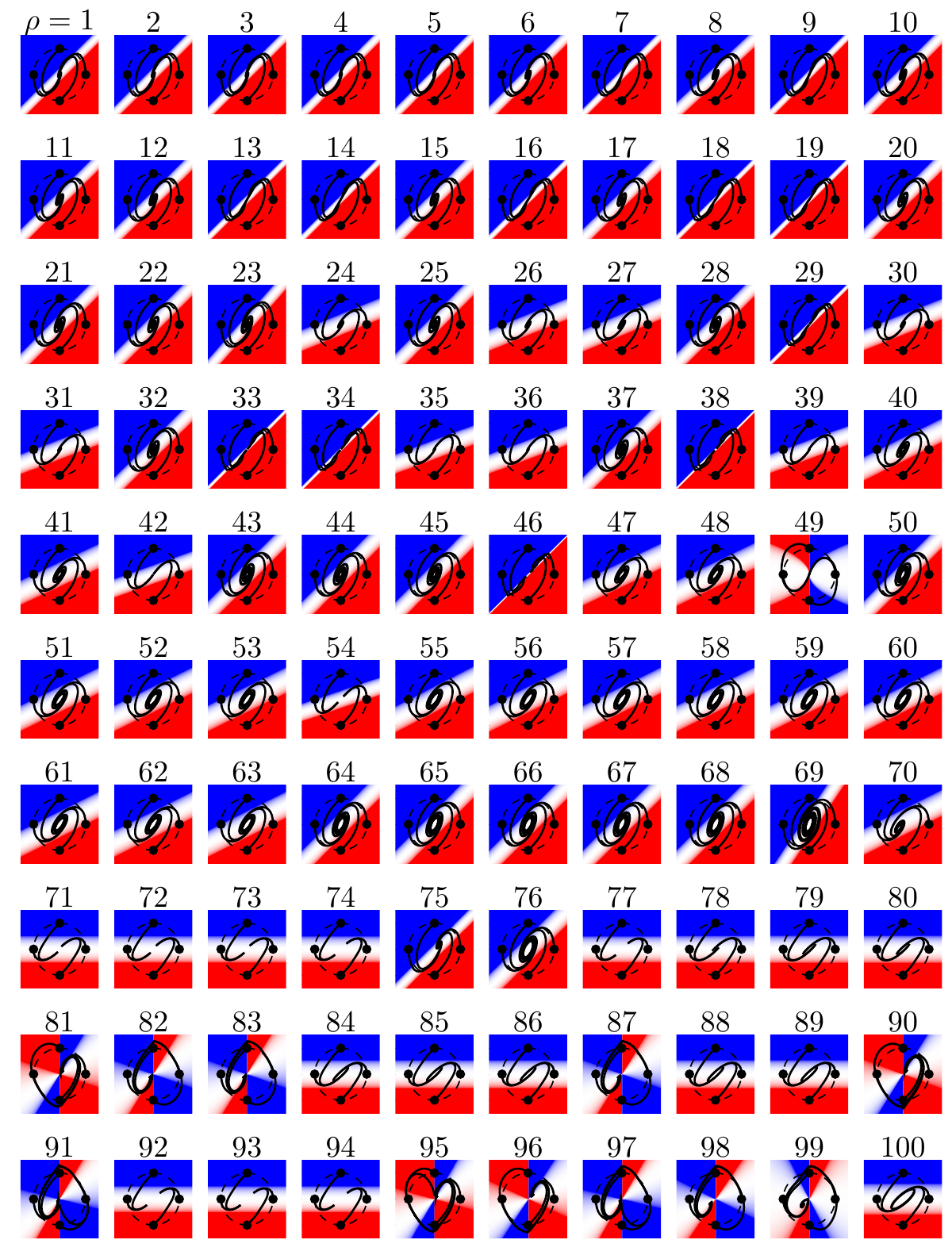}
  \caption{Visualization of the best control laws for each one of the hundred 100 Monte Carlo realizations.
  The index $\rho$ denotes the different realizations.
  The realizations are sorted by the cost of their best individual.
  1 being the best realization and 100 the worst.
  For simplicity, the trajectories for the four initial conditions are only plotted for the two first periods only.
  }
  \label{fig:MC_best_each_real}
\end{figure}
Figure~\ref{fig:MC_best_each_real} displays the best individuals of the hundred realizations.
We notice that the first twenty-three control laws are all similar.
They are similar to the optimal linear control with different band width.
Starting from the 33$^{\rm rd}$ run, the best individuals takes more than two periods to reach the fixed point.
We remark a majority of linear-like solutions, especially among the best ones.
The first nonlinear solution is ranked 49$^{\rm th}$ and the other ones are ranked beyond the 80$^{\rm th}$ realizations.
We also notice that those nonlinear solutions still present a symmetry in their actuation map.
Among the least performing realizations, there are 15 control laws that are equivalent to $-a_2$ to within one multiplicative positive constant, resulting on horizontally symmetric solutions.
We can expect that the genetic operators in the following sections combines appropriately $a_1$ and $a_2$ for as faster stabilization.

The study shows that a Monte Carlo sampling is able to build a good performing control laws for such problem.
However, only $1/3$ of all realizations achieved  a stabilization in less than 2 periods.
We note in particular that the best control law of the median realization is far from an efficient solution as the solution takes many periods to reach the fixed point.
Monte Carlo sampling lacks reproducibility on the ability to build efficient solutions.
In the next sections, crossover and mutation are included progressively in the learning process, showing that they allow much better performances in terms of speed, solution and reproducibility.

% ########################### END MONTE CARLO ############################

% ############################### CROSSOVER ##############################
\subsection{Exploitation with crossover}
In this section, we improve the Monte Carlo sampling by introducing the crossover genetic operator.
We follow the algorithm described in figure~\ref{fig:LGPC_algo} but only considering the crossover operator.
For this, we set the genetic operators probabilities to 0 except for the crossover, set to 1.
As we rely on a evolution process, we also set elitism to 1 to assure that the best individual is saved throughout the generations.
\begin{table}[htb]
  \centering
  \begin{tabular}{>{\centering}p{2.5cm}>{\centering}p{6cm}>{\centering\arraybackslash}p{2cm}}
  Parameter & Description & Value\\
  \midrule
   $\Nps$ & Population size & $100$\\
   $\Ng$ & Number of generations & $10$\\
   $\Ntour$ & Tournament size & $7$\\
   $\Ne$ & Elitism & $1$\\
   $\Pcros$ & \textbf{Crossover probability} & 1\\
   $\Pmut$ & Mutation probability & 0\\
   $\Prep$ & Replication probability & 0\\
\end{tabular}

  \caption{Parameters for MLC with crossover only.}
  \label{tab:crossover_parameters}
\end{table}
The parameters related to the evolution process are detailed in \ref{tab:crossover_parameters}.
The remaining parameters are to same as the one used for the Monte Carlo sampling, see table~\ref{tab:search_space_param}.
\begin{figure}[htb]
  \centering
  \includegraphics[width=0.5\linewidth]{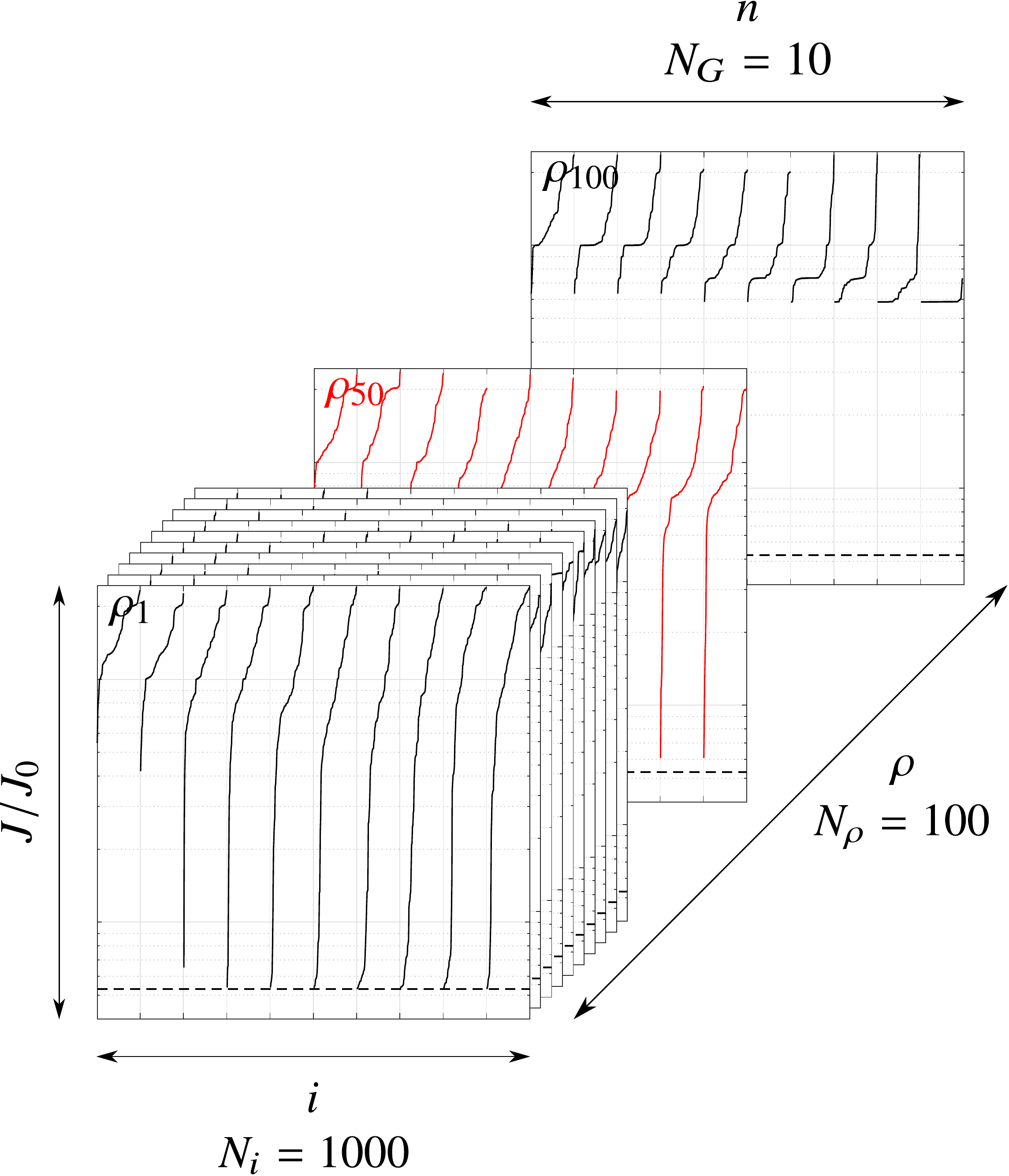}
  \caption{Conceptual figure of the $N_{\rho}=100$ realizations of MLC.
  Each graph represents a realization including $\Nps=10$ generations of 100 individuals.
  The median realization is denoted $\rho=50$ and colored in red.
}
  \label{fig:concept_C}
The resulting algorithm is an exploitative-only or crossover-only MLC.
To have a fair comparison between Monte Carlo and MLC, we evaluate $\Ni=1000$ individuals distributed in a population of $\Nps=100$ individuals that evolves 9 times for a total of $\Nrho=10$ generations.
\end{figure}
Like for the Monte Carlo sampling, $N_{\rho}=100$ realizations of the crossover-only MLC are performed.
Each realization consist of a population of $\Nps = 100$ individuals evolving through $\Ng=10$ generations for a total of $\Ni=1000$ evaluations.
Figure~\ref{fig:concept_C} summarizes the parameters for this study.

\begin{figure}[htb]
  \centering
  \includegraphics[width=0.5\textwidth]{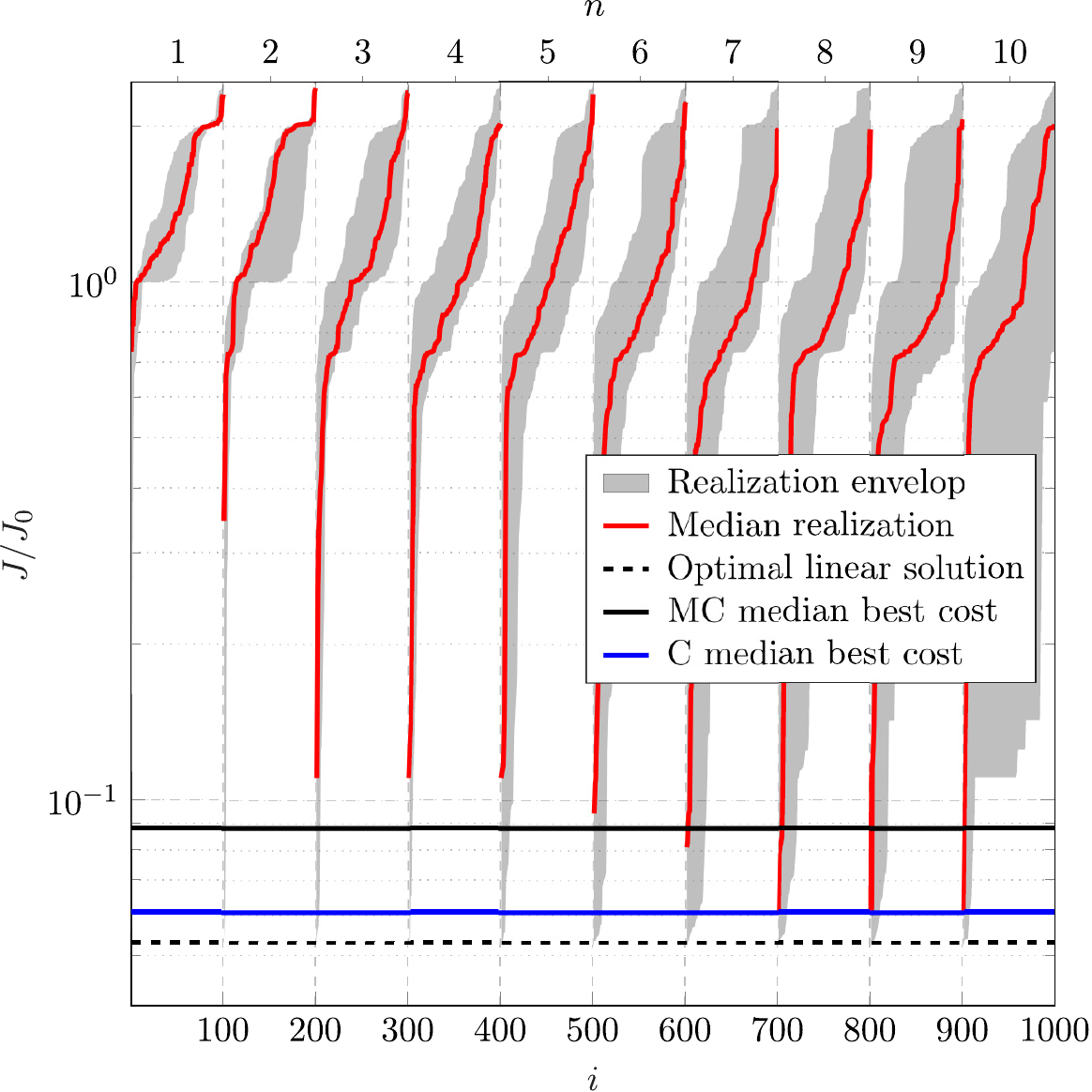}
  \caption{Costs of individuals for 100 realizations of crossover-only MLC.
  The gray region represents the envelop of the 100 realizations.
  The individuals are divided in 10 groups of 100 individuals representing the 10 generations.
  In a given generation the individuals are sorted by their cost.
  The median realization is depicted in red.
  The black horizontal line represents the median cost for the Monte Carlo sampling.
  The blue horizontal line represents the median cost ($\Delta J/J_0=93.94\%$) for the crossover-only MLC.
  The black dotted line represents the cost of the optimal linear control law.
  The vertical axis is in $\log$ scale.}
  \label{fig:C_am}
\end{figure}
Like for Monte Carlo sampling, the 100 realizations are collapsed in a single figure~\ref{fig:C_am}.
First, we note that the median realization of crossover-only MLC performs better than the Monte Carlo sampling one. 
Second, the convergence speed is improved,
indeed already after 700 evaluations, the cost of the best individual is lower compared to the Monte Carlo sampling.
This shows that a genetic programming algorithm only including recombination with crossover is already able to stabilize the Landau oscillator and performs better than Monte Carlo sampling.

\begin{figure}[htb]
\centering
\subfloat[]{\label{fig:C_median}\includegraphics[width=0.45\textwidth]{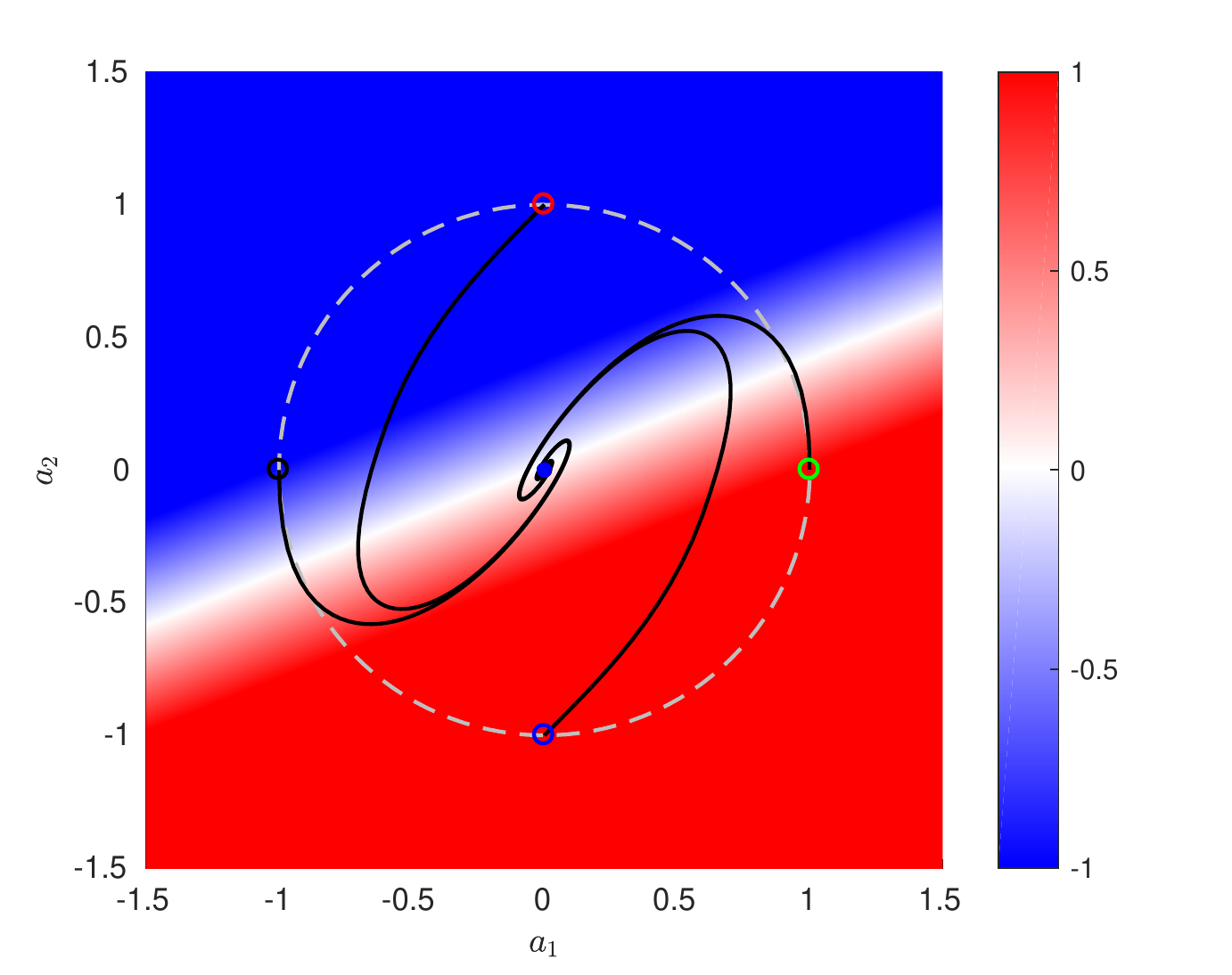}}%
\hfil
\subfloat[]{\label{fig:C_best}\includegraphics[width=0.45\textwidth]{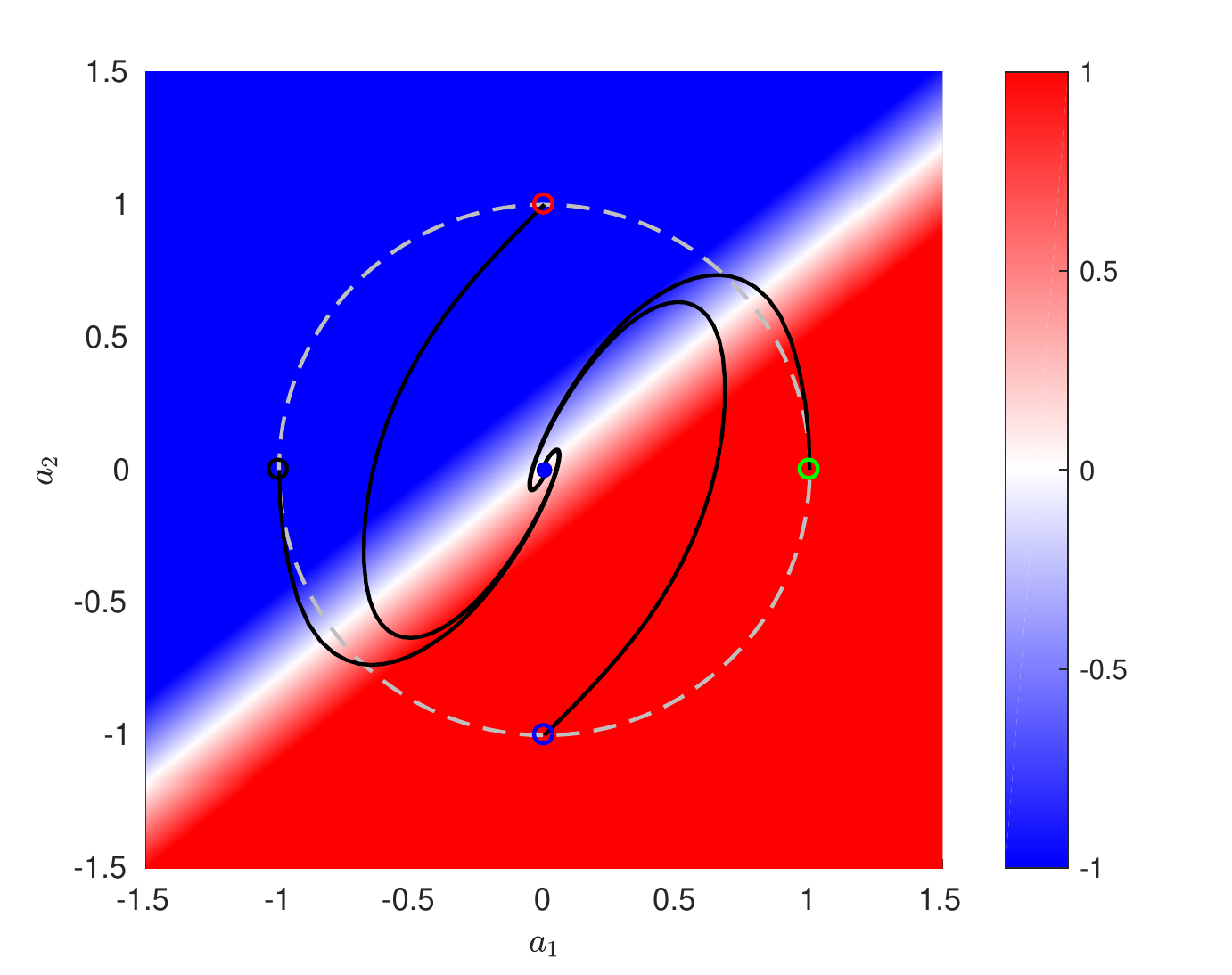}}%
\caption{Visualization of control laws in the phase space.
(a) Best control law of the median crossover-only MLC realization \eqref{Eq:Cmedian}.
(b) Best control law of the best crossover-only MLC realization \eqref{Eq:Cbest}.
The limit cycle is depicted with a dashed line.
}
\end{figure}
The best control law of the median crossover-only MLC is
\begin{equation}
  b_{\rm C, median }=a_1 -2.4876a_2
  \label{Eq:Cmedian}
\end{equation}
and the best control law of the best realization is 
\begin{equation}
  b_{\rm C, best} = 0.63361(4a_1 - 5a_2);
  \label{Eq:Cbest}
\end{equation}
Their associated cost reduction are $\Delta J/J_0=93.94\%$ and $\Delta J/J_0=94.69\%$ respectively.
After simplification of the expressions, both control laws are actually linear with a cost reduction close to the optimal linear solution ($\Delta J_{\rm opt}/J_0=94.70\%$).
We note, in particular, in figures~\ref{fig:C_median} and \ref{fig:C_best} that the separatrix angle is different for the two solutions leading to a faster convergence for the best realization.
\begin{figure}[htb]
  \centering
  \includegraphics[width=0.7\linewidth]{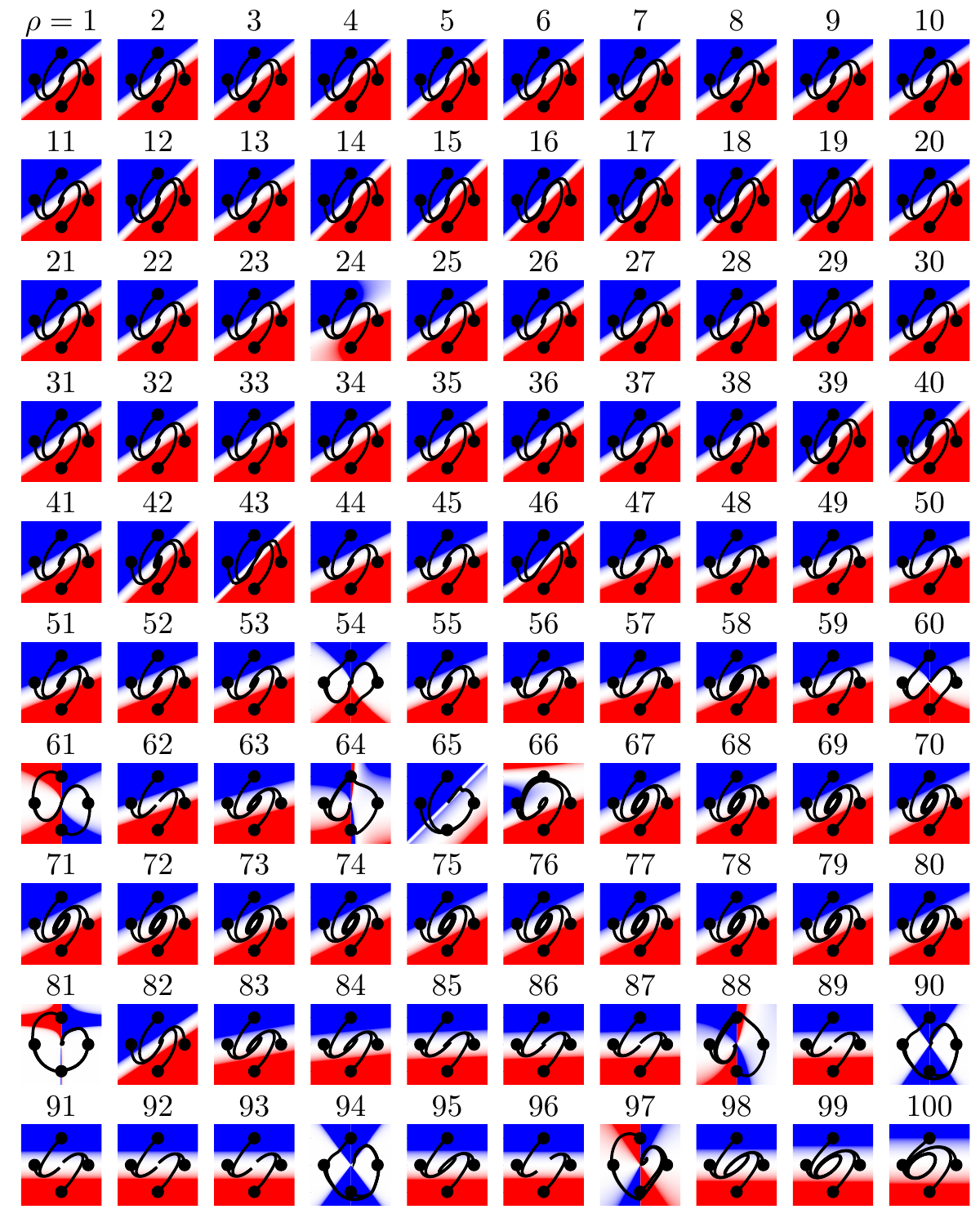}
  \caption{Same as \ref{fig:MC_best_each_real} but for crossover-only MLC.
  }
  \label{fig:C_best_each_real}
\end{figure}
Moreover, figure~\ref{fig:C_best_each_real} shows that 59 of the 100 realizations optimized control laws able to stabilize the oscillator in less than 2 periods.
Most of the best control laws derived are linear and close to the optimal linear one, except for realizations 24 and 54 that presents nonlinear behaviors.
The note that there is around the same number of horizontally symmetric solutions (13) compared to Monte Carlo sampling.
An explanation may be a poor first generation of individuals that are not able to build better control laws with crossover only.
The explorative power of mutation is expected to generate new individual and more complex individuals to enrich the population.

The exploitative-only or crossover-only MLC shows that only with one genetic operator genetic programming is more performing than Monte Carlo sampling.
Crossover-only MLC builds better control laws are optimized with less evaluations and has a better reproducibility.
Adding an explorative operator is expected to improve the results.
% ########################### END CROSSOVER ##############################

% ######################### CROSSOVER MUTATION ############################
\subsection{Exploitation and exploration with crossover and mutation}\label{Sec:MLC_CM}
\begin{table}[htb]
  \centering
   \begin{tabular}{>{\centering}p{2.5cm}>{\centering}p{6cm}>{\centering\arraybackslash}p{2cm}}
  Parameter & Description & Value\\
\midrule
   $\Nps$ & Population size & $100$\\
   $\Ng$ & Number of generations & $10$\\
   $\Ntour$ & Tournament size & $7$\\
   $\Ne$ & Rlitism & $1$\\
   $\Pcros$ & \textbf{Crossover probability} & $0,0.1,...,1$ \\
   $\Pmut$ & \textbf{Mutation probability}  & $1-\Pcros$ \\
   $\Prep$ & Replication probability & 0\\
\end{tabular}

  \caption{Parameters for crossover-mutation MLC }
  \label{fig:parameters_CM}
\end{table}
In this section, we progressively introduce exploration in the optimization process by tuning the crossover and mutation probabilities.
$\Pcros$ goes from 1 to 0 by steps of 0.1 and $\Pmut$ increases accordingly.
Thus, for $(\Pcros,\Pmut)=(1,0)$ MLC is purely exploitative, like in the previous section, and for $(\Pcros,\Pmut)=(0,1)$ MLC is purely explorative.
Table~\ref{fig:parameters_CM} summarizes the parameters employed for this section.

\begin{figure}[htb]
\centering
\subfloat[]{\label{fig:config_CM_MAD}\includegraphics[width=0.45\textwidth]{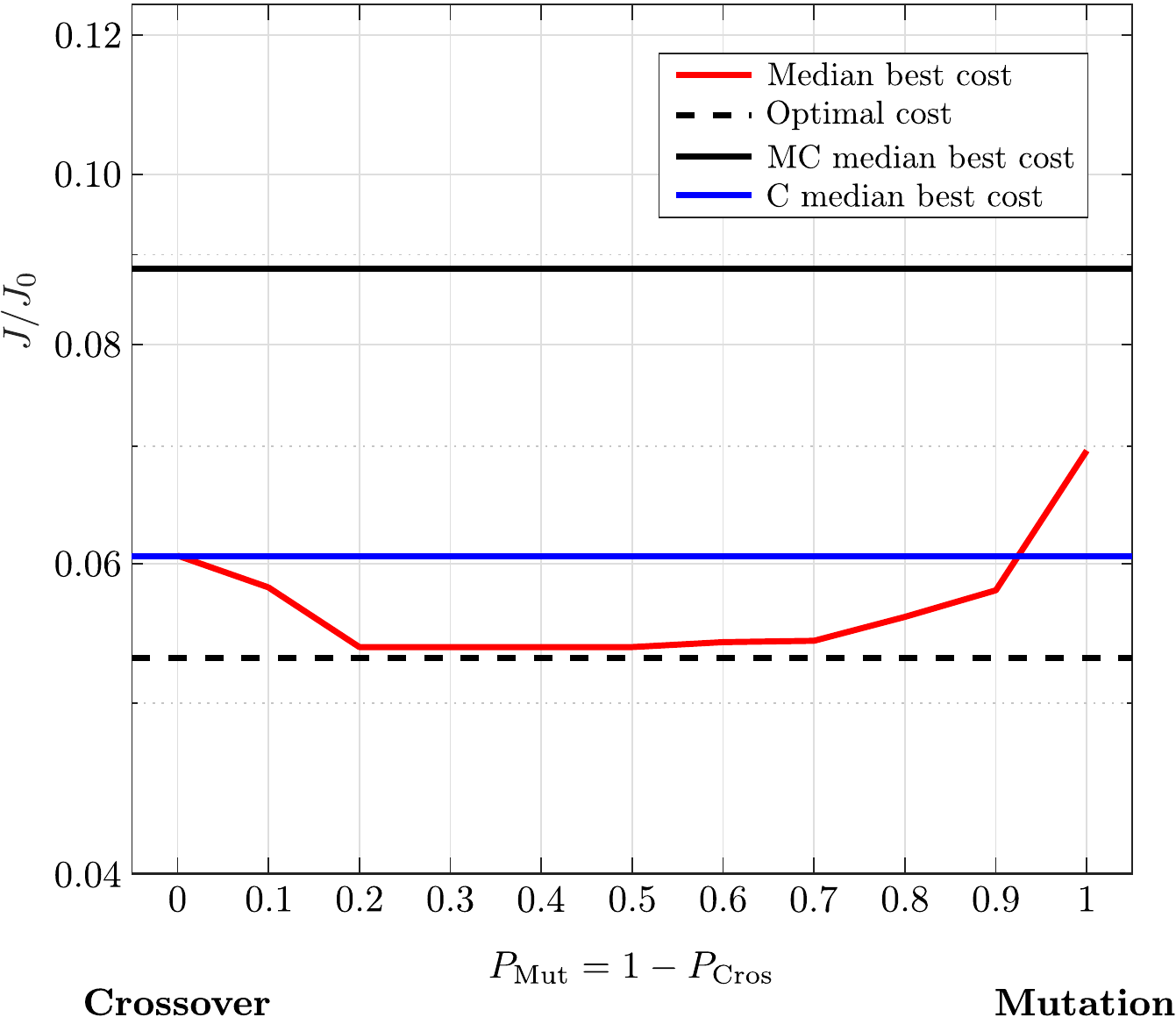}}%
\hfil
\subfloat[]{\label{fig:CM_am}\includegraphics[width=0.45\textwidth]{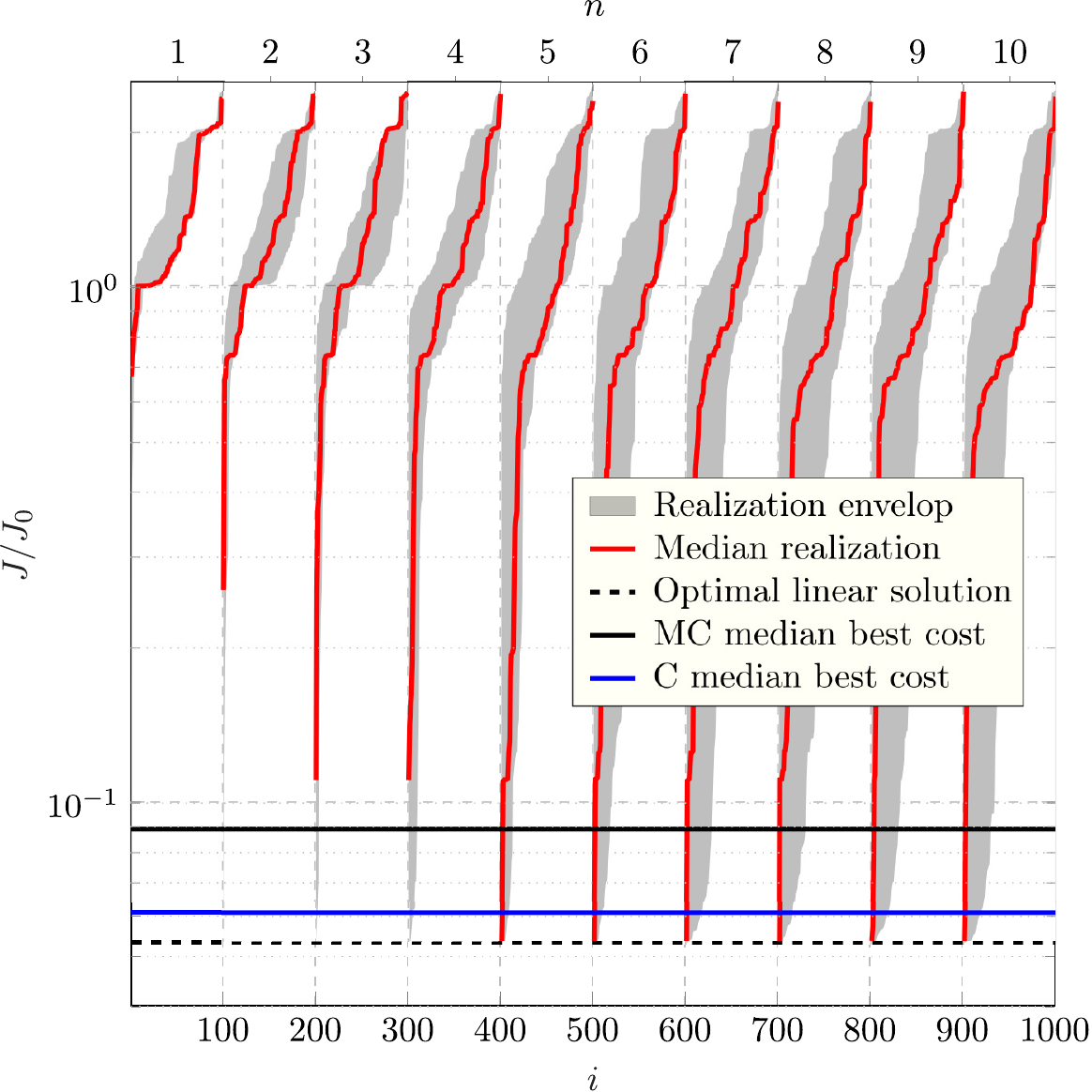}}%
\caption{
(a) Performance of MLC regarding the parameters $\Pmut=1-\Pcros$.
The most cost reduction ($\Delta J_{\rm}/J_0=94.62\%$) is achieved for configurations $\Pmut=0.2$ to $\Pmut=0.5$.
  The vertical axis is in log-scale.
(b) Same figure as \ref{fig:C_am} but for the $(\Pcros,\Pmut)=(0.7,0.3)$ configuration of MLC.
The best median best cost for crossover-mutation MLC is not displayed as it value would overlap with the cost of the optimal linear solution.
}
\end{figure}
In total 11 configurations are tested. 
For each couple of $(\Pcros,\Pmut)$ 100 realizations have been carried out.
Figure~\ref{fig:config_CM_MAD} summarizes the performances of the median realization for each configuration.
We recall that in this study we define the median as the $50^{\rm th}$ realization.
First, we note that an exploration-only MLC based only on mutation performs worse than any other combination but still better than Monte Carlo sampling.
This shows that random `mutation' of the best individuals throughout the generations is able to improve the performance of individuals and is better than just a random sampling.
However, exploration-only MLC is less performing than exploitation-only MLC.
This may be explained by the fact that crossover recombines individuals that are known to be good
whereas mutation still relies on randomness to improve its best individuals.
Second, we note that any combination of crossover and mutation
performs better than crossover and mutation only.
In particular, they are 4 configurations $\Pmut=0.2, 03, 04, 0.5$ that reduce the cost to the same level
and close to the performance of the optimal linear control solution.
The two best ones being $\Pmut=0.3$ and $\Pmut=0.4$.
Figure~\ref{fig:CM_am} shows the learning process of the $(\Pcros,\Pmut)=(0.7,0.3)$ configuration.
We note in particular that compared to crossover-only MLC, this configuration learns better control laws and with less evaluations.
Indeed, already after 500 evaluations, an individual performing better than crossover-only MLC has been found.

\begin{figure}[htb]
  \centering
  \includegraphics[width=0.7\linewidth]{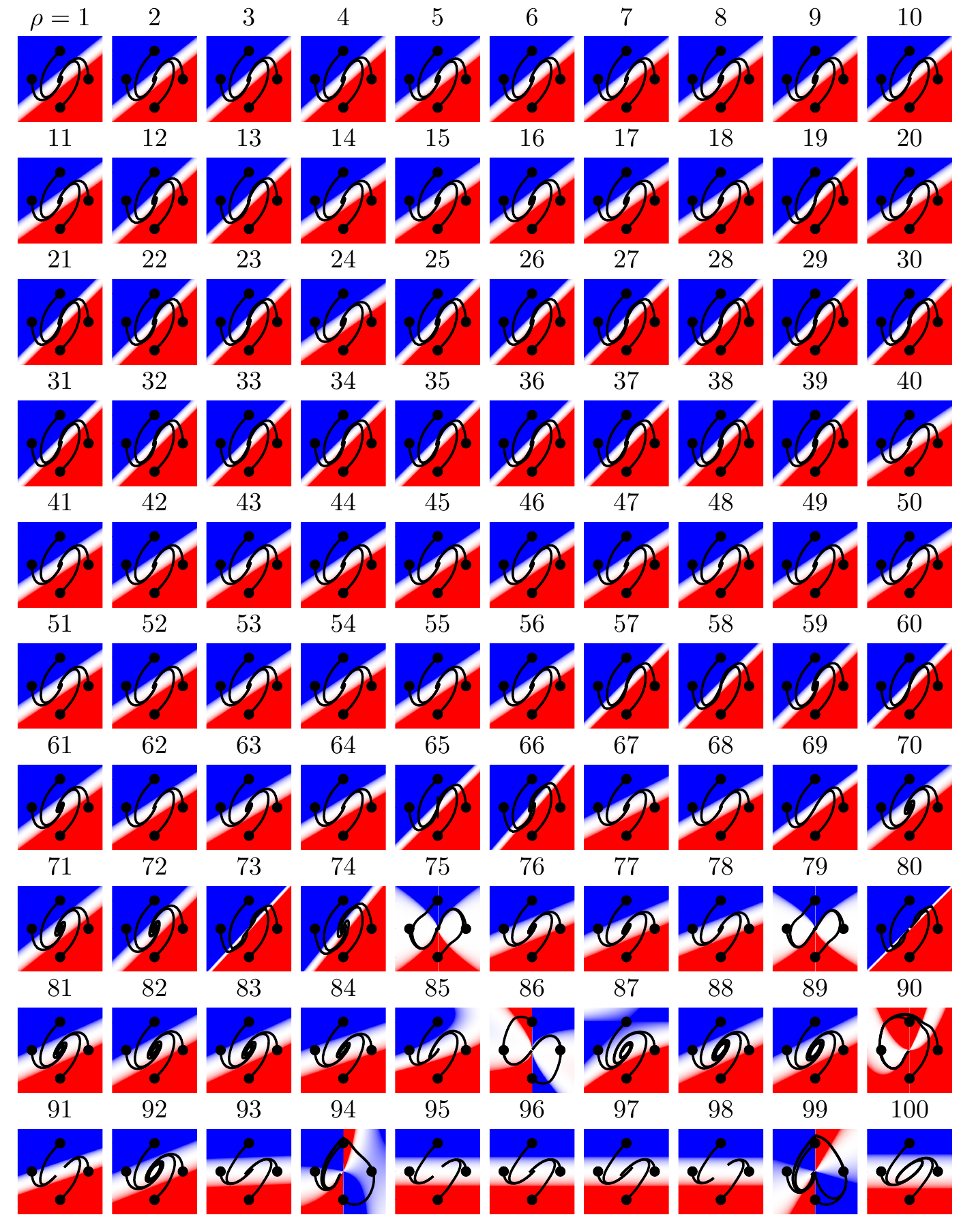}
  \caption{Same as \ref{fig:C_best_each_real} but for the $(\Pcros,\Pmut)=(0.7,0.3)$ MLC configuration.
  }
  \label{fig:CM_best_each_real}
\end{figure}
\begin{figure}[htb]
\centering
\subfloat[]{\label{fig:CM_median}\includegraphics[width=0.45\textwidth]{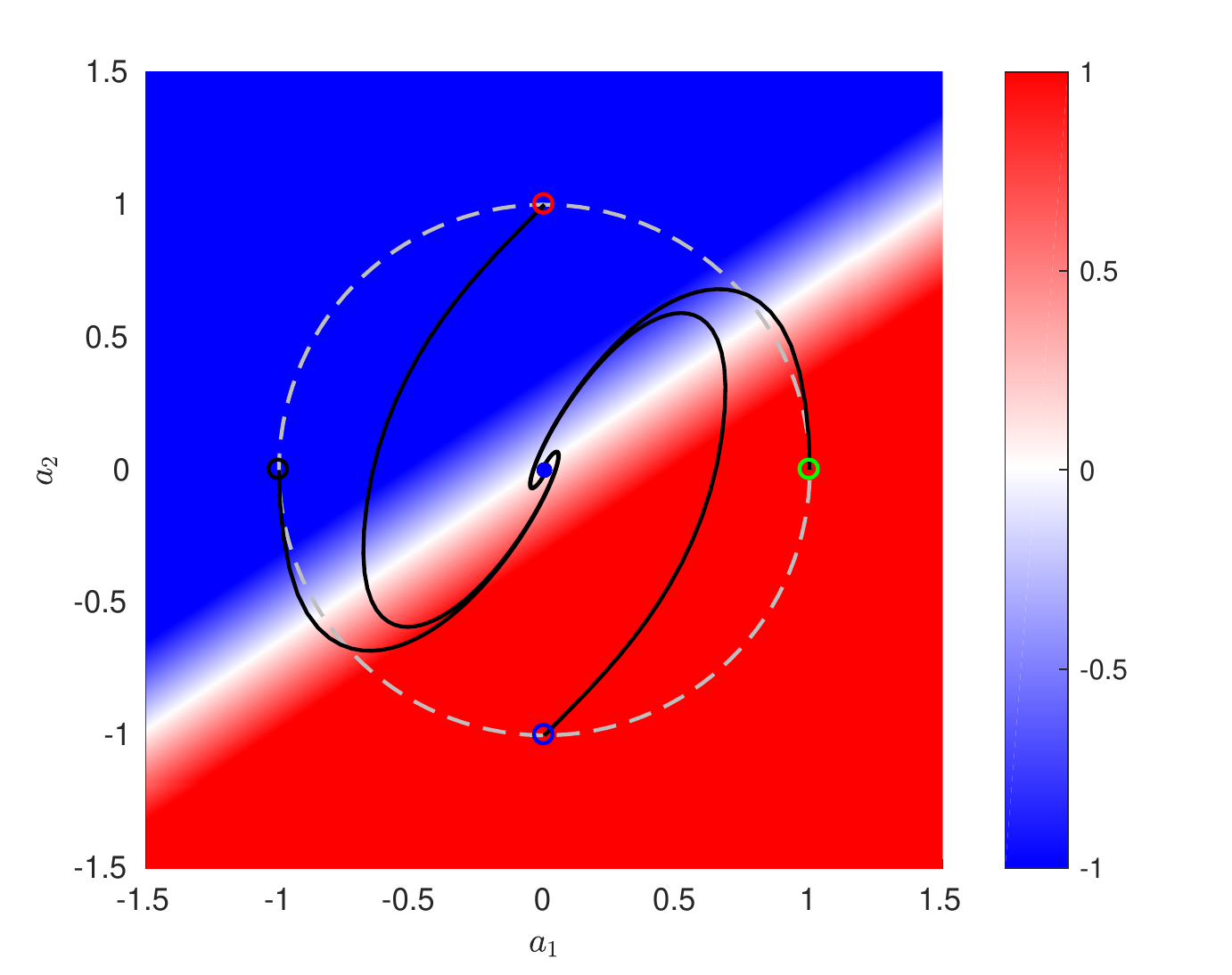}}%
\hfil
\subfloat[]{\label{fig:CM_best}\includegraphics[width=0.45\textwidth]{Figures/AppA/C_best_best_Visu.pdf}}%
\caption{Visualization of control laws in the phase space.
(a) Best control law of the median crossover-mutation MLC realization \eqref{Eq:CMmedian}.
(b) Best control law of the best crossover-mutation MLC realization \eqref{Eq:CMbest}.
The limit cycle is depicted with a dashed line.
}
\end{figure}

Figure~\ref{fig:CM_best} shows that for the $(\Pcros,\Pmut)=(0.7,0.3)$ configuration, 79 of the 100 realizations optimized control laws able to stabilize the oscillator in less than 2 periods.
Again, like for crossover-only MLC, most of the best control laws derived are linear and close to the optimal linear one, except for realizations 75 and 79 that presents nonlinear behaviors.
Moreover, 6 realizations optimized horizontally symmetric control laws, which is half the number of crossover-only MLC and Monte Carlo sampling.
This shows that the introduction of mutation improves indeed the exploration capability of MLC.

Equation~\eqref{Eq:CMmedian} is the control law that has been found by the median realizations of the four best configurations in figure~\ref{fig:config_CM_MAD}.
\begin{equation}
  b_{\rm CM, median }=2a_1 - 3a_2
  \label{Eq:CMmedian}
\end{equation}
This same control law has been in the four configuration.
The coefficients have been built by additions and subtraction of $a_1$ and $a_2$ only,
these realizations ignored the random constants in the registers.
The best control law of the best realization of $(\Pcros,\Pmut)=(0.7,0.3)$ MLC configuration is:
\begin{equation}
  b_{\rm CM, best}= 2.4932a_1 -3.1001a_2.
  \label{Eq:CMbest}
\end{equation}
Again, it is a linear control but whose coefficients have been build from the random constants.
The cost reduction associated to the median and best realization are, respectively, ($\Delta J/J_0=94.62\%$) and ($\Delta J/J_0=94.69\%$).
Figures~\ref{fig:CM_median} and~\ref{fig:CM_best} displaying their corresponding visualizations.
The only difference noticeably is the angle of the separatrix.

We've shown that combining crossover and mutation allows better performances than crossover and mutation alone.
Control laws performing as well as the optimal linear control are built with less evaluations.
The reproductivity of the results have also been improved as there are more realizations being able to quickly stabilize the oscillator.
In the next section, we analyze the full-fledge MLC, i.e., including crossover, mutation and replication.

 % ################## CROSSOVER MUTATION REPLICATION #######################
\subsection{MLC Optimization with crossover, mutation and replication}
\begin{table}[htb]
  \centering
   \begin{tabular}{>{\centering}p{2.5cm}>{\centering}p{6cm}>{\centering\arraybackslash}p{3.5cm}}
Parameter & Description & Value\\
\midrule
   & Function library & $ F_1= \{+,-,\times,\div\}$\\
   $\boldsymbol s$ & Controller inputs & $a_1$, $a_2$ \\
   $\Nvar$ & Number of variable registers & $3$ \\
   $\Ncst$ & Number of constant registers & $3$ \\
   $\Ninstrmax$ & Max. number of instructions & $ 5$\\
   $\gamma$ & Penalization parameter & 1\\
  \hline
   $\Nps$ & Population size & $100$\\
   $\Ng$ & Number of generations & $10$\\
   $\Ntour$ & Tournament size & $7$\\
   $\Ne$ & Elitism & $1$\\
   $\Pcros$ & \textbf{Crossover probability} & $\Pcros$ \\
   $\Pmut$ & \textbf{Mutation probability} & $\Pmut$ \\
   $\Prep$ & \textbf{Replication probability} & $1-\Pcros-\Pmut$ \\
\end{tabular}

  \caption{\xMLC for the optimization. The operators probability $(\Pcros,\Pmut,\Prep)$ are studied in this section.}
  \label{tab:parameters_CMRE}
\end{table}
In this section, we study the influence of the operators probability $(\Pcros,\Pmut,\Prep)$.
For this, we run MLC optimizations with different combinations of $(\Pcros,\Pmut,\Prep)$.
We start from $(\Pcros,\Pmut,\Prep)=(1,0,0)$, full crossover, and we increase and decrease each parameter with a step of 0.1.
We recall that $\Pcros+\Pmut+\Prep=1$.
There are in total 66 combination of parameters to test.
We run $\Nrho=100$ realizations for each combination of probabilities.
The parameters are summarized in table~\ref{tab:parameters_CMRE}.

\begin{figure}[htb]
  \centering
  \includegraphics[width=0.6\linewidth]{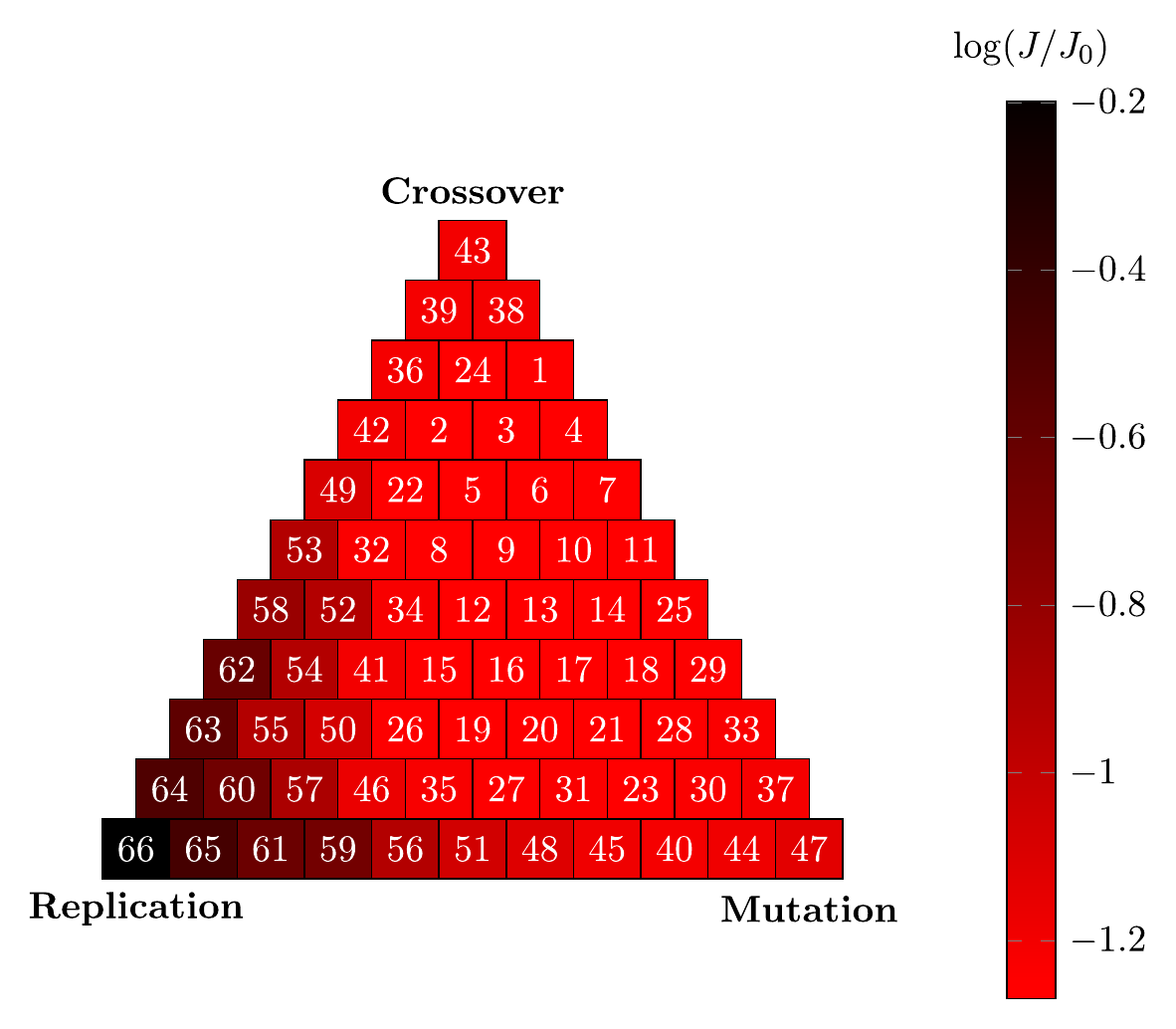}
  \caption{Performance of MLC regarding the parameters $(\Pcros,\Pmut,\Prep=1-\Pcros-\Pmut)$.
  Each block of the triangle corresponds to one probability configuration.
  A neighbor block means a variation of $0.1$ of the genetic operator probabilities.
  Top block is crossover-only $(1,0,0)$, bottom-left block is replication-only $(0,0,1)$ and bottom-right block is mutation-only $(0,1,0)$.
  For each configuration 100 realizations have been carried out.
The color code symbolize the performance of the best individual for the median realization of each configuration.
  The configurations are ranked following their relative performance, $1$ being the best configuration and $66$ the worst.
The first 21 configurations have all the same minimum value corresponding to a cost reduction of $\Delta J/J_0=94.62\%$.
Configurations $52$ to the $66$ perform worse than Monte Carlo sampling.
  }
    \label{fig:TriangleMedian}
\end{figure}
Figure~\ref{fig:TriangleMedian} presents the results of the 66 configurations.
The replication-only is, as expected, the worse configuration.
Indeed, this is equivalent to run a Monte Carlo sampling with only $\Ni=100$ individuals.
The addition of crossover and mutation progressively improves the replication-only optimization.
We notice that both crossover-only and mutation-only MLC are less efficient than most of the combinations of crossover, mutation and replication.
Also, it is the combination of crossover and mutation that gives the best results.
We note that from configuration $(\Pcros,\Pmut,\Prep)=(0.4,0.1,0.5)$, ranked $52^{\rm nd}$, the median costs are lower than the median cost of Monte Carlo sampling.

It is worth noting that crossover-only MLC (ranked $43^{\rm rd}$) is better than mutation-only MLC (ranked $47^{\rm th}$).
This indicates that crossover and exploration has a better learning potential than the exploitative power of mutation.

Also, we highlight that the median cost of the 21 first configurations are all equal.
Indeed, the median realizations of these 21 configurations all manage to build the same controller, in 13 different ways:
\begin{equation*}
b = 2a_1-3a_2
\end{equation*}
that reduces the cost by $\Delta J/J_0=94.62\%$, see Sec.\ \ref{Sec:MLC_CM} and figure~\ref{fig:CM_median}.
Surprisingly, even for such simple configuration, the replication operator also plays a non-negligible part as the 21 first configurations are all equivalent.
We can conclude that memory is beneficial to explore a search space even when it is expected to have few minima.

\begin{figure}[htb]
  \centering
  \includegraphics[width=0.5\textwidth]{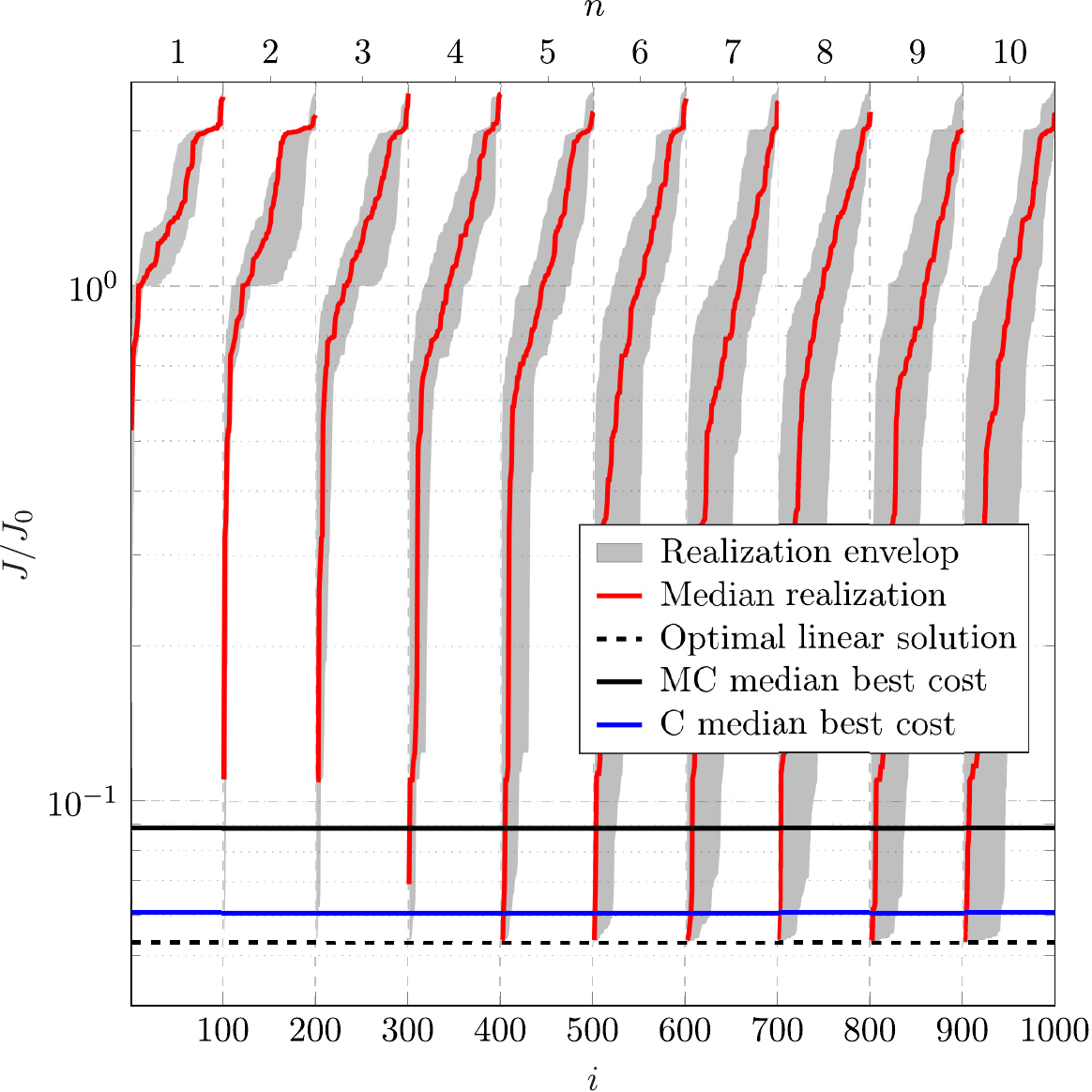}
  \caption{Same figure as \ref{fig:C_am} but for full-fledge MLC with the configuration $(\Pcros,\Pmut,\Prep)=(0.6,0.3,0.1)$.}
  \label{fig:CMRE_am}
\end{figure}
To differentiate the configurations, we look at the cost of the best individual for their $25^{\rm th}$ realization.
The five best configurations have similar costs and are in the same region: $(\Pcros,\Pmut,\Prep)=(0.6,0.2,0.2)$, $(0.7,0.2,0.1)$, $(0.6,0.3,0.1)$, $(0.5,0.2,0.3)$, $(0.5,0.4,0.1)$.
We choose to study the configuration at the `center' of them all $(\Pcros,\Pmut,\Prep)=(0.6,0.3,0.1)$.
Figure~\ref{fig:CMRE_am} shows the learning process of this configuration for the median realization.
Like for crossover-mutation $(\Pcros,\Pmut,\Prep)=(0.7,0.3,0)$ MLC, the learning is accelerated compared to crossover only and the best solution reaches reduces the cost similarly to the optimal linear solution.

\begin{figure}[htb]
  \centering
  \includegraphics[width=0.7\linewidth]{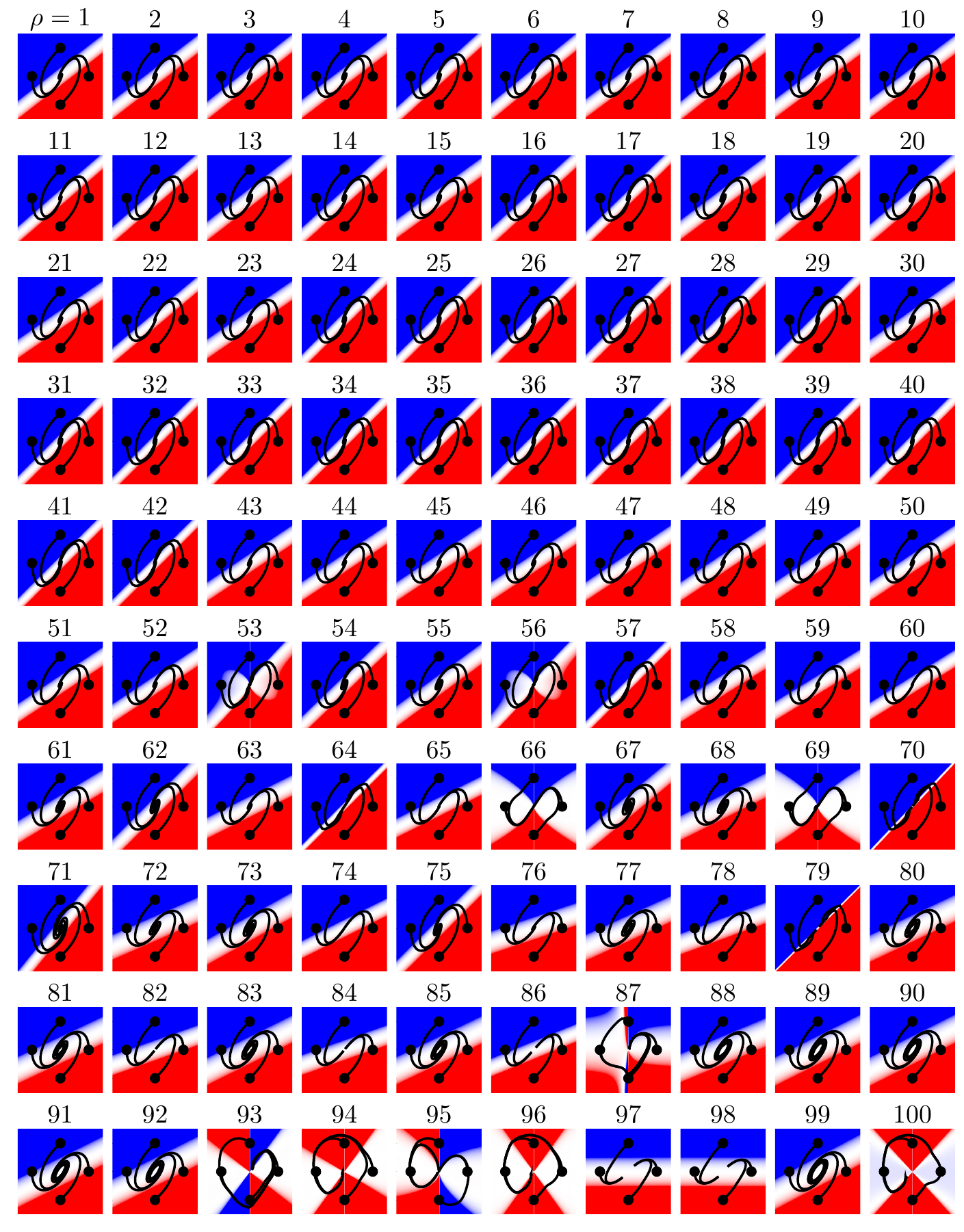}
  \caption{Same as \ref{fig:CM_best_each_real} but for full-fledged MLC.
  }
  \label{fig:CMRE_best}
\end{figure}
Figure~\ref{fig:CMRE_best} shows the best control laws of all realizations of $(\Pcros,\Pmut,\Prep)=(0.6,0.3,0.1)$ MLC.
We note, first that 78 realizations manage to stabilize the fixed point in less than two periods.
This number is one less than the crossover-mutation $(\Pcros,\Pmut,\Prep)=(0.7,0.3,0)$ MLC.
This is not surprising as the control laws are less recombined due to the 0.1 replication probability.
However, we note that there are only 2 horizontally symmetric solution, suggesting that memory benefits in building complex control laws.

% ------------ Best control law --------------
The best control law all probability combinations and realizations combined is found for the configuration $(\Pcros,\Pmut,\Prep)=(0.5,0.4,0.1)$ and reads after simplification:
\begin{equation*}
    b_{\rm best} = \left(\frac{a_1-2a_2}{a_1}\right)^2 (a_1-a_2).
\end{equation*}
It allows a cost reduction of $\Delta J_{best}/J_0=94.78\%$, better than the optimal linear control, revealing that the linear control is in fact not the global minimum of the problem.
It is worth noting that $b_{\rm best}$ is built only from $a_1$ and $a_2$.
No constants have been used to `adjust' the control law.
This may be explained by the fact that tuning constants from the initial random ones is costly in terms of instructions.
\begin{figure}[htb]
\centering
\subfloat[Phase space, actuation command, instantaneous cost function and radius for the controlled case.]{\includegraphics[width=0.45\textwidth]{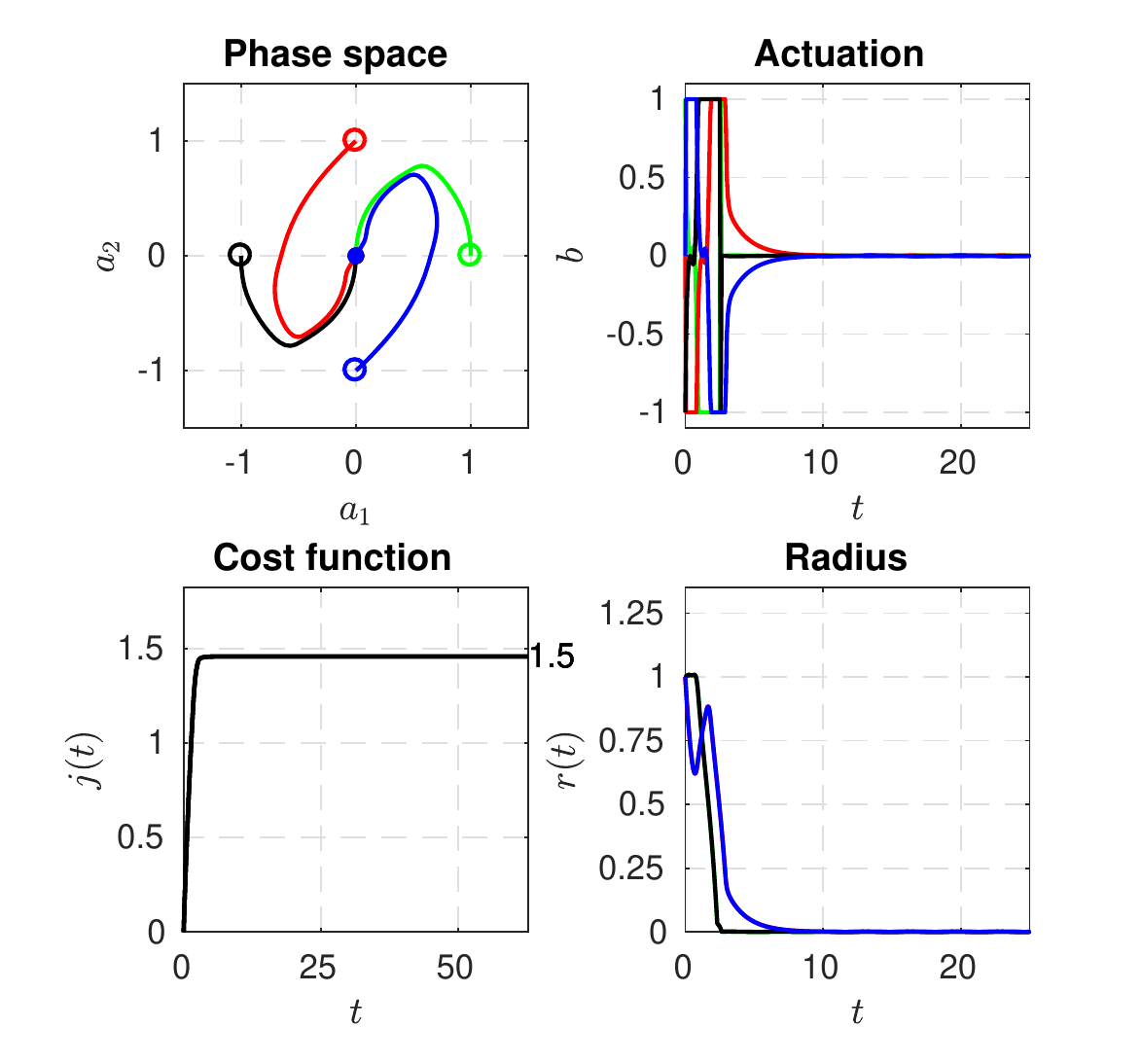}}%
\hfil
\subfloat[Visualization of the control law in the phase space.]{\includegraphics[width=0.53\textwidth]{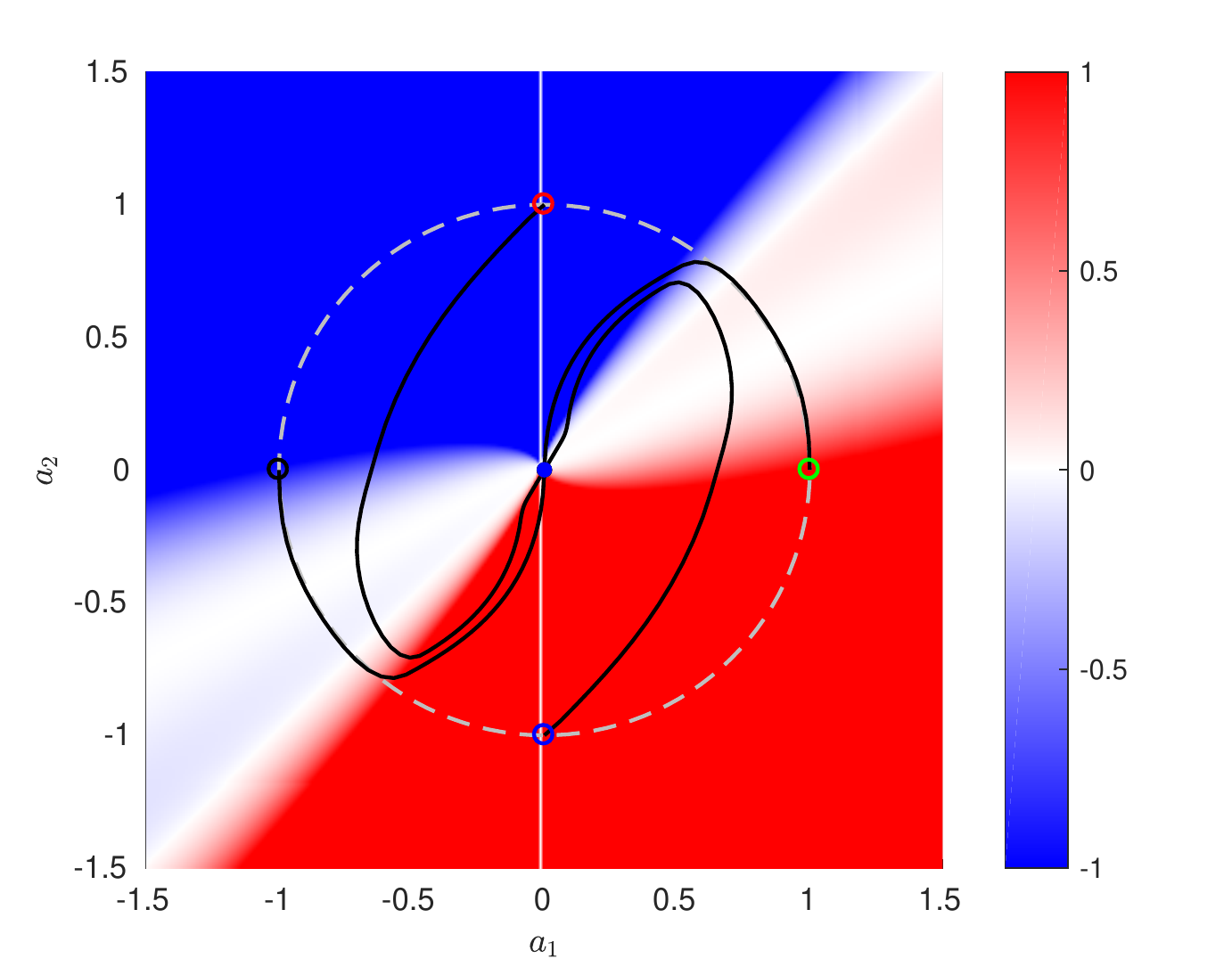}}%
\caption{\label{Fig:BestOfAll} Figures characterizing the best control law $b_{\rm best}$ derived thanks to \xMLC with the $(\Pcros,\Pmut,\Prep)=(0.5,0.4,0.1)$ combination.}
\end{figure}
Figure~\ref{Fig:BestOfAll} shows the phase portrait of the controlled oscillator with the best control found $b_{\rm best}$.
The actuation map displays the nonlinear effect of the control.
The actuation is globally similar to the linear control, however the control intensifies near the fixed point.
An absence of control in the diagonal direction prevents the system to go beyond the limit cycle unlike the optimal linear control and thus converges even faster.
The curvature of the lobes that delimit the control must play a role in the fast convergence of the system towards the fixed point.
We note that the oscillator converges linearly when close enough to the fixed point.
Thus, \xMLC manages to find an unexpected structure for the control law that performs better than the optimal linear control.
We can expect that this nonlinear control law can be improved with adequate constant tuning.

If we take a look at the best individual of the $(\Pcros,\Pmut,\Prep)=(0.6,0.3,0.1)$ combination, all realizations combined, we notice that the best control law is not only linear:
\begin{equation*}
    b_{(0.6,0.3,0.1)\rm, best} = 2.4353 a_1- 3.1238a_2
\end{equation*}
but the associated cost reduction $\Delta J/J_0=94.70\%$ is similar to the `optimal' linear control found with \texttt{fminsearch}.
This shows that the solution (equation~\eqref{eq:opt_lin}) is, in fact only a local minimum of the space of linear control laws.
When we run a \texttt{fminsearch} with this new control law as initial condition, we do not observe any improvement thus we can assume that this new solution is another local minimum in the space of linear control laws.
One explanation of the presence of several minima in such simple dynamical system may be the nonlinear saturation of the actuation command.
We do not display the phase portrait of the best linear control found with the $(\Pcros,\Pmut,\Prep)=(0.6,0.3,0.1)$ combination as it is identical to figure~\ref{Fig:OptLinSolution} and no differences can be seen with the naked eye.

Here, we shall stop the analysis of the influence of the probability operators $(\Pcros,\Pmut,\Prep)$ to focus on other meta-parameters.
In the following, we employ MLC with the $(\Pcros,\Pmut,\Prep)=(0.6,0.3,0.1)$ configuration as it is one that have the lowest median cost value.

\section{Influence of population size and number of instructions}\label{sec:mlc_parametric_study}
In this section, we present the influence of other parameters on the learning process of \texttt{xMLC}.
In particular, we investigate the role of the population size $\Nps$, the maximum number of instructions $\Ninstrmax$ and the choice of the function library.

\begin{table}[htb]
  \centering
   \begin{tabular}{>{\centering}p{2cm}>{\centering}p{6cm}>{\centering\arraybackslash}p{5cm}}
Parameter & Description & Value\\
\midrule
   & \multirow{2}{*}{\textbf{Function library}} & $ F_1= \{+,-,\times,\div\}$\\
   &  & $F_2 = F_1 \cup \{\mathrm{exp},\tanh,\sin,\cos,\mathrm{log}\}$\\
   $\boldsymbol{s}$ & Controller inputs & $a_1$, $a_2$ \\
   $\Nvar$ & Number of variable registers & $3$ \\
   $\Ncst$ & Number of constant registers & $3$ \\
   $\Ninstrmax$ & \textbf{Max. number of instructions} & [2,5,10,20,50,100,200,500,1000]\\
  \hline
   $\Nps$ & \textbf{Population size} & [10,20,50,100,200,500,1000]\\
   $\Ng$ & \textbf{Number of generations} & [100,50,20,10,5,2,1]\\
   $\Ntour$ & Tournament size & $7$\\
   $\Ne$ & Elitism & $1$\\
   $\Pcros$ & Crossover probability & 0.6 \\
   $\Pmut$ & Mutation probability & 0.3 \\
   $\Prep$ & Replication probability & 0.1 \\
\end{tabular}

  \caption{Parameters for the parametric study of MLC.
  The studied parameters are in bold.
  Parameters are separated in two sets: (top) parameters that define the control laws, (bottom) parameters that define the learning process.}
  \label{tab:LGPC_param_study}
\end{table}
For a fair comparison between the parameters, we run all MLC optimizations with $\Ni=1000$ individuals.
Seven population sizes are tested: 10, 20, 50, 100, 200, 500, 1000 with the adequate number of generations: 100, 50, 20, 10, 5, 2, 1.
The 1000 individual run is equivalent to a Monte Carlo optimization,
Also, we vary the maximum number of instructions $\Ninstrmax$, nine values are tested: 2, 5, 10, 20, 50, 100, 200, 500, 1000.
This means that for $\Ninstrmax=2$, the instruction matrix only contains two rows, and one thousand rows for $\Ninstrmax=1000$.
Finally, we look at the impact of the function library in the optimization process.
We investigate two libraries: $F_1 =\{+,-,\times,\div\}$ and $F_2 =\{+,-,\times,\div,\mathrm{exp},\tanh,\sin,\cos,\mathrm{log}\}$.
The rest of the parameters are the same as the previous section and are summarized in table~\ref{tab:LGPC_param_study}.
The operator probabilities $(\Pcros,\Pmut,\Prep)$ are chosen according to Sec.\ \ref{Sec:ParamStudyGenOp}.

To have a relevant estimation of the performance of each combination of parameters, we realize $\Nrho=100$ runs for each combination of parameters.
 
\subsection{$F_1 =\{+,-,\times,\div\}$}
\begin{figure}[htb]
\centering
\subfloat[]{\label{fig:PI1Med}\includegraphics[width=0.53\textwidth]{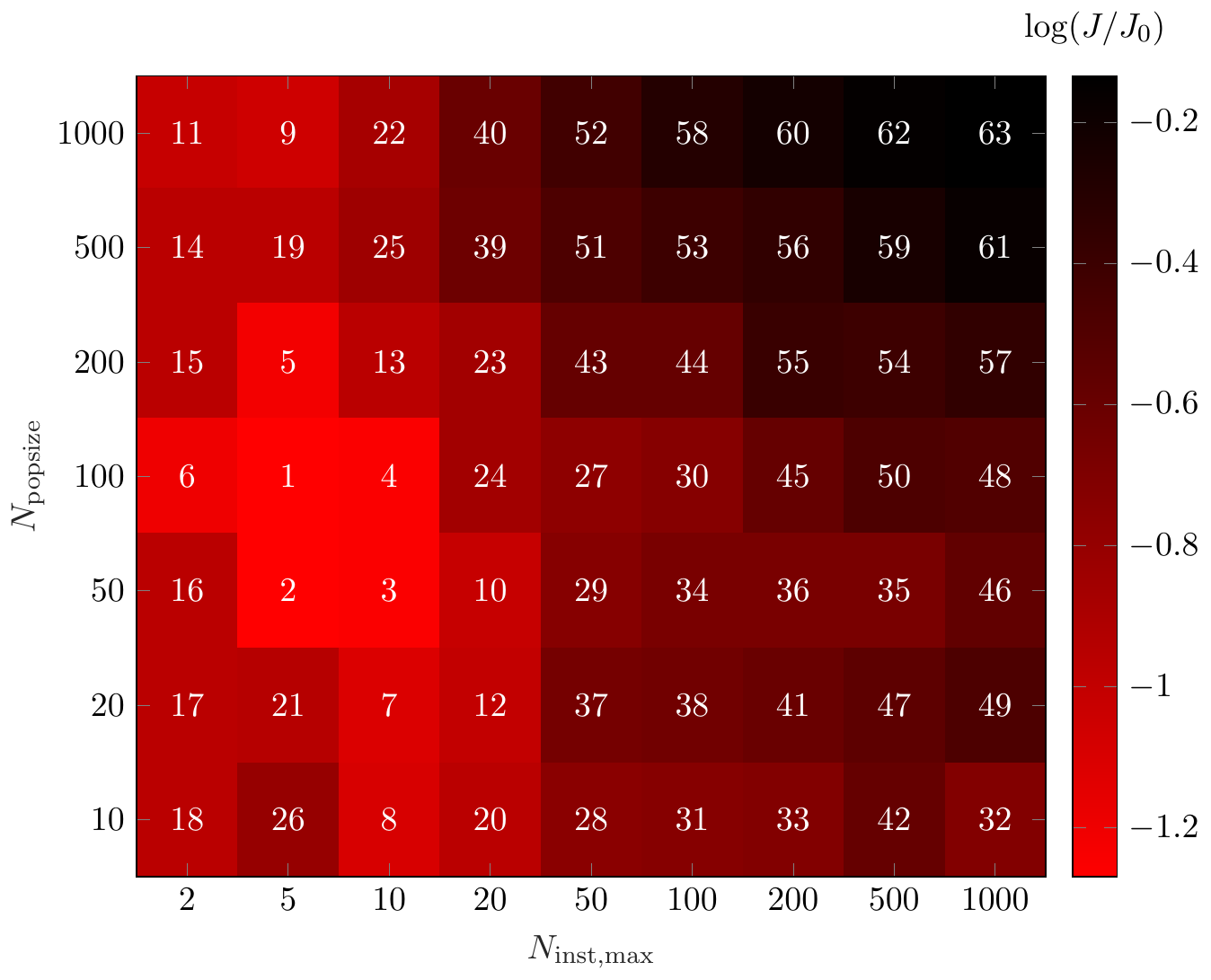}}%
\hfil
\subfloat[]{\label{fig:PI1Val}\includegraphics[width=0.45\textwidth]{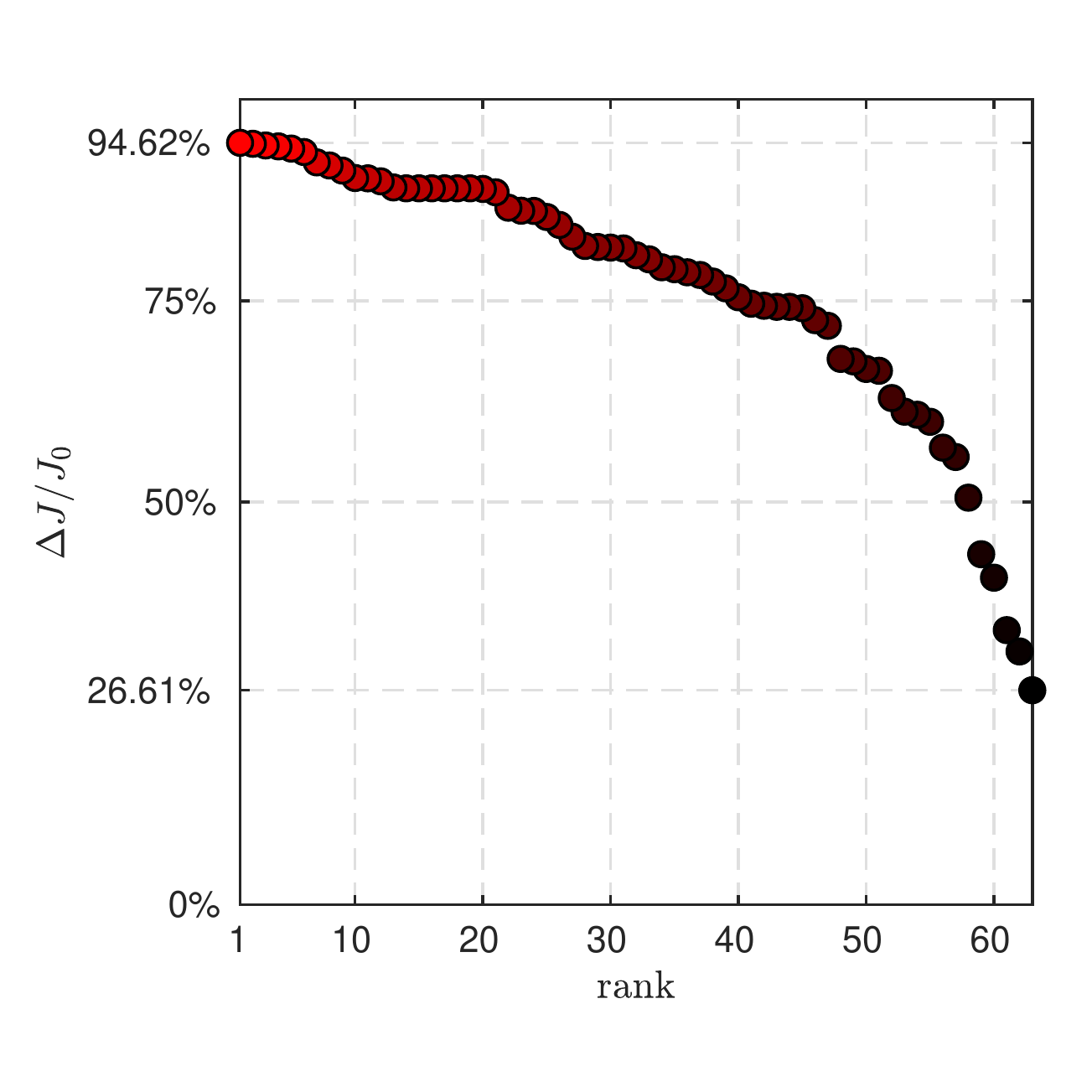}}%
\caption{\label{fig:PI1Median} (a) Performance of MLC regarding population size and maximum number of instructions for the function library $F_1=\{+,-,\times,\div\}$.
  The combination of parameters are ranked following the performance of their median realization.
  The combination ranked first is $(\Ninstrmax,\Nps)=(5,100)$.
(b) Cost reduction of the median realization of the different combinations.
The horizontal axis represent the rank of the configuration described in (a).
(a) and (b) share the same color code.}
\end{figure}
Let's first look at the results for the library employed in Sec.\ \ref{Sec:ParamStudyGenOp}.
In figure~\ref{fig:PI1Median}, we notice, first, that there is a favored region in the parameter space for a best median performance.
This region includes $\Ninstrmax \in [5,10]$ and $\Nps \in [50,100]$ and the performance decreases smoothly beyond this region.
The best configuration of all is $(\Ninstrmax,\Nps)=(5,100)$.

We remark that for a fixed population size, performance decreases when we increase the maximum number of instructions.
This can be explained by the fact that as we allow more instructions, the search space becomes all the more bigger.
Indeed, with more instructions, more and more complex control laws can be built, with the possibility to have a high level of function nesting.
When a large number of instructions is employed, a bigger number of generations is preferred.
Indeed, as the matrices have many more rows, several rounds of crossover and mutation are needed to shape the control law by selecting the adequate operators, registers and avoiding that further instructions destroy the intermediate expressions.
This also shows that too much instructions can be a hindrance for the learning process.
For the `simple' problem that is the stabilization of the oscillator where rather simple control laws are expected to work, a small number of instructions seems to be favored.
However, to chose too few instructions is not advantageous as we lose too much in complexity of the control laws.
The maximum number of instructions should be large enough to include relevant solutions but not too much otherwise, the search space becomes too large to be effectively explored in a reasonable number of generations.

Concerning the population size, we notice that population sizes around 50 and 100 are favored.
Also, some Monte Carlo samplings, especially with $\Ninstrmax=2$ and 5, perform better than MLC with $\Nps=10$.
This is because small populations are more vulnerable to random variations of the population due to mutation; a given structure can more easily take over the population even if it is not the best one, discarding relevant structures and leading the population into a local minimum.
This effect is also referred in biology as \emph{genetic drift}.

\begin{figure}[htb]
\centering
\subfloat[Phase space, actuation command, instantaneous cost function and radius.]{\includegraphics[width=0.405\textwidth]{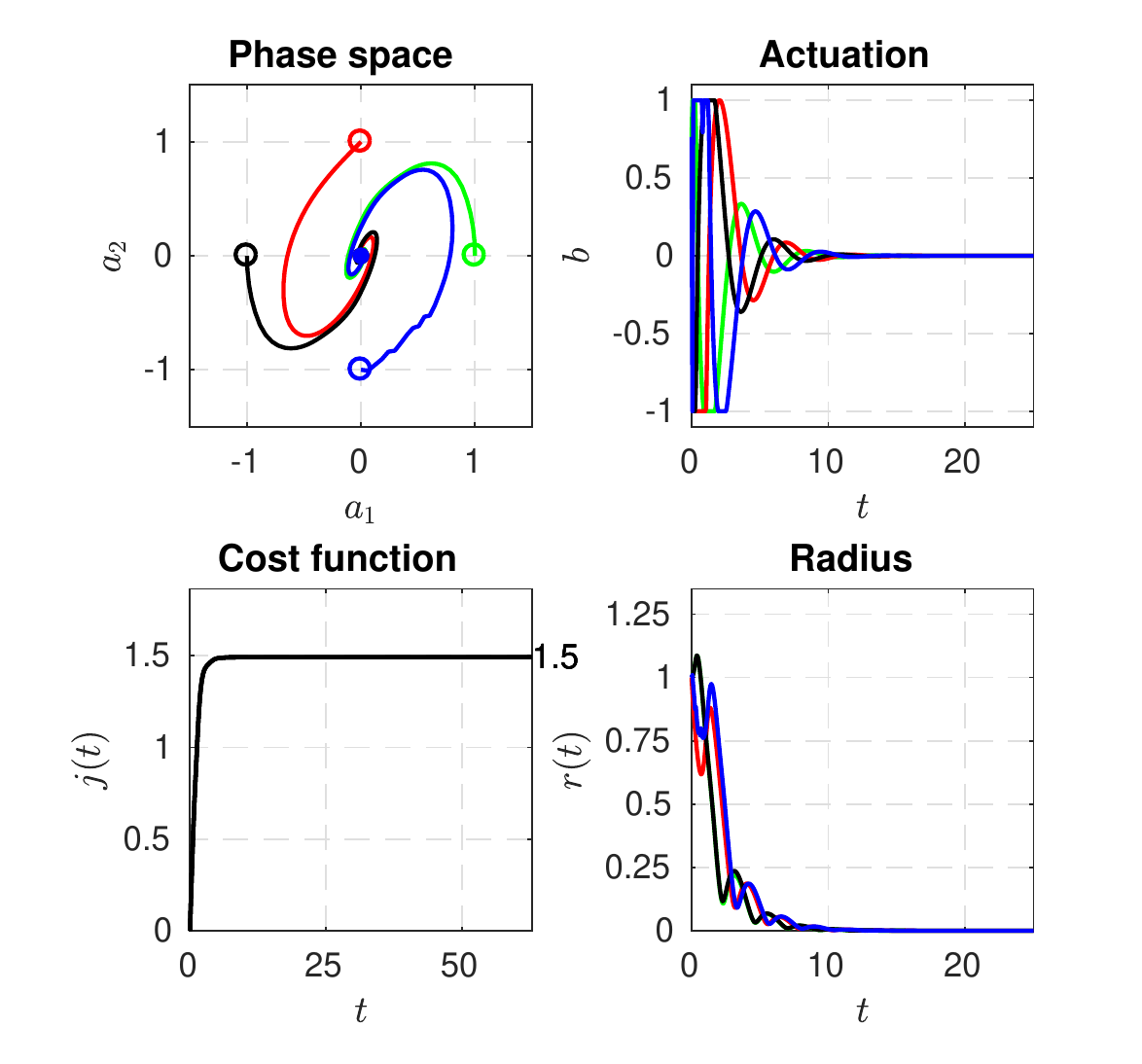}}%
\hfil
\subfloat[Visualization of the control law in the phase space. The limit cycle is depicted with a dashed line.]{\includegraphics[width=0.477\textwidth]{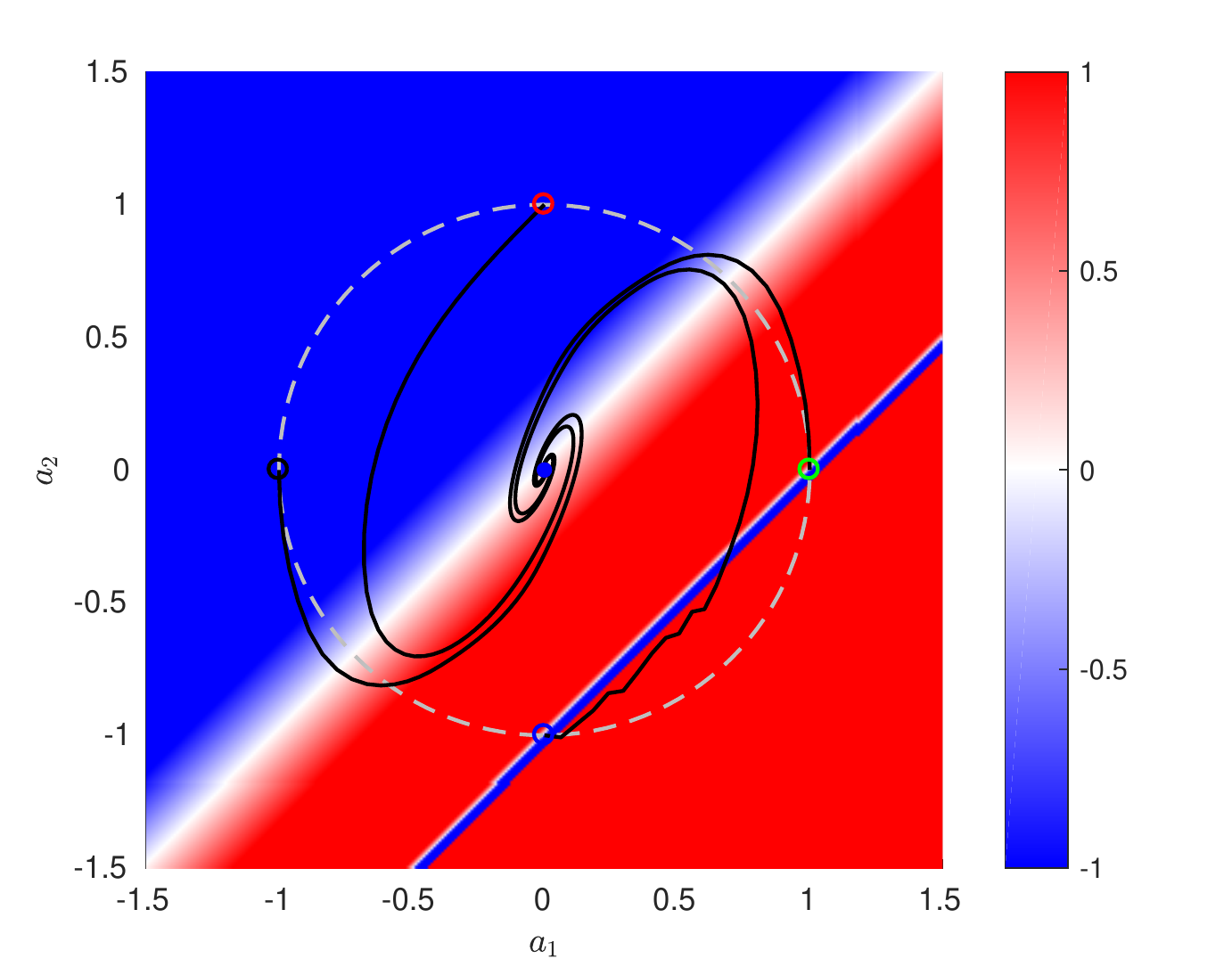}}%
\caption{\label{fig:PI1_best} Visualization of the controlled Landau oscillator with the best control law found for the function library $F_1=\{+,-,\times,\div\}$.
}
\end{figure}
The best control law among all realizations is:
\begin{equation*}
b_{F_1\rm ,best}=\cfrac{ (a_2 - a_1)^2}{
    \cfrac{
    \cfrac{(a_2 - a_1)}{-0.69096}
     - 5(a_2 - a_1)
    }{
    (a_2 - a_1)			}
     - 6(a_2 - a_1)
    }
     - 3(a_2 - a_1)
\end{equation*}
and is depicted in figure~\ref{fig:PI1_best}.
This complex expression, nesting divisions, has been found with the combination $(\Ninstrmax,\Nps)=(20,200)$ and reduces the cost by $\Delta J / J_0=95.50\%$.
Even though a limit a 200 instructions was needed to find it, we note that the matrix of this control law has in fact 23 rows.
Figure~\ref{fig:PI1_best} shows that the convergence of the oscillator towards the fixed point is not as fast as previous solution, indeed, we can still distinguish oscillations after the second period.
However, this solution seems to favor less intense actuation and thus reducing the $J_b$ component of the cost function.
Indeed, the cost of the linear optimal control is as follows: ${J_a}_{\rm opt}/J_0 = 2.16\% ,\; {J_b}_{\rm opt}/J_0 = 3.14\%$ and the cost of $b_{F_1\rm,best}$ is: $J_a/J_0 = 1.76\% ,\; J_b/J_0 = 2.74\%$

\subsection{$F_2 = F_1 \cup \{\mathrm{exp},\tanh,\sin,\cos,\mathrm{log}\}$}
\begin{figure}[htb]
\centering
\subfloat[]{\label{fig:PI2Med}\includegraphics[width=0.53\textwidth]{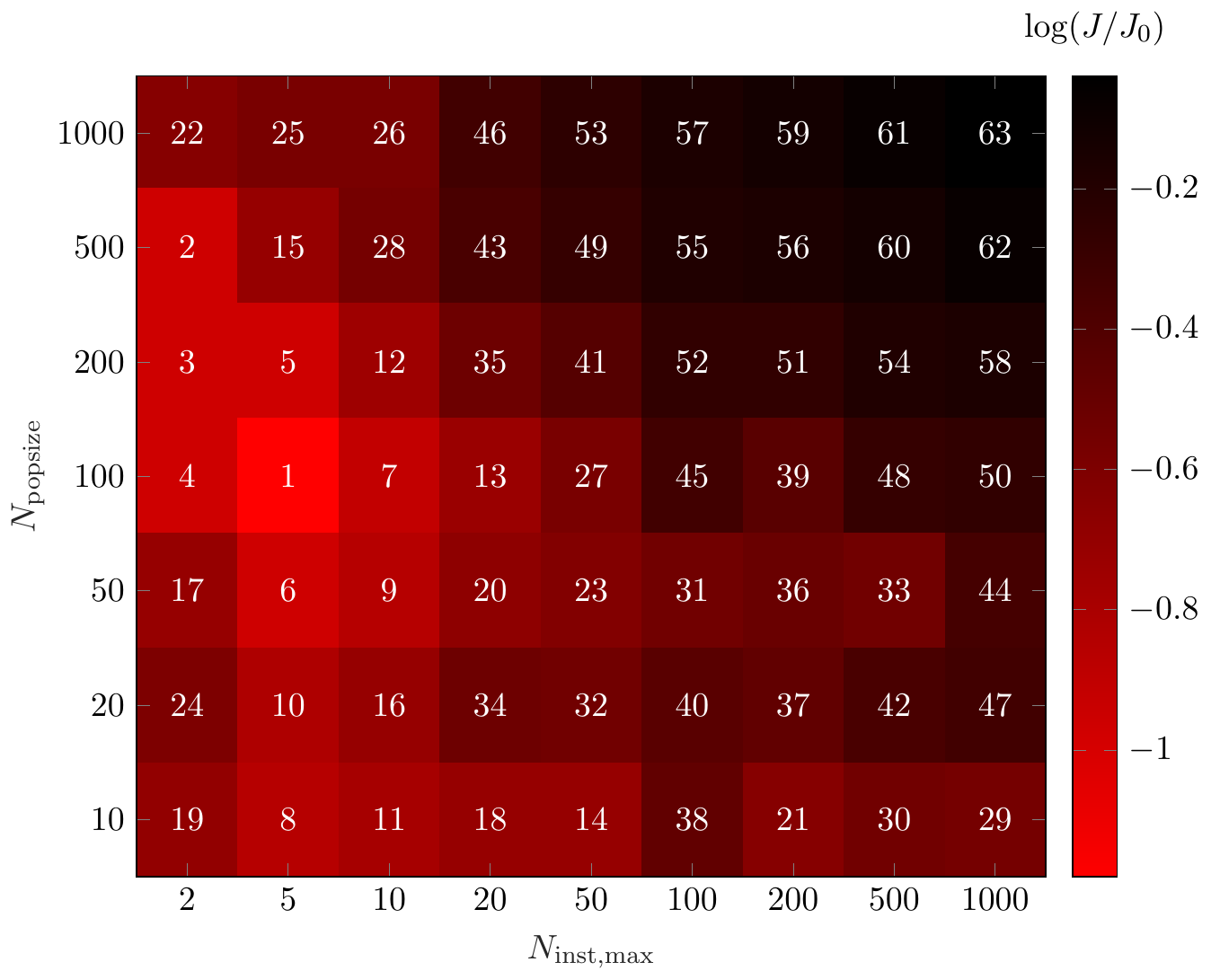}}%
\hfil
\subfloat[]{\label{fig:PI2Val}\includegraphics[width=0.45\textwidth]{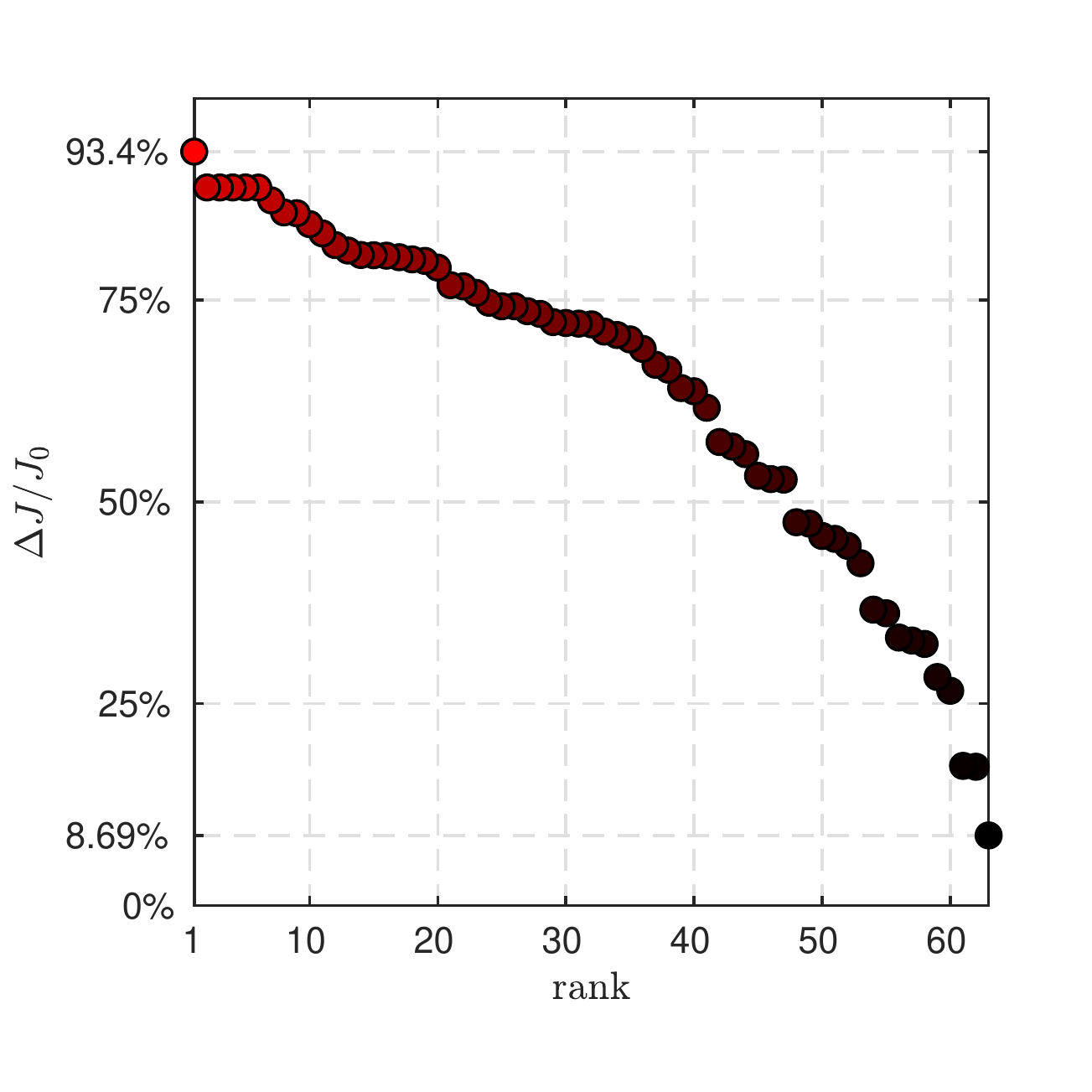}}%
\caption{\label{fig:PI2Median} (a) Performance of MLC regarding population size and maximum number of instructions for the function library $F_2 = F_1 \cup \{\mathrm{exp},\tanh,\sin,\cos,\mathrm{log}\}$.
  The combination of parameters are ranked following the performance of their median realization.
  The combination ranked first is $(\Ninstrmax,\Nps)=(5,100)$.
(b) Cost reduction of the median realization of the different combinations.
The horizontal axis represent the rank of the configuration described in (a).
(a) and (b) share the same color code.}
\end{figure}
We now add nonlinear functions in the function library, allowing oscillations with $\cos$ and $\sin$, exponential and logarithm growth with $\exp$ and $\log$ and saturation with $\tanh$.
As we allow more complex expressions, we can expect to find complex and more efficient solutions than the linear optimal control.
Figure~\ref{fig:PI2Median} show similar tendencies for the median realization as for the $F_1$ library.
We note however that the performances have dropped in general.
This can be explained by the fact, as we augment the function library, the search space becomes bigger and more individuals are needed to explore more control law structures.
The combination $(\Ninstrmax,\Nps)=(5,100)$ stands out from the other combinations as the best one with a reduction of $\Delta J /J_0=93.40\%$ of the unforced cost.

The best configuration for the $F_2$ library is $(\Ninstrmax,\Nps)=(5,100)$ as it is ranked first in terms of median cost.
 
The all time best control law has been found for the combination $(\Ninstrmax,\Nps)=(20,50)$ and reduces the cost by $\Delta J/J_0=97.08\%$ with a matrix containing 28 rows.
We shall not show the control law as it is exceedingly complex.
It comprises 215 instances of $\cos$, 217 instances of $\div$, 216 instances of $(a_1-a_2)$ and two times the same constant.
Following the number of occurrences of the operators, we can assume that the same pattern has been repeated and nested to shape the control law.
However, we cannot trust this solution as it appears to exploit the rather coarse time step chosen for integration.
Indeed, the dynamics and cost of this control law vary considerably when the integration time step decreases.

The success of MLC is that it managed to detect and exploit the most receptive state of the oscillator to efficiently control it with minimum actuation energy.
The most difficult task for MLC may have been to build the transient part of the control, to guide the system towards the state of interest.
Indeed, the best strategy found by MLC has been to let the system evolve freely until it gets to the region where the forcing is the most efficient, see figure~\ref{Fig:BestOfAll}.
Such strategy balances effectively stabilization and actuation power.
It has been built thanks to nonlinear functions and
we can assume that augmenting the function library with more nonlinear functions, such as inequality signs ($>$ or $\leq$) and boolean operators, can help the learning of better control laws.
However, one must select the function library carefully since a too diverse library lowers the general performance.

\chapter{Define user's problem}\label{AppMyPlant}
The definition of a MLC problem and its parameters is done by two files: \texttt{MyProblem\_problem.m} and \texttt{MyProblem\_parameters.m} respectively.
In Sec.\ \ref{AppA_parameters}, the \texttt{Plant/MyPlant/MyProblem\_parameters.m} file is detailed.
This file should be used as a template to build one's own parameter file.
A template for the definition of a MATLAB problem is also given in Sec.\ \ref{AppA_problem}.
To adapt the parameter file to her/his own problem, the user is invited to create a new folder in \texttt{Plant/}  with a name related to her/his problem.
There is no restriction on the name of the folder as it is not used in the code.

\section{User's parameter file}\label{AppA_parameters}
The \texttt{Plant/MyPlant/MyProblem\_parameters.m} is a basis to define one's own parameter file.
In this section, we provide comments and suggestions on how to choose the parameters.
The parameters displayed are just examples,
they are coherent between them but do not represent any real problem.
	
\lstset{numbers=left, numberstyle=\tiny, stepnumber=1, numbersep=2pt,
keywordstyle=\color{blue}, 
}

\begin{lstlisting}[name=MyParameters,firstnumber=1]
function parameters = MyPlant_parameters()
	% Template file to write one's own parameter file.
	% Every instance of "MyPlant" should be replaced by an adequate name.
	% Lines ending with a "*" should not modified.
	%
	% Guy Y. Cornejo Maceda, 2022/07/01
	%
	% See also default_parameters.

	% Copyright: 2022 Guy Cornejo Maceda (gy.cornejo.maceda@gmail.com)
	% The MIT License (MIT)

%% Options
	parameters.verbose = 2;
	
\end{lstlisting}

The parameter file is divided in four main sections.
First, the general parameters define the type of problem and how the \xMLC code handles it.
\begin{lstlisting}[name=MyParameters,firstnumber=16]
%% General parameters
    parameters.Name = 'TestRun'; % Name used to save the MLC object.
    % Name of the problem file. This name corresponds to the first part of
    % the MyPlant_problem.m file.
    % Of course, one can have several parameter files referring to the same
    % problem.
    parameters.EvaluationFunction = 'MyPlant'; 
    % Following if the plant is a MATLAB file or an external
    % solver/experiment:
    %       - 'MATLAB'  : when there is a *_problem.m file.
    %       - 'external': should be used when a numerical solver is
    %       employed.
    %       - 'LabView' : should be used when xMLC is interfaced with an
    %       experiment using LabView.
    % For the last options, the EvaluationFunction parameter does not
    % matter.
    parameters.ProblemType = 'MATLAB'; % 'MATLAB', 'external' or 'LabView'
    % Path for external evaluation.
    % When the code is interfaced with an external solver or LabView,
    % files are exchanged via the folder in the PathExt path.
    % For 'MATLAB' type problems, this path is useless.
    parameters.PathExt = '/Costs'; % Folder to exchange information.
    
\end{lstlisting}

Secondly, the problem parameters define the inputs and outputs of the controller, and some parameters related to the system to control.

\begin{lstlisting}[name=MyParameters,firstnumber=39]
%% Problem parameters    
    % Inputs and outputs definition
        % The inputs and outputs are considered from the controller point
        % of view. Thus, outputs corresponds to the actuation commands and 
        % the inputs are the sensors signals or time dependent functions.
        % Outputs---Control laws
        ProblemParameters.OutputNumber = 2; % Number of control laws
        % Inputs---Sensors and time dependent functions
            % si(t)---Sensor signals
            ProblemParameters.NumberSensors = 2;
            % Name of the variables in the solver.
            % Here 'a1' and 'a2' represent state vectors.
            ProblemParameters.Sensors = {'a1','a2'}; 
            % hi(t)---Time dependent functions
            ProblemParameters.NumberTimeDependentFunctions = 2; 
            % Include the list of time dependent functions in the
            % MATLAB/Octave syntax. The time variable should be 't'.
            ProblemParameters.TimeDependentFunctions = ...
                {'sin(2*pi*1.5*t)','sin(2*pi*2.5*t)'}; 
            % Include the list of time dependent functions one more time in
            % the solver's syntax (Fortran, MATLAB, LabView, ...).
            % If no time dependent functions are employed then comment.
            ProblemParameters.TimeDependentFunctions(2,:) = ...
                {'sin(2*pi*1.5*t)','sin(2*pi*2.5*t)'};
        % Consistency test.                                              %*
        ProblemParameters.InputNumber = ...                              %*
            ProblemParameters.NumberSensors+...                          %*
            ProblemParameters.NumberTimeDependentFunctions;              %* 
        % Control syntax definition.                                     %*
        Sensors = cell(1,ProblemParameters.NumberSensors);               %*
        TDF = cell(1,ProblemParameters.NumberTimeDependentFunctions);    %*
        for p=1:ProblemParameters.NumberSensors                          %*
            Sensors{p} = ['s(',num2str(p),')'];                          %*
        end                                                              %*
        for p=1:ProblemParameters.NumberTimeDependentFunctions           %*
            TDF{p} = ['h(',num2str(p),')'];                              %*
        end                                                              %*
        ControlSyntax = horzcat(Sensors,TDF);                            %*
\end{lstlisting}

The plant parameters are divided in two groups:
The parameters that are essential because used in the code,
and additional parameters employed in the problem file.
The user is free to add any parameters she/he wants it this part.
All parameters are sent to the problem for the evaluation of one individual.
        
\begin{lstlisting}[name=MyParameters,firstnumber=99]
    % Plant parameters.
        % Essential problem parameters used in the code.
            ProblemParameters.T0 = 0; % Not always used
            ProblemParameters.Tmax = 10; % 
            % Actuation limitation : [lower bound,upper bound]
            ProblemParameters.ActuationLimit = ...
                [-2,2;... % Boundaries for the first actuator.
                -1,1];    % Boundaries for the second actuator.
            % The evaluation of the cost function and the control points
            % will be rounded to the 'RoundEval' parameter.
            ProblemParameters.RoundEval = 6;        
            % Estimate performance
            % When a control law is evaluated several times, its
            % performance can be computed by taking the average of all the
            % evaluations.
            % Other options are also possible if there is a drift, such as:
            %   - 'last' : takes the cost of the evaluation
            %   - 'worst': takes the cost of the worst evaluation
            %   - 'best' : takes the cost of the best evaluation
            ProblemParameters.EstimatePerformance = 'mean'; 

        % Problem parameters used in the *_problem.m file.
            % Maximum evaluation time otherwise returns a bad value
            ProblemParameters.TmaxEv = 5;
            % Penalization parameter
            % It is a vector that multiplies the secondary components of
            % the cost function such as:
            % J = Ja + gamma(1)*Jb + gamma(2)*Jc
            ProblemParameters.gamma = [0.01,10];
            % Other parameters used by the solver or normalization of the
            % cost function.
            ProblemParameters.ReynoldsNumber = 100; % For example.
            ProblemParameters.J0 = 1;               % For example.
            ProblemParameters.Jmin = 0;             % For example.
            ProblemParameters.Jmax = inf;           % For example.
            % ...
        
    % Assignation
    parameters.ProblemParameters = ProblemParameters; %*
    
\end{lstlisting}

Then, the control law parameters define the space of the matrices, i.e., the space of control laws.
The following parameters define the maximum number of lines (or instructions) in the matrix, the range of each column, i.e., the mathematical operations, and the registers (constant and variable).

\begin{lstlisting}[name=MyParameters,firstnumber=118]
%% Control law parameters
    % Number of instructions.
    % InitMin and InitMax define the range of number of instructions when
    % the matrices are generated randomly.
    % The number of instructions for each matrix is then chosen randomly 
    % with an uniform distribution.
    % Max defines the maximum number of instructions during the learning
    % process. This is used during the crossover operation as the number of
    % instructions can grow larger than the initial number.
        ControlLaw.InstructionSize.InitMax=20;
        ControlLaw.InstructionSize.InitMin=1;
        ControlLaw.InstructionSize.Max=20;
        
    % Operators.
        ControlLaw.OperatorIndices = [1:5,7:9];
            %  implemented:  - 1  addition       (+)
            %                - 2  subtraction    (-)
            %                - 3  multiplication (*)
            %                - 4  division       (%) - protected (my_div)
            %                - 5  sinus         (sin)
            %                - 6  cosines       (cos)
            %                - 7  logarithm     (log) - protected (my_log)
            %                - 8  exp           (exp)
            %                - 9  tanh          (tanh)
            %                - 10 square        (.^2)
            %                - 11 modulo        (mod)
            %                - 12 power         (pow)
        % Precision of the evaluation of the control laws.
        % The default value is 6.
        % If this value is changed, one must also update the protected
        % operators: my_div and my_log.
        ControlLaw.Precision = 6; 
        
    % Definition of the registers.
        % Number of variable registers.
        % The minimum number is the sum between the number of inputs and
        % outputs. 
        VarRegNumberMinimum = ProblemParameters.OutputNumber+...         %*
            ProblemParameters.InputNumber;                               %*
        % More variable registers can be added if necessary.
        ControlLaw.VarRegNumber = VarRegNumberMinimum + 3; 
        % Number of constant registers.
        % Here, we only set random constants in the constant registers but 
        % one can also include sensor signals.
        ControlLaw.CstRegNumber = 4;
        % Range of the random constants constants.
        % In this example, the range of the four constants is [-1;1].
        ControlLaw.CstRange = repmat([-1,1],ControlLaw.CstRegNumber,1);
        % Total number of registers.
        ControlLaw.RegNumber = ...         % Total number of operands     *
            ControlLaw.VarRegNumber + ...  % Number of variable registers *
            ControlLaw.CstRegNumber;       % Number of constant registers *   
        % Register initialization (recommended)                                       
            NVR = ControlLaw.VarRegNumber;                               
            RN = ControlLaw.RegNumber;                                   
            r{RN}='0';                                                   
            r(:) = {'0'};                                                
            % Variable registers                                         
            for p=1:ProblemParameters.InputNumber                        
                r{p+ProblemParameters.OutputNumber} = ControlSyntax{p};  
            end                                                          
            % Constant registers                                         
            minC = min(ControlLaw.CstRange,[],2);                        
            maxC = max(ControlLaw.CstRange,[],2);                        
            dC = maxC-minC;                                              
            for p=NVR+1:RN                                               
                r{p} = num2str(dC(p-NVR)*rand+minC(p-NVR));              
            end                                                          
        % Assignation                                                    
        ControlLaw.Registers = r;                                        %*
\end{lstlisting}
            
To accelerate the learning, individuals with originally different mathematical expression but exactly equal after simplification are searched to be removed.
For this, the control laws are evaluated on random sensor samplings.
Those samplings are referred as `control points' in the code.
Random time samples are also generated for the time dependent functions.
            
\begin{lstlisting}[name=MyParameters,firstnumber=189]
        % Control law estimation.
        ControlLaw.ControlPointNumber = 1000;
        % Sensor range to define the sampling points.
        % In this case, the range of all sensors is [-2,2].
        ControlLaw.SensorRange = repmat([-2 2],...
            ProblemParameters.NumberSensors,1);
        % Definition of random sampling points (ControlPoints) and random
        % time samplings for the control law estimation
            Nbpts = ControlLaw.ControlPointNumber;                       %*
            Rmin = min(ControlLaw.SensorRange,[],2);                     %*
            Rmax = max(ControlLaw.SensorRange,[],2);                     %*
            dR = Rmax-Rmin;                                              %*
        ControlLaw.EvalTimeSample = rand(1,Nbpts)*ProblemParameters.Tmax;%*
        ControlLaw.ControlPoints = ...                                   %*
            rand(ProblemParameters.NumberSensors,Nbpts).*dR+Rmin;        %*

    % Assignation
    parameters.ControlLaw = ControlLaw;                                  %*
    
\end{lstlisting}

Finally, the MLC parameters define the optimization process.
In addition to the genetic operators probabilities parameters, there are also some screening of the individuals options.

\begin{lstlisting}[name=MyParameters,firstnumber=208]
%% MLC parameters
    % Population size
    parameters.PopulationSize = 10;
    % Optimization options
        % Optimization of the first generation only (Monte Carlo).
        % It removes the duplicated individuals, redundant, etc.
    parameters.OptiMonteCarlo = 1; 
    % Remove individuals whose evaluation failed (bad)
    parameters.RemoveBadIndividuals = 1; 
    % Remove already evaluated individuals
    parameters.RemoveRedundant = 1; 
    % Remove the individuals if they have already been evaluated in a
    % earlier generations.
    parameters.CrossGenRemoval = 1;
    % Evaluate the initial condition of the registers (here:b=0)
    parameters.ExploreIC = 1; 
    % The crossover and mutation operators are until new individuals are
    % generated or MaxIterations iterations is reached.
    parameters.MaxIterations = 100;
    % Reevaluation of the individuals for stochastic problems.
    % This parameter implicitly set to 1 for experiments.
    %   - 0: No revaluation
    %   - 1: Force the reevaluation of the replicated and "elitism" indiv.
    %   - n: Each individual is reevaluated n times.
    parameters.MultipleEvaluations = 0;
    % Selection parameters.
    parameters.TournamentSize = 7;
    parameters.p_tour = 1;
    % Genetic operator parameters
    parameters.Elitism = 1;
    parameters.CrossoverProb = 0.6;
    parameters.MutationProb = 0.3;
    parameters.ReplicationProb = 0.1;
    % Other genetic parameters
        % How many mutations per matrix (MutationType)?
        %   - 'classic': depends on the mutation rate. The mutation rate is
        %   the probability that a given line changes.
        %   - 'at_least_one': the mutation rate is set such as there is at
        %   least one mutation per matrix.
        %   - 'number_per_chromosome': the mutation rate is set such as
        %   there is at least MutationNumber mutations per matrix.
        %   least one mutation per matrix.
    parameters.MutationType = 'number_per_matrix';
    parameters.MutationNumber = 1;
    parameters.MutationRate = 0.05;
    % Number of cuts in the matrices for the crossover operation.
    parameters.CrossoverPoints = 1;
    % If CrossoverMix=1, the offsprings are built by alternating their
    % parents sections, if CrossoverMix=0, the sections are selected
    % randomly.
    parameters.CrossoverMix = 1; 
    % The crossover operation gives two offsprings by crossover operation.
    parameters.CrossoverOptions = {'gives2'};
    % Other parameters
    % Cost value for individuals whose evaluation failed.
    parameters.BadValue = 10^36;
    % Remove the individuals that do not satisfy a given test.
    % The removed individuals are referred as "wrong" individuals.
    parameters.Pretesting = 0;  

%% Constants
    parameters.PHI = 1.61803398875;

%% Other parameters
    % Parameter that stores the list of saving times.
    % Can help to keep track of the latest version.
    parameters.LastSave = '';

end
\end{lstlisting}

\section{User's MATLAB problem file}\label{AppA_problem}
In this section, we describe the typical structure of MATLAB problem file.
We detail in particular, the types of the inputs and outputs.

The input of the function, \texttt{Arrayb},  is a $\Nb \times1$ `cell' array, where $\Nb$ is the number of components of the control law.
The elements of the array are string of characters that define the control law.

\begin{lstlisting}[name=MyProblem,firstnumber=1]
function  J_out = MyPlant_problem(Arrayb,parameters,visu)
  % MYPLANT_PROBLEM is a template file to define one's own problem file to
  % be solved with MATLAB.
  % The user can also include a calling to his solver in this file.
  % Several initial conditions can be used.
  % To help understand the commands, we illustrate the commands with
  % Arrayb set to {'1.34','a1+a2*sin(2*pi*1.5*t)'}, for example.
  % This example could correspond to the case where the individual is:
  % {'1.34','s(1)+s(2)*h(1)'}.
  % The values of s(1), s(2) and h(1) are replaced before the evaluation by
  % the values set in the parameter file, see MyPlant_parameters.m
  %
  % Guy Y. Cornejo Maceda, 2022/07/01
  %
  % See also LandauOscillator_problem.

  % Copyright: 2022 Guy Cornejo Maceda (gy.cornejo.maceda@gmail.com)
  % The MIT License (MIT)

%% Parameters
    ActuationLimit = parameters.ProblemParameters.ActuationLimit;
    ActMin = ActuationLimit(:,1);
    ActMax = ActuationLimit(:,2);
    gamma = parameters.ProblemParameters.gamma;
    BadValue = parameters.BadValue;
    RoundEval = parameters.ProblemParameters.RoundEval;
    
%% Control law synthesis
    % Bound the actuation with a clip function.
    BoundArrayb = limit_to(Arrayb,ActuationLimit); 
    % The results of this command is then:
    %   2x1 cell array
    % 
    %     'clip(1.34,-2,2)'
    %     'clip(a1+a2*sin(2*pi*1.5*t),-1,1)'
    
    % Control law
    % The i-th control law is the i-th element of Arrayb.
    bx1=BoundArrayb{1};
    bx2=BoundArrayb{2};
    % Definition
    eval(['b = @(t,a1,a2)[',bx1,';',bx2,'];']);

%% Objective
% Stabilization of the fixed point (example)

%% Some problem parameters
    % Parameter A
    A = 1;
    % Parameter B
    B = 1;
    % Initial conditions
    initial_conditions = [0;0.1];

%% Resolution parameters
    % Solver
        solver = 'ode5';
    % Time discretization
        % Number of time steps
        N = 500;
        T0 = parameters.ProblemParameters.T0;
        Tmax = parameters.ProblemParameters.Tmax; % frequency=1
        time = linspace(T0,Tmax,N+1);
        TmaxEv = parameters.ProblemParameters.TmaxEv;
\end{lstlisting}

In this example, a solver with fixed size step is employed to integrate the ODE,
however when a solver with a variable size step is chosen (like \texttt{ode45}) the integration time can significantly increase.
Thus, we add a function \texttt{T\_maxevaluation} to limit the integration time.
This function returns an error when the integration time is larger than the variable \texttt{TmaxEv},
triggering the exception in the \texttt{try} function.

\begin{lstlisting}[name=MyProblem,firstnumber=67]
%% Equation resolution
    % Equations
    % Unforced dynamical system (example)
     DynSys = @(t,a)[(A-a(1).^2-a(2).^2).*a(1)-B*a(2);...
                   (A-a(1).^2-a(2).^2).*a(2)+B*a(1)];
	
    % Controlled dynamical system
 	ConDynSys = @(t,a) (DynSys(t,a) + b(t,a(1),a(2)) +...
        T_maxevaluation(TmaxEv,toc)*[0;0]);

%% Time integration
% A try function is employed to treat the case where the evaluation fails.
try
    % Resolution
        tic
        y = feval(solver,ConDynSys,time,initial_conditions);

%% Cost function
    % Ja (example)
    ja = y(:,1).^2+y(:,2).^2;
    Ja = mean(ja);
    % Jb (example)
    % Evaluate the control law -> cell structure
    b_cell = arrayfun(b,time',y(:,1),y(:,2),'UniformOutput',false);
    % Convert in matrix
    b_matrix = cell2mat(transpose(b_cell));
    jb = sum(b_matrix.^2,2);
    Jb = mean(jb);
    % Jc (example)
    Jc = rand;
    J = Ja+gamma(1)*Jb+gamma(2)*Jc;

catch err
    J_out = {BadValue,BadValue,BadValue,BadValue};
    fprintf(err.message);
    fprintf('\n');
    return
end
\end{lstlisting}

The output, \texttt{J\_out}, is also a cell array.
Its dimension is the number of components of the cost function $+1$.
It this example, $J$ is composed of three elements, thus the output is: ${J,J_a,J_b,J_c}$.
A cell type is chosen for the output to allow cost function of different types (boolean, string of characters, etc.).

\begin{lstlisting}[name=MyProblem,firstnumber=105]
%% Output
    J_round = round(J,RoundEval);
    Ja_round = round(Ja,RoundEval);
    Jb_round = round(Jb,RoundEval);
    Jc_round = round(Jc,RoundEval);
    J_out = {J_round,Ja_round,Jb_round,Jc_round};

%% Plot
if nargin > 2 && visu
   figure
   % Include your figure here.
end
end
\end{lstlisting}

\chapter[Interfacing with a solver or an experiment]{Interfacing with an external numerical solver or an experiment}\label{AppInterface}
In this appendix, we give some guidance on how to interface \xMLC with an external numerical solver or an experiment.

\section{External numerical solver}
For external numerical solver, e.g., based on Fortran or Python, there is mainly two ways to interface the codes.
If the evaluation time of each individual takes a few seconds, the simplest way is to call the solver directly from the problem file along with the adequate parameters.
When the computation of one individual takes several dozens of minutes or even a few hours in a single processor, it is recommended to evaluate the individuals \emph{in parallel} on a computer cluster.
\xMLC includes several scripts that help the management of the control laws to evaluate and the cost files once the computation is over.

The first script is \texttt{MLC\_tools/External\_evaluation\_START.m}:
\begin{lstlisting}[name=START,firstnumber=1]
    % EXTERNAL_EVALUATION_START starts the run.
    % This script generates the first generation of individuals and
    % creates a file Gen1population.mat , a cell array containing the
    % control laws to evaluate.
    % The file is locate in save_runs/RUNNAME/Populations.
    %
    % Guy Y. Cornejo Maceda, 2022/07/01
    %
    % See also external_evaluation_CONTINUE, External_evaluation_END.

    % Copyright: 2022 Guy Cornejo Maceda (gy.cornejo.maceda@gmail.com)
    % The MIT License (MIT)

Initialization;
%% Start
    mlc=MLC('MyExternalPlant');
    mlc.parameters.PathExt='';
    
%% Generate population
    mlc.generate_population;
    
%% Save
    mlc.save_matlab('Gen0');
\end{lstlisting}
This script creates a MLC object with the  `MyExternalPlant' problem.
This problem is similar to the `MyPlant' one, except that the \texttt{Name} and \texttt{ProblemType} parameters are set to \texttt{ExternalTestRun} and \texttt{`external'}, respectively.
The script creates a MAT file (Gen1population.mat) containing the control laws to be evaluated in cell array.
This file is located in \texttt{save\_runs/MyExternalPlant/} in this example.
The script also saves the MLC object generated under the name \texttt{Gen0}.
The index `0' indicates that the first generation has not been evaluated yet.
Once this script is executed, the evaluation of the individuals on a computer cluster can start.

The expected output of each individual evaluation is a file containing the components $J_k$  of the cost function.
The files should be in ASCII format and named \texttt{GenXIndY.dat} where \texttt{X} is the generation number and \texttt{Y} the creation order of the individual.
Here is an example of such file for the evaluation of the 14$^{\rm th}$ individual of the 2$^{\rm nd}$ generation.
\begin{lstlisting}[numbers=none]
1.345 2.050 0.767
\end{lstlisting}
All these files should be located in the folder defined by the variable \texttt{PathExt}; its value is modified on line~17 of \texttt{External\_evaluation\_START.m}.
No cost file is expected for individuals whose evaluation failed.

Once the evaluation the evaluation of all individuals is done, to continue the optimization process, one needs to run the function \texttt{MLC\_tools/External\_evaluation\_CONTINUE.m} with the generation that has been evaluated.
\begin{lstlisting}[firstnumber=1]
function External_evaluation_CONTINUE(gen)
    % EXTERNAL_EVALUATION_CONTINUE continues the run.
    % To be used after the evaluation of the individuals of generation GEN.
    % Retrieves the cost information and makes the population evolve.
    % New control laws are generated in the Population folder and ready to 
    % be evaluated.
    %
    % Guy Y. Cornejo Maceda, 2022/07/01
    %
    % See also External_evaluation_END, External_evaluation_START.

    % Copyright: 2022 Guy Cornejo Maceda (gy.cornejo.maceda@gmail.com)
    % The MIT License (MIT)

%Evolve_population_script
Initialization;
mlc=MLC('MyExternalPlant');
%% Load
    mlc.load_matlab('ExternalTestRun',['Gen',num2str(gen-1)]);

%% Complete
    matJ = External_build_matJ(mlc.parameters,gen);
    complete_evaluation(mlc,gen,matJ);

%% Evolve
    evolve_population(mlc);

%% Save
    mlc.save_matlab(['Gen',num2str(length(mlc.population)-1)]);
\end{lstlisting}
The script loads the previous save of the MLC class object, loads the cost files, generates the next generation of individuals and create the associated MAT file containing the control laws.
The function that loads the cost files is \texttt{External\_build\_matJ} (line 22), its output is a matrix containing the components of the cost function (the columns) for each individual (the rows).
One is free to adapt the code and especially \texttt{External\_build\_matJ} to her/his own file system.
The function \texttt{External\_evaluation\_CONTINUE} needs then to be iterated to advance the generations.

For the final generation the \texttt{External\_evaluation\_END} function needs to be run with the final generation as input.
\texttt{External\_evaluation\_END} does exactly the same as \texttt{External\_evaluation\_CONTINUE} except generating the next generation.
The optimization process can then be resumed by running the \texttt{External\_evaluation\_CONTINUE} function.

In practice, these three scripts (`START', `CONTINUE' and `END') are meant to be ran directly at the start of MATLAB.
Once they are executed, the MATLAB session can be ended.
The best way to use them is in shell script comprising all the commands.
Here is an example of such script in bash:
\lstset{
  language={bash},}
\begin{lstlisting}[numbers=none]
#!/bin/bash

# Parameters
GenMax=10
 
# --- Start ---
matlab -r "External_evaluation_START;"
# Evaluation
# ...

# Initialization of the generation counter
Gen=1

# --- Optimization loop ---
while [ "$Gen" != "GenMax" ]
do
	matlab -r "External_evaluation_CONTINUE($Gen)"
	# Evaluation
	# ...
	
	# Increment Gen
	Gen=$(($Gen+1))
done

# --- Complete the final generation
matlab -r "External_evaluation_END($Gen)
\end{lstlisting}

\section{Experiment}
The \xMLC code, running on MATLAB, has been interfaced with several experiments via LabVIEW and dSPACE/Simulink-based platforms.
For an experiment, there are essentially two main loops:
\begin{itemize}
\item The fast evaluation loop, that runs at the sampling frequency. It contains the plant, sensors, actuators and real time controller.
\item The slow learning loop, that updates the control laws once they are evaluated. Its characteristic frequency is much lower than the evaluation loop. It contains the real time controller and the MATLAB session.
\end{itemize}

The first step to interface the code with an experiment is to be able to run a real-time control loop for any control law or mathematical expression.
Here are some elements the user needs to take into account to run a real-time control loop and to ease the interface with \xMLC:
\begin{itemize}
\item The graphic model of the experiment shall include a clip function for the actuation command sent to the actuators. This assures the actuators are not harmed by out of range inputs.
\item The sampling frequency is well chosen such as the frequencies of interest are well resolved. The Nyquist theorem requires the sampling frequency to be at least twice the largest frequency.
\item Automatic saving of the time series under a unambiguous. 
\end{itemize}

Once any control law can be evaluated, there are essentially two ways to interface the \xMLC with the experiment:
\begin{itemize}
\item Update the set of control law at each generations.
This approach has been employed when interfacing \xMLC with dSPACE/Simulink.
The \texttt{external} option needs to be chosen as \texttt{ProblemType} in the parameters.
The idea is to have a Simulink model including a list of control laws that is compiled and sent to the real time controller. The list of control laws is a function that takes as input the sensor signals, an integer and a gives back the actuation command as output.
The integer  is used to `select' the control law in the list.
In \xMLC, this function can be automatically generated with the method \texttt{@MLC/expe\_create\_control\_select.m}. 
The function is named \texttt{ControlLawSelect.m} and is located in \texttt{save\_runs/tmp/}.
This approach needs a recompilation of the Simulink once every generation.
Once, the generation is evaluated and the time series saved, the cost of each individual can be computed and sent to \xMLC to create the generation.
This approach is the more complex as it requires an extra Python or Bash layer to command the real time controller and the MATLAB session.
\item Another approach is to update the control laws one by one.
This approach has been employed when interfacing with a LabVIEW system.
The \texttt{LabView} option needs to be chosen as \texttt{ProblemType} in the parameters for this approach.
Instead of sending the control law to a MATLAB function, \xMLC creates a script, \texttt{LabViewControlLaw.txt}, containing the control law and readable by LabVIEW.
The \texttt{LabViewControlLaw.txt} file is created in the external path defined by \texttt{PathExt} in the parameters.
This approach does not require any compilation as it employs a control law parser.
Once the evaluation is done and the time series saved, the costs can be computed and send to \texttt{xMLC}.
The process will then continue automatically.
This approach is the most simple but an extra step of control law simplification may be needed if the parsing takes too much time compared to sampling time.
The user is free to modify:
\begin{itemize}
\item The \texttt{MLC\_tools/CreatefunctionLabview.m} function to create the \texttt{LabViewControlLaw.txt} file that suits her/his needs;
\item The \texttt{@MLCind/evaluate\_ind.m} method in the `LabView' section to send the control law to LabVIEW and compute the cost from the time series.
\end{itemize}
\end{itemize}
We shall not detail the interfacing any longer as it often depends on the experiment.
For more information on the interfacing of the code with LabVIEW, dSPACE/Simulink or any other platform please contact Guy Y. Cornejo Maceda at: \href{mailto:Yoslan@hit.edu.cn}{Yoslan@hit.edu.cn}.

\chapter{Accelerating the learning}\label{sec:accelerators}
In this appendix, we describe some accelerators that manage to increase the learning rate of MLC.
The main idea of these accelerators is to avoid redundant evaluations, meaning to prevent the evaluation of the same control laws.
First, we need to detect equivalent control laws.
However, this is not an easy task as simplification of two mathematical expressions is not guaranteed to give the same results because of the commutative operators.
Moreover, another complication is the protection of the $\div$ and $\log$ operators.
Because of the protection, these functions present a discontinuous behavior near 0 that prevents from simplification.
Thus, we propose another way to detect equivalent expressions.
To detect if two control laws are similar, we evaluate them on random sample inputs and we compare the control outputs.
These methods allows to detect mathematical equivalent control laws even though their expressions are very different.
Indeed, thanks to such test, control laws that are beyond the actuation thresholds will also be ruled out.
For example, the control laws $b=1.01$ and $b=10^23$ will be considered as the same control laws if the threshold is set to 1.
Moreover, the random samples are taken in the range of the sensors signals making the comparison between the control laws even more meaningful.
In practice, each time a new control law is built, we evaluate it over 1000 random sample inputs and store the result in the database.
Each new control law is then compared to the database to verify if it has not been already evaluated.

Equivalent control laws can then be filterer out in two ways:
\begin{description}
\item[In the population:] if a new individual is equivalent to an individual already present in the population, then it is discarded and a new individual is built (randomly if it is the first generation or with one of the genetic operators).
\item[In the database:] if a new individual is equivalent to a previous individual in the current or past generations, then it is discarded and a new individual is built.
\end{description}
Of course the second option includes the first one.
Those rules do not apply for individuals generated thanks to replication and elitism as their role is to emulate memory through the generations.
Those filtering assures that the population of individuals is always moving towards unexplored regions of the control landscape and backward steps are not possible.

\bibliographystyle{plainnat} % Same style as the xROM book.
\bibliography{bibliography}

\backmatter
%\addcontentsline{toc}{chapter}{Index}
%\printindex
%-----------------------------------------------------------------------
% The asterisk excludes chapter from the table of contents.
\chapter*{Authors}
%-----------------------------------------------------------------------
\begin{wrapfigure}[7]{r}[0pt]{0.15\textwidth} \vspace*{-08mm}
	\begin{flushright}
		\includegraphics[width=0.14\textwidth]{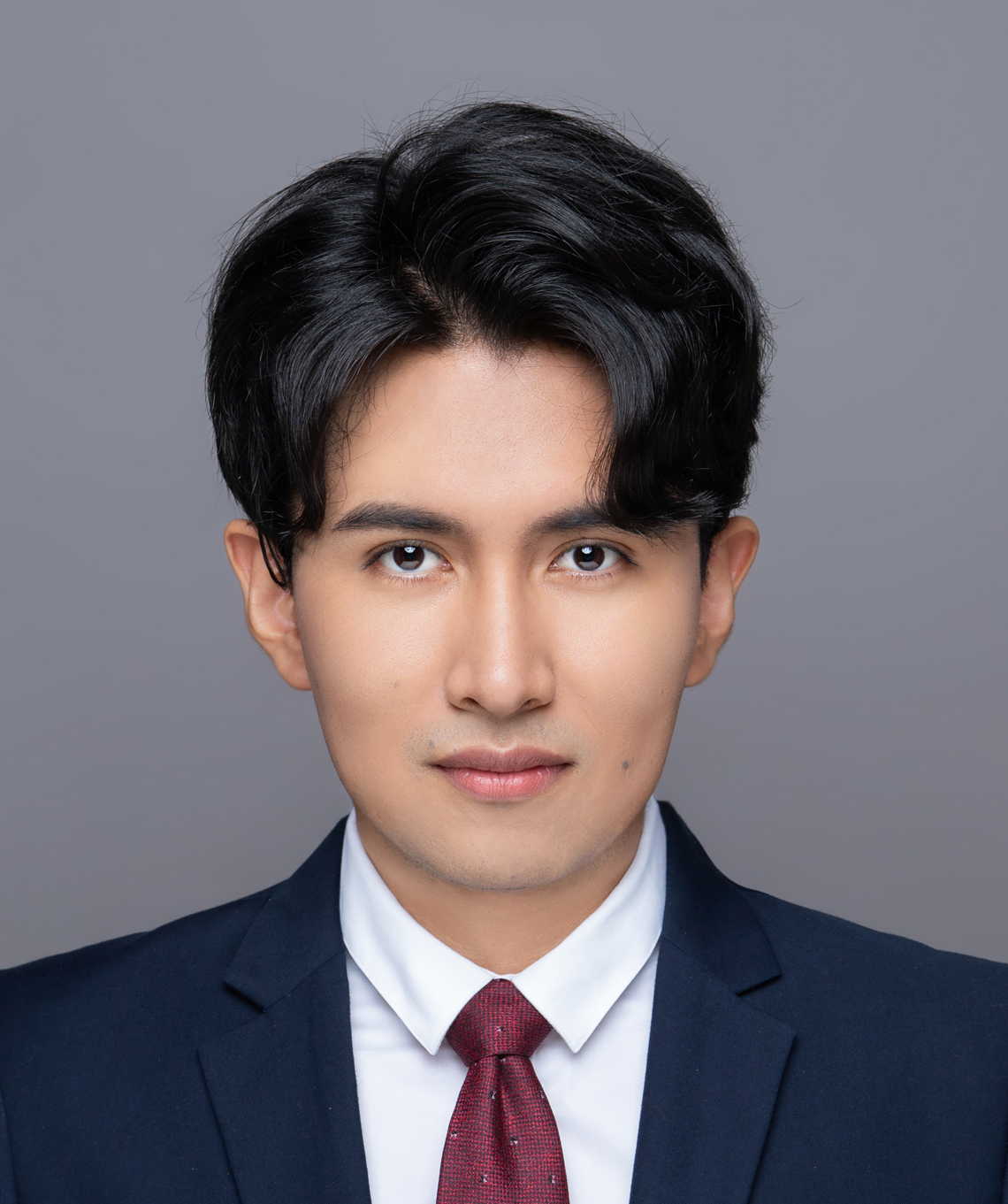}
	\end{flushright}
\end{wrapfigure}
\textbf{Guy Y.\ Cornejo Maceda} is a postdoctoral researcher at Harbin Institute of Technology at Shenzhen, 
working on turbulence control with machine learning methods for a fast learning of robust, feedback control laws and their visualization.
He obtained his Ph.D. in 2021 at Paris-Saclay University (France) under the supervision of Profs. Bernd R. Noack and François Lusseyran 
with whom he accelerated the learning of feedback control laws by a factor 10 compared to previously employed machine learning control. 
His methodology has been demonstrated on numerical and experimental MIMO plants.
Examples include
 drag reduction and stabilization of the fluidic pinball, 
 stabilization of the open cavity flow,
 lift increase and drag reduction of a high-Reynolds number airfoil,
 drag reduction and side-force mitigation of a yawed truck model and 
even the challenging feedback smart-skin separation control over a smooth ramp 
with 30 actuators and 56 sensors.

%-----------------------------------------------------------------------
\bigskip
\begin{wrapfigure}[7]{r}[0pt]{0.15\textwidth} \vspace*{-08mm}
\begin{flushright}
	\includegraphics[width=0.14\textwidth]{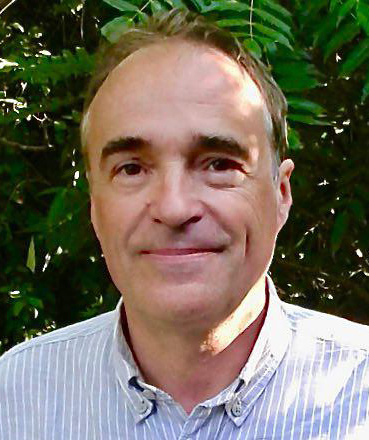}
\end{flushright}
\end{wrapfigure}
\textbf{Fran\c{c}ois Lusseyran} is currently Emeritus Research Director at the CNRS in a laboratory of Paris-Saclay University, LISN (France). His career as a fluid mechanics experimentalist has always followed the thread of the expression of nonlinearities in the space-time dynamics of various flow contexts. This began with the creation of vortex circulation at the Max Planck Institut f\"{u}r Str\"{o}mungsforschung in 1978 in G\"{o}ttingen (Germany) and now contributes to the control of vortex structures, developing in wakes and resonant flows, by searching for control laws with the help of machine learning algorithms. Other research topics have included, for example, the dynamic relationship between the heart and its arterial system (LMFE Orsay-Paris, France), transitions between gas-liquid two-phase flow regimes (LEMTA, Nancy, France, University of Illinois -- Urbana-Champagne, USA), model reduction in turbine mixers and in jets with transverse flow (LEMTA, Helixor, Rosenfeld, Germany) dynamics of open cavities in incompressible regime and moderate Reynolds numbers (LISN ex-LIMSI -- LFD, FIUBA-UBA, Buenos Aires, Argentina), many years of collaboration on metrological issues and data analysis (IPPT PAN, Warsaw, Poland) and reduction of temporal dynamics forced by the so-called coherent structures (CORIA, Rouen, France -- LFD -- LadHyX, Palaiseau, France).
%\newpage

%-----------------------------------------------------------------------
%-----------------------------------------------------------------------
\bigskip
\begin{wrapfigure}[7]{r}[0pt]{0.15\textwidth} \vspace*{-08mm}
	\begin{flushright}
   \includegraphics[width=0.14\textwidth]{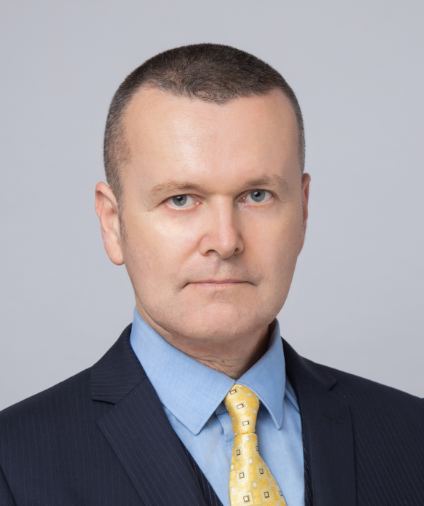}
	\end{flushright}
\end{wrapfigure}
\textbf{Bernd R.\ Noack} is National Talent Professor
at the Harbin Institute of Technology, Shenzhen,
and Honorary Professor and Chair of Turbulence Control at TU Berlin.
Until 2020, he was Research Director CNRS at LIMSI, Paris-Saclay 
and Professor at TU Braunschweig and TU Berlin.
Past affiliations include the United Technologies Research Center,
Max-Planck Society, German Aerospace Center and University of G\"ottingen.
He develops closed-loop turbulence control solutions
for cars, airplanes and transport systems
in an interdisciplinary effort with leading groups in China, Europe and USA/Canada.
His team is advancing the frontiers of nonlinear control-oriented reduced-order models and machine learning control,
an automated learning of control laws in experiment and simulation.
He has co-authored over 250 refereed publications, including 2 patents, 2 textbooks, 3 other books, 3 review articles
and over 120 international journal articles.
His work has been honored by numerous awards,
e.g., a Fellowship of the American Physical Society,
a CNRS Scientific Excellence award,
a Senior ANR Chair of Excellence in France,
and the von Mises Award of International Association of Applied Mathematics and Mechanics.
He is awarded outstanding faculty at HIT and 
a highly cited researcher (top 1\% in Mendeley/Stanford list).

%-----------------------------------------------------------------------
\end{document}